%% file: 00_main.tex
\documentclass[twocolumn,tighten,times]{aastex62}

\newcommand{\herschel}{\textit{Herschel}}
\newcommand{\spitzer}{\textit{Spitzer}}
\newcommand{\hst}{\textit{HST}}

\usepackage{savesym}
\savesymbol{tablenum}
\usepackage[binary-units]{siunitx}
\restoresymbol{SIX}{tablenum}

\renewcommand\micron{\mbox{\si{\micro}{m}}}

\renewcommand\arcdeg{\mbox{$^\circ$}}% 
\renewcommand\arcmin{\mbox{$^\prime$}}% 
\renewcommand\arcsec{\mbox{$^{\prime\prime}$}}% 
\newcommand\ha{H$\mathrm{\alpha}$}
\newcommand{\nii}{[\ion{N}{2}]}
\newcommand\msun{\mbox{\si{M_\odot}}}
\newcommand\lsun{\mbox{\si{L_\odot}}}
\newcommand\smpy{\mbox{\si{M_\odot.yr^{-1}}}}
\newcommand\sbstar{$\Sigma_\mathrm{star}$}
\newcommand\sbfir{$\Sigma_\mathrm{IR}$}
\newcommand{\textred}[1]{{#1}}
\newcommand{\stkout}[1]{\ifmmode\text{\sout{\ensuremath{#1}}}\else\sout{#1}\fi}
\graphicspath{{./}{figures/}}

\received{September 18, 2020}
\revised{December 9, 2020}
\accepted{December 24, 2020}
\submitjournal{ApJ}

\shorttitle{ALMA 1.3mm Survey of Bright Lensed SMGs}
\shortauthors{Sun et al.}
\begin{document}

\title{ALMA 1.3\,mm Survey of Lensed Submillimeter Galaxies (SMGs) Selected by \textit{Herschel}: \footnote{{\it Herschel} is an ESA space observatory with science instruments provided by European-led Principal Investigator consortia and with important participation from NASA.} \\
Discovery of Spatially Extended SMGs and Implications
}

\correspondingauthor{Fengwu Sun}
\email{fengwusun@email.arizona.edu}

\author{Fengwu Sun}
\affiliation{Steward Observatory, University of Arizona, 933 N. Cherry Avenue, Tucson, 85721, USA}

\author{Eiichi Egami}
\affiliation{Steward Observatory, University of Arizona, 933 N. Cherry Avenue, Tucson, 85721, USA}

\author{Timothy D.\ Rawle}
\affiliation{ESA, Science Operations Department, STScI, Baltimore, MD 21218, USA}

\author{Gregory L.\ Walth}
\affiliation{The Observatories of the Carnegie Institution for Science, 813 Santa Barbara Street, Pasadena, CA 91101, USA}
\affiliation{Steward Observatory, University of Arizona, 933 N. Cherry Avenue, Tucson, 85721, USA}

\author{Ian Smail}
\affiliation{Centre for Extragalactic Astronomy, Department of Physics, Durham University, South Road, Durham, DH1 3LE, UK}

\author{Miroslava Dessauges-Zavadsky}
\affiliation{Observatoire de Gen\`eve, Universit\'e de Gen\`eve, 51, Ch. des Maillettes, 1290 Versoix, Switzerland}

\author{Pablo G.\ P\'erez-Gonz\'alez}
\affiliation{Centro de Astrobiolog\'ia, Departamento de Astrof\'isica, CSIC-INTA, Cra.\ de Ajalvir km.4 E-28850--Torrej\'on de Ardoz, Madrid, Spain}

\author{Johan Richard}
\affiliation{Universit\'e Lyon, Universit\'e Lyon1, ENS de Lyon,
CNRS, Centre de Recherche Astrophysique de Lyon UMR5574, Saint-Genis-Laval, France.}

\author{Francoise Combes}
\affiliation{Sorbonne Universit\'e, Observatoire de Paris, Universit\'e PSL, CNRS, LERMA, 75014 Paris, France}
\affiliation{Coll\`ege de France, 11 Place Marcelin Berthelot, 75231 Paris, France}

\author{Harald Ebeling}
\affiliation{Institute for Astronomy, University of Hawaii, 2680 Woodlawn Drive, Honolulu, HI 96822, USA}

\author{Roser Pell\'o}
\affiliation{Aix Marseille Universit\'e, CNRS, CNES, LAM (Laboratoire d’Astrophysique de Marseille), UMR 7326, 13388, Marseille, France}

\author{Paul Van der Werf}
\affiliation{Leiden Observatory, Leiden University, P.O. Box 9513, NL-2300 RA Leiden, The Netherlands}

\author{Bruno Altieri}
\affiliation{European Space Astronomy Centre, ESA, Villanueva de la Ca\~nada, 28691, Madrid, Spain}

\author{Fr\'ed\'eric Boone}
\affiliation{CNRS, IRAP, 9 Av.\ colonel Roche, BP 44346, 31028, Toulouse Cedex 4, France}

\author{Antonio Cava}
\affiliation{Department of Astronomy, University of Geneva, 51 Ch. des Maillettes, 1290 Versoix, Switzerland}

\author{Scott C.\ Chapman}
\affiliation{Department of Physics and Astronomy, University of British Columbia, 6225 Agricultural Road, Vancouver, BC V6T 1Z1, Canada}
\affiliation{National Research Council, Herzberg Astronomy and Astrophysics, 5071 West Saanich Road, Victoria, BC V9E 2E7, Canada}
\affiliation{Department of Physics and Atmospheric Science, Dalhousie University, Halifax, NS B3H 4R2, Canada}

\author{Benjamin Cl\'ement}
\affiliation{Institute of Physics, Laboratory of Astrophysics, Ecole Polytechnique Fdrale de Lausanne (EPFL), Observatoire de Sauverny, 1290 Versoix, Switzerland}

\author{Alexis Finoguenov}
\affiliation{Department of Physics, University of Helsinki, Gustaf H\"allstr\"omin katu 2, 00014 Helsinki, Finland}

% \author {Daniel P.\ Marrone}
% \affiliation{Steward Observatory, University of Arizona, 933 N. Cherry Avenue, Tucson, 85721, USA}

\author{Kimihiko Nakajima}
\affiliation{National Astronomical Observatory of Japan, 2-21-1 Osawa, Mitaka, Tokyo 181-8588, Japan}

\author{Wiphu Rujopakarn}
\affiliation{Department of Physics, Faculty of Science, Chulalongkorn University, 254 Phayathai Road, Pathumwan, Bangkok 10330, Thailand}
\affiliation{National Astronomical Research Institute of Thailand (Public Organization), Don Kaeo, Mae Rim, Chiang Mai 50180, Thailand}
\affiliation{Kavli Institute for the Physics and Mathematics of the Universe (WPI),The University of Tokyo Institutes for Advanced Study, The University of Tokyo, Kashiwa, Chiba 277-8583, Japan}

\author{Daniel Schaerer}
\affiliation{Observatoire de Gen\`eve, Universit\'e de Gen\`eve, 51, Ch. des Maillettes, 1290 Versoix, Switzerland}

\author{Ivan Valtchanov}
\affiliation{European Space Astronomy Centre, ESA, Villanueva de la Ca\~nada, 28691, Madrid, Spain}

% \author{...}



%\input{00_abstract_fsun}
\input{00_abstract_v4}



\section{Introduction} 
\label{sec:01_intro}
\input{01_intro.tex}

\section{Observations and Data} 
\label{sec:02_obs}
\input{02_observation.tex}

\section{Results} 
\label{sec:03_res}
\input{03_results}

\section{Physical Properties of the Detected Lensed SMGs} 
\label{sec:04_ana}
\input{04_analyses}

\section{Discussion}
\label{sec:05_dis}
\input{05_discussion.tex}

\section{Summary}
\label{sec:06_con}
\input{06_sum.tex}

% \citet{2015ApJ...805...23C} provides a example of how the citation in the
% article references the external code at
% \doi{10.5281/zenodo.15991}.  Unfortunately, bibtex does
% not have specific bibtex entries for these types of references so the
% ``@misc'' type should be used.  The Repository tutorial explains how to
% code the ``@misc'' type correctly.  The most recent aasjournal.bst file,
% available with \aastex\ v6, will output bibtex ``@misc'' type properly.

\acknowledgments

E.E.\ would like to thank Observatoire de Lyon and European Space Astronomy Centre (ESAC) Faculty Council for hosting and supporting his long-term visits, during which this project was originally started. 
I.R.S.\ acknowledges support from the Science and Technology Facilities Council (grant number ST/T000244/1). 
\textred{H.E.\ gratefully acknowledges financial support from Space Telescope Science Institute for program SNAP-12884.}
W.R.\ is supported by the Thailand Research Fund/Office of the Higher Education Commission Grant Number MRG6280259 and Chulalongkorn University's CUniverse.

% ALMA: 
This paper makes use of the following ALMA data: ADS/JAO.ALMA\#2015.1.01548, \#2016.1.00372 and \#2017.1.01658.S. ALMA is a partnership of ESO (representing its member states), NSF (USA) and NINS (Japan), together with NRC (Canada), MOST and ASIAA (Taiwan), and KASI (Republic of Korea), in cooperation with the Republic of Chile. The Joint ALMA Observatory is operated by ESO, AUI/NRAO and NAOJ.

% spitzer 
This work is based on observations made with the \spitzer\ Space Telescope, obtained from the NASA/ IPAC Infrared Science Archive, both of which are operated by the Jet Propulsion Laboratory, California Institute of Technology under a contract with the National Aeronautics and Space Administration.

% SPIRE 
\herschel\ is an ESA space observatory with science instruments provided by European-led Principal Investigator consortia and with important participation from NASA.
SPIRE has been developed by a consortium of institutes led by Cardiff University (UK) and including Univ. Lethbridge (Canada); NAOC (China); CEA, LAM (France); IFSI, Univ. Padua (Italy); IAC (Spain); Stockholm Observatory (Sweden); Imperial College London, RAL, UCL-MSSL, UKATC, Univ.\ Sussex (UK); and Caltech, JPL, NHSC, Univ.\ Colorado (USA). This development has been supported by national funding agencies: CSA (Canada); NAOC (China); CEA, CNES, CNRS (France); ASI (Italy); MCINN (Spain); SNSB (Sweden); STFC, UKSA (UK); and NASA (USA).

% HST
This research is based on observations made with the NASA/ESA Hubble Space Telescope obtained from the Space Telescope Science Institute, which is operated by the Association of Universities for Research in Astronomy, Inc., under NASA contract NAS 5–26555. These observations are associated with program SNAP 12884, CLASH Treasury Program (GO 12065) and RELICS Treasury Program (GO 14096).


% \vspace{5mm}
\facilities{ALMA, \spitzer\ (IRAC), \herschel\ (PACS and SPIRE), \hst\ (ACS and WFC3-IR)}


\software{
astropy \citep{astropy:2013,astropy:2018},  
SExtractor \citep{sex},
CASA \citep{casa},
emcee \citep{emcee},
MAGPHYS \citep{magphys,battisti19},
\textred{GALFIT \citep{galfit}}
          }



% \clearpage
\appendix
% \section{Images of lensed SMGs}

\input{98_appendix.tex}




% Bibliography by bibtex:
% \clearpage
\bibliography{00_main}


% \listofchanges

\end{document}

%% file: 00_abstract_v4.tex
\begin{abstract}

We present an ALMA 1.3\,mm (Band 6) continuum survey of lensed submillimeter galaxies (SMGs) at $z=1.0\sim3.2$ with an angular resolution of $\sim0\farcs2$. 
These galaxies were uncovered by the \textit{Herschel} Lensing Survey (HLS), and feature exceptionally bright far-infrared continuum emission ($S_\mathrm{peak} \gtrsim 90$\,mJy) owing to their lensing magnification.
We detect 29 sources in 20 fields of massive galaxy clusters with ALMA. 
% Two sources are spatially extended at 1.3\,mm with a low surface brightness ($\lesssim 0.1$\,\si{mJy.arcsec^{-2}}) and only detectable in \textit{uv}-tapered maps.
Using both the \textit{Spitzer}/IRAC (3.6/4.5 \micron) and ALMA data, we have successfully modeled the surface brightness profiles of 26 sources in the rest-frame near- and far-infrared.
Similar to previous studies, we find the median dust-to-stellar continuum size ratio to be small ($R_\mathrm{e,dust}$/$R_\mathrm{e,star}$\,=\,0.38$\pm$0.14) for the observed SMGs, indicating that star formation is centrally concentrated.  
This is, however, not the case for two spatially extended main-sequence SMGs with a low surface brightness at 1.3\,mm ($\lesssim 0.1$\,\si{mJy.arcsec^{-2}}), in which the star formation is distributed over the entire galaxy ($R_\mathrm{e,dust}$/$R_\mathrm{e,star}$\,$>$\,1).
As a whole, our SMG sample shows a tight anti-correlation between ($R_\mathrm{e,dust}$/$R_\mathrm{e,star}$) and far-infrared surface brightness ($\Sigma_\mathrm{IR}$) over a factor of $\simeq$\,1000 in $\Sigma_\mathrm{IR}$.
% This indicates that SMGs with less vigorous star formation (i.e., lower $\Sigma_\mathrm{IR}$) are likely to show a lower specific star-formation rate at their center because the star formation is distributed over the whole galaxy %rather than concentrated in the center (i.e., larger $R_\mathrm{e,dust}$/$R_\mathrm{e,star}$).
This indicates that SMGs with less vigorous star formation (i.e., lower $\Sigma_\mathrm{IR}$) lack central starburst and are likely to retain a broader spatial distribution of star formation over the whole galaxies (i.e., larger $R_\mathrm{e,dust}$/$R_\mathrm{e,star}$).
The same trend can be reproduced with cosmological simulations as a result of central starburst and potentially subsequent ``inside-out" quenching,
which likely accounts for the % {\irs\sout{dissipation of intense star formation in SMGs as well as the}} 
emergence of compact quiescent galaxies at $z\sim2$.
\end{abstract}

%% file: 01_intro.tex
% Overview Submillimeter galaxies 
As the most vigorous stellar nursery in the Universe, submillimeter galaxies (SMGs\footnote{Observed 1.3\,mm flux density at $\gtrsim 1\,$mJy in this work. See discussion of definition in \citet{hodge20}.}) are discovered in abundance at $z>1$, contributing $\sim 20\%$ of the cosmic star formation rate density up to $z\sim 4$ \citep[e.g.,][]{swinbank2014,casey14}. 
Due to the critical role played by dust grains in the interstellar medium (ISM), commonly produced by asymptotic giant branch (AGB) stars and supernovae (SNe), these galaxies are observed to be highly dust-obscured in the rest-frame UV/optical bands \citep[e.g.,][]{whitaker17,dudzevic19}.
In the far-infrared (FIR), dust grains, heated up by the intense star formation, emit thermal continuum radiation accessible through submillimeter/millimeter observations.
These galaxies are found to host massive gas reservoirs with relatively short gas depletion time scales ($t_\mathrm{dep}\sim10^2$\,Myr; e.g., \citealt{tacconi08,bothwell13,miettinen17}).
After the truncation of sufficient gas supply, SMGs are believed to evolve towards compact quiescent galaxies (cQs) seen at lower redshift  \citep[e.g.,][]{toft14, simpson14}, which will eventually become massive elliptical galaxies in the local Universe potentially through additional gas-poor mergers  \citep[e.g.,][]{vd05, oogi13}.

% Formation and quenching of SMGs
The trigger mechanism of SMGs remains a subject of debate.
As the local analogs of SMGs, ultra-luminous infrared galaxies (ULIRGs) are observed to result from major mergers, with compact and prominent star-forming regions in their nuclei \citep[e.g.,][]{sanders88}.
At $z>1$, similar scenarios have been proposed by certain galaxy evolution theories \citep[e.g.,][]{narayanan10,mcalpine19}, but minor mergers \cite[e.g.,][]{gomez18} and secular starburst \citep[e.g.,][]{dave10} are other possible physical explanations supported by either observational evidence or theoretical frameworks.

%%% following sentence:
% Although more efforts are clearly required to clarify how SMGs formed, the quenching fates of these extreme starbursts have been probed by various methodologies. 
Most recently, powerful ground-based interferometers like ALMA have started to reveal the compact dust continua of SMGs by high-resolution imaging \citep[e.g.,][]{simpson15, hodge16, hodge19, elbaz18, puglisi19, gullberg19, tadaki20}.
With a typical half-light radius of $\sim$1--2\,kpc, these intense star-forming regions are still larger than those in the local (U)LIRGs, while their sizes do match with cQs at slightly lower redshift \citep[e.g.,][]{vandokkum08, vdw14}.
In addition, the number density and clustering properties of SMGs also coincide with those of cQs, indicating an underlying evolutionary connection \citep{hickox12,simpson14, an2019, dudzevic19}.
Therefore, understanding how star formation ceases in SMGs (i.e., quenching) holds important clues to connecting actively star-forming galaxies and their (possible) red and dead descendants.

%%%%%%%%%%%%%%%%%%%%%%%%%%%%%%%%%%%%%%%
%%% Details on galaxy quenching scenario

% Observations in the past few decades have already shown that the quenching efficiency of star formation is related to the galaxy mass, environment and morphology \citep[e.g.,][]{peng10,martig09}. 
% At $z\gtrsim1$, galaxy mass could be the predominant factor that correlates with the quenching rate. 
Spatially resolved studies have suggested that the massive star-forming galaxies at $z \simeq 1 - 2$ exhibit a rising specific star-formation rate (sSFR; SFR per unit stellar mass) profile from their center to the outskirts, indicating that the fade-out of star formation commences from the galactic center \citep[e.g.,][]{tacchella2015, tacchella18, nelson16, spilker19}.
Such an ``inside-out" process of quenching can also be reproduced within the context of cosmological simulation \citep{tacchella2016}.
% With the peak of surface star-formation rate located at the galactic center, it is not difficult to derive a profile of gas depletion time scale ($t_\mathrm{dep}$) rising toward the outskirts under the assumption of the Kennicutt-Schmidt law \citep{ks98}.
% However, this may not be the whole story in their quenching lifetime,
Various physical mechanisms have been proposed to interpret the quenching process, including the gas consumption by star formation, gas outflows driven by stellar and supermassive black hole feedbacks \citep{ds86,dm05}, as well as suppression of exterior gas supply through shocking heating due to gravitationally infalling gas \citep{db06,db08}. 

% How to probe? The uniqueness of this survey 
Although emerging observational clues suggest an inside-out fashion of stellar mass assembly and gas depletion, it is yet to be confirmed as the standard process in the evolution of SMGs with compact and powerful star-forming regions.
The morphological modeling of SFR and stellar mass profile requires high-resolution imaging of SFR tracers (e.g., \ha, UV/FIR continuum) and stellar component (the rest-frame optical/near-infrared continuum).
However, the accuracy of measurements at shorter wavelength is clearly subject to the strong dust extinction in the center of SMGs \citep[e.g.,][]{simpson17, lang19}.
% \citet{lang19} applied a dust-extinction correction to pixel-to-pixel mass-to-light ratios over SMGs using a rest-frame optical color, and the derived stellar mass profile is more concentrated than inferred from single band rest-frame optical imaging alone.
Meanwhile, \ha-based SFR profile is also sensitive to the active galactic nuclei (AGN) contribution, since the AGN fraction is higher in compact star-forming galaxies compared to more extended ones at similar redshifts \citep{barro13}.
In a nutshell, it is necessary to develop novel modeling techniques for stellar mass and SFR that are less sensitive to dust extinction and AGN contribution. 

To address this issue, one possible solution is to observe at longer wavelength. 
Since most of the star formation in SMGs is obscured by dust, FIR surface luminosity can represent the surface SFR with a sufficient accuracy.
% Additionally, CO observations can help constrain the distribution of molecular gas, providing an independent probe of the gas fraction distribution \citep[e.g.,][however, the spatial variance of mass-to-light ratio was not considered there]{spilker19}. 
In order to avoid the heavy dust obscuration of stellar continuum in SMGs, it is also better to sample the rest-frame near-infrared (NIR) bands rather than the optical ones  (e.g., \hst/WFC3-IR F160W samples the rest-frame $\sim$\,$V$ band for $z=2$ galaxies).
However, current sensitive mid-infrared imaging instrument (e.g., \spitzer/IRAC) cannot allow such a study with ALMA-like angular resolution as required.

In this regard, cluster-lensed SMGs \citep[e.g.,][]{smail97,swinbank10} can be useful targets to provide the morphological evidence.
% disentangle the quenching process. 
Magnified by gravitational lensing provided by a foreground cluster, these targets are sufficiently bright at multiple wavelengths, and their angular sizes are also stretched significantly (note, however, that the lensing effect conserves surface brightness). 
Compared to galaxy-lensed SMGs with a bright lensing galaxy always in the front, cluster-lensed ones are often free from blending with foreground galaxies, ensuring simplicity of the morphological modeling on their stellar component.  
Strong magnification gradients on the $\simeq$kpc scales relevant for resolving galaxies are also much less a concern than in galaxy-lensed cases.
During the course of our \herschel\ Lensing Survey (HLS, \citealt{egami10}; Egami et al., in prep.), we uncovered a substantial number of lensed SMGs with exceptionally bright FIR continuum ($S_\mathrm{peak} \gtrsim 90\,$mJy).
We then carried out observations of their dust continua in the ALMA Band 6 at 1.3\,mm, as well as stellar continua with \spitzer/IRAC at 3.6/4.5\,\micron.
In this work, we present the observations and analyses of the stellar and dust components in cluster-lensed SMGs at both integrated and spatially resolved scales. 
% We intend to show a trend of sSFR suppression at the center of SMGs compared to the whole plane during the recession of their star formation activeness.
% This result suggests the inside-out quenching as a common mode of transition from SMGs to cQs.

% Layout of paper
This paper is arranged as follows: 
Section~\ref{sec:02_obs} introduces all the obtained ALMA and \spitzer/IRAC data and corresponding data reduction techniques, with several ancillary data from various sources.
Section~\ref{sec:03_res} presents the fundamental analysis of our data, including detection, photometry and surface brightness profile modeling.
In Section~\ref{sec:04_ana} we perform spectral energy distribution (SED) modeling and show the statistical results of galaxy properties from both the integrated and spatially-resolved analyses.
We discuss the underlying physics and make necessary comparison with both observational evidence and theoretical predictions in Section~\ref{sec:05_dis}.
The conclusions and broader implication can be found in Section~\ref{sec:06_con}.
Throughout this paper, we assume a flat $\Lambda$CDM cosmology with $h=0.7$ and $\Omega_m = 0.3$. 
The AB magnitude system \citep{abmag} is used to express source brightnesses in the optical and NIR.

%% file: 02_observation.tex
% HLS survey
\subsection{The Sample}
\label{ss:02_sample}

% For the purpose of discovering and studying SMGs with the facilitation of strong gravitational lensing, 
To discover and study a significant sample of gravitationally lensed SMGs, we have conducted an extensive imaging survey of massive galaxy clusters in the FIR using the \herschel\ Space Observatory \citep{herschel10}, known as the HLS (\citealt{egami10}).  The target clusters were selected mainly from the samples produced by the following three surveys: (1) the \textit{ROSAT} All-Sky Survey (RASS) with the X-ray-luminous cluster sample tabulated by H.\ Ebeling (private communication), (2) the COnstrain Dark Energy with X-ray (CODEX) survey, which utilizes the combination of the RASS X-ray and Sloan Digital Sky Survey (SDSS) optical data \citep{finoguenov20}, and (3) the South Pole Telescope (SPT) survey, which selected clusters via the Sunyaev-Zel'dovich (SZ) effect (the SPT-SZ survey; \citealt{bleem15}).  The full HLS cluster sample will be presented and described by the forthcoming survey paper (Egami et al., in prep.).

%\todo{Selection of clusters --EE}

With a substantial number of SMG detections at $z_\mathrm{phot}\gtrsim 1$, we specifically selected a subset of exceptionally bright sources and obtained ALMA follow-up observations.
The selection criteria used were:
(1)~SPIRE color $S_{500}/S_{250}>0.4$ to ensure the selection of $z\gtrsim1$ sources,
(2)~FIR continuum peak ($S_\mathrm{peak}$) brighter than 90\,mJy if the source is within 1\arcmin\ from the cluster center, 
(3)~$S_\mathrm{peak}>$150\,mJy or with a spectroscopic redshift if the source is beyond 1\arcmin\ from the cluster center, 
% (4)~spectroscopically confirmed previously if the source is beyond 1\arcmin\ from the cluster center, 
and (4)~observable with ALMA (Dec.\,$<+30^\circ{}$).
The second criterion is designed to select bright sources at $L_\mathrm{IR} \gtrsim 10^{13.1}$\,\si{\mu^{-1}.L_\odot} at $z\sim 2$.
The third criterion ensures that the resultant sample includes the brightest sources in the HLS data even if some of them may be boosted by a galaxy component on the line of sight.
Due to the coarse resolution of SPIRE, the $S_\mathrm{peak}$ quoted here is the sum of all sub-mm sources within a radius of $\sim15$\arcsec. 
Multiple source systems will be decomposed later using the ALMA data, and these individual sources will not necessarily satisfy the same \herschel\ selection criteria.

We eventually constructed a sample of 20 sources based on the criteria listed above.\footnote{One source in the cluster field SPT\,J0345-6419 was observed with ALMA but was later identified as a low-redshift IR-bright galaxy rather than a $z\gtrsim1$ SMG. We therefore removed it from our sample.}
There are also several other HLS sources satisfying the same brightness criteria, but we do not incorporate them here since they were not observed by ALMA or lie at different redshift range ($z \sim 5$; e.g., HLS0918, \citealt{combes12}, \citealt{rawle14}; HLS0257, Sun et al.\ in prep.; HLS2043, \citealt{zavala15}; Walth et al.\ in prep.).
% Finally, we conducted ALMA Band-6 continuum observations of submillimeter sources discovered in 21 cluster field at 1.3\,mm with an angular resolution of $\sim$0.2\arcsec.
% As a parallel effort, we also performed \spitzer/IRAC imaging of their stellar continua at 3.6/4.5\,\micron.   
% To improve the SED modelling of several sources, we also included \hst\ and UKIRT data obtained through both public and private access in the analysis.

% A basic summary of the obtained \herschel, ALMA and \spitzer\ data is presented in Table~\ref{tab:01_log}.

\subsection{\herschel}
\label{ss:02a_hershel}
\input{02a_herschel.tex}

\subsection{ALMA}
\label{ss:02b_alma}
\input{02b_ALMA.tex}

\subsection{\spitzer/IRAC}
\label{ss:02c_irac}
\input{02c_IRAC.tex}

\subsection{Other ancillary data}
\label{ss:02d_other}
\input{02d_other.tex}

%% file: 02a_herschel.tex
% As a part of HLS, we have performed \herschel/PACS \citep{pacs10} and SPIRE \citep{spire10} 
The HLS has performed far-infrared imaging observations of 581 massive galaxy clusters with a total observing time of 418.7\,hours at two typical depths.
% deep mode:
The HLS-deep survey imaged 54 clusters deeply with PACS \citep{pacs10} and SPIRE \citep{spire10} through an open-time key program (44 targets; PI: Egami; PID: KPOT\_eegami\_1) and an open-time Cycle 2 program (10 targets; PI: Egami; PID: OT2\_eegami\_5).
Three of the clusters studied here (MACS\,J1115.8+0129, Abell\,2813 and Abell\,3088) were observed by HLS-deep, and therefore have a five-band coverage with both PACS (100/160\,\micron) and SPIRE (250/350/500\,\micron).
All of the PACS 100 and 160\,\micron\ observations consist of two orthogonal scan maps, each comprising 18–22 repetitions of 13 parallel 4-arcmin scan legs.
The SPIRE observations for the two Abell clusters were performed with 20 repetitions in the large scan map mode, each with two 4\arcmin\ scans and cross-scans (1.6\,h scan for three bands simultaneously).   
MACS1115 was observed through 11-repetition small scan maps, and each repetition consisted of one scan and one cross-scan of 4\arcmin\ length (0.4\,h scan).

% snapshot mode (rms noise as 9.6, 8.9 and 10.8 for one macs cluster; use 10mJy as typical value in text)
The HLS-snapshot survey obtained shallower SPIRE-only data for 527 clusters through two open-time programs during Cycle 1 and 2 (PI: Egami; PID: OT1\_eegami\_4, OT2\_eegami\_6), providing SPIRE data for the remaining 17 clusters studied here.
With shallow observations ($3-8$\,min scan) in the small scan map mode, HLS-snapshot provides nearly confusion-limited images in all three SPIRE bands (typical RMS noise $\sim$\,10\,\si{mJy.beam^{-1}} at 250, 350 and 500\,\micron, compared with the confusion noise levels of 5.8, 6.3 and 6.8\,\si{mJy.beam^{-1}} measured by \citealt{nguyen10}). 

% \todo{Need descriptions on basic data reductions, and statistics of data quality}
Our SPIRE images were produced via the standard reduction pipeline in \textsc{HIPE} v12.2 \citep{hipe}, and the processing routine was detailed in \citet{rawle16} for \herschel\ coverage of the \hst\ Frontier Fields (HFF, \citealt{lotz17}).
The observation ID (OBSID) and total scan time of each obtained \herschel/SPIRE observation are summarized in Table~\ref{tab:01_log}.
HLS-deep PACS images were generated with \textsc{UniMap} \citep{piazzo15} with a pixel scale of 1\farcs0 at 100\,\micron\ and 2\farcs0 at 160\,\micron, also detailed in \citet{rawle16}. 
%for A2813 and A3088 is not included here to ensure the uniformity of the analysis over the 20-source sample.
% With all the image products, we extract an initial source catalog with local maxima finding, small-aperture photometry and appropriate aperture correction under the assumption of point-like sources. 
% A merged catalog is then constructed for the convenience of planning ALMA and \spitzer\ observations.

% Since we are aware that at least 4 sources exhibit extended structure in SPIRE images, we request mosaic observations for these special cases. 
% SPIRE fluxes are finally re-extracted using ALMA detection information, detailed in Section~\ref{sss:03bi_spire}.

%% file: 02b_ALMA.tex
ALMA Band 6 observations were carried out through project 2015.1.01548, 2016.1.00372 and 2017.1.01658 (PI: Egami) between April 30, 2016 and September 30, 2018. 
Since four sources exhibit extended structures in the SPIRE images (FWHM$>$20\arcsec\ at 250\,\micron), we requested multi-pointing observations for these special cases. 
We observed all of our 20 targets in one of two spectral window settings. 
For 15 sources without previous spectroscopic redshift determination, we performed continuum-only observations with a central frequency at 233\,GHz (corresponding to 1.287\,mm).
For 5 sources with prior spectroscopic redshift information, we acquired both dust continuum and at least one CO line spectrum, and thus the final effective frequencies of these continuum products range from 224 to 238\,GHz.
\textred{The diameter of the ALMA field of view (FoV) at the requested frequencies is 25\arcsec.
}
A brief summary of our ALMA observations is also presented in Table~\ref{tab:01_log}.

The data were taken in various weather conditions with a median precipitable water vapor (PWV) of 0.77\,mm, with the 16th to 84th percentile ranging from 0.56 to 1.84\,mm.
% Median on-source integration time is 5\,minutes, with $\pm 1\sigma$ distribution between 3 and 12\,minutes.
A median angular resolution of 0\farcs26\ was achieved, and the median maximum recoverable scale was 1\farcs7.
%A median angular resolution of 0.26\arcsec\ was achieved, and the median largest angular resolution is 1.7\arcsec.

All the ALMA data were reduced with \textsc{CASA} \citep{casa} with the pipelines v4.7.2 and v5.4.1 for observations obtained in different cycles.
Before the formal reduction work, we first checked the combined continuum image and spectral cube of each target, delivered by the ALMA archive. 
% If no obvious source was detected above a 4$\sigma$ significance, we will only produced UV-tapered continuum images rather than natural or Briggs weighting images.
If any obvious source was detected above a 4$\sigma$ significance, we would examine the spectral cubes, searching for possible spectral line features, and both line and continuum would be imaged in the natural/Briggs weighting and \textit{uv}-tapered modes separately. 
If undetected, we would only produce the continuum images. % with the same settings of weighting.
We performed continuum imaging at four different levels of synthesized beam size: Briggs weighting (\textsc{robust}=0.5), natural weighting (\textsc{robust}=2), 1\arcsec\ \textit{uv}-tapering and 2\arcsec\ \textit{uv}-tapering. 
These settings were used for visualizing both compact and extended emission structures in the SMGs.
Interactive cleaning was performed during each imaging process.
The noise level of the final continuum products is $0.11^{+0.04}_{-0.05}$\,\si{mJy.beam^{-1}} with a 1\arcsec-tapered beam, which we used to obtain photometry for most of the targets.

%% file: 02c_IRAC.tex
% \spitzer/IRAC data are utilized for constraining the stellar properties of SMGs.
We obtained \spitzer/IRAC Channel 1 (3.6\,\micron) and Channel 2 (4.5\,\micron) images through various programs. 
The majority of our data were from Program 12095 and 90218 (both PI: Egami), covering 13 of 20 clusters.
We also included other data with public access on the \spitzer\ Heritage Archive\footnote{\href{https://sha.ipac.caltech.edu/}{https://sha.ipac.caltech.edu/}}. 
This includes Programs 80168, 90213 (PI: Bouwens), 12005, 14281 (PI: Bradac), 60099, 80012 (PI: Brodwin), 60034 (PI: Egami), 30344 (PI: Jarvis), 80162, 90233 (PI: Lawrence), 70149 (PI: Menanteau), 80066 (PI: Rawle), 61061 (PI: Sheth), 12123 (PI: Soifer), 40370, 80096 (PI: Stanford), 14061, 60194 (PI: Vieira). 
These programs provide 3.6/4.5\,\micron\ coverages for all the 20 clusters.
All of these fields are observed with cycling sub-pixel dithering patterns with four or five dithering points at least.
% except for 0345-6419 which is finally identified as a low-redshift LIRG rather than $z\gtrsim1$ SMG.

We started our IRAC data processing from the archive-delivered level 1 (BCD) products.
A standard and automatic \textsc{MOPEX} reduction routine was applied with an output pixel size of 0\farcs6\,pixel$^{-1}$.
We registered the output frames with the \textit{GAIA} DR2 \citep{gaiadr2} using \textsc{SExtractor} (for catalog extraction; \citealt{sex}) and \textsc{SCAMP} (for astrometric computation; \citealt{scamp}).
This achieved a final astrometric error of  $\lesssim 0\farcs1$ with the produced IRAC images.
The median $5\sigma$ point-source depth in each field, estimated from the variance of sky background, is presented in Table~\ref{tab:01_log}.

%% file: 02d_other.tex
For a number of clusters, we also included other ancillary data to improve the quality of analysis, mainly for the optical SED fitting. 
This provides more accurate extinction and stellar mass estimates, compared with IRAC-only analyses:

MACS\,J0553.4-3342--- this cluster was observed with the \hst\ treasury program RELICS \citep{coe19}, and therefore we used its 7-band \hst\ data for photometry and SED fitting (ACS/F435W, F606W, F814W and WFC3-IR/F110W, F125W, F140W, F160W).

MACS\,J1115.8+0129--- this cluster was observed with the \hst\ treasury program CLASH \citep{postman12}, and therefore its 9-band \hst\ data were utilized (ACS/F435W, F606W, F775W, F814W, WFC3-IR/F105W, F110W, F125W, F140W, F160W).
We directly used the processed data of MACSJ0553 and MACSJ1115 available on MAST\footnote{Mikulski Archive for Space Telescopes (MAST),   \href{https://archive.stsci.edu/}{https://archive.stsci.edu/}}.

RXC\,J2332.4-5358--- this cluster was observed through the \hst\ SNAP program 12884 (PI: Ebeling). 
The WFC3-IR/F110W and F140W images were obtained with an integration time of  706\,s per filter. 
We reduced the data with \textsc{Drizzlepac} v2.1.17 \citep{2012drzp} under the \textsc{PyRAF} environment with an output pixel size of 0\farcs06\,pixel$^{-1}$. 

Aperture photometry of SMGs in \hst\ images is conducted with \textsc{SExtractor}. We do not obtain flux measurement of HLS0553-C (blended with a star $\sim5^\mathrm{m}$ brighter than the SMG in the F814W band) and HLS2332-C (blended with an irregular galaxy). 

RXC\,J1314.3-2515--- This cluster was observed with WFCAM on UKIRT in both the $J$ and $K$ bands (PI: Walth), and here we only use the NIR photometry of the lensed SMG in this cluster field.
%detailed analysis on this source will be presented in Walth et al., in prep.

No additional optical/NIR data was included for the analysis of remaining sources.
The \herschel\ sources in RXC\,J1314.3-2515, MACS\,J0455.2+0657 and MACS\, J0600.1-2008 were observed with JCMT/SCUBA-2 at 850\,\micron\ \citep{cheale19}, and here we quote the 850\,\micron\ flux densities to improve the quality of far-IR SED modeling.

%% file: 03_results.tex
% In this section, we perform basic analysis on our combined dataset described in Section~\ref{sec:02_obs}.
% We first obtained secure SMG detections on ALMA maps, and then acquired multi-band photometry using these ALMA priors. 
% We also modeled the source angular sizes in both rest-frame near-IR (IRAC 3.6/4.5\,\micron) and FIR (ALMA Band 6).

\subsection{ALMA detection and photometry}
\label{ss:03a_detect}
\input{03a_detection.tex}

\subsection{Multi-wavelength photometry}
\label{ss:03b_photo}
\input{03b_photometry.tex}

\subsection{Quantitative morphological modeling}
\label{ss:03c_mor}
\input{03c_morphology.tex}

\subsection{Summary of ALMA and IRAC counterparts}
\label{ss:03d_sum}

\input{03d_summary.tex}

% \subsection{Dust-to-stellar continuum size ratio}
% \label{ss:03f_size}
% \input{03f_size.tex}

%% file: 03a_detection.tex
To obtain reliable and complete ($\gtrsim1$\,mJy) ALMA detections of lensed SMGs in all 20 observed cluster fields, we used \textsc{SExtractor} v2.19.5 \citep{sex} to derive uniform source extraction in all ALMA continuum image products (without primary beam correction).
Based on a quick visual inspection of natural-weighted image products without \textit{uv}-tapering (median beam FWHM $=0\farcs28$), we found that all of the obvious sources detected in our data were spatially resolved. 
Since \textit{uv}-tapering can increase the detectability of extended structures, though at the expense of resolution, we performed the source extraction with the 1\arcsec-tapered image products.
% Compared with the natural/briggs-weighted images, the tapered images do increase the signal-to-noise (S/N) ratio of several low-surface-brightness targets (e.g., HLS0043-B, HLS0840, HLS2155-B) without significantly deteriorating the S/N of relatively compact sources. 

We ran \textsc{SExtractor} for source detection %at S/N threshold =3.5
and automatic photometry with Kron-like elliptical apertures (\textsc{PHOT\_AUTOPARAMS} values of 1.8 and 2.5) in the primary-beam-uncorrected maps \textred{where the primary beam response is greater than 0.2}. 
We estimated the photometric errors ($\sigma_\mathrm{aper}$) based on aperture-to-beam size ratio ($\Omega_\mathrm{aper}/\Omega_\mathrm{beam}$) and continuum RMS ($\sigma_\mathrm{RMS}$) according to the following equation:

\begin{equation}
\label{eq:aper}
\sigma_\mathrm{aper} = {\sigma_\mathrm{RMS}}
\cdot \sqrt{\frac{(\Omega_\mathrm{aper}/\Omega_\mathrm{beam})^2}{1+(\Omega_\mathrm{aper}/\Omega_\mathrm{beam})/2}}
\end{equation}
%%% ({\bf\color{red} EE: Describe how you came up with this equation with an additional factor in the denominator.})
which is derived from a simulation of applying random apertures (enclosed area as $\Omega_\mathrm{aper}$) on Gaussian-blurred (kernel area as $\Omega_\mathrm{beam}$) Gaussian white noise maps.
%%%% At a median aperture-to-beam size ratio of 4.4 for sources in our sample, the derived $\sigma_\mathrm{aper}$ is very close to $\sigma_\mathrm{RMS}\cdot\sqrt{\Omega_\mathrm{aper}/\Omega_\mathrm{beam}}$.
We also evaluated the photometric uncertainty by directly applying random apertures on source-free regions in ALMA primary-beam-uncorrected maps.
We measured the standard deviation of flux densities enclosed within the apertures of identical size, and the results are consistent with the prediction by Equation~\ref{eq:aper}.
We then corrected the flux densities and their errors for the gain of primary beam.

Based on this method, we detected 77 sources at S/N$>$3.0 in all ALMA images.
Here the S/N is defined as the ratio between the aperture-photometry flux density and its uncertainty ($f/\sigma_\mathrm{aper}$).
To eliminate possible false detections, we studied the false detection rate in Appendix~\ref{sec:app_1}, and found that an S/N cut at 4.0 would ensure the total false detection number to be $\lesssim 1$. % in all of our ALMA images.
% We find that 
We therefore detected 28 sources at S/N$>$4.0 based on the 1\arcsec-tapered images. 
\textred{All of the sources were detected in the area where the primary beam response is greater than 0.5, except for HLS2155-A.
}

We also included another S/N$>$4.0 detection in 2\arcsec-tapered images, namely HLS0840. 
This source is exceptionally extended with an ALMA-measured 1.3 mm effective radius $R_\mathrm{e, ALMA}$ of $3\farcs7\pm1\farcs7$ with a relatively low surface brightness. 
It is split into multiple S/N$\sim$3 components in our 1\arcsec-tapered image but remains as a single S/N=4.5 source in the 2\arcsec-tapered map.
\textred{We also detected a point-like $0.5\pm0.1$\,mJy source at its center in the 0\farcs2-resolution image. This might represent the core of the galaxy, although it only contributes $13\pm4$\% of the total flux density in the 2\arcsec-tapered map.
}
We included this source because of the robustness of the detection in the 2\arcsec-tapered images, and no other similar example was found in our data.

We also studied the completeness of our source extraction in Appendix~\ref{sec:app_1}.  We conclude that the completeness of point-like sources at S/N=5.0 in 1\arcsec-tapered map (0.79\,mJy, under the assumption of a median continuum RMS and a primary beam correction of $\sqrt{2}$) is 80$\pm$4\%.

In one cluster field, CODEX\,52909, we did not detect any significant source.
% SPIRE source in 0345-6419 is confirmed as a low-redshift one, thus its continuum strength is below our detection limit.
The SPIRE source in CODEX\,52909 is extended, and it can be a composite of several ALMA sources at S/N$\simeq$3--3.5 with reddened IRAC counterparts. 
To avoid any confusion of $\sim3\sigma$ fake sources in this mosaic ALMA FoV, we did not analyze this field any further.
%simply treat this cluster field as no significant SMG detection. 

Because of the noise fluctuation in the ALMA maps, the fluxes of sources at low S/N tend to be overestimated, known as flux boosting effect \citep[e.g.,][]{geach17, stach19}.
We examined this effect for sources detected at relatively low S/N (HLS0546, HLS0840, HLS0043-B and HLS2155-A; Table~\ref{tab:02_phot}).
% The flux boosting is a consequence of the tendency of low signal-to- noise ratio sources to have their measured fluxes preferentially increased by noise fluctuations in the maps.
Assuming the surface brightness profiles measured in Section~\ref{ss:03c_mor}, we simulated the visibility data for these sources 10--15 times per each with \textsc{CASA}, and the RMS noise of mock data was controlled to match with that of our observations. 
We then applied the same imaging and source extraction routine, measured the median output flux densities and compared them with those of input models.
Based on our simulations, no conspicuous flux boosting effect was identified.

%% file: 03b_photometry.tex
\begin{figure*}[!tb]
\centering
\includegraphics[width=0.495\linewidth]{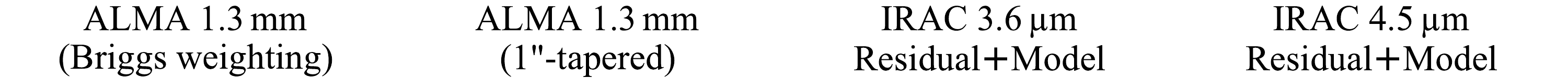}
\includegraphics[width=0.495\linewidth]{figures/thumbnail/header.pdf}
\includegraphics[width=0.495\linewidth]{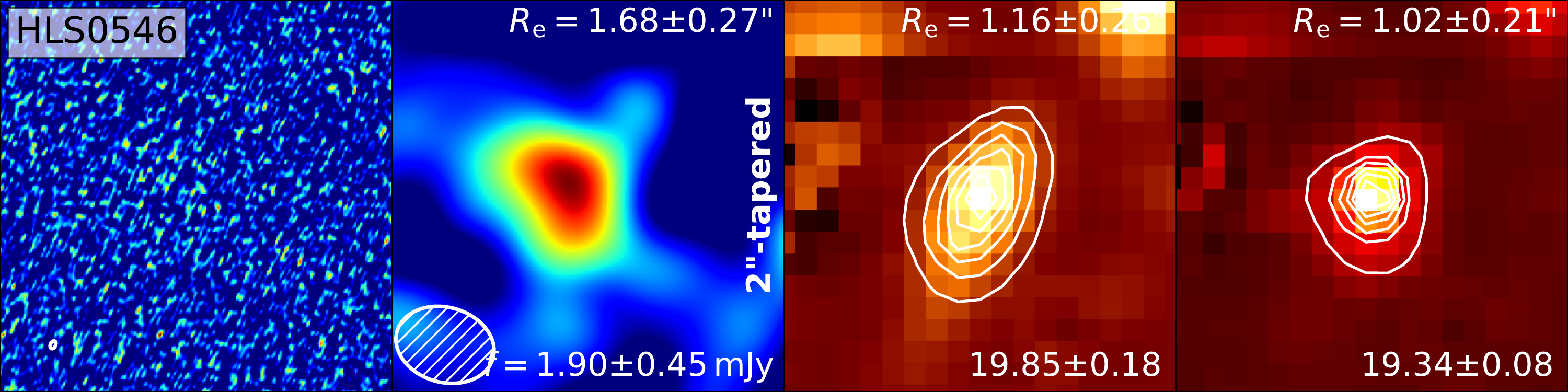}
\includegraphics[width=0.495\linewidth]{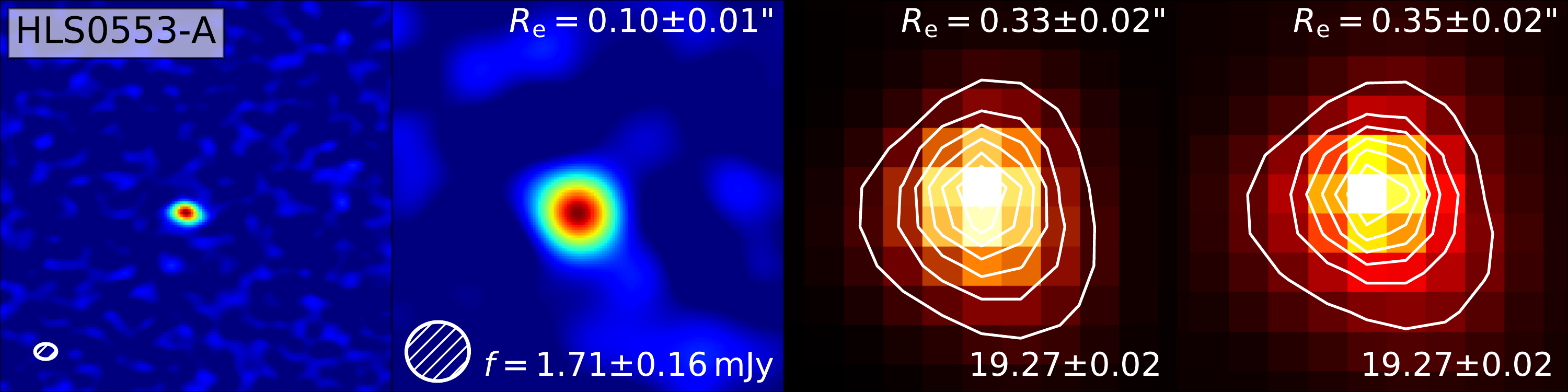}
\includegraphics[width=0.495\linewidth]{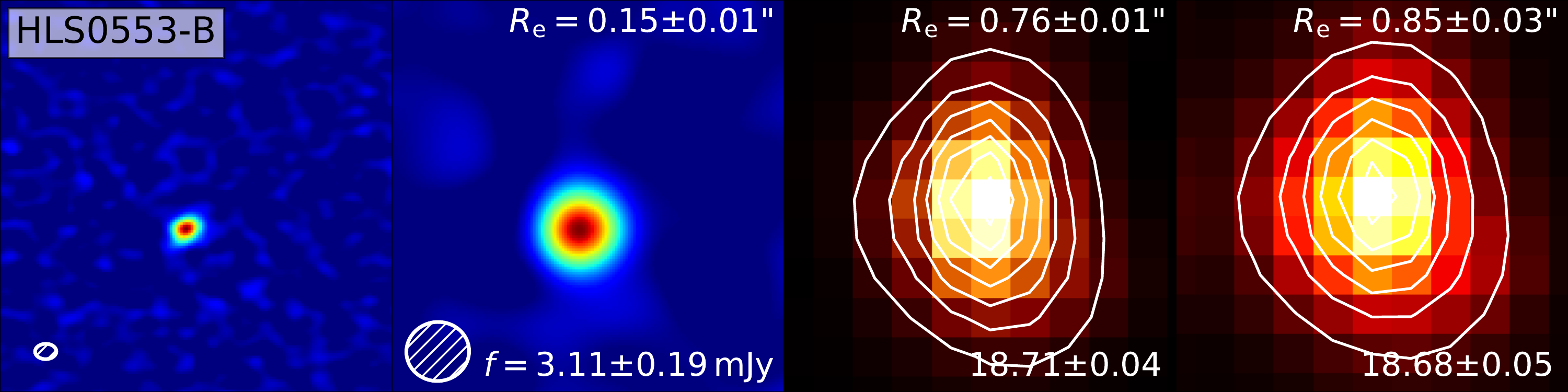}
\includegraphics[width=0.495\linewidth]{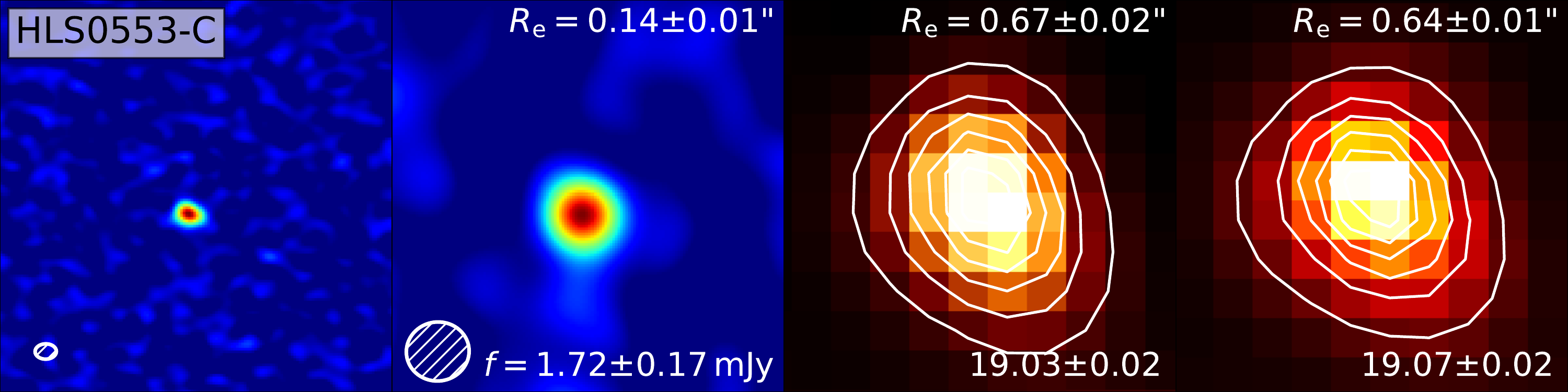}
\includegraphics[width=0.495\linewidth]{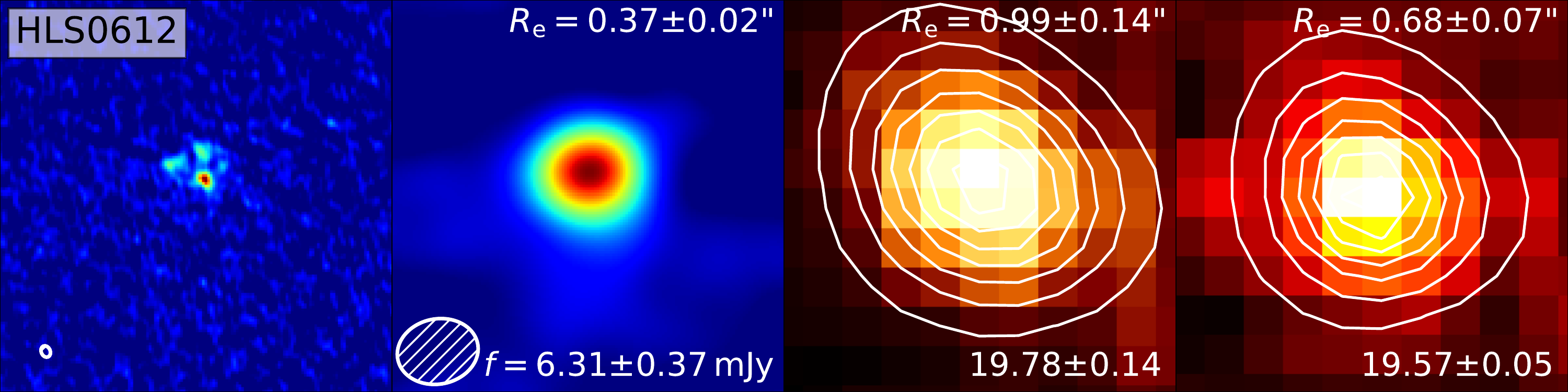}
\includegraphics[width=0.495\linewidth]{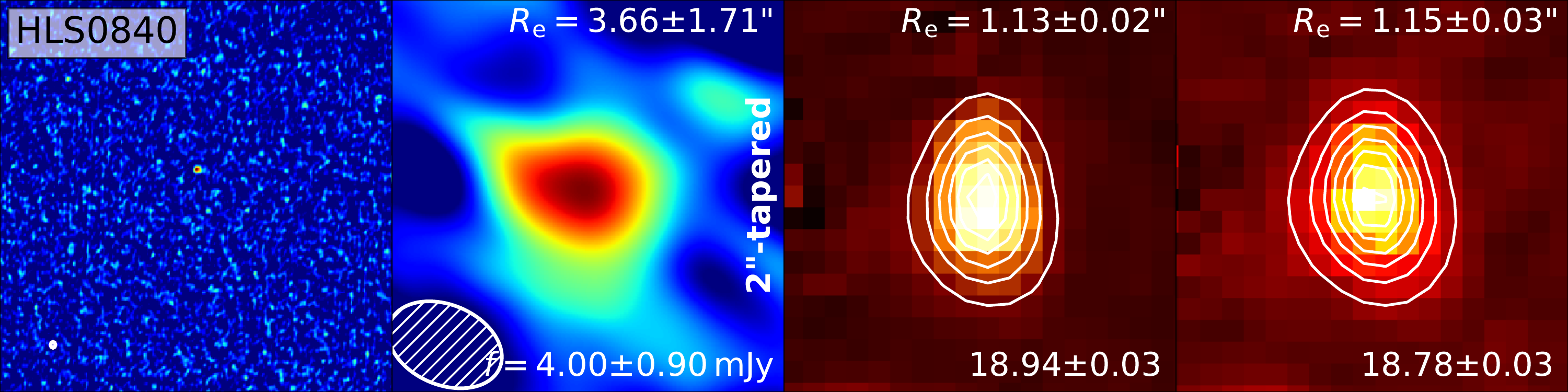}
\caption{Postage stamp images of six cluster-lensed SMGs in this work, and the remaining ones are shown in Figure~\ref{fig:apd_01}. 
The galaxies are at the center of each image. 
For each source, we show their ALMA 1.3\,mm Briggs-weighted map in the first column (resolution $\Delta \theta \sim 0\farcs2$), and \textit{uv}-tapered map in the second column ($\Delta \theta \sim 1$\arcsec; $2$\arcsec-tapered for HLS0546 and HLS0840). 
The synthesized ALMA beam is shown as a hatched ellipse at the lower-left corner, and the source sizes and fluxes are noted in the tapered images.
IRAC 3.6/4.5\,\micron\ maps after the neighborhood subtraction with \textsc{GALFIT} are shown in the third/forth columns. 
The best-fit models, convolved with the PSFs, are overlaid as white contours. 
The source sizes and magnitudes are noted in the upper-right and lower-right corners of and IRAC images.
}
\label{fig:3_cutout}
\end{figure*}

% Based on the SMG detections by ALMA, we performed multi-band photometry on these sources.
% This includes \herschel/SPIRE flux re-extraction and \spitzer/IRAC photometry. 
% Table~\ref{tab:02_phot} presents the summary of these photometric results.

\subsubsection{\herschel}
\label{sss:03bi_spire}
\herschel/PACS 100 and 160\,\micron\ flux densities were measured with an aperture radius of 5\arcsec, and the uncertainty was inferred from the variance of sky background.
We adopted the aperture correction factors suggested in PACS data handbook.
HLS1115 was observed to be 20.7$\pm$3.1\,mJy at 100\,\micron\ and 50.6$\pm$3.8\,mJy at 160\,\micron.
HLS0307-28-A was observed to be 50.7$\pm$2.0\,mJy at 100\,\micron\ and 110.1$\pm$4.7\,mJy at 160\,\micron.
HLS0307-28-B was blended with a low-redshift source and remained undetected in PACS images, and the two sources in A2813 fell outside of the PACS footprint.

We measured source flux densities using PSF photometry in \herschel/SPIRE 250, 350 and 500\,\micron\ images. 
We assumed point-source models for the majority of ALMA-detected SMGs, except for two extended sources ($R_\mathrm{e}>1\farcs 6$), HLS0840 and HLS0546, which were fit with a 2D Gaussian model.
PSF photometry was conducted using \textsc{GALFIT} \citep{galfit}.
We used ALMA source coordinates as prior source positions in the SPIRE maps. 
For single-source cases, we floated the source position in the SPIRE images due to the coarse spatial sampling.
If multiple sources exist in a field, the relative positions of their models are fixed. 
In several cluster fields, we also included other PACS (A2813, A3088) or \textit{WISE}-detected (RXCJ2155) sources to optimize the fitting through multi-component decomposition.

28 out of 29 ALMA-detected sources were successfully extracted from all three SPIRE bands. 
%A2813B$_1$ 
HLS0043-B was undetected in the SPIRE 350 and 500\,\micron\ images, and therefore we only present its $3\sigma$ upper limit ($\sim$11\,mJy).

\subsubsection{\spitzer/IRAC}
\label{sss:03bii_irac}

IRAC photometry was performed with two main methods. 
Since our targets are located in cluster fields, some of them may be blended with foreground cluster members, decreasing the accuracy of aperture photometry. 
We use \textsc{GALFIT} to model the source brightness and morphology of 24 cluster-lensed SMGs.
The IRAC warm-mission PSFs released by \citet{hora12} for Channel 1 and 2 were used for model convolution.
The orientation angle of spacecraft was also considered in our modeling routine.

S\'ersic and PSF source models were assumed for different sources in IRAC maps.
\textred{We adopted a S\'ersic model for the majority of SMGs, and a range of S\'ersic index ($n$) between 0.2 and 4.0 was allowed.} 
If the best-fit $n$ is beyond this range, we fixed it at 0.2, 0.5 (as Gaussian), 1.0 (as exponential) or 1.5, depending on the goodness of fit.
\textred{15 sources have well constrained S\'ersic indices in at least one IRAC band, and nine sources were fit with fixed $n$.}
We also applied a 2D Gaussian model for all the SMGs.
Due to the large PSF size (FWHM$\sim$1\farcs8) and relative small source size (\textred{the median half-light radius along the semi-major axis is} $R_\mathrm{e}=0\farcs8$), the degree of freedom for \textsc{GALFIT} modeling was limited, and thus Gaussian fits do not show significant deviations in magnitude or source size from S\'ersic fits, as also mentioned by \citet{puglisi19}.

% A2813B$_1$ and A3088$_1$ 
HLS0043-B and HLS0307-B were fit with PSF models, since their faintness ($\sim$\,21 mag) and heavy blending with bright foreground sources ($\sim$\,17 mag) led to a divergence in the Gaussian/S\'ersic modeling. 
The upper limit of their $R_\mathrm{e}$ is estimated by the minimum measurable deviation from PSF size ($\sim$0\farcs4). 
% and thus 2.5\,kpc at $z_\mathrm{med}=1.8$. 
HLS1314 is observed as a lensed arc in the NIR images (UKIRT/WFCAM) and poorly modeled with a single Gaussian/S\'ersic profile. Therefore, we performed aperture photometry of this source after subtracting nearby sources.
Since three clumps were seen in HLS1314 in the 1\arcsec-tapered ALMA map, we also modeled its morphology with triple S\'ersic profiles at the positions of the ALMA clumps.
% The summed flux matches the aperture-photometry one in each IRAC band.
When the photometric measurements of the three clumps were combined, the total magnitudes derived in the IRAC images were recovered.

Another five sources, namely HLS1124-A/B, HLS0455, HLS0505 and HLS0600,
exhibited irregular morphologies in the ALMA continuum maps, revealing galaxy-lensed rings or multiple components at a resolution of $\sim 0\farcs 2$.
All of these sources are blended with nearby sources, and their morphologies cannot be well quantified through Gaussian or S\'ersic models.

To perform reliable IRAC photometry on these five sources, we adopted their ALMA continuum images as their morphological models in the IRAC bands.
We clipped their Briggs-weighted, native Band-6 continua at $3\sigma$ and then convolved them with the corresponding PSFs as the source models.
Based on the $i$-band optical images from the Pan-STARRS DR1 and DES DR1 \citep{ps1dr1, desdr1}, we set up 2D Gaussian models for their nearby sources, which were also convolved with the IRAC PSFs.
We then used an MCMC routine \citep[\textsc{emcee};][]{emcee} to fit in the IRAC images the brightnesses of the ALMA sources, the effective radii and brightnesses of the optical sources, and the sky background.
Photometry of the ALMA sources was then performed with the residual maps after the best-fit models for the nearby sources were subtracted.
Figure~\ref{fig:02_sunfit} shows the ALMA images of the SMGs, optical images of nearby sources, and IRAC 3.6/4.5\,\micron\ images, before and after this MCMC neighborhood subtraction routine.

\begin{figure}[!tb]
\centering
\includegraphics[width=\linewidth]{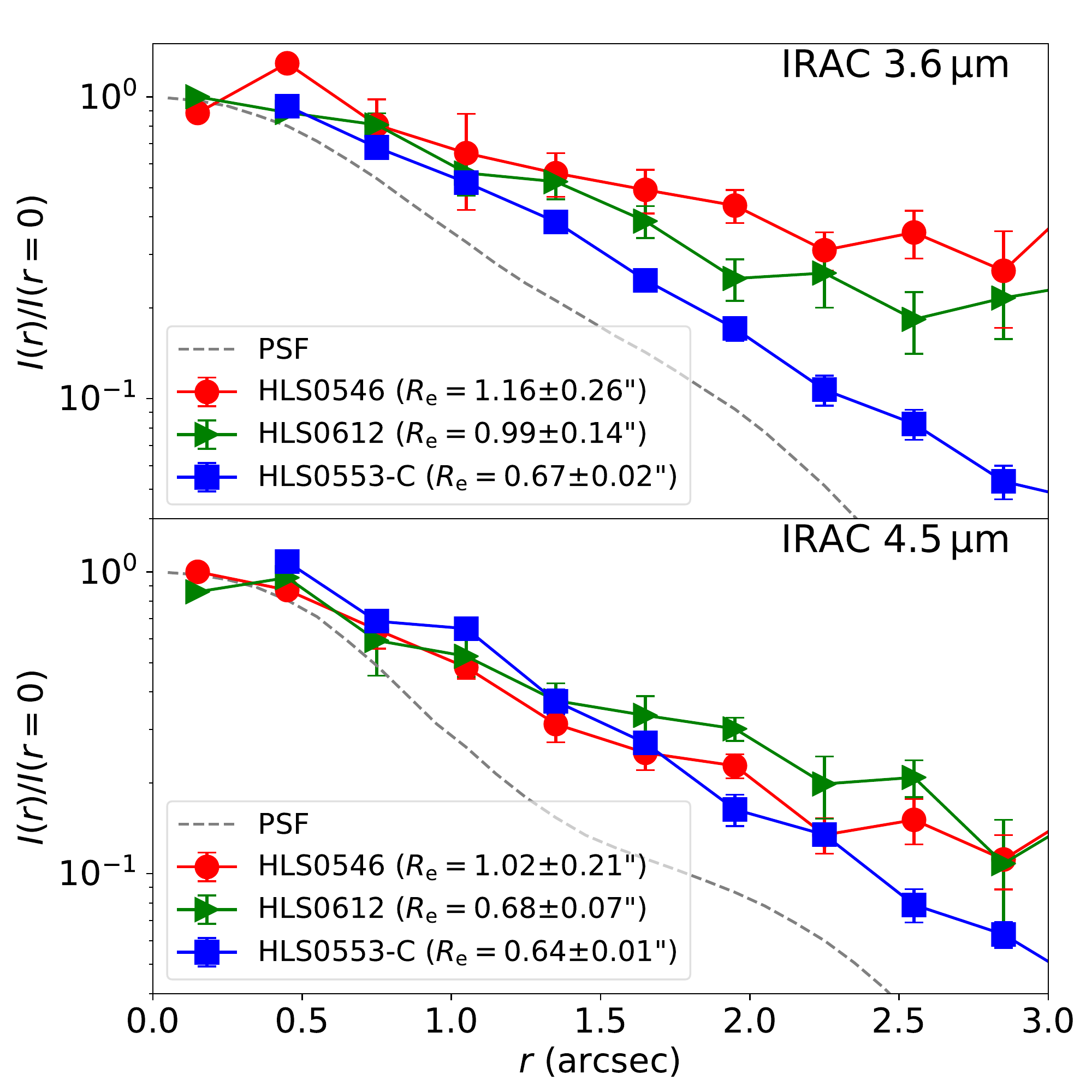}
\caption{IRAC 3.6\,\micron\ (upper panel) and 4.5\,\micron\ (lower panel) radial surface brightness profiles of three representative sources in this study (postage stamp images shown in Figure~\ref{fig:3_cutout}). 
Source names with their $R_\mathrm{e}$, obtained with \textsc{GALFIT}, are labeled in the legends. 
Radial profiles of IRAC PSFs \citep{hora12} are shown as grey dashed lines.
}
\label{fig:03_cog_irac}
\end{figure}

%% file: 03c_morphology.tex
\begin{figure}[!t]
\centering
\includegraphics[width=\linewidth]{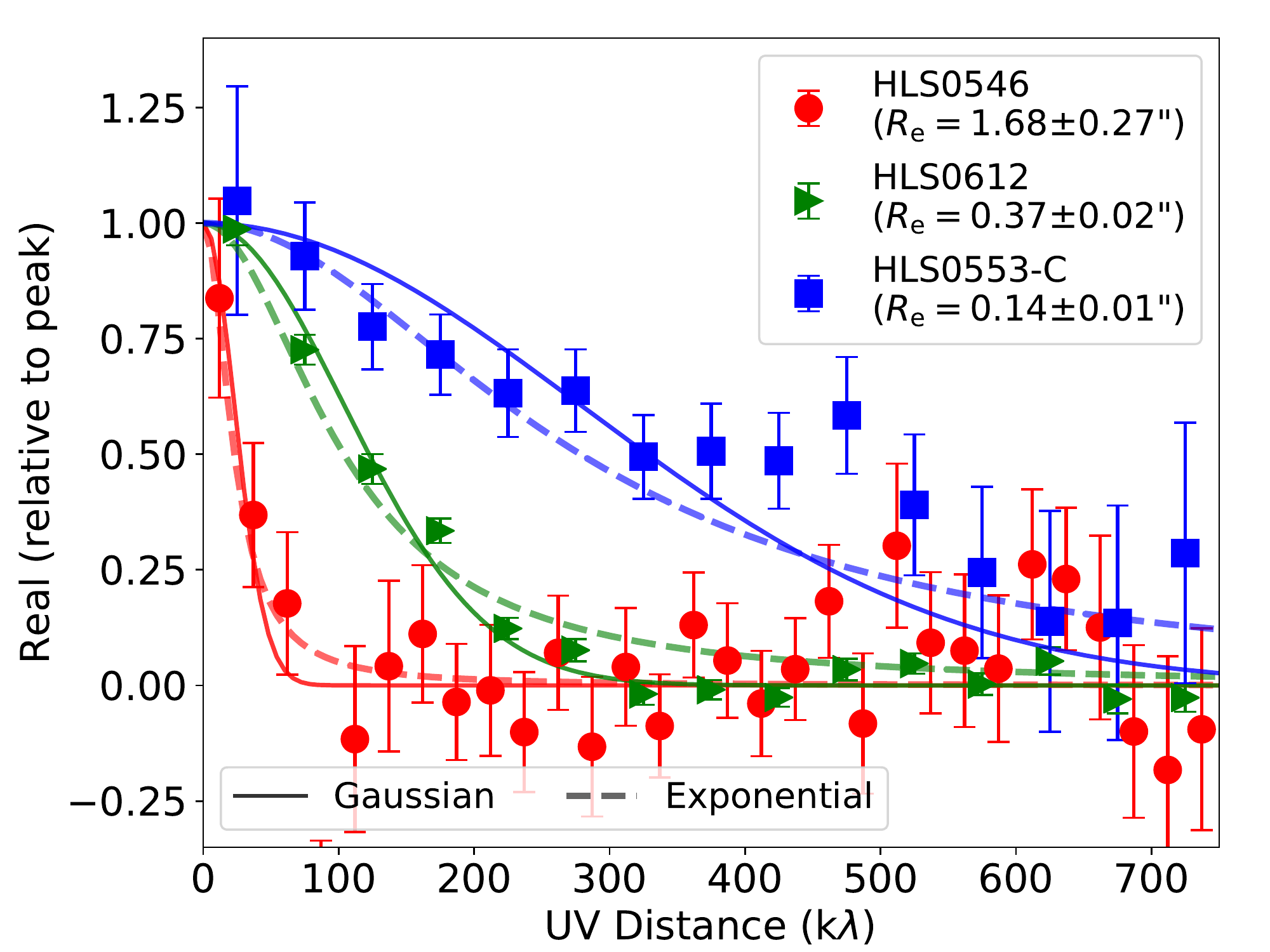}
\caption{Visibility profiles of three representative sources in this study (same as the ones in Figure~\ref{fig:03_cog_irac}). 
Source names with their $R_\mathrm{e}$, obtained through a \textsc{CASA} \textsc{uvmodelfit} routine, are labeled in the upper-right legend.
Solid lines present the best-fit Gaussian profiles, and dashed lines indicate the corresponding exponential profile with the same $R_\mathrm{e}$.
All points and lines are normalized according to their peak flux density.
}
\label{fig:03_uvdist}
\end{figure}

Morphological modeling of our sources in the IRAC images was performed along with photometry using \textsc{GALFIT}, as detailed in Section~\ref{ss:03b_photo}. 
For 24 of our sources with successful IRAC surface brightness profile modeling, we measured a median S\'ersic index of 0.9 and 1.0 at 3.6 and 4.5\,\micron, which is close to $n_\mathrm{med}=1.2\pm0.3$ reported by \citet{chen15} for $z\simeq 1 - 3$ SMGs in the \hst/WFC3-IR F160W band. % and $n_\mathrm{med}=1.0\pm0.1$ by \citet{gullberg19}, for $z=1\sim3$ SMGs in the \hst/WFC3-IR F160W band.
Among the six sources which we display their postage stamp images in Figure~\ref{fig:3_cutout}, 
the IRAC 3.6/4.5\,\micron\ radial surface brightness profiles of three representative sources are shown in Figure~\ref{fig:03_cog_irac}.
The median IRAC effective radius is around $0\farcs8$, corresponding to 6.5\,kpc at $z=2$ %$z_\mathrm{med}=1.8$
in physical scale before lensing correction. 
For the five sources with heavy blending and complex morphology (Figure~\ref{fig:02_sunfit}), we did not measure their structural profiles in either the IRAC or ALMA images.
Here, we concentrate on the modeling of structural profile of the remaining 24 lensed SMGs at 1.3\,mm.
As already described, we continue to decompose HLS1314 as three sources, so 26 sources are studied in this subsection.

We obtained structural parameters of our sources in the ALMA data, modeling their visibility data using \textsc{uvmodelfit} (for single-source case) and \textsc{uvmultifit} (for multiple-source case; \citealt{uvmultifit}), both under \textsc{CASA} environment.
We assumed Gaussian models for the surface brightness profile of all sources\textred{, consistent with the modeling procedure applied by \citet{puglisi19} and \citet{tadaki20}}.
We note that \citet{hodge16} measured a median S\'ersic index of $n=0.9\pm0.2$ for $z\sim2.5$ SMGs, and \citet{gullberg19} measured $n=1.0\pm 0.1$ at 870\,\micron. 
These S\'ersic indices are closer to exponential disk profiles ($n=1$) rather than Gaussian ones ($n=0.5$). 
Since the $R_\mathrm{e}$ modeled by a Gaussian profile is smaller than exponential one by only $\sim0.045$\,dex in \citet{hodge16}, less than the median uncertainty of $R_\mathrm{e}$ in this work ($\sim0.063$\,dex), we keep this Gaussian profile assumption for SMGs in the ALMA data. 
We used ALMA source positions, flux densities and morphological parameters, obtained through \textsc{SExtractor} photometry, as the initial guess of source models in the \textit{uv}-plane. 
Figure~\ref{fig:03_uvdist} shows three representative visibility profiles of lensed SMGs with different values of $R_\mathrm{e}$ (length of semi-major axis) shown in Figure~\ref{fig:03_cog_irac}.
There is no clear difference between the $\chi^2$ of the best-fit circular Gaussian and exponential models for HLS0546 and HLS0553-C, and the Gaussian profile fits HLS0612 better.

There were two cases that required special attention for multiple-source modeling on the \textit{uv}-plane. 
HLS0612 was observed to have two minor nearby components in the ALMA high-resolution image (Figure~\ref{fig:3_cutout}), while \textsc{uvmultifit} favored a single-source model.
Therefore, we discarded the three-source fitting results.
On the contrary, the goodness of a three-source fit with HLS1314 was better than that of a single-source one, so we continued to decompose it into HLS1314-A, B and C, consistent with the structural fitting in the IRAC bands.

The effective radii and their uncertainty of all the 26 sources (24 SMGs and HLS1314 with three components) are presented in Table~\ref{tab:03_irm}. 
We also compared the IRAC (average of 3.6/4.5\,\micron) and ALMA source centroids, and did not find any detectable offset (Appendix~\ref{sec:app_2}).

\begin{figure}[!tb]
\centering
\includegraphics[width=\linewidth]{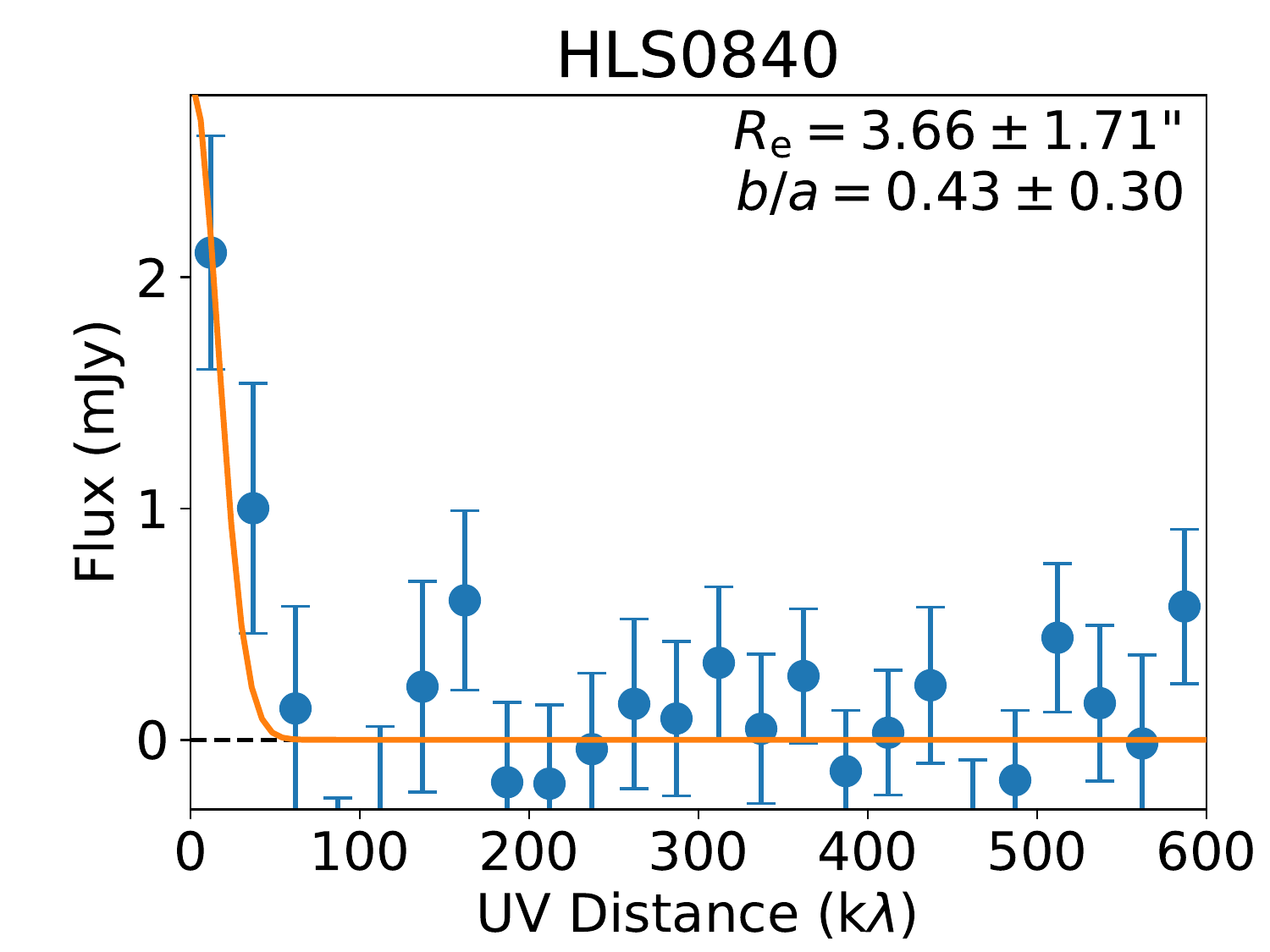}
\caption{Visibility profile of HLS0840 (blue filled circles), the source with the most extended 1.3\,mm continuum in our sample. 
Best-fit morphological parameters obtained with \textsc{uvmodelfit} are noted in the upper-right corner.
The best-fit Gaussian profile is indicated as the orange solid curve.
}
\label{fig:0840}
\end{figure}

\textred{
All of our measured effective radii at 1.3\,mm are smaller than their corresponding maximum recoverable angular scales for given ALMA antenna configurations, except for HLS0840 (Figure~\ref{fig:0840}). 
As a result, the Re measurement of HLS0840 has a large error bar (1\farcs7).
To examine the quality of \textit{uv}-profile modeling for sources with extended $R_\mathrm{e}$ or at relatively low S/N, we also simulated and modeled the visibility data for the same sources as described in Section~\ref{ss:03a_detect} for flux boosting evaluation.
The median $R_\mathrm{e}$ measured from the 10--15 sets of simulated visibility data for each source % with the same \textsc{uvmodelfit} routine 
is consistent with that of the input model at the $1\sigma$ confidence level, demonstrating the overall validity of our source-size measurements with ALMA.
}

%% file: 03d_summary.tex
% fidelity 
We successfully identified the IRAC counterparts of all the 29 sources discovered with ALMA at $\mathrm{S/N}>4.0$, indicating that our sample is free from the contamination by any false detection.
\textred{Our Monte-Carlo simulation also suggests a low probability ($<2$\%) of random association between ALMA and IRAC sources.}
We display all the ALMA and IRAC images of sources in our sample in Figure~\ref{fig:3_cutout}, \ref{fig:02_sunfit} and~\ref{fig:apd_01}.
We notice that HLS0840 and HLS0546 are two spatially extended sources ($R_\mathrm{e,ALMA}>R_\mathrm{e,IRAC}\sim 1$\arcsec) at 1.3\,mm with exceptionally low surface brightness ($\lesssim 0.1$\,\si{mJy.arcsec^{-2}}) and only detectable in \textit{uv}-tapered ALMA maps instead of those at native resolution (Figure~\ref{fig:3_cutout}).
The visibility profiles of these two sources are displayed in Figure~\ref{fig:03_uvdist} and~\ref{fig:0840}.

% cluster/galaxy lensed
Although all of the sources were discovered in lensing cluster fields, we identified five sources that are subject to additional boosting by foreground galaxies, i.e., HLS1124-A/B, HLS0455, HLS0505 and HLS0600 as shown in Figure~\ref{fig:02_sunfit}.
Among the remaining cluster-lensed cases, HLS0553 and HLS2332 are two spectroscopically confirmed triply-imaged systems behind the cluster fields of MACSJ0553 and RXCJ2332, respectively.
HLS0553-A/B/C have been reported at $z=1.14$ by \citet{ebeling17}, and HLS2332-A/B/C were reported at $z=2.73$ by \citet{greve12}.
We do not find any other multiply-imaged SMGs in our sample.

For the sources with known spectroscopic redshifts (e.g., from a CO search with the IRAM 30-m telescope; Dessauges-Zavadsky et al., in prep.), our ALMA Band-6 observations also targeted CO lines when possible.  We confirmed the redshifts of HLS0553 and HLS2332, 
and obtained/confirmed redshifts for additional seven sources (Sun et al., in prep): HLS0455 ($z=2.93$; \citealt{zavala15}), HLS1314 ($z=1.45$), HLS1124-A/B ($z=1.80$), HLS0011-A/B ($z=2.27$) and HLS0600 ($z=2.89$).  
We did not detect any spectral line feature in the ALMA data of the remaining 16 sources, and thus their redshifts remain unknown.

% intrinsically multiple sources
\textred{Among the full sample of 29 ALMA sources, we found 16 are potentially isolated sources with no detectable companion brighter than 0.6\,mJy at 1.3\,mm. %\todo[inline]{IRS: Within what area and what is amplification gradient on that scale?} 
We also identified six SMGs that exhibit close companions at similar redshifts, namely HLS1124 (observed as 4 components and grouped as 2 in Table~\ref{tab:02_phot}; hereby noted as 4/2 and same later), HLS0111 (2/2), HLS0600 (2/1), HLS0612 (3/1), HLS1314 (3/1) and HLS2104 (2/2).}
Such grouping is determined by the angular separation and \textsc{SExtractor} deblending threshold ($\sim$3\arcsec) on 1\arcsec-tapered maps.
% All of these pairs of sources are spectroscopically confirmed within a maximum velocity separation of 800\,\si{km.s^{-1}} of each other except for HLS2104 whose spectroscopic redshift is unknown. 
\textred{Source groups HLS0111, HLS0600, HLS1124 and HLS1314 are spectroscopically confirmed within a maximum velocity separation of 800\,\si{km.s^{-1}} of each other. }

% \begin{figure}[!htb]
% \centering
% \includegraphics[width=\linewidth]{figures/thumbnail/header.pdf}
% \includegraphics[width=\linewidth]{figures/thumbnail/0546-5345_0_thumbnail.pdf}
% \includegraphics[width=\linewidth]{figures/thumbnail/0612-4317_0_thumbnail.pdf}
% \includegraphics[width=\linewidth]{figures/thumbnail/MACSJ0553_2_thumbnail.pdf}
% \caption{}
% \label{fig:04thumb}
% \end{figure}

%% file: 04_analyses.tex
% Based on the measurements and modeling efforts described in Section~\ref{sec:03_res}, we modeled the SEDs of all sources and performed our analysis at the level of both integrated (i.e., unresolved) and spatially-resolved scales. 
% Our results on the integrated scale include how observed color properties are affected by redshift and galaxy properties, 
% whether there is an intrinsic offset between stellar and dust continua, 
% how intense the ongoing star formation is in our lensed SMG sample, and what are the amount and behaviour of dust.
% At the spatially-resolved scale, which is the very unique point for this study due to the lensing magnification, we study the surface density distributions of both star and dust (star formation), and compare their relative spatial distributions with the strength of star formation in the galaxies.

\subsection{SED fitting and photometric redshift}
\label{ss:04a_sed}
\input{04a_SEDfitting}

\subsection{Multi-wavelength color and redshift}
\label{ss:04b_color}
\input{04b_color}

% \subsubsection{Spatial offset between stellar and dust continua}
% \label{sss:04aii_off}
% \input{04b_2offset}
% \subsection{Integrated galaxy properties}

\subsection{Star formation in lensed SMGs}
\label{ss:04c_sf}
\input{04c_SF}

\subsection{Dust to stellar mass ratio}
\label{ss:04d_dust}
\input{04d_dust}

% resolved study: 
% \subsection{Resolved study of dust and stellar continua}

\subsection{Dust-to-stellar continuum size ratio}
\label{ss:04e_size}
\input{04e_size}

\subsection{IR surface luminosity and dust temperature}
\label{ss:04f_sbfir}
\input{04f_sbfir.tex}

\subsection{Dust-to-stellar size ratio versus IR surface luminosity}
\label{ss:04g_main}
\input{04g_main}

\subsection{Stellar surface density}
\label{ss:04h_svigstar}
\input{04h_sigstar}

%% file: 04a_SEDfitting.tex
% \subsubsection{Photometric Redshift}

% With the 4-band FIR photometry including ALMA 1.3\,mm, we computed the photometric redshift ($z_\mathrm{phot}$) using LIRG templates compiled in \citet{rieke09}.
% We first assumed that LIRG SED templates at several total IR luminosities ($\log(L_\mathrm{TIR}/\mathrm{L}_\odot)$ = 10.75,\ 11.00,\ 11.25,\ 11.50,\ 11.75) have an equal possibility to match the intrinsic SED shape of any given SMG.
% We then calculated the probability distribution of SMG redshift through a Bayesian approach.
% % special case of CODEX
% CODEX35646$_0$ exhibits a significant smaller 1.3\,mm flux density compared to the prediction by the 3-band SPIRE SED, and thus we suggest that the ALMA source cannot account for all the SPIRE flux density measured for this source (further discussed in Section~\ref{ss:04a_sed} and Figure~\ref{fig:04_color}).
% For this special case, we did not use the 1.3\,mm flux density for the $z_\mathrm{phot}$ determination.  
% The most probable redshift is adopted as $z_\mathrm{phot}$ shown in Table~\ref{tab:02_phot} with the dagger symbol, with a typical uncertainty of $\Delta z = 0.1 (1+z)$.
% When we derive the galaxy properties later, we do not include the uncertainty of redshift for the error analysis.

% \subsubsection{SED fitting}

% To interpret our multi-wavelength photometry of 29 lensed SMGs (MACSJ1314 as a whole), 
We perform SED modeling with the high-$z$ extension of \textsc{magphys} \citep[][]{magphys,dacunha15}.
We also use the photo-$z$ extension of \textsc{magphys} \citep{battisti19} to estimate the photometric redshift ($z_\mathrm{phot}$) and physical properties simultaneously when the spectroscopic redshift ($z_\mathrm{spec}$) is unknown.
\textsc{magphys} assumes a Chabrier IMF \citep{chabrier03}, a continuous delayed exponential star-formation history (SFH), a two-component dust absorption law \citep{cf00} and energy balance between dust absorption in the UV and re-emission in the infrared.
Other key components of the model assumptions in \textsc{magphys} have been detailed in \citet{martis19}. 

We adopt our observed photometric measurements to feed the SED fitting routine without applying any lensing magnification correction. 
This is due to a lack of detailed lens models for several of the clusters in this study.
Note, therefore, that all the derived physical quantities are those including the effects of lensing.
We also assume that there is no differential magnification effect (i.e., effective magnification factors change as function of the source size and therefore as function of wavelength in general).
% Here we simply ignore the differential lensing effect (i.e.\ different parts of the SMG is magnified at different scale), and the derived physical quantity is lensing-boosted.

For the majority of our sample, their SEDs are modeled using six-band photometry from 3.6\,\micron\ to 1.3\,mm. 
We also utilize the ancillary data described in Section~\ref{ss:02d_other} to improve the fitting, especially to constrain the amount of rest-frame optical dust extinction and therefore stellar mass. 
All the individual best-fit SEDs are displayed in Figure~\ref{fig:apd_03}, and a summary of the best-fit galaxy properties is presented in Table~\ref{tab:04_prop}.

The 16--50--84th percentile of the redshift distribution for our sample is 1.23--1.93--2.73, and the highest and lowest photometric redshifts are $z_\mathrm{phot}=0.96^{+0.17}_{-0.45}$ (HLS1623) and 3.23$^{+0.48}_{-0.59}$ (HLS0043-B).
Here the $z_\mathrm{phot}$ is the median of likelihood distribution of redshift for each source.
We assess the uncertainty of photometric redshift estimate based on the likelihood distribution obtained with \textsc{magphys}. 
The typical uncertainty of derived $z_\mathrm{phot}$ is $\Delta z = 0.16^{+0.05}_{-0.04} \, (1+z_\mathrm{phot})$. 
HLS1115 exhibits a small $\Delta z$ of 0.03 because of the existence of 9-band \hst\ data.
We also evaluate the far-IR $z_\mathrm{phot}$ by matching with the LIRG templates in \citet{rieke09}, and the derived redshifts are consistent with those by \textsc{magphys} within $\sim 1\sigma$ confidence interval.

We also study the dust temperature with far-IR data over a rest-frame wavelength of 50\,\micron, following \citet{greve12}. 
We fit the dust continua of all SMGs with modified black-body (MBB).
The dust absorption coefficient is assumed to be $\kappa = 0.040\times (\nu/250\,\mathrm{GHz})^\beta$ in the unit of \si{\meter^{2}\,\kilogram^{-1}}, where $\nu$ is the frequency in GHz in the rest-frame.
We assume a fixed dust emissivity of $\beta=1.8$, which is widely adopted in previous studies \citep[e.g.,][]{ds17,dudzevic19}.
We also incorporate the uncertainty of redshift %($\Delta z /(1+z)$) 
in that of the dust temperature if the $z_\mathrm{spec}$ is unknown.
The dust masses derived from this MBB fitting are consistent with those from \textsc{magphys}, listed in Table~\ref{tab:04_prop}, with a $1\sigma$ dispersion of 0.08\,dex.

HLS1623 was observed to be much fainter at 1.3\,mm compared to the prediction from the SPIRE SED.
This may indicate that the SPIRE fluxes are contributed by certain ALMA-undetected components in the same FoV. %or the possible low-quality of ALMA data.
Here, we use its SPIRE-only SED to assess the dust temperature, but we only use its IRAC and ALMA flux densities for \textsc{magphys} SED fitting.

%% file: 04b_color.tex
\begin{figure*}[t]
\centering
\includegraphics[width=0.47\linewidth]{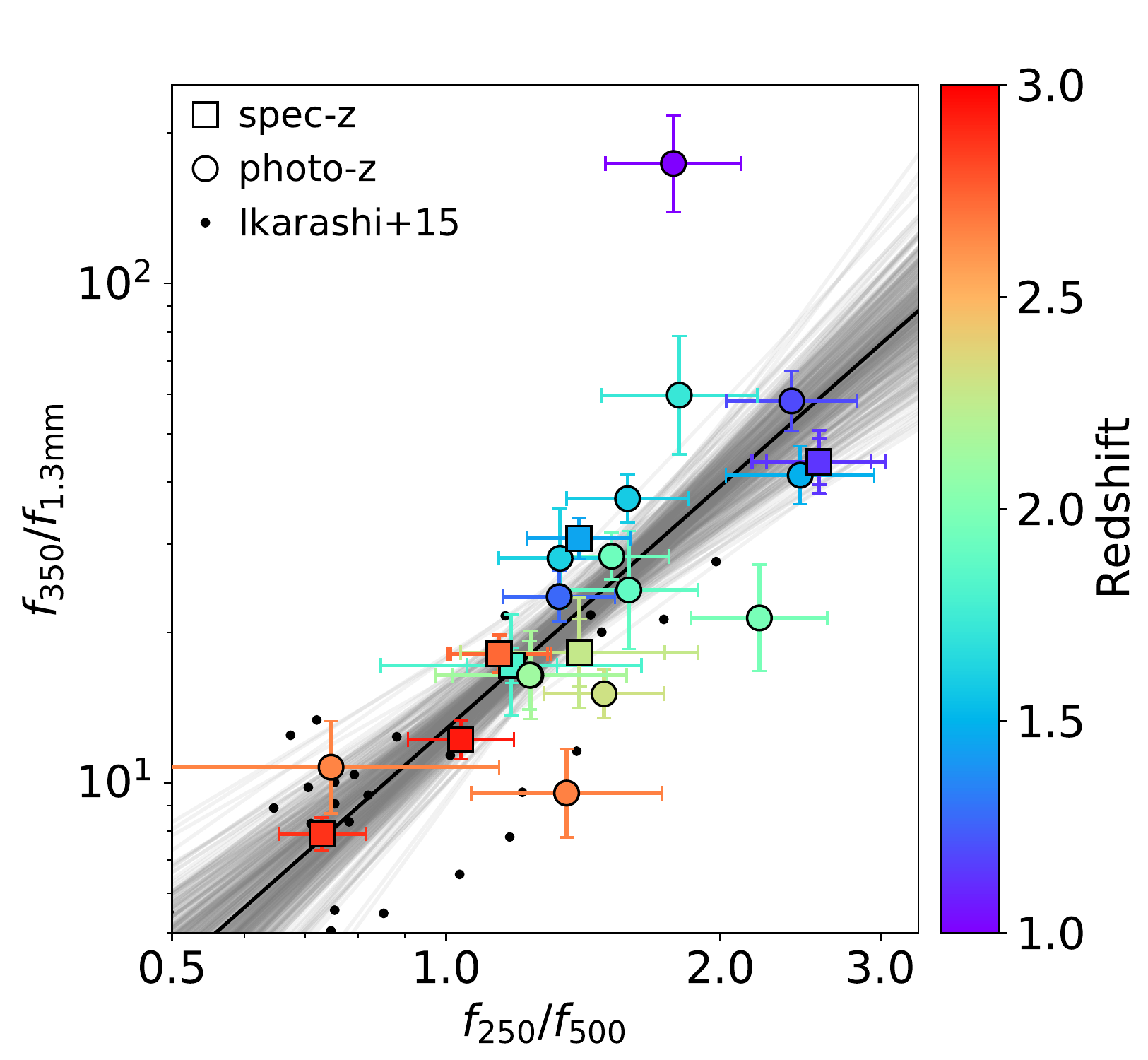}
\includegraphics[width=0.47\linewidth]{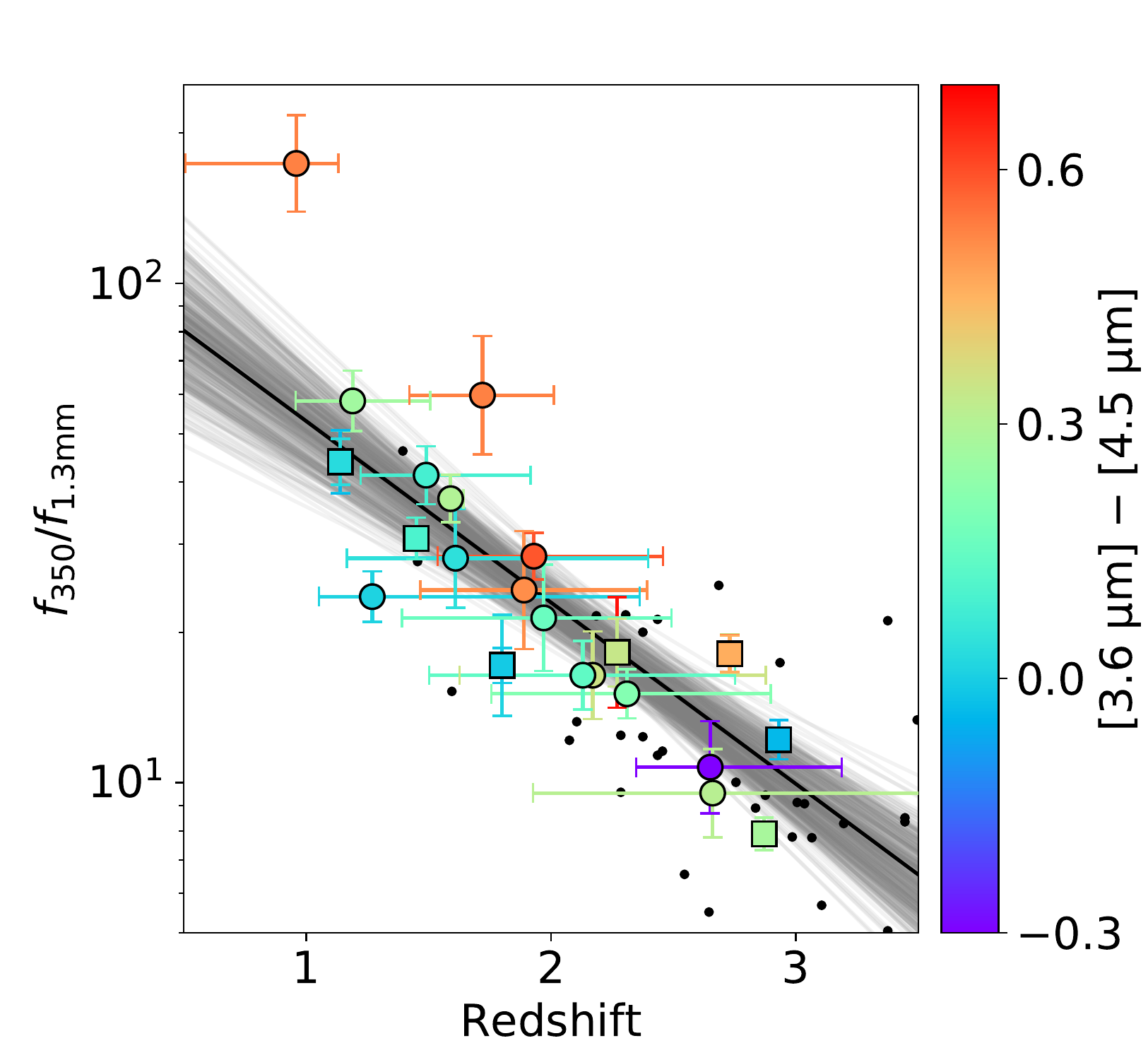}
\caption{\textit{Left}: FIR color-color diagram of all lensed SMGs in this work, color-coded with their redshifts. 
The squares denote $z_\mathrm{spec}$-confirmed sample and the circles denote $z_\mathrm{phot}$ ones.
Unlensed SMGs presented in \citet{ikarashi15} are shown as black dots.
A linear fitting through MCMC is plotted as black solid line with grey line groups indicating the uncertainty.
\textit{Right}: FIR color ($f_{350}/f_{1.3\si{mm}}$) versus redshift of all our sources, color-coded with their IRAC color.
Markers stand for the same in left panel.
Similar MCMC linear fitting is plotted as black solid line with grey line groups for its uncertainty.
HLS1623 shows exceptionally blue $f_{350}/f_{1.3\si{mm}}$ color ($\sim 170$), indicating the 1.3\,mm source detection in this cluster field may be incomplete.
}
\label{fig:04_color}
\end{figure*}

% Color indices obtained in multiple wavelengths are strong indicator of the redshift and physical properties of galaxies.
% The color of the \spitzer/IRAC 3.6/4.5\,\micron\ bands is sensitive to redshift, stellar age and dust absorption, and FIR colors are mainly affected by redshift, dust temperature ($T_\mathrm{dust}$) and emissivity ($\beta$; assumed as fixed value $=1.5$ in this work).

We first compare the FIR colors, namely  $f_{250}/f_{500}$ and  $f_{350}/f_{1.3\si{mm}}$, of all galaxies detected in our survey (Figure~\ref{fig:04_color}; left). 
Here we do not use the color of  $f_{250}/f_{350}$ or  $f_{350}/f_{500}$ due to their narrow ranges of distributions.
%%%%%
We find the majority of our galaxy exhibit a positive linear correlation between these two color indices.
An MCMC fitting through \textsc{emcee} suggests the power-law relation as $\log(f_{350}/f_{1.3\si{mm}}) = (1.507 \pm 0.247) \times \log(f_{250}/f_{500}) + (1.13 \pm 0.05)$.
This is consistent with the unlensed SMG sample presented in \citet{ikarashi15}, for which we apply a conversion from 1.1\,mm flux densities to 1.3\,mm by a factor of 0.53.
This indicates red $f_{250}/f_{500}$ and red $f_{350}/f_{1.3\si{mm}}$ colors will occur simultaneously, basically controlled by source redshift and dust temperature.

We then compare $f_{350}/f_{1.3\si{mm}}$ versus source redshifts, and an exponential relation can be identified against both the spectroscopic and photometric redshift samples (Figure~\ref{fig:04_color}; right).
Similar MCMC routine implies an underlying relation as $\log(f_{350}/f_{1.3\si{mm}}) = -(0.376  \pm 0.044) \times z + (2.102 \pm 0.095)$, with a standard dispersion of 0.16\,dex for the measured $f_{350}/f_{1.3\si{mm}}$ ratio.
This trend is also consistent with the unlensed sample in \citet{ikarashi15} at similar redshift range.
Such a correlation suggests that the $f_{350}/f_{1.3\si{mm}}$ color of SMGs, at least in this study, is only weakly affected by dust temperature and mainly reflects the redshift.
% Sources below this exponential relation do show higher dust temperature, while this effect is minor and $f_{350}/f_{1.3\si{mm}}$ can be considered as an estimator of redshift, though precision is limited when 250 and 500\,\micron\ measurements are ignored.

On the other hand, the IRAC $[3.6\,\micron]-[4.5\,\micron]$ color of our SMG sample seems to show a larger dispersion at any given redshift (see the color coding in the right panel of Figure~\ref{fig:04_color}), and thus no substantial correlation can be derived between the IRAC color and redshift.
This reflects the complexity of stellar age and dust absorption among our SMG sample.
However, we find that at a given redshift, a red IRAC color is likely to occur simultaneously with a blue $f_{350}/f_{1.3\si{mm}}$ color.
The red IRAC color can be a signature of high dust extinction, and thus a high dust column density and high IR surface brightness ($\Sigma_\mathrm{IR}$).
Since $\Sigma_\mathrm{IR}$ is observed to be correlated with dust temperature at various redshift ranges \citep[e.g.,][]{ds17,spilker16},
this would result in a bluer $f_{350}/f_{1.3\si{mm}}$ color as we have shown.

Several $z\sim 1$ galaxies were observed with a red IRAC color ($\sim$0.5), namely HLS0307-28-A, HLS1623, which suggests high dust extinction in these systems ($A_V\gtrsim7$) through \textsc{magphys} SED modeling.
Such a high $A_V$ is not seen in AS2UDS sources with secure optical/NIR detection at $\lambda_\mathrm{obs} \lesssim 2.2$\,\micron\ \citep[but are founded in sources with IRAC-only detections;][]{dudzevic19}.
Due to the lack of deep rest-frame UV and optical data, it is not clear whether the determination of such high $A_V$ values is reliable.
%such high $A_V$ cannot be confirmed and are likely to be overestimated.
In the case of HLS0553 and HLS2332, where \hst\ photometry exists for two of the three triplet lensed images, the \hst\ photometry obtained at $\lambda\leq1.6$\,\micron\ slightly reduces the estimated $A_V$ (by $\sim 0.4$) and thus the best-fit stellar mass (by $\sim 0.2$\,dex; note that the median uncertainty is $0.32$\,dex if \hst\ data is excluded).
This underscores the difficulty of deriving dust absorption in galaxies with only two-band IRAC observations of their stellar continua.

%% file: 04c_SF.tex
\begin{figure}[tb!]
\centering
\includegraphics[width=\linewidth]{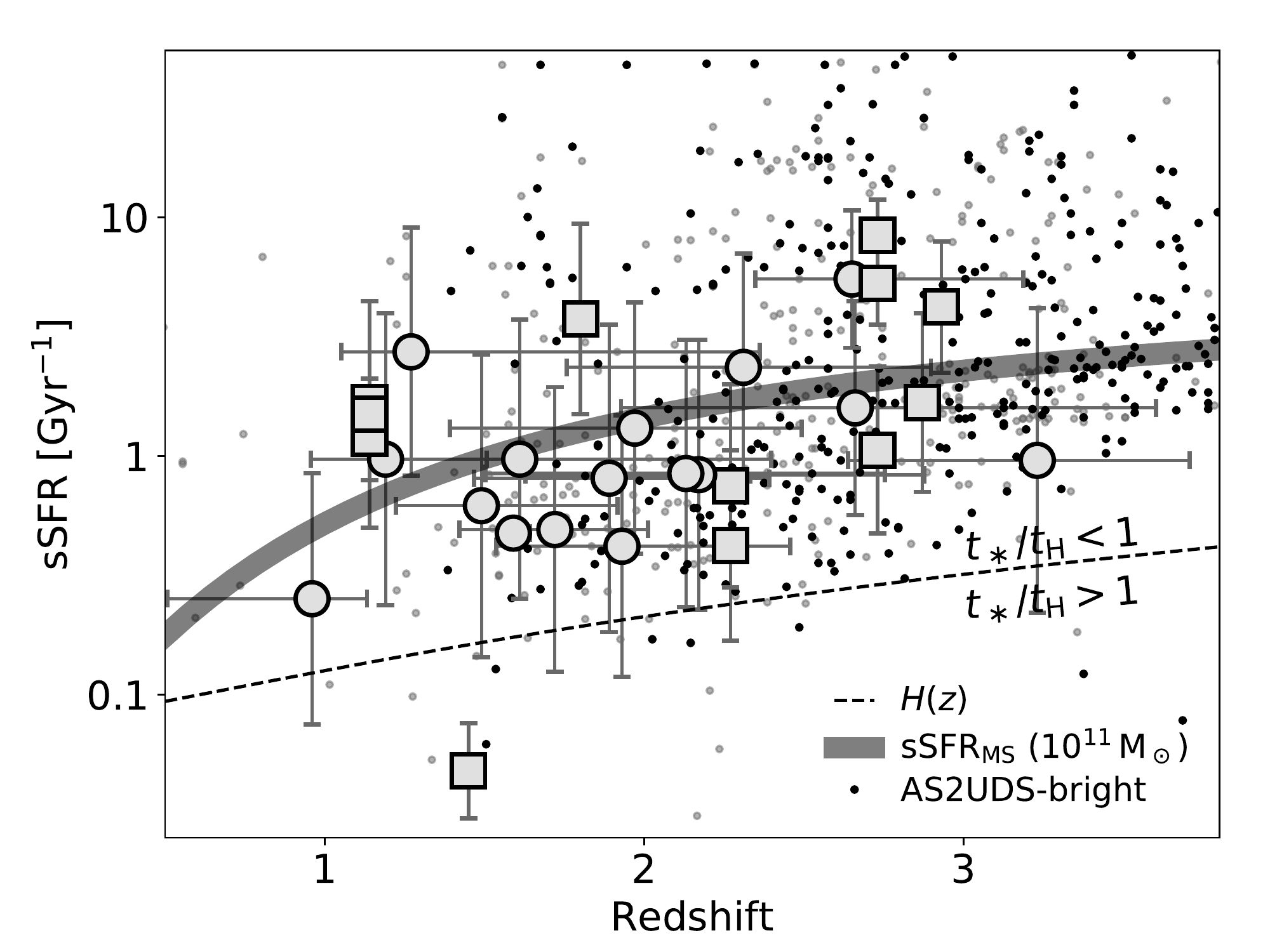}
\caption{sSFR (SFR$\cdot M_\ast^{-1}$) versus redshift of all lensed SMGs in this work.
Squares stand for $z_\mathrm{spec}$ confirmed sample and circles stand for $z_\mathrm{phot}$ ones.
We compare the distribution of our sources with AS2UDS sample (black dots for sources brighter than 3.6\,mJy at 850\,\micron, i.e., the completeness limit of the SCUBA-2, and grey dots for fainter ones which are below that limit; \citealt{dudzevic19}), the main sequence of star-forming galaxies with a stellar mass of $10^{11}$\,\si{M_\ast} \citep[grey solid line;][]{speagle14}, and Hubble parameter $H(z)$ (black dashed line). }
\label{fig:ssfr}
\end{figure}

The dust-obscured fraction of SFR in a galaxy has been claimed to be correlated with the stellar mass with no conspicuous evolution found with this correlation from redshift 2.5 to 0 \citep{whitaker17}.
For a galaxy with $M_\ast = 10^{10}$\,\msun, 80\% of the total star formation is obscured, and for typical $z\sim2$ SMGs with a stellar mass of $\sim10^{11}$\,\msun\ (e.g., \citealt{hainline11,dudzevic19}), this fraction is higher than 95\%.
Without appropriate lensing correction, we cannot accurately determine the intrinsic stellar mass of our lensed SMG sample.
However, the median value is $10^{11.8}\,\mu^{-1}$\msun, still well above $10^{10}\,$\msun\ if a $\mu=5$ magnification factor correction is applied.
Therefore, it is reasonable to assume that the majority of star formation in our sample is dust-obscured, and that the observed IR luminosity (and emitting regions) adequately represent the total SFR (and star-forming regions).

We measure a median total IR luminosity of $L_\mathrm{IR} = 10^{12.92\pm 0.07}\,\mu^{-1}$\lsun\ and thus a median SFR of 552$\pm$93 $\mu^{-1}$\smpy\ without lensing corrections through SED fitting.
% Note that \textsc{magphys} applies a decomposition of thermal dust emission ($L_\mathrm{IR}$) originating from both the diffuse ISM ($f_\mu$) and the stellar birth cloud ($1-f_\mu$), and we find that the value of $f_\mu$ is $0.56\pm0.11$ among our sample.
% The typical uncertainty of $f_\mu$ for each source is $\Delta f_\mu = 0.10$. 
% \citet{martis19} showed that \textsc{magphys} only considers the IR luminosity from birth clouds as SFR indicator through a \citet{ke12} conversion, and in dusty star-forming galaxies, a significant amount of dust emission is caused by heating from evolved stellar population rather than ongoing star formation.
% Our study also suggest the similar scenario, and the SFR we derived is 0.31$\pm$0.07\,dex lower than the direct conversion from the total IR luminosity to SFR, without distinguishing the origin of emission.
Figure~\ref{fig:ssfr} displays the sSFR versus redshift of all the sources in our sample.
The distribution of sSFR in redshift space is consistent with those of galaxies on the so-called star-forming ``main sequence'' (MS) with a stellar mass of $10^{11}$\,\si{M_\odot} \citep{speagle14}, which have a median sSFR of 1.1\,\si{Gyr^{-1}} and a $1\sigma$ dispersion of 0.46\,dex.   
If the differential magnification is negligible, sSFR will conserve by lensing, and hence the distribution of sSFR should be the same among lensed and unlensed SMGs.

We also compare the stellar mass doubling time scale ($t_\ast=$1/sSFR) with the Hubble time ($t_H(z)=1/H(z)$) at the redshifts of SMGs in this work.
The ratio between these two quantities ($t_\ast/t_H$) can in principle act as an indicator of whether a galaxy is undergoing a significant star-formation event \citep[e.g.,][]{tacchella18}.
We find a median ratio of 0.18$\pm$0.02, suggesting that the majority of our sample are vigorously star forming galaxies.
This value is consistent with the median of AS2UDS SMGs at $1<z<3$ \citep{dudzevic19}, regardless of whether the unlensed SMG is brighter than the single-dish SCUBA-2 detection limit at 850\,\micron\ (3.6\,mJy) or not.
Except for HLS1314 (the square point at the bottom of Figure~\ref{fig:ssfr}), no other galaxy in this study shows $t_\ast/t_H\geq1$.  
% {\color{red} (Do you want to explain why HLS1314 is an exception here?)}
% Additionally, a direct \citet{ke12} conversion of total IR luminosity to SFR will imply $t_\ast/t_H<1$ for all the 29 galaxies in this study, with a median ratio of 0.09.

%% file: 04d_dust.tex
\begin{figure}[!t]
\centering
\includegraphics[width=\linewidth]{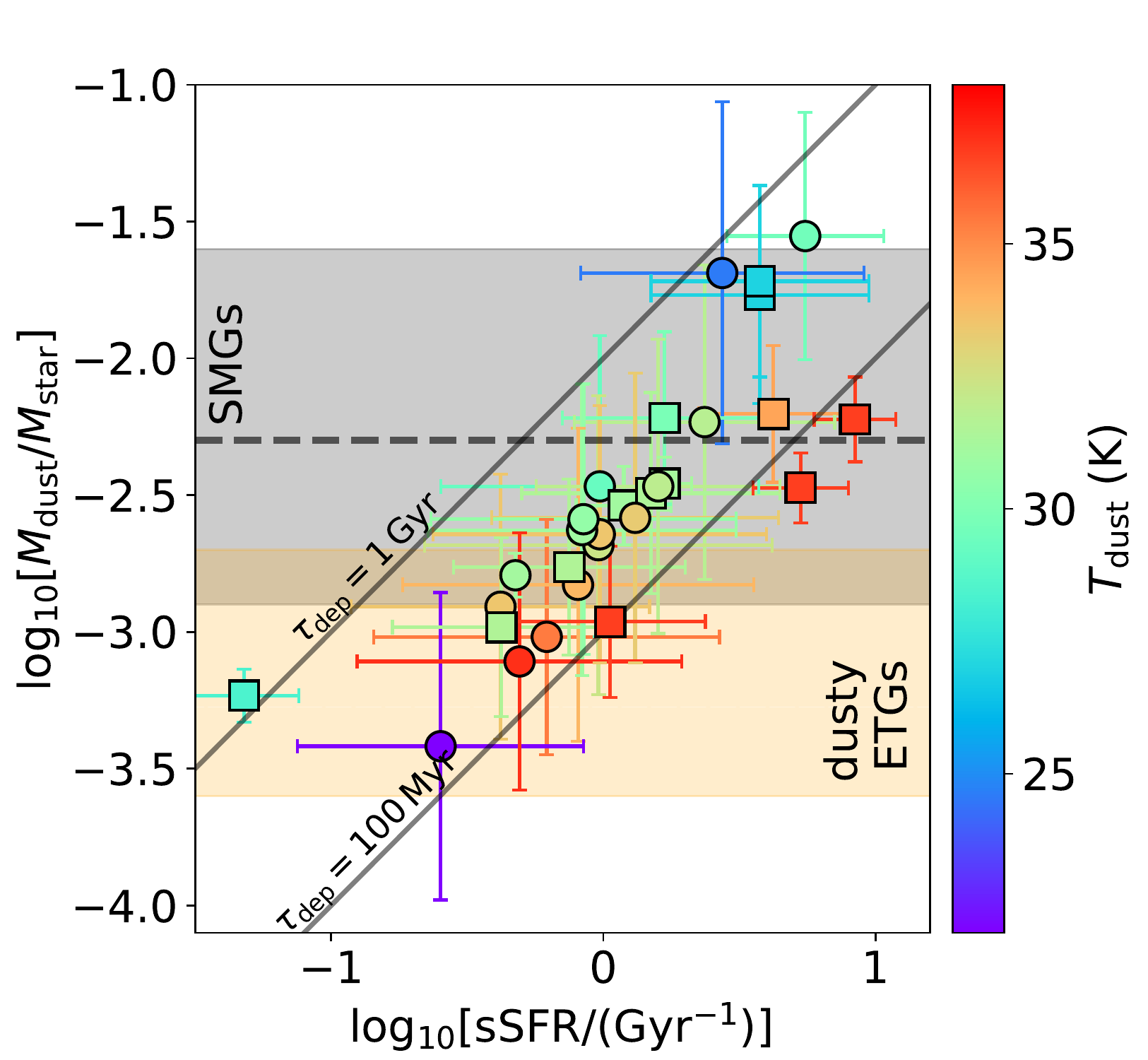}
\caption{Dust-to-stellar mass ratio versus sSFR, color-coded with dust temperature, of all 29 SMGs in this work. 
Squares represent sources with confirmed $z_\mathrm{spec}$ while circles for $z_\mathrm{phot}$-only ones.
Grey-shaded region represents $1\sigma$ distribution range for 24 SMGs in \citet{santini10} at $z_\mathrm{med}=2$ (dust mass recomputed by \citealt{calura17}), with grey dashed line for the median $\log(M_\mathrm{dust} /M_\ast)$.
Orange-shaded region represents $1\sigma$ range for dusty ETGs at $z<0.06$ \citep{agius13,rowlands15}. 
The two grey solid lines indicate the cases in which the gas depletion time scale is identical to 1\,Gyr or 100\,Myr, assuming a constant gas-to-dust-ratio of 100.
}
\label{fig:05_dsr}
\end{figure}

Similar to sSFR, the dust-to-stellar mass ratio ($M_\mathrm{dust}/M_\ast$; also referred to as the specific dust mass) and dust temperature are key observables of the properties of star-forming galaxies that are conserved in lensing.   
% should be conserved quantities if no differential magnification effect is at work.
Since dust is produced through the process of star formation, the ratio between dust and stellar mass should be closely related to the sSFR.
Figure~\ref{fig:05_dsr} plots the dust-to-stellar mass ratio versus sSFR in all 29 SMGs, color-coded with dust temperature.
A positive linear relation can be found between these two quantities, although the normalization is subject to $T_\mathrm{dust}$. % \citep[also shown in][]{martis19}.

Although such a correlation is not a surprise, the wide range of the dust-to-stellar mass ratio and sSFR of this 29-SMG sample is remarkable. 
Distributed between $-3.4$ and $-1.5$, the $\log (M_\mathrm{dust}/M_\ast)$ span of this sample is similar to that of low-redshift galaxies, whose SFRs are distributed between $10^{-1.5}$ and $10^{2}$\,\smpy\ \citep{dacunha10}. 
The upper end of our $M_\mathrm{dust}/M_\ast$ distribution matches with previous SMG literature \citep[e.g.,][]{santini10,dudzevic19}, while our sample does have a significant excess at $\log(M_\mathrm{dust} /M_\ast) \lesssim -3$.
We perform a Kolmogorov--Smirnov (K-S) test of $\log(M_\mathrm{dust}/M_\ast)$ between our sample and AS2UDS SMGs in % both \citet[24 SMGs at a median redshift of 2]{calura17} and
\citet[454 SMGs at $1 < z < 3$]{dudzevic19},
and a null hypothesis that SMGs in these two samples share the same $M_\mathrm{dust}/M_\ast$ distribution can be rejected ($p$-value less than 0.01). %at a significance of $a=0.05$.
The lower end of our $M_\mathrm{dust}/M_\ast$ distribution matches with that of low-redshift dusty early-type galaxies \citep[ETGs;][]{agius13}, while still higher than that of dust-poor early-type galaxies \citep{rowlands12, rowlands15}.

The number excess at the lower end of the $M_\mathrm{dust}/M_\ast$ distribution can actually be a signature of evolved systems with lower gas fraction and dust destruction in the post-starburst stage \citep[e.g.,][]{rowlands15, li19}. 
\textred{Applying a \citet{scoville16} conversion from luminosity at a rest frame of 850\,\micron\ ($L_{\nu,850}$) to the molecular gas mass ($M_\mathrm{gas}$), we find a low gas fraction ($f_\mathrm{gas}=M_\mathrm{gas}/(M_\mathrm{gas}+M_\mathrm{star})$) of $0.17\pm0.08$ for four sources at 
$M_\mathrm{dust}/M_\ast<10^{-3}$.}
% With the help of cluster-lensing magnification, our survey was probably able to detect dusty starbursts undergoing the period of ceasing star formation and thus dust destruction and expulsion.
We further discuss this issue of dust-to-mass ratio and its implication in Section~\ref{sss:05c_iievolution}.
% This scenario can also be testified through spatially resolved study in following sections.

Assuming a gas-to-dust ratio of 100, we calculate the gas depletion time scale ($\tau_\mathrm{dep} = M_\mathrm{gas}/\mathrm{SFR}$) for our sample, and the median value with 1$\sigma$ dispersion is $226_{-73}^{+196}$\,Myr. 
This is consistent with the median value of AS2UDS sample ($\sim 320$\,Myr) recomputed with similar method \citep{dudzevic19}.

%% file: 04e_size.tex
\begin{figure}[!t]
\centering
\includegraphics[width=\linewidth]{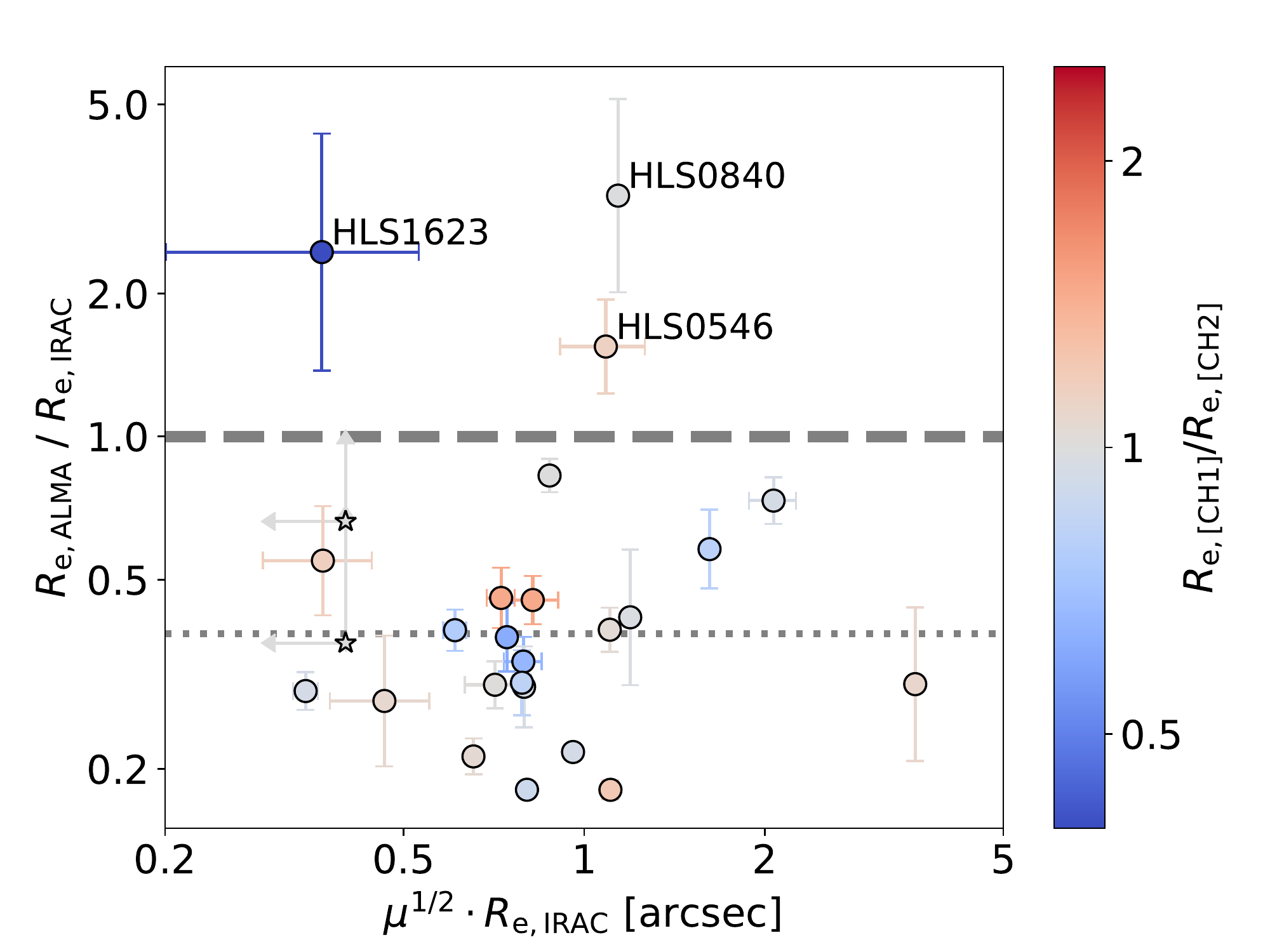}
\caption{Effective radius ($R_\mathrm{e}$) ratio between ALMA and IRAC counterparts, versus $R_\mathrm{e}$ measured in the IRAC bands (unit: arcsec; before lensing correction).
The circles present sources modeled with S\'ersic profiles in the IRAC bands, while stars present the two sources modeled with PSF profiles. % (HLS0043-B and HLS0307-28-B).
The symbols are color-coded with their $R_\mathrm{e}$ ratio between 3.6 and 4.5\,\micron, usually close to 1.
The grey dashed line indicates the case in which the two effective radii are identical, while the dotted line shows the median value of $R_\mathrm{e}$ ratios (0.38).
% A histogram of source size ratio ($R_\mathrm{e,ALMA}/R_\mathrm{e,IRAC}$) is also shown in the right panel.
}
\label{fig:07_size}
\end{figure}

SMGs are generally found to host dust continua that are more compact than the stellar ones \citep[e.g.,][]{hodge16, hodge19, lang19, gullberg19}.
This can be interpreted as an evolutionary connection from SMGs to cQs at slightly lower redshift: 
after star formation ceases in a $\sim$1\,kpc-scale region at the galaxy center, a cusp of stellar component will remain in this region, which can be observed to be compact and quiescent \citep[e.g.,][]{toft14,simpson15,barro16,lang19}.
% In this part, we dedicatedly study the difference of structural profile between dust and stellar components.

We adopt the half-light radii measured at 1.3\,mm ALMA \textit{uv}-plane ($R_\mathrm{e,ALMA}$) as the effective radius of dust continuum (noted as $R_\mathrm{e,dust}$) and thus the star-forming region.
We note that this ignores any radial gradients of dust temperature or opacity that could alter the measured size of dust continuum.
Based on the FIRE-2 simulation, \citet{cochrane19} predicted that the effective radius of dust emission goes up with the observed wavelength in the submillimeter/millimeter.
This is because observations at longer wavelengths are more sensitive to the cooler gas and dust components in the outer region. % but this effect is expected to be small with our sample.
With an effective rest-frame wavelength of $440_{-95}^{+137}\,$\micron\ for our sample, the measured $R_\mathrm{e,dust}$ is not expect to vary by more than 10\% due to the difference in the sampled wavelength. 
% Note that we do not find any obvious dependence of $R_\mathrm{e,ALMA}$ on redshift or dust temperature in our sample.

% Figure~\ref{fig:07_size} displays the ratio of $R_\mathrm{e}$ measured in the ALMA and IRAC bands.
% The median value of $R_\mathrm{e,ALMA}/R_\mathrm{e,IRAC}$ is observed as $0.38_{-0.10}^{+0.35}$.
% Except for HLS1623, HLS0546 and HLS0840, 23 out of the 26 sources exhibit compact ALMA dust continua  relative to their stellar components.
With the IRAC data, we define the effective radius as the geometric mean of the effective radii measured at 3.6/4.5\,\micron\ ($R_\mathrm{e,IRAC} = R_\mathrm{e,[CH1]}^{1/2} \cdot R_\mathrm{e,[CH2]}^{1/2}$).
At $z_\mathrm{med}=1.9$, IRAC Channel 1/2 samples the rest-frame $J/H$ bands with a similar angular resolution.
Adopting the \citet{calzetti00} extinction law, dust extinction in the rest-frame $J/H$ bands ($A_J$ and $A_H$) will only be 0.3/0.2 of $A_V$.
For a median $A_{V,\mathrm{med}} = 2.9$ in this study, the dust extinction in the IRAC 3.6/4.5\micron\ bands is 0.9/0.6, which is only $\sim$8\%\ of the extinction seen in \hst\ $J_{125}/H_{160}$ bands for $z\simeq 2$ SMGs in \citet{lang19}.
Therefore, the intrinsic stellar distribution should be close to the light profile in the IRAC bands, without significant overestimate of the effective radius of stellar mass distribution due to the concentration of dust extinction at the galaxy center.
We measure the effective radius ratio between 3.6 and 4.5\,\micron\ as $0.99^{+0.14}_{-0.19}$ for our sample, and such a consistency between the radii seen in the two bands also suggests the weak influence of dust extinction on light profiles. 
Therefore, we directly adopt the $R_\mathrm{e,IRAC}$ as a representation of the effective radius of stellar mass (noted as $R_\mathrm{e,star}$).

One caveat is that the measured $A_V$ does not account 
for the stars which are fully dust-obscured, and thus the $A_V$ derived from an energy-balanced SED fitting code does not necessarily represent the extinction to the full stellar mass component \citep[e.g.,][]{casey14b}.
As pointed out in \citet{lang19}, it is possible that a compact and obscured stellar component remains undetected at the ALMA/IRAC continuum centroid \citep[e.g.,][]{simpson17}, resulting in an even more compact configuration of stellar mass than the IRAC light profile.

The comparison of the effective radii of dust and stellar continua is shown in Figure~\ref{fig:07_size}.
Except for three sources (HLS1623, HLS0546 and HLS0840), 23 out of the 26 sources have more compact ALMA dust continua relative to their stellar components.
The median dust-to-stellar continuum size ratio ($R_\mathrm{e,dust}/R_\mathrm{e,star}$) is 0.38, with a 1$\sigma$ distribution from 0.28 to 0.73. 
\citet{lang19} presented a $R_\mathrm{e,870\micron} / R_\mathrm{e,star}$ of $0.6\pm0.2$, similar to our measurement.
This is much smaller than the dust-to-stellar size ratio of local spiral galaxies ($\sim 1.0$, e.g., \citealt{hunt15}).

%% file: 04f_sbfir.tex
\begin{figure}[!t]
\centering
\includegraphics[width=\linewidth]{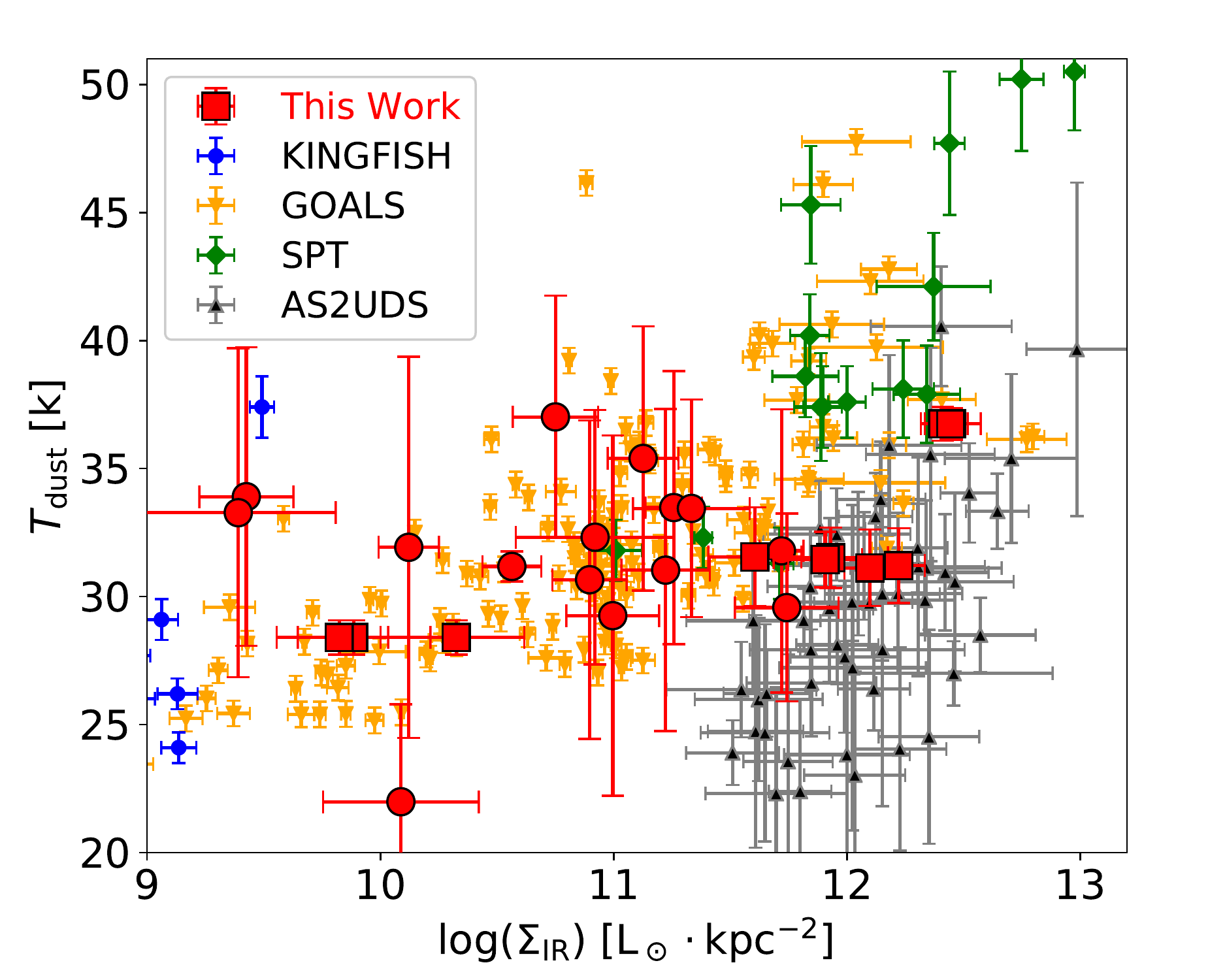}
\caption{Dust temperature versus IR surface luminosity of SMGs in this work (red squares for sources with confirmed $z_\mathrm{spec}$ and circles for $z_\mathrm{phot}$-only ones).
We also compare our data with galaxy sample from GOALS (local LIRG/ULIRGs in orange; \citealt{ds17}), AS2UDS ($z_\mathrm{med}=2.7$ SMGs in black; \citealt{dudzevic19,gullberg19}),
SPT ($z_\mathrm{med}=4.3$ SMGs in green; \citealt{spilker16,strandet16}) 
and KINGFISH (nearby galaxies in blue; \citealt{skibba11,hunt15}).
}
\label{fig:06_tsb}
\end{figure}

Since all of the 26 sources (except for heavily-blended galaxy-lensed cases) are resolved in the ALMA 1.3\,mm maps, we are able to obtain the surface luminosity of their dust continua.
This should also be a conserved quantity independent of lensing magnification.
We hereby define the IR surface luminosity as $\Sigma_\mathrm{IR} = L_\mathrm{IR}/(2\pi R_\mathrm{e,ALMA}^2)$, and thus $\Sigma_\mathrm{IR}$ is the average IR surface luminosity within a radius of $R_\mathrm{e,ALMA}$.
Note that at a rest-frame wavelength of $\sim$440\,\micron, the $R_\mathrm{e,ALMA}$ is expected to trace the size of cold dust component and star-forming region better than the far-IR emission \citep[e.g., 70\,\micron\ continuum size used in][]{ds17}.
Due to the effect of radial $T_\mathrm{dust}$ gradient or dust optical depth, the $R_\mathrm{e}$ at the wavelength of FIR SED peak would be smaller than the $R_\mathrm{e,ALMA}$ \citep[e.g.,][]{cochrane19}, and thus the $\Sigma_\mathrm{IR}$ might be underestimated. 
The conversion factor between IR and 1.3\,mm surface brightness ($\Sigma_\mathrm{IR} / \Sigma_\mathrm{1.3mm}$) is $10^{10.65\pm0.20}$\,\si{mJy.arcsec^{-2}.L_\odot^{-1}.kpc^{2}} for sources in our sample,
and the factor between surface SFR density and IR luminosity ($\Sigma_\mathrm{SFR}/\Sigma_\mathrm{IR}$) is $10^{-10.17\pm0.07}$\,\si{M_\odot.yr^{-1}.L_\odot^{-1}}.

We plot the dust temperature versus IR surface luminosity in Figure~\ref{fig:06_tsb}, comparing the distribution with those of the GOALS (local LIRG/ULIRGs; \citealt{ds17}), AS2UDS ($z_\mathrm{med}\simeq 2.7$ SMGs; \citealt{gullberg19,dudzevic19}), SPT ($z_\mathrm{med}\simeq 4.3$ SMGs; \citealt{spilker16,strandet16}) and KINGFISH (nearby galaxies; \citealt{skibba11,hunt15}) galaxies.
Galaxies in this work exhibit a wide 3\,dex range of $\Sigma_\mathrm{IR}$ from $10^{9.4}$ to $10^{12.4}$\,\si{L_\odot \cdot kpc^{-2}}.
Such a range of $\Sigma_\mathrm{IR}$ does resemble that of the local (U)LIRGs in the GOALS sample.

We find that the dust temperature of SMGs in this work is barely correlated with the IR surface luminosity.
At $\Sigma_\mathrm{IR} < 10^{11.5}$\,\si{L_\odot \cdot kpc^{-2}}, the lensed SMGs exhibits comparable dust temperature as local LIRGs \citep[][also assumed $\beta=1.8$]{ds17}. 
This indicates that the $\Sigma_\mathrm{IR} - T_\mathrm{dust}$ relation for LIRGs may not evolve with redshift up to $z\sim 2$.
\citet{symeonidis13} analyzed \herschel-selected LIRGs at $z\lesssim 1$ and suggested no significant evolution of dust temperature at a constant IR luminosity ($L_\mathrm{IR} \lesssim 10^{11.5}\,$\lsun) across $z=0\sim 1$.  
However, ULIRGs ($L_\mathrm{IR} \geq 10^{12}\,$\lsun) at $z\gtrsim 1$ are much cooler than their local analogs or more optically thick.
% \citet{dudzevic19} analyzed 707 SMGs and reported no evolution in dust temperature at a constant IR luminosity across $z=1.5\sim 4$.
Assuming that $\Sigma_\mathrm{IR}$ is well correlated with the intrinsic $L_\mathrm{IR}$ \citep[reported at various redshifts, e.g.,][]{wiphu11,lutz16,fujimoto17}, our result is consistent with \citet{symeonidis13} but extending out to $z\sim 2$.

At $\Sigma_\mathrm{IR}> 10^{11.5}$\,\si{L_\odot \cdot kpc^{-2}}, the lensed SMGs in our sample seem to show lower dust temperature than both local ULIRGs and $z\sim 4$ SPT sources (biased towards galaxy-lensed cases), but consistent with $z_\mathrm{med}=2.7$ SMGs in AS2UDS sample ($\Sigma_\mathrm{IR}$ based on 870\,\micron\ continuum size; \citealt{gullberg19,dudzevic19}).
However, such a comparison is limited to the sample size since only two independent sources in our sample (HLS0553 and HLS2332) are at $\Sigma_\mathrm{IR}> 10^{12}$\,\si{L_\odot \cdot kpc^{-2}}.

%where most of the galaxies in our sample are spectroscopically confirmed, the dust temperature of SMGs in this work is well correlated with the FIR surface luminosity, which is also seen with the GOALS and SPT galaxies in completely different redshift ranges.
% This indicates that the $\Sigma_\mathrm{IR} - T_\mathrm{dust}$ relation for the most intense starburst galaxies barely evolves with redshift.

% However, such a trend cannot be confirmed for low-$\Sigma_\mathrm{IR}$ sources.
% We find that sources with low IR surface luminosity generally exhibit 
% Three fairly extended sources, namely MACSJ0840.$_0$, MACSJ1314.$_\Sigma$ and MACSJ2155.$_1$, shows $\Sigma_\mathrm{IR}<10^{10.5}$\,\si{L_\odot \cdot kpc^{-2}} but peculiarly high dust temperature of $T_\mathrm{dust}>30$\,\si{K}.
% Although these sources are offset from the general trend of the $\Sigma_\mathrm{IR} - T_\mathrm{dust}$ relation, 
% local analogs of these low-$\Sigma_\mathrm{IR}$ and high-$T_\mathrm{dust}$ galaxies do exist in KINGFISH sample, such as the barred starburst galaxy NGC\,2146 ($\Sigma_\mathrm{IR}=10^{9.5}$\,\si{L_\odot \cdot kpc^{-2}}, $T_\mathrm{dust}=37$\,\si{K}). %

% The relatively high $T_\mathrm{dust}$ could be interpreted with their lower $A_V$ ($\sim1.9$ compared to $A_{V,\mathrm{med}} = 3.5$ of the whole sample) since their dust grain can be exposed to harder interstellar radiation field than those in a heavier dust extinction cases \citep{rr13}. 
% Such a scenario can be intrinsically related to the low-metallicity ISM picture \citep{madden06} as observed in local dwarf galaxies. 

%% file: 04g_main.tex
\begin{figure*}[!ht]
\centering
\includegraphics[width=0.495\linewidth]{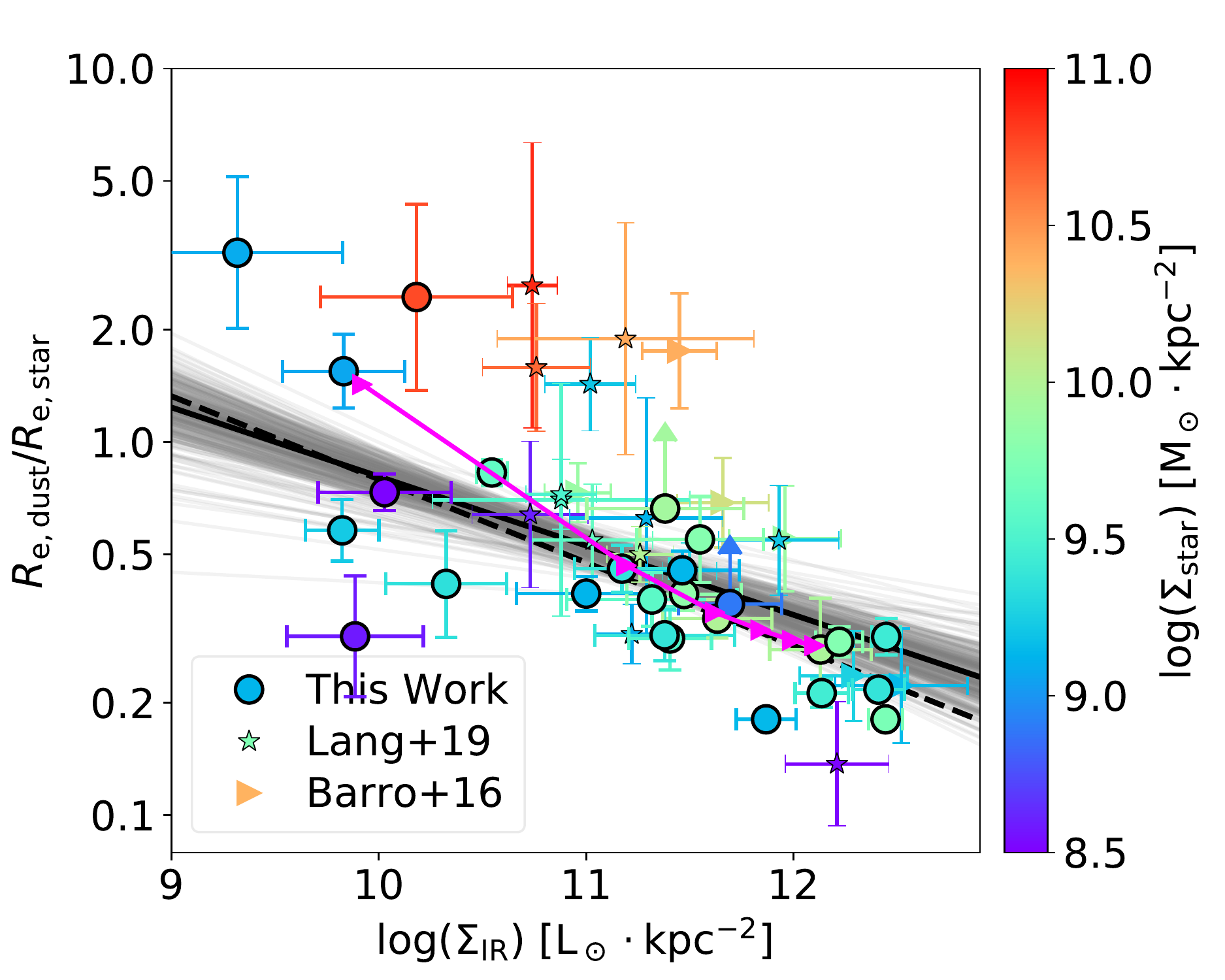}
\includegraphics[width=0.495\linewidth]{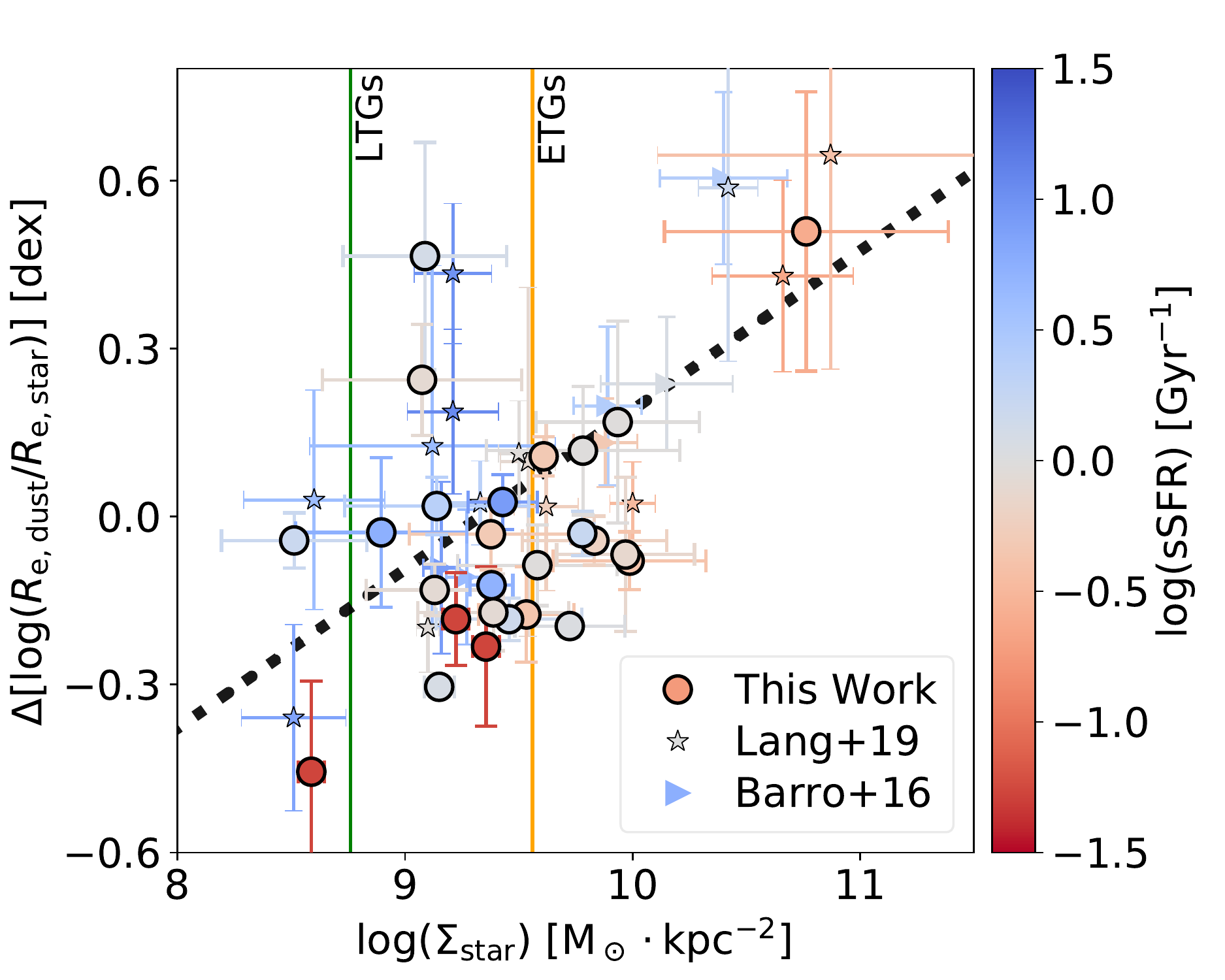}
\caption{\textit{Left}: Effective radius ratio between dust and star, versus FIR surface luminosity density of 26 sources in this work (circles with bold black edges), 14 SMGs from \citet[][stars with slim black edges]{lang19}, 6 cSFGs from \citet[][rightward triangles]{barro16}. 
Symbols are color-coded with their surface stellar-mass densities.
The best-fit linear relation for 26 sources in this work is shown as the black dashed line, and the best-fit relation of the combined sample is shown as the black solid line, with the grey lines indicating the uncertainty. 
The solid magenta line shows the track of the two-component model discussed in Section~\ref{sss:05_two-component1}.
\textit{Right}: The deviation of effective radius ratio from the best-fit $R_\mathrm{e,dust} / R_\mathrm{e,star} - \Sigma_\mathrm{IR}$ relation (i.e., the black solid line in the left panel),
versus surface stellar-mass density $\Sigma_\mathrm{star}$.
The symbols are the same as in the left panel but color-coded with galaxy-wide sSFR obtained through SED fitting.
The best-fit linear regression (the black dotted line) suggests a positive correlation between the two parameters, although the intercept might be a function of sSFR.
Median $\Sigma_\mathrm{star}$ of $M_\mathrm{star}=10^{11}$\,\msun\ early and late-type galaxies (ETGs/LTGs) at $z=1.75$ are plotted as the orange and green solid vertical lines \citep{vdw14}.
}
\label{fig:08_main}
\end{figure*}

We then investigate how the dust-to-stellar continuum size ratio is correlated with other physical quantities.
Similar to \citet{lang19}, we do not find any obvious correlation between the size ratio and sSFR on integrated-galaxy scales. However, we do find that this continuum size ratio is well correlated with FIR surface luminosity, plotted in the left panel of Figure~\ref{fig:08_main}.
We find that with the increase of $\Sigma_\mathrm{IR}$, the dust-to-stellar continuum size ratio decreases slowly.
We fit a linear relation through \textsc{emcee}, and the best-fit relation for our 26 sources is shown as follows:
\begin{equation}
\label{eq:01_hls}
\log(\frac{R_\mathrm{e,dust}}{R_\mathrm{e,star}}) = (-0.23 \pm 0.03) \cdot \log(\frac{\Sigma_\mathrm{IR}}{10^{10}}) - (0.10 \pm 0.04)
\end{equation}
where $\Sigma_\mathrm{IR}$ is in the unit of \si{L_\odot\cdot kpc^{-2}}.

We also incorporate six compact star-forming galaxies (cSFGs) presented in \citet{barro16} and 14 SMGs presented in \citet{lang19} into a combined dataset of 46 sources.
All of these sources are around $z=2.2\pm0.3$ with accurate ALMA image-plane morphology modeling at 870\,\micron.
The $R_\mathrm{e,star}$ of the six cSFGs in \citet{barro16} was measured using \hst/WFC3 F160W images without any correction for dust extinction ($A_{V, \mathrm{SED}}$ is as low as $1.3\sim1.6$).
$R_\mathrm{e,star}$ of 14 SMGs in \citet{lang19} was also based on F160W imaging but corrected with a pixel-to-pixel $A_V$ map.
A similar MCMC fitting to these 46 sources shows the following $R_\mathrm{e,dust}/R_\mathrm{e,star} - \Sigma_\mathrm{IR}$ relation:
\begin{equation}
\label{eq:02_all}
\log(\frac{R_\mathrm{e,dust}}{R_\mathrm{e,star}}) = (-0.19 \pm 0.04) \cdot \log(\frac{\Sigma_\mathrm{IR}}{10^{10}}) - (0.09 \pm 0.04)
\end{equation}
which is consistent with the fitting of 26 HLS sources within $1\sigma$.
Such a linear relation is relatively tight over the 3.1\,dex range of \sbfir\ distribution.
We measure a standard deviation of the 46 sources from the best-fit correlation as 0.21\,dex, and the typical uncertainty of effective radius ratio is $0.11\pm0.06$\,dex.

We also fit the data with least-squares method and bootstrapping. 
The derived relation and its uncertainty are consistent with those by MCMC (slope is $-0.27 \pm 0.04$).
If $L_\mathrm{IR}$ and $R_\mathrm{e,star}$ are (\romannumeral1) random variables with a narrow distribution and (\romannumeral2) independent of $R_\mathrm{e,dust}$ distributed over a wide range, then one should expect to derive a linear relation between $\log(R_\mathrm{e,dust}/R_\mathrm{e,star})$ and $\log(\Sigma_\mathrm{IR})$ with a slope of $-0.5$.
We also generate mock dataset of $L_\mathrm{IR}$, $R_\mathrm{e,dust}$ and $R_\mathrm{e,star}$ that satisfies our selection criteria (i.e., above certain thresholds of $L_\mathrm{IR}$ and $\Sigma_\mathrm{IR}$) and matches our measurements with respect to the variance, and the resultant slope is $-0.46$. 
However, our fittings suggest that such a slope can be ruled out at a confidence of $>5\sigma$, indicating that the observed relation is physical and not a direct consequence of the relatively wider distribution range of $R_\mathrm{e,dust}$.

% 
% We further the discussion on this $R_\mathrm{e,dust} / R_\mathrm{e,star} - \Sigma_\mathrm{IR}$ diagram in Section~\ref{ss:05a_ssfr}, and we find the inside-out quenching mechanism can well explain the derived empirical relation.

%% file: 04h_sigstar.tex
One of the remarkable properties of $z\sim2$ compact quiescent galaxies (cQs) is their inferred ultra-high stellar surface density ($\gtrsim10^{10}$\,\si{M_\odot \cdot kpc^{-2}} within effective radius; e.g., \citealt{vandokkum08,vdw14}). 
Such a compact configuration of the stellar component indicates a previous star-formation history at a similarly compact scale.
% or forms much earlier when galaxy size is relativily small \citep[e.g.,][]{williams17}
This could be achieved through a nuclear starburst, since galaxy interaction/merger and disk instability can cause gas inflows and thus trigger the central stellar density enhancement \citep[e.g.,][]{hopkins08}.
% In this subsection, we study the stellar surface density (\sbstar) of our SMG sample.

We define \sbstar\ as the average stellar mass within $R_\mathrm{e,star}$, namely $\Sigma_\mathrm{star} = M_\ast/(2\pi R_\mathrm{e,star}^2)$, similar to the definition of \sbfir.
The \sbstar\ of our SMG sample ranges between $10^{8.6}$ and $10^{11.0}$, with a median value of $10^{9.4\pm 0.1}$ (unit: \si{M_\odot.kpc^{-2}}), similar to that seen for SMGs by \citet{hodge16}. 
Just like \sbfir, \sbstar\ is a conserved quantity with respect to gravitational lensing.

%%% stop here 

We first compare the \sbstar\ of our sample with the general galaxy population in similar redshift and mass ranges.
\citet{vdw14} showed that for early- and late-type galaxies at $z\sim1.75$ with a stellar mass of $10^{11}$\,\msun, which is close to that of our SMG sample, the median \sbstar\ should be $10^{9.6\pm0.3}$ and $10^{8.8\pm0.4}$ \si{M_\odot.kpc^{-2}}, respectively.
Therefore, the majority of galaxies in our sample do match the stellar surface density of early-type galaxies rather than the late-type, although a test on the $M_\ast - R_\mathrm{e}$ plane \citep[e.g.,][]{hodge16} cannot be performed due to the incompleteness of lensing magnification.

We then color-code the $R_\mathrm{e,dust} / R_\mathrm{e,star} - \Sigma_\mathrm{IR}$ plot with \sbstar\ as shown in the left panel of Figure~\ref{fig:08_main}.
Although a tight linear relation can be found between the two quantities, we notice that the deviation of continuum size ratios from the best-fit relation %(i.e., the vertical distance to the black solid line) 
seems to be correlated with \sbstar.
We then plot this residual of $R_\mathrm{e,dust} / R_\mathrm{e,star}$ versus \sbstar\ in the right panel of Figure~\ref{fig:08_main}.
An MCMC linear fitting under the assumption of equal weighting 
of all data points suggests a positive relation between the two quantities as:
\begin{equation}
\label{eq:03_residual}
\Delta[\log(\frac{R_\mathrm{e,dust}}{R_\mathrm{e,star}})] = (0.24 \pm 0.07) \cdot \log(\frac{\Sigma_\mathrm{star}}{10^{10}}) + (0.17 \pm 0.05)
\end{equation}
where $\Sigma_\mathrm{star}$ is in the unit of \si{M_\odot.kpc^{-2}}.

This indicates that the relation shown as Equation~\ref{eq:02_all} holds for the majority of $z\sim2$ SMGs with a stellar mass surface density of roughly $10^{9} \sim 10^{10}$\,\si{M_\odot.kpc^{-2}}. 
With the enhancement of the central stellar mass surface density above $\sim10^{10}$\,\si{M_\odot.kpc^{-2}}, SMGs will show larger $R_\mathrm{e,dust} / R_\mathrm{e,star}$ ratio than the regular relation.
This may reflect not only a newly-formed cuspy stellar profile in the galaxy center, but also potentially the quenching of concentrated star formation and dissipation of dust remnant through multiple physical process (e.g., stellar or SMBH feedback), which we discuss further in Section~\ref{sss:05b_iiiquench}.
% may indicate a compact star-forming progenitor, or previous nuclear starburst driven by interaction/merger \citep{hopkins08}.  

We also note that the vertical scatter in the right panel of Figure~\ref{fig:08_main} is related with the sSFR at the galaxy-integrated scale, as color-coded in the diagram. 
At a given \sbstar, the increase of galaxy-wide sSFR will lead to the growth of $R_\mathrm{e,dust}/R_\mathrm{e,star}$, which can be a trivial result of the fact that since sSFR is proportional to FIR luminosity and thus $R_\mathrm{e,dust}$ when \sbfir\ conserves.
% We continue the discussion of $\Sigma_\mathrm{star} - \mathrm{sSFR}$ in Section~[\#ref].

%% file: 05_discussion.tex
% In this section, we discuss the underlying physical mechanism that drives the observable shown in Section~\ref{sec:04_res}.
% This include the comparison of sSFR at galaxy-wide with their center, and its implication to an inside-out quenching scenario with necessary comparison with theoretical predictions.
% We also study the offset between the dust and stellar component to investigate the trigger mechanism of $z\sim2$ SMGs.
% The evolutionary connection between SMGs and cQs is then further discussed, and we also testify how the effect of differential lensing will affect our presented results.

% Finally, we explain other possible physical pictures accounting for the observed tight relation between $R_\mathrm{e,dust}/R_\mathrm{e,star}$ and $\Sigma_\mathrm{FIR}$.

\subsection{Discovery of spatially extended SMGs with low surface brightness}
\label{sss:05a_ext}
\input{05a_extended}

\subsection{Central starburst vs.\ galaxy-wide star formation}
\label{sss:05_two-component1}

\input{05_two-component1}
\subsection{Structural evolution of SMGs}
%\subsection{Inside-out quenching of SMGs}
\subsubsection{Main sequence offset}
\label{sss:05b_idms}
\input{05b_1sSFR}

\subsubsection{MS offset versus dust-to-stellar size ratio}
%\subsubsection{Indications of inside-out quenching}
\label{sss:05b_iidms_xx}
\input{05b_2model}

\subsubsection{Further test of the evolutionary sequence}
\label{sss:05b_ivqorb}
\input{05b_3qorb}

% \subsubsection{The effect of differential magnification}
% \label{sss:05a_iiidifflens}
% \input{05b_3difflens}

%%% general evolution picture, from triggering to quenching
\subsection{The evolutionary picture of SMGs}

\subsubsection{Trigger mechanism of SMGs}
\label{sss:05c_itrigger}

\input{05c_1trigger}

\subsubsection{The connection between SMGs and cQs}
\label{sss:05c_iievolution}
\input{05c_2evolution}

%% file: 05a_extended.tex
One major discovery of this work is the existence of spatially extended SMGs with a low IR surface luminosity ($\Sigma_\mathrm{IR}< 10^{10}$\,\si{L_\odot\cdot kpc^{-2}}), e.g., HLS0546 and HLS0840.
These galaxies exhibit extended dust continua compared with the stellar ones, and their FIR surface luminosities are $\gtrsim$1\,dex lower than those of typical SMGs at $z=1\sim3$ ($\Sigma_\mathrm{IR} \gtrsim 10^{11}$\,\si{L_\odot\cdot kpc^{-2}}; e.g., \citealt{hodge16,gullberg19} and this work).

A galaxy with a similar $\Sigma_\mathrm{IR}$ has been reported by the ALMA Frontier Fields Survey (e.g., A2744-ID05, $\Sigma_\mathrm{IR} = 10^{9.9}$\,\si{L_\odot\cdot kpc^{-2}}; \citealt{gl17,laporte17}). 
However, the dust continuum of this galaxy is relatively compact ($R_\mathrm{e,dust} = 1.3$\,kpc) even compared with the size of stellar continuum measured with \hst/WFC3-IR ($R_\mathrm{e,star}=4.0$\,kpc).
Such a small dust-to-stellar size ratio is different from the extended nature of the SMGs reported here.
HLS0546 and HLS0840 are also different from the spatially extended SMG at $z=2.8$, SMM\,J02399-0136 ($\Sigma_\mathrm{IR} = 10^{10.2}$\,\si{L_\odot\cdot kpc^{-2}}; \citealt{genzel03,ivison10}) because of even lower $\Sigma_\mathrm{IR}$ and a lack of merger feature for our sample.
Most recently, \citet{tadaki20} reported the discoveries of \hst-selected massive ($M_\mathrm{star}>10^{11}$\,\msun) star-forming galaxies at $z \sim 2$ with $R_\mathrm{e,dust}\simeq 5$\,kpc.
The dust-to-stellar size ratio and IR surface luminosity of these sources are generally consistent with those of spatially extended SMGs reported here.
% Therefore, this galaxy is more likely a 

It should be noted that currently spatially extended $z\sim 2$ SMGs with a low IR surface luminosity density ($\Sigma_\mathrm{IR} \simeq 10^{9.5}$\,\si{L_\odot\cdot kpc^{-2}}) can only be detected and resolved through ALMA observations with deep integration (e.g., pointed observations such as \citealt{tadaki20}; blank-field surveys such as 1.2\,mm ASPECS, \citealt{gl20,aravena20}) or of lensing-cluster fields (this work or ALMA Cycle 6 large program ALCS, Kohno et al., in prep.).
Previous ALMA Band-6 surveys in cosmological deep fields (e.g., HUDF, \citealt{dunlop17}; GOOD-S, \citealt{franco18,hatsukade18}) can only reach a 4$\sigma$ depth of $\gtrsim 0.24$\,\si{mJy\,arcsec^{-2}} at 1.3\,mm, which is not sufficient to detect these low-surface-brightness galaxies ($\lesssim 0.1$\,\si{mJy\,arcsec^{-2}}).
Without lensing, these extended SMGs would remain barely resolved with a $\sim$1\arcsec\ beam, and therefore further \textit{uv}-tapering of the data would hardly improve the detectability of these galaxies. 
Galaxy-lensing usually results in difficulty of source-plane reconstruction, especially in the rest-frame optical due to the existence of a bright foreground lensing galaxy, introducing large uncertainties into multi-wavelength comparison at spatially-resolved scales.
In contrast, cluster-lensed SMGs, often with a magnification factor of $\gtrsim 5$ and reduced contamination from foreground objects, are significantly stretched spatially ($R_\mathrm{e}\gtrsim 1$\arcsec).
\textred{Therefore, the detectability of cluster-lensed intrinsically extended SMGs can be substantially improved by \textit{uv}-tapering (as shown in Figure~\ref{fig:3_cutout}) or high-sensitivity facilities with a larger beam size and recoverable angular scale.}

%% file: 05_two-component1.tex
Qualitatively, one simple way to explain the observed correlation between $R_\mathrm{e,dust} / R_\mathrm{e,star}$ and $\Sigma_\mathrm{FIR}$ seen in Figures~\ref{fig:08_main} (the left panel) is to assume a model in which a compact dust component ($R_\mathrm{e,dust}\sim 1$\,kpc) with a varying surface luminosity ($\Sigma_\mathrm{IR\ }$\,$\simeq$\, $0-10^{12.5}$\,\si{L_\odot.kpc^{-2}}) is superposed on an extended component ($R_\mathrm{e,dust}\sim 5$\,kpc, $\Sigma_\mathrm{IR}\sim 10^{9.5}$\,\si{L_\odot.kpc^{-2}}).  
In this model, the former corresponds to central starburst while the latter corresponds to galaxy-wide star formation (e.g., a star-forming disk).  
Such a two-component model was also suggested by \citet{gullberg19} based on the morphological evidence from an SMG stacking analysis.  
In this model, as the compact component (i.e., the central starburst) increases its brightness, $\Sigma_\mathrm{IR}$ increases while $R_\mathrm{e,dust}/R_\mathrm{e,star}$ decreases, and the trend will reverse when the compact component fades.

In the left panel of Figure~\ref{fig:08_main}, we overplot the behavior of this two-component model (the solid magenta line), using a set of representative values for various parameters.
More specifically, the effective radii of the compact and extended dust components were fixed to 1 and 5 kpc.  
The infrared luminosity of the extended component was fixed to $10^{12.1}$\,L$_{\sun}$ while that of the compact component was varied between 0 and $10^{12.9}$\,L$_{\sun}$ to mimic the rise/decline of the central starburst.  
$R_{\rm e,dust}$ was then measured for the combined source.  
% With this set of parameters, the resultant 1.3 mm surface brightness will range from 0.2 to 30 mJy arcsec$^{-2}$, matching our data as well as that of \citet{gullberg19}. 
For the stellar component, $R_{\rm e,star}$ and $M_{*}$ were assumed to be 3.5 kpc and $10^{11}$\,M$_{\sun}$.
% Note that in this figure, the magnification factors cancel out.
As Figure~\ref{fig:08_main} shows, this simple two-component model reproduces the observed trend well.  

%% file: 05b_1sSFR.tex
% \begin{figure*}[!ht]
% \centering
% \includegraphics[width=0.9\linewidth]{figures/sSFRratio_vs_sSFR_z_cc.pdf}
% \caption{The ratio between the total and central sSFR of SMGs ($R_\mathrm{e,dust}^2 / R_\mathrm{e,star}^2$ by our definition; see Section~\ref{sss:05a_issfr} for justification), versus the galaxy-wide total sSFR (left) and the central sSFR (right). 
% 26 sources in this work are shown as filled circles, and 20 literature samples \citep{lang19,barro16} are shown as stars and triangles, respectively. 
% All SMGs are color-coded with their redshifts. In the right panel, we also fit the trends of SMGs in two redshift bins, namely $z>2$ bin (26 sources at $z_\mathrm{med}=2.31$, solid red line) and $z<2$ bin (20 sources at $z_\mathrm{med}=1.60$, solid line in light coral). }
% \label{fig:09_ssfr}
% \end{figure*}

% {\color{blue} \sout{The observed correlation between $R_\mathrm{e,dust} / R_\mathrm{e,star}$ and $\Sigma_\mathrm{FIR}$ in SMGs (Section~\ref{ss:04g_main}) indicates that the strength of central star-forming activity may be related to the dust and stellar continuum size ratio at the galaxy scale.
% To examine whether this relation can be produced by an evolutionary effect, }}
To evaluate the star-forming properties of the observed SMGs further,
we calculate the offset between the observed SFR and that of the expected on the star-forming main sequence, which is a function of redshift and stellar mass \citep[e.g.,][]{speagle14,schreiber15}.
This quantity is commonly referred as the main-sequence offset ($\Delta\mathrm{MS} = \log[\mathrm{SFR} / \mathrm{SFR}_\mathrm{MS} (M_\mathrm{star}, z)]$), and here the SFR on the main sequence ($\mathrm{SFR}_\mathrm{MS}$) is calculated using the formula given by \citet{speagle14}.
Following \citet{aravena20}, we also define the boundaries of the main sequence as $\Delta\mathrm{MS}=\pm0.4$, and classify sources above/below the main sequence as starburst/passive galaxies.

The accurate derivation of $\Delta\mathrm{MS}$ requires the knowledge of intrinsic SFR and stellar mass and thus the magnification factor ($\mu$). 
Based on published cluster mass models of MACS1115 and MACS0553 \citep{oguri10,zitrin15,ebeling17}, we derive a median magnification factor of $\sim 5$ for HLS1115 and HLS0553-A/B/C.
% Similar magnification can also be estimated from a comparison of median lensed $R_\mathrm{e,star}$ and $M_\mathrm{star}$ with mass--radius relation of early-type galaxies at similar redshift \citep{vdw14} \ref{ss:04h_svigstar}. 
Therefore, we assume a lensing magnification of $\mu = 5$ for all the 26 sources uniformly with morphological measurements.
We show that for a $z\sim 2$ lensed SMGs with an intrinsic stellar mass of $10^{11}$\,\msun, an uncertainty of $0.5$\,dex (i.e., a factor of $\sim 3$) with a lensing magnification factor would lead to an error of only 0.12\,dex with $\Delta\mathrm{MS}$.
This is smaller than the uncertainty of observed, lensing-boosted $M_\mathrm{star}$ and SFR.
Therefore, $\Delta\mathrm{MS}$ is a relatively robust quantity against the uncertainty of magnification factor.

\begin{figure}[!tb]
\centering
\includegraphics[width=\linewidth]{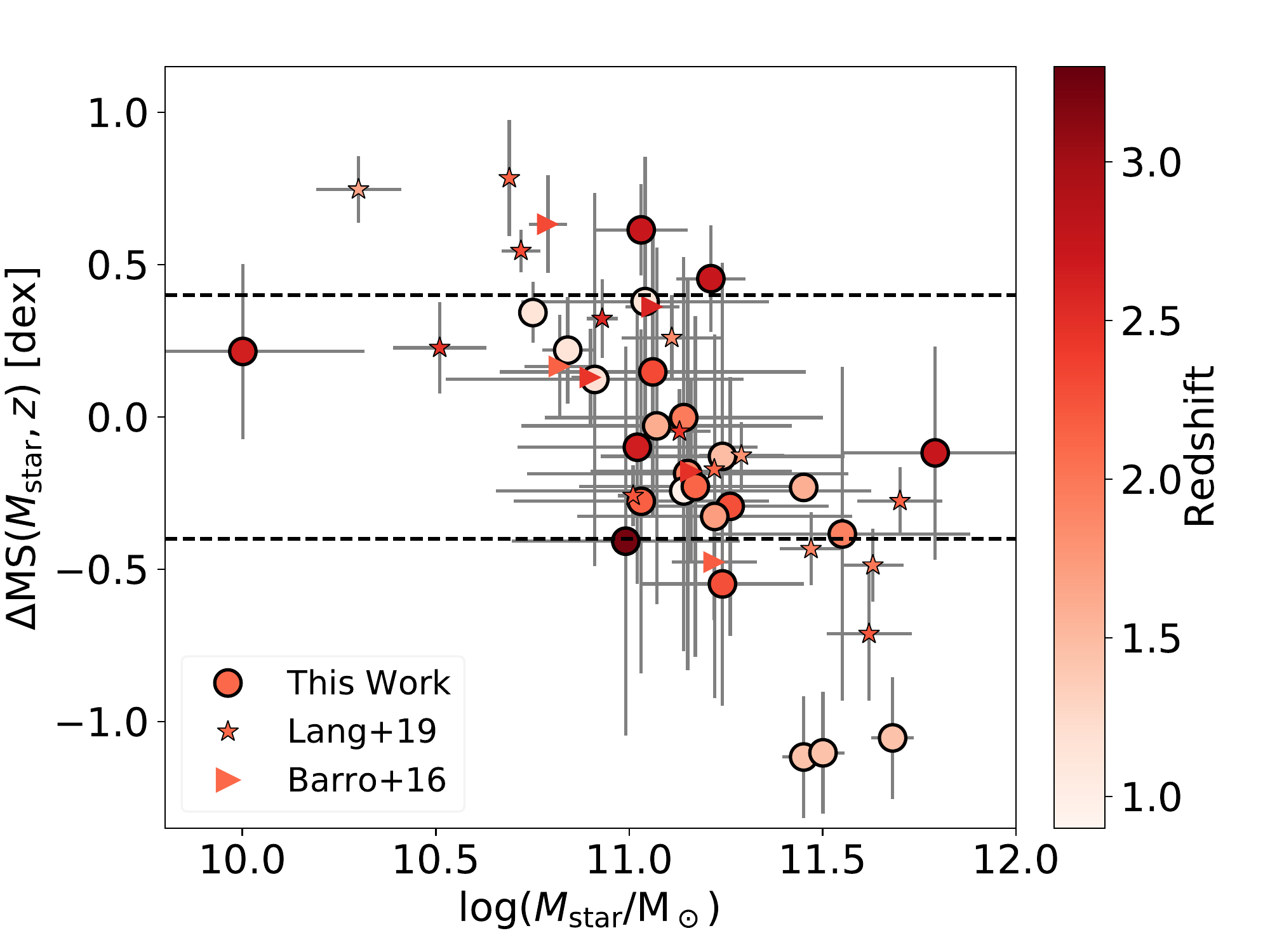}
\caption{Main sequence offset ($\Delta\mathrm{MS}$, defined as $ \log[\mathrm{SFR}/\mathrm{SFR}_\mathrm{MS}(M_\mathrm{star}, z)]$) versus stellar mass ($M_\mathrm{star}$) of the joint 46-source sample. 
The symbols are the same as Figure~\ref{fig:08_main} though color-coded with their redshifts.
We assumed a uniform magnification factor of $\mu=5$ for 26 sources presented in this work.
$\Delta\mathrm{MS}=\pm 0.4$ are specially noted with dashed lines which differentiate the regions for passive, main-sequence and starburst galaxies.
}
\label{fig:dms_ms}
\end{figure}

Figure~\ref{fig:dms_ms} displays the distribution of main-sequence offsets and stellar masses for the joint 46-source sample. 
A general agreement between the lensed and unlensed sample is clear, justifying the use of $\mu=5$ as a representative magnification factor for the lensed sample.
One may notice an anti-correlation between the $\Delta\mathrm{MS}$ and $M_\mathrm{star}$. 
However, this could be the consequence of a selection bias against relatively low-mass ($M_\mathrm{star} \lesssim 10^{10.8}\,$\msun) galaxies without starburst.

Among the full sample, six (nine) sources can be classified as starburst (passive) galaxies. 
The remaining 31 sources are therefore galaxies on the main sequence.
K-S tests suggest no significant difference among the redshift distribution of the three subsamples.
% This is also the same for the MS and passive sample.
% However, the redshifts of starburst sample could be higher than those of passive sample, since the $p$-value for the same redshift distribution is only 0.14.

% Such an anti-correlation between the $\mathrm{sSFR}_\mathrm{t}/\mathrm{sSFR}_\mathrm{c}$ and $\mathrm{sSFR}_\mathrm{c}$ %at various redshifts 
% implies that the galaxy-wide $\mathrm{sSFR}_\mathrm{t}$ has a delayed response to the sharp change of central star-forming activities. 
% If a $z\sim 2$ galaxy is in its terminating phase of star formation, then the rapid decline of its central sSFR, for example due to AGN feedback, will not lead to a simultaneous and dramatic decrease of sSFR over the whole galaxy.
% This is also true for the bursting phase, since the central starburst is known to emerge at a relatively compact scale. % ($\sim1-2$\,kpc).
% Such a scenario is commonly referred to as ``inside-out'' quenching, supported by both theoretical \citep[e.g.,][]{tacchella2016} and observational \citep[e.g.,][]{tacchella2015, spilker19} evidence.
% We then further compare this observed trend with theoretical prediction in the following Section~\ref{ss:05b_quench}.

\textred{
Figure~\ref{fig:10_quench} (the top panel) shows the relation between the IR surface brightness and main-sequence offset.
There is a positive correlation between these two quantities, and an MCMC fitting to the joint 46-source sample suggests $\log(\Sigma_\mathrm{IR}) = (1.10\pm0.19)\Delta\mathrm{MS} + (11.42 \pm 0.10)$, although the dispersion is considerable (0.62\,dex) for the measured IR surface luminosity.
Such a large dispersion was also seen in \citet{elbaz18}, indicating that central starbursts may or may not be present in galaxies which apparently fall near the main sequence.
This relation demonstrates that sources with larger $\Delta\mathrm{MS}$ are generally galaxies with more vigorous central star-forming activities and thus higher surface density of IR luminosity.
Note, however, that the spatially extended SMGs reported in Section~\ref{sss:05a_ext} are on the star-forming MS, indicating that extended dust continua reflect active star formation over the whole galaxy.
}

%% file: 05b_2model.tex
\begin{figure}
\centering
\includegraphics[width=\linewidth]{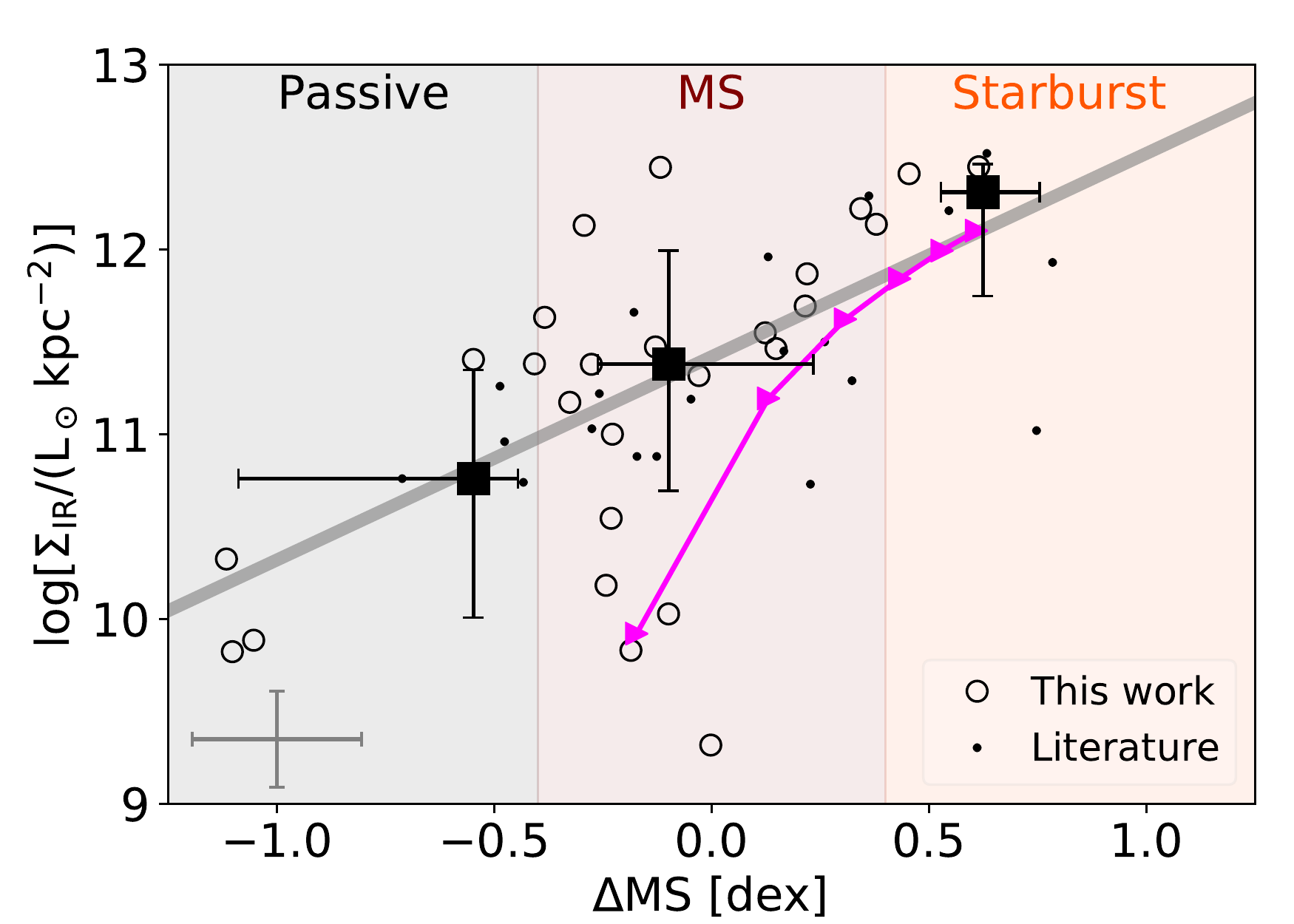}
\includegraphics[width=\linewidth]{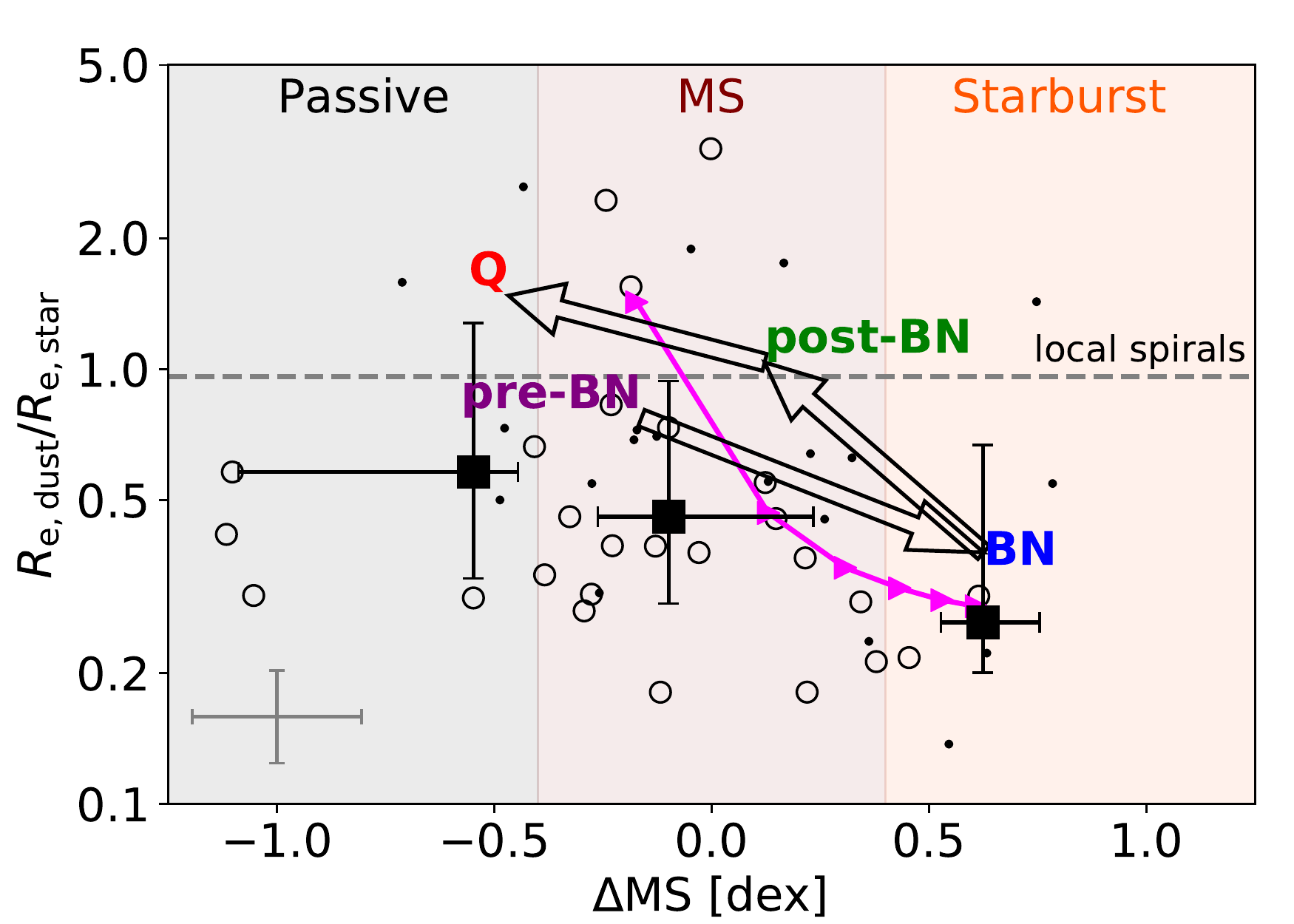}
\caption{\textit{Top}: Infrared surface luminosity ($\Sigma_\mathrm{IR}$) versus main-sequence offset ($\Delta$MS).
The empty circles denote the 26 sources in this work, and the black dots denote the 20 sources reported in \citet{lang19} and \citet{barro16}.
Typical uncertainty of each source is shown as the grey error bars at the lower-left corner.
The median value of all the sources in the three bins of $\Delta$MS (i.e., considered as passive, main-sequence and starburst galaxies) are shown as black squares with error bars denoting the 16-84th percentile of distribution.
Best-fit linear relation is shown as the grey solid line.
\textit{Bottom}: Dust-to-stellar source size ratio ($R_\mathrm{dust}/R_\mathrm{star}$) versus main-sequence offset.
Symbols are the same as the top panels.
The black empty arrows denote the theoretical evolutionary path of a $M_\mathrm{star}\sim 10^{10}$\,\msun\ galaxy, computed from the models of \citet{tacchella2016}.
Four proposed evolutionary phases, namely pre-blue nugget (pre-BN), blue nugget (BN), post-blue nugget (post-BN) and quenching phase (Q), are labelled in colored bold texts although we stress that the term ``blue'' does not apply to these highly obscured systems.
Typical $R_\mathrm{dust}/R_\mathrm{star}$ of local spiral galaxies is shown as the horizontal dashed line \citep{hunt15}.  
In both panels, the solid magenta line shows the track of the two-component model discussed in Section~\ref{sss:05_two-component1}.
}
\label{fig:10_quench}
\end{figure}

According to Figure~\ref{fig:08_main} and Section~\ref{ss:04g_main}, one should further expect an anti-correlation between the $\Delta\mathrm{MS}$ and $R_\mathrm{e,dust}/R_\mathrm{e,star}$. % given the $\Sigma_\mathrm{IR} - \Delta\mathrm{MS}$ relation presented above.
Figure~\ref{fig:10_quench} seems to show a weak anti-correlation between the dust-to-stellar size ratio and main-sequence offset.
A least-squares linear fitting with bootstrapping suggests the slope is less than zero but only at a $2.4\sigma$ significance.
Similar conclusion has been made in \citet{lang19} where the authors claimed no conspicuous relation was found between sSFR and $R_\mathrm{e,dust}/R_\mathrm{e,star}$.

\input{05_two-component2}

\subsubsection{Comparison with theoretical evolutionary tracks}
\label{sss:05b_iiiquench}

To better understand the observed trends and underlying properties of the observed SMGs, we further investigate the distribution of  $R_\mathrm{e,dust}/R_\mathrm{e,star} - \Delta\mathrm{MS}$ in the lower panel of Figure~\ref{fig:10_quench} by comparing with theoretical predictions from cosmological simulations.
By stacking the simulated galaxy profiles of different evolutionary phases, \citet{tacchella2016} suggested that the radial profile of sSFR declines from center to outskirts during the central starburst phase, resulting in a concentrated star-forming region with a smaller radius when compared with that of evolved stellar continuum.
% However, this sSFR profiles rises as the fade-out of star formation, which leads to a larger ratio between the radii of star-forming region and stellar component.
However, after the gas compaction and central enhancement of SFR, quenching then starts from the center as a combined effect of gas depletion due to star formation, feedback and truncation of further gas inflow.
Compared with the galaxy center, the outskirts can still retain a ring-like star-forming region, increasing the ratio of $R_\mathrm{e,SFR}/R_\mathrm{e,star}$.
% A cartoon illustration of this process is shown as Figure~\ref{fig:cartoon}.

Through this so-called compaction and ``inside-out'' quenching scenario, the main-sequence offset (and thus IR surface luminosity) of an SMG can increase and subsequently decline, and the dust-to-stellar size ratio will decline first and then rise.  
Such behaviors of a galaxy evolution model are generally consistent with the trends as we see in Figure~\ref{fig:08_main} and \ref{fig:10_quench}.

To conduct a quantitative comparison with the theoretical predictions, we utilize the radial profiles of $\Sigma_\mathrm{SFR}$ and $\Sigma_\mathrm{star}$ for 26 simulated galaxies presented in \citet{tacchella2016}.
The median stellar mass of these galaxies is $\sim 10^{10}$\,\msun\ at $z = 2$, and \citet{tacchella2016} stacked the radial profiles into four evolutionary phases, namely pre-blue nugget (pre-BN),
%\footnote{{\color{blue}The term ``blue nugget'' may be a misnomer for SMGs dominated by compact central starburst because the amount of dust extinction would be significant, which would redden the color.}},
blue nugget (BN), post-blue nugget (post-BN), and quenching (Q) phase with increasing time. 
Here, the BN refers to a massive compact star-forming galaxy of high central density in stellar mass, following \citet{zolotov15}, although the blue stellar population in our systems is heavily dust reddened and therefore would appear red, making ``blue'' nugget somewhat a misnomer.
Using the stacked $\Sigma_\mathrm{SFR}(r)$ and $\Sigma_\mathrm{star}(r)$ profiles, we compute the main-sequence offset and the ratio between the effective radii of star-forming region and stellar component in these four phases, and overlay the evolutionary trend with arrows in the bottom panel of Figure~\ref{fig:10_quench}. 
We assume the BN phase occurs at $t_{z=2}$, i.e., the age of the universe at $z=2$ in the unit of Gyr. 
The remaining three phases are assumed to occur at $t=t_{z=2} - 0.4$, $+0.3$ and $+1.0$ Gyr, consistent with the time range reported in \citet{tacchella2016}.
% (same as the right panel of Figure~\ref{fig:09_ssfr}). 

We find that the observed distribution of $R_\mathrm{e,dust}/R_\mathrm{e,star}$ and $\Delta\mathrm{MS}$ generally matches the theoretical evolutionary tracks. % prediction, especially in terms of the slope and intercept.
The slopes of these arrows are less than zero, consistent with the tentative negative slope reported in Section~\ref{sss:05b_iidms_xx}, and the vertical offset between the pre-BN\,$\rightarrow$\,BN and post-BN$\rightarrow$Q trend can also explain the large dispersion of $R_\mathrm{e,dust}/R_\mathrm{e,star}$ seen around the main sequence.

% {\irs \sout{Furthermore, we also notice a potential lack of main-sequence SMGs at $0.9<R_\mathrm{e,dust}/R_\mathrm{e,star}<1.5$ in the joint sample. % corresponding to the typical ratio for local spiral galaxies \citep{hunt15}. 
% This could be tentatively explained by the gap between the bursting (pre-BN\,$\rightarrow$\,BN) and quenching (post-BN\,$\rightarrow$\,Q) trend in simulations.
% The intermediate BN\,$\rightarrow$\,post-BN phase has a shorter timescale and half of the vector is in the $\Delta\mathrm{MS}$ bin for starburst galaxies, which may make it hard to detect main-sequence SMGs at this radius ratio.
% However, K-S test suggests that the dust-to-stellar size ratios of main-sequence SMGs still satisfy a normal distribution ($p$-value as 0.79), and therefore no strong statement could be made.}}

We find that the five main-sequence SMGs with more extended dust continua (i.e., $R_\mathrm{e,dust}/R_\mathrm{e,star} > 1$; including the two extended SMGs reported in Section~\ref{sss:05a_ext}) hold a higher stellar mass surface density (mean $\Sigma_\mathrm{star} = 10^{9.9\pm0.4}$\,\si{M_\odot.kpc^{-2}}) than the 26 main-sequence compact sources at $R_\mathrm{e,dust}/R_\mathrm{e,star}<1$ (mean $\Sigma_\mathrm{star} = 10^{9.4\pm0.1}$\,\si{M_\odot.kpc^{-2}}).
This indicates that SMGs with more extended dust continua are likely at a later evolutionary phase than those compact ones.
% We also find a sharp decline of sSFR ($0.9\pm0.2$\,dex) from the starburst subsample to the extended main-sequence subsample, consistent with the quenching scenario.}} 

One minor mismatch is that the compact/extended SMGs exhibit a even lower/higher $R_\mathrm{e,dust}/R_\mathrm{e,star}$ than the theoretical predictions.
This result is tentative because the SMG sample is biased to a higher stellar mass (median $M_\mathrm{star} = 10^{11.1}\,$\msun). 
By dividing the simulated galaxies at $M_\mathrm{star}=10^{10.2}\,$\msun\ into two mass bins, \citet{tacchella2016} reported that galaxies with higher stellar masses show a more noticeable evolutionary pattern of the $\mathrm{SFR}(r)$ profile. 
This will likely result in a broader distribution of $R_\mathrm{e,SFR} / R_\mathrm{e,star}$ at various evolutionary phases.
Therefore, we suggest that our observations are consistent with the theoretical evolutionary track of galaxies in a gas compaction and potentially subsequent ``inside-out'' quenching process, which can be the driver of the tight $R_\mathrm{e,dust} / R_\mathrm{e,star} - \Sigma_\mathrm{IR}$ correlation seen in Figure~\ref{fig:08_main}.

% One caveat, however, is that the Pre-BN$\rightarrow$BN and BN$\rightarrow$Post-BN tracks are degenerate in Figure~\ref{fig:10_quench}, not allowing us to distinguish bursting galaxies from quenching ones on the basis of this figure alone.  We will discuss in Section~\ref{sss:05b_iiiqorb} how we may be able to break this degeneracy with more observational data.  

% We then further this discussion in Section~\ref{ss:05f_qorb}.

% thus we will observed a larger effective radius of dust-o 

% \begin{figure}
% \centering
% \includegraphics[width=0.9\linewidth]{figures/cartoon.pdf}
% \caption{Cartoon of the structural profile evolution during the ``inside-out'' quenching \citep{tacchella2016} of SMGs. 
% Profiles of stars (red) and star formation (blue) are shown in arbitrary scale as a function of galactic radius ($r$). 
% Through the evolution from bursting (BN) to quenching (Q) phase, we should expect to observe a fade-out of central starburst (i.e., lower $\Sigma_\mathrm{IR}$) and the broadening of $R_\mathrm{e,dust} / R_\mathrm{e,star}$ due to both the bulge formation and dust depletion/dispersal, exactly as the trend we present in the bottom panel of Figure~\ref{fig:10_quench}.}
% \label{fig:cartoon}
% \end{figure}

%% file: 05_two-component2.tex
In Figure~\ref{fig:10_quench}, we also plot the two-component model discussed in Section~\ref{sss:05_two-component1} (the solid magenta line).  
Unlike the left panel of Figure~\ref{fig:08_main}, this simple model does not produce a good fit in either plot.
This is because in these plots, the observed SMGs exhibit a large spread in $\Sigma_\mathrm{IR}$ and $R_\mathrm{e,dust}/R_\mathrm{e,star}$, reflecting the diversity of apparently main-sequence SMGs in the sense that some of them are dominated by central starburst while others are dominated by galaxy-wide star formation (as noted
by \citealt{puglisi19}).  
Since the two-component model, as defined in Section~\ref{sss:05_two-component1}, assumes a transition of SMGs from those dominated by central starburst to those dominated by galaxy-wide star formation as sSFR decreases from the starburst to main-sequence range, it fails to reproduce the main-sequence SMGs dominated by central starburst.  
In order to reproduce this population of compact main-sequence SMGs with this type of two-component model, it would be
necessary to decrease the luminosity of galaxy-wide star formation further.

%% file: 05b_3qorb.tex
Future observations of spatially extended SMGs may testify whether they are undergoing a starburst and subsequent quenching process.
This can be performed through NIR spectroscopy by providing tighter constraints on instantaneous SFR (e.g., through \ha\ spectroscopy, sensitive to star formation in previous $\sim3-10$\,Myr; \citealt{ke12}; although this is subject to dust obscuration) and comparing with the dust-based one which traces the star formation at a longer duration ($\sim100$\,Myr).
If the properly dust-corrected \ha\ SFR is significantly smaller than the dust-based one, then it is very likely that the galaxy is in a quenching phase, since the \ha\ emission can also be contributed by AGN activity. 
% \todo[inline]{IRS: Not really.  Just means dust correction isn't precise enough.  I don't think the H-alpha SFR is a robust measure to compare against as this will depend on the precision of the reddening correction and even then it won't show anything other than some SFR is very obscured if the H-alpha still falls short of the FIR.   if you wanted to you could replace this with something on Pa-alpha using JWST - which might be more reasonable?}
In addition to this, NIR spectroscopy can also probe the stellar age (e.g., through 4000\,\si{\AA} break; \citealt{barro16ha, newman15, newman18}) and thus the evolutionary stage of the galaxy (but only for the less obscured stellar populations).

Previous works also reported a stronger \nii~$\lambda$6585 flux compared with that of \ha\ in a $z= 1.67$ fast-quenching galaxy \citep{barro16ha} and $z \gtrsim 2$ cluster-lensed cQs \citep{newman18}. 
A high value of the \nii/\ha\ ratio is typical of the LINER galaxies in the BPT diagram \citep{bpt81}, suggesting a suppressed star formation, which has been observed in low-redshift red-sequence or post-starburst galaxies \citep[e.g.,][]{yan06}, although possible AGN contamination will be an issue with LINER-like spectra in general.
For most of the targets in this work, such an analysis is inaccessible due to the general lack of $J/H$-band photometry and spectroscopy, and thus the true SFH can only be determined accurately with more observational data.

% However, our $H$-band spectroscopy, together with $J$ and $K$-band photometry, on MACSJ1314.$_\Sigma$, does suggest a low instantaneous SFR in contrast with FIR-traced one, which can be significantly contaminated by evolved stars ($\sim100-200$\, Myr; \citealt{ke12}). 
% We also find MACSJ1314.$_\Sigma$ exhibit stronger [N II] flux compared with \ha, consistent with the $z\sim 1.7$ fast-quenching galaxy in \citet{barro16ha} and $z \gtrsim 2$ cluster-lensed cQs in \citet{newman18}. 
% Such a high value of the [N II]/\ha\ ratio is typical of LINER galaxies in BPT diagram \citep{bpt81}, suggesting suppressed star formation since similar line ratio is seen in low-redshift red-sequence or post-starburst galaxies \citep[e.g.,][]{yan06}.
% Detailed analysis on this SMG will be presented in Walth et al., in prep., and here we only carefully conclude that the star-formation phase (i.e., bursting or quenching) of SMGs can be reliably determined though further NIR observation.

%% file: 05c_1trigger.tex
Our ALMA observations suggest that late-phase wet-wet major merger could be the triggering mechanism for \textred{$\sim$27\% (6 out of 22 cases)} of the SMGs studied in this work because of the existence of companions 
\textred{(angular separation of $2\farcs5 _{-1.8} ^{+5.7}$ in the image plane) at similar redshift (Section~\ref{ss:03d_sum}; e.g., \citealt{bournaud11}).
% We observed comparable 1.3\,mm flux densities between the different components in each SMG group,
% This indicates that major merger could act as a possible triggering mechanism for SMGs \citep[e.g.,][]{bournaud11}. 
Our derived major-merger pair ratio ($\gtrsim$27\%)} is consistent with \citet[][as 27\%]{fujimoto17} and \citet[][as 22\%]{an2019}, but note that we do not have a uniform survey of companions due to the cluster lensing effect, and therefore our measurement should be a lower limit.

% Our observations detected no major-merger companion for the majority of SMGs in this lensed sample. 
\textred{The remaining 16 sources in this work} are potentially isolated SMGs with no companion brighter than $\sim 0.6$\,mJy at 1.3\,mm, a quarter of the median ALMA flux density we measured for the primary sources.
Since a stellar mass ratio as $4:1$ between the major and minor components is widely adopted to distinguish major and minor merger \citep[e.g.,][]{lotz11, man16}, under the assumption that FIR flux ratio equals to stellar mass ratio for galaxies in a merging system, our observations suggest no evidence of major-merger companion for \textred{$\lesssim$73\% of the SMGs in this lensed sample}.
Note that we cannot rule out the possibility that some SMGs are in the late phase of major merger (e.g., separation is $\lesssim 1$\,kpc between various components), which should be further tested through high-resolution observations of gas kinematics \citep[e.g.,][]{litke19, neeleman19}.

% As illustrated in Section~\ref{sec:01_intro}, the triggering mechanism of SMGs is currently under debated.
Based on the companion search, the co-centered distribution of dust/stellar continua and the evolutionary trends presented in this work, we conclude that SMGs could be triggered by a variety of mechanisms including major merger, minor merger and secular burst, consistent with the conclusions in \citet{fujimoto17}, \citet{wipu19}, \citet{lang19} \textred{and \citet{jimnez20}.}
Such a diversity of triggering mechanism is different from what we have seen in galaxies of comparable infrared luminosities in the local Universe, i.e., ULIRGs, in which the major merger is the prevalent mode of igniting a circumnuclear starburst \citep[e.g.,][]{sanders88}.
% \todo[inline]{IRS: But I thought we were claiming most of these are not starbursts (i.e., on the main-sequence)?\\ EE: Added some text to clarify the statement.}

We find no clear difference of the central and total sSFR \textred{between the 27\% SMGs with close FIR-bright companions and the rest of our sample}.
This suggests that other mechanisms like minor mergers \citep[e.g.,][]{gomez18}, mergers with gas-poor companions and secular inflow of gas \citep[e.g.,][]{dekel14} may trigger the star formation in the core of SMGs at comparable intensity as the major merger. % \todo[inline]{IRS: or mergers with gas-poor companions.}
However, we also stress that there are a number of caveats in our analysis of the merging fraction, most notably being the need for companions to be FIR-bright and to lie close enough in the source plane to suffer comparable amplification, which may weaken these conclusions.

% \citet{chen15} pointed out that the star-forming timescale of ALESS SMGs ($t_\ast\sim$100\,Myr) is shorter than the infall timescale ($\sim$250\,Myr) that \citet{dekel14} predicted, raising a timescale tension of gas inflow as trigger mechanism of SMGs.
% For our sample with lower $\Sigma_\mathrm{SFR}$, the gas depletion timescale based on the global Kennicutt-Schmidt law \citep{ks98} is comparable with the infall one.
% This consistency can relieve such a reported timescale tension and allow the sufficient compaction of infall gas.

% In addition to these, the theoretical picture of ``inside-out'' quenching in \citet{tacchella2016} also assumes minor merger and smooth gas inflow as the drivers of gas compaction and the triggering of ``blue-nugget'' phase \citep{zolotov15}. 
% Since our observations are consistent with such a theoretical prediction, it is possible that a significant number of SMGs in our sample are driven by interactions less intense than major merger.

%% file: 05c_2evolution.tex
Previous studies have suggested that SMGs are linked to the cQs seen at slightly lower redshift. 
The evidence includes the matched number densities \citep[e.g.,][]{simpson14,dudzevic19}, clustering properties \citep[e.g.,][]{hickox12, an2019} and size/mass similarities \citep[e.g.,][]{barro16, lang19}.
% Due to a lack of high-resolution optical/NIR images for the majority of SMGs in this work, we do not have accurate lens models to make a similar comparison.
% Therefore, 
In this subsection, we discuss about the possible evolutionary connection between SMGs and cQs using those physical quantities that are independent of lensing effects.

%  Evidence 1: Dust-to-stellar mass ratio

\textred{We show that a subset of SMGs in our sample exhibits similar dust-to-stellar mass ratios (four at $M_\mathrm{dust} / M_\mathrm{star} < 10^{-3}$) as early-type galaxies at low redshift (Figure~\ref{fig:05_dsr}).
}
% \textbf{We show that a subset of SMGs in our sample exhibit low dust-to-stellar mass ratios (four sources at $M_\mathrm{dust} / M_\mathrm{star} < 10^{-3.0}$ and seven sources at $M_\mathrm{dust} / M_\mathrm{star} < 10^{-2.9}$), comparable to those of  early-type galaxies at low redshift (Figure~\ref{fig:05_dsr}).}
Such a low dust-to-stellar mass ratio is consistent with the stacked $z\sim 1.8$ quiescent galaxies \citep{gobat18}.
Assuming a gas-to-dust ratio of 100, these galaxies also match with the $z\sim 0.7$ post-starburst (PSB) galaxies reported in \citet{suess17}. 
One caveat is that the lack of NIR data may lead to an overestimate of dust extinction and thus stellar mass through an energy-balance approach in the SED fitting (e.g., HLS1623), but we show that for HLS1314 with accurate $J/K$-band photometry, its $A_V$ is tightly constrained and thus its $\log(M_\mathrm{dust} / M_\mathrm{star})$ can be determined as $-3.25\pm0.10$, similar to the values for PSBs and early-type galaxies quoted above.
% Detailed spatially-resolved analysis of the dust, gas and stellar component of this fast-quenching object will be presented in Walth et al., in prep.

% Stellar surface densuty
We also show that the stellar surface density ($\Sigma_\mathrm{star}$) of SMGs in this work match better with early-type than late-type galaxies at similar redshift \citep{vdw14}, in the right panel of Figure~\ref{fig:08_main}.
%Since SMGs generally exhibit smaller star-forming region compared with their present spatial stellar distribution, 
Since the spatial distribution of star-forming regions in SMGs is typically more compact than those of stars, the ongoing intense star formation will lead to an even higher $\Sigma_\mathrm{star}$ in later phases, increasing the difference from the typical value of late-type galaxies at that cosmic age.
Such a comparison has been conducted by \citet{barro13}, \citet{hodge16} and \citet{lang19} via \hst\ imaging and interpreted as structural consistency between the stellar components of SMGs and quiescent galaxies, suggesting a possible evolutionary link after the cessation of star formation in SMGs.

\begin{figure}[tb!]
\centering
\includegraphics[width =\linewidth]{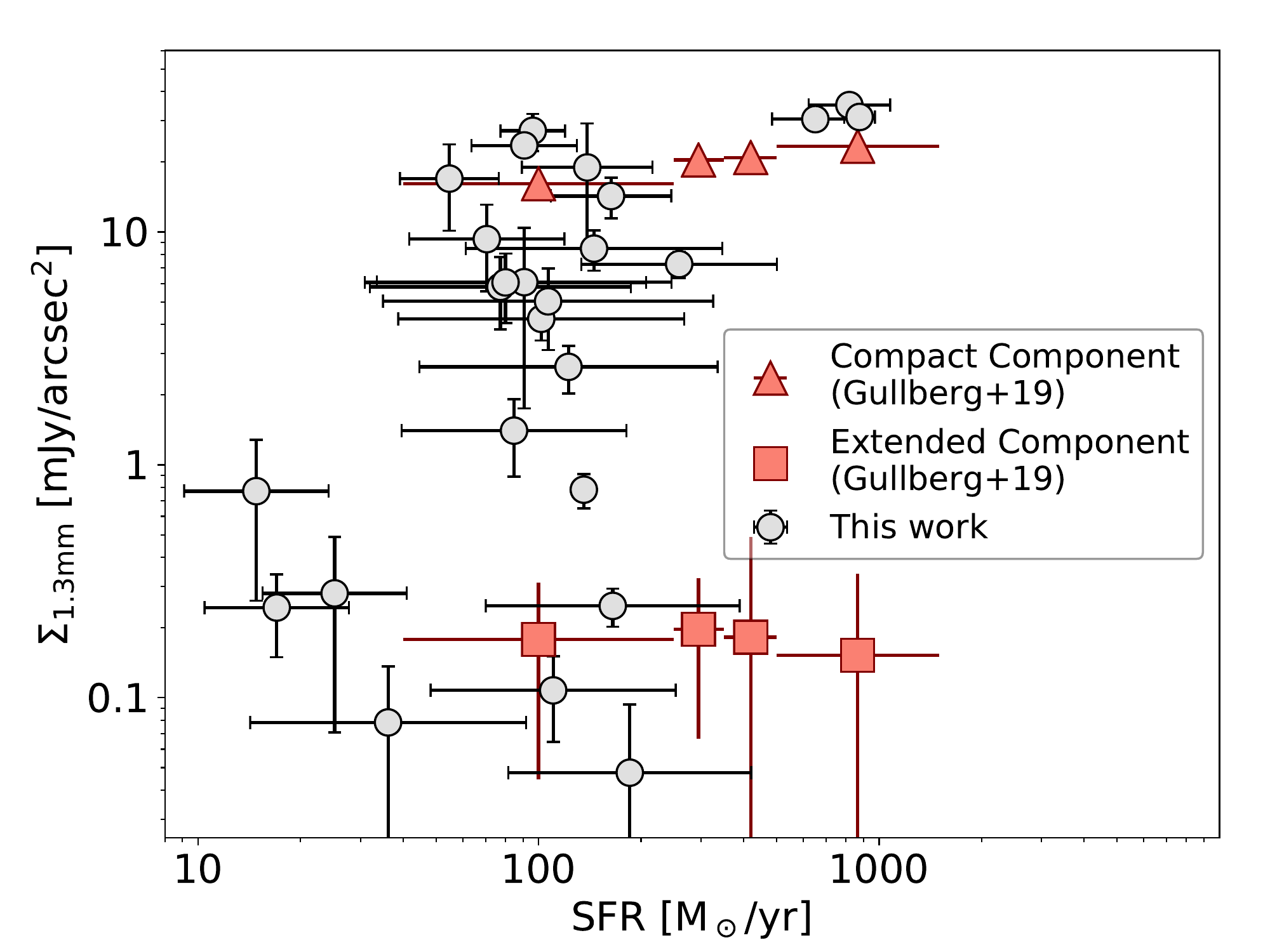}
\caption{1.3\,mm surface brightness ($\Sigma_{1.3\,\si{mm}}$) versus intrinsic SFR (assuming a lensing magnification of $\mu=5$) of resolved SMGs in our sample (grey circles with black edges). \citet{gullberg19} reported the existence of both compact (shown as triangles) and extended dust components (squares) of $z\sim 3$ SMGs through stacking. 
Assuming typical SED of a SMG at $z=3$, we convert the 870\,\micron\ surface brightness in \citet{gullberg19} to that at 1.3\,mm by dividing a factor of 2.7, which is $\sim 0.2$\,\si{mJy.arcsec^{-2}} for extended components.
This value matches those of low-surface-brightness SMGs in our sample ($\Sigma_{1.3\,\si{mm}} \sim 0.1$\,\si{mJy.arcsec^{-2}}).
}
\label{fig:gullberg}
\end{figure}

% Extended structure
We further show that the IR surface luminosity and spatial extent of extended %low-sSFR$_\mathrm{c}$ 
SMGs in this work may match with SMGs in a transitional phase to quiescent galaxies. 
\citet{gullberg19} showed the existence of an extended dust component ($R_\mathrm{e,dust} \sim 4$\,kpc) of typical $z\sim 3$ SMGs through stacking 153 of them in the ALMA Band 7.
Such an extended component is reported to contribute to $\sim 13\%$ of the total emission at 870\,\micron, and the corresponding surface luminosity is $\Sigma_\mathrm{IR} \sim 10^{9.9}$\,\si{L_\odot.kpc^{-2}}, under the assumption of typical SED of an SMG at $z\sim3$.
% original (1st draft) is 10^9.7 for z~2 (Band7), revised
Furthermore, in the sample of \citet{gullberg19}, the FIR surface brightness of the compact dust component ($R_\mathrm{e,dust}\sim 1$\,kpc) decreases with the declining of total $L_\mathrm{IR}$ while it remains the same for the extended component (Figure~\ref{fig:gullberg}).

In this work, we do discover SMGs with an IR surface brightness comparable to the faint and extended component of SMGs reported by \citet{gullberg19}, and such a comparison is presented in Figure~\ref{fig:gullberg}.
Taking HLS0546 ($\Sigma_\mathrm{IR}= 10^{9.8\pm0.3}$\,\si{L_\odot.kpc^{-2}}) as an example, assuming a reasonable lensing magnification of $\mu \simeq 5$, the physical effective radius of the dust continuum will be $6.3\pm 1.0$\,kpc.
Consider the radial gradient of dust temperature, %and a possibly larger lensing magnification, 
this result could be consistent with the size of the extended dust continuum of a typical SMG observed at rest-frame 220\,\micron\ \citep{gullberg19}.
% This could be consistent with the $\sim4$\,kpc size of the extended dust continuum of a typical SMG observed at rest-frame 220\,\micron\ if the radial gradient of dust temperature is considered and actual lensing magnification is greater than 5. 
This magnification assumption will also lead to a de-magnified stellar mass of $M_\ast = 10^{11.0\pm0.4}$\,\msun\ and $R_\mathrm{e,star}=3.5\pm0.6$\,kpc, within a 1$\sigma$ distribution of early-type galaxies seen at the given redshift \citep{vdw14}.  
Therefore, it is possible that after the dissipation of central star formation on a compact physical scale ($1-2$\,kpc), SMGs will maintain a low-level IR surface brightness over a more extended galaxy structure ($\gtrsim 4$\,kpc), as suggested by the evolutionary tracks in Section~\ref{sss:05b_iiiquench}.  
At the same time, the previous concentrated star formation would lead to the formation of
\textred{compact ($\sim2$\,kpc) quiescent and spheroidal galaxies within a timescale of $\sim 300$\,Myr.
}
Therefore, spatially extended SMGs with large dust-to-stellar radii ratio in this study may provide possible evidence of evolutionary connection between typical SMGs and compact quiescent galaxies at slightly lower redshift.

% the low-sSFR$_\mathrm{c}$ SMGs 

% and stellar mass density
% and Figure~\ref{fig:08_main} right panel

%% file: 06_sum.tex
We have obtained and analysed ALMA 1.3\,mm, \herschel/SPIRE 250/350/500\,\micron\ and \spitzer/IRAC 3.6/4.5\,\micron\ data of 29 lensed SMGs in 20 cluster fields.
These SMGs were discovered by the \herschel\ Lensing Survey as exceptionally bright sources in the far infrared ($S_\mathrm{peak} \gtrsim 90$\,mJy).
We have carried out modeling of their structural profiles in both the IRAC and ALMA bands, as well as their SEDs from NIR to millimeter wavelengths.
Since accurate lens models are not yet available for many of the observed SMGs, we focus our discussion on quantities that are independent of lensing effects, such as surface brightness and size ratios.  
When necessary, we also assumed a canonical magnification factor of $\mu \simeq 5$. 
The main results of this study are the following:
\begin{enumerate}
\item 
29 sources were detected in our \textit{uv}-tapered ALMA 1.3\,mm maps at S/N$>$4.0. % with total false detection number no more than one. 
Five sources are identified as galaxy-lensed or highly-blended cases, and their IRAC fluxes were successfully decomposed using ALMA and optical priors.
The remaining 24 sources are cluster-lensed SMGs, and all of their dust continua were spatially resolved (HLS1314 is resolved as three components).

\item 
% At the assist of strong gravitational lensing,
Because of gravitational lensing,
we were able to resolve the structural profile of stellar continuum in 24 out of 26 SMGs in \spitzer/IRAC bands.
% The remaining two sources were resolved by ALMA, but their heavy blending with nearby bright sources prevented surface brightness profile modelling in the IRAC images.
88\% of the SMGs in this study show smaller half-light radii in dust continua compared with the stellar ones.
The medium $R_\mathrm{e.dust}/R_\mathrm{e,star}$ was found to be 0.38$\pm$0.14.
Two sources (HLS0840 and HLS0546) are discovered as spatially extended main-sequence SMGs with low surface brightness ($\lesssim 0.1$\,\si{mJy.arcsec^{-2}} at 1.3\,mm) and only detectable in \textit{uv}-tapered ALMA maps.

\item
We fit the SED of all the observed SMGs with \textsc{MAGPHYS}, deriving their physical properties such as SFR and stellar/dust mass.
The dust-to-stellar mass ratio of SMGs is correlated with sSFR (at galaxy scale; subjected to dust temperature $T_\mathrm{dust}$), and \textred{four of them exhibit $\log(M_\mathrm{dust}/M_\mathrm{star}) < -3$}, resembling dusty early-type galaxies rather than typical SMGs at $z\sim2$.

\item
We find that the IR surface luminosity ($\Sigma_\mathrm{IR}$) of SMGs in our sample spans over a wide range of 3\,dex, similar to that covered by local (U)LIRGs in the GOALS sample \citep{ds17}. 
At a given $\Sigma_\mathrm{IR}$ below $ 10^{11.5}$\,\si{L_\odot.kpc^{-2}}, our SMGs show consistent dust temperature as local LIRGs, indicating no significant evolution of $\Sigma_\mathrm{IR} - T_\mathrm{dust}$ relation for LIRG-like galaxies from $z\sim 2$ to the present Universe.
At $\Sigma_\mathrm{IR} > 10^{11.5} $\,\si{L_\odot.kpc^{-2}}, the $T_\mathrm{dust}$ of ULIRGs at $z\sim 2$ are lower than those of the local ones as reported previously \citep[e.g.,][]{symeonidis13,dudzevic19}.

\item
We find that the IR surface luminosity of SMGs is anti-correlated with the dust-to-stellar source size ratio ($R_\mathrm{e,dust}/R_\mathrm{e,star}$). % which is not an artifact of lensing effect.
% This indicates that SMGs with less vigorous star formation are likely to show lower specific star-formation rate at their center, in contrast to the whole galaxy plane.
% This is consistent with the features of galaxies in cosmological simulation that are experiencing central starburst and subsequent inside-out quenching \citep{tacchella2016}.
Compared with a simple analytic model and cosmological simulations \citep{tacchella2016}, this relation could be interpreted by the morphological evolution of SMGs through central starburst and potentially subsequent inside-out quenching:
the central starburst leads to an increase in $\Sigma_\mathrm{IR}$ and compaction of star formation (i.e, smaller $R_\mathrm{e,dust}$), and in a following quenching phase, SFR quickly declines from the center (smaller $\Sigma_\mathrm{IR}$) but retains a considerable level at the outskirts, resulting in a larger $R_\mathrm{e,dust} / R_\mathrm{e,star}$.
% These processes can result in the enhancement and decline of central star formation (i.e., $\Sigma_\mathrm{IR}$), as well as the compaction and broadening of (i.e., $R_\mathrm{e,dust} / R_\mathrm{e,star}$)

\item
The distribution of spatial offset between the dust and stellar continua in SMGs does not support any intrinsic offset rather than astrometric uncertainty. 
\textred{$\gtrsim 27\%$ of the SMGs} can be visually identified as late-stage wet-wet major-merger pair \citep[consistent with][]{fujimoto17,lang19}. 
Therefore, it is possible that SMGs could be triggered through a variety of mechanisms, including major/minor merger and secular evolution, although lensed samples are not well suited to identify early-stage mergers due to the spatially varying magnification.

\item
Our observations are consistent with the hypothesis that SMGs are progenitors of the compact quiescent galaxies at slightly lower redshift. This is because  
(1) low-sSFR SMGs show dust-to-stellar mass ratios ($M_\mathrm{dust}/M_\mathrm{star}$) comparable to those of compact early-type or post-starburst galaxies,
(2) the high stellar surface densities of SMGs match those of early-type galaxies rather than late-type ones at similar redshift,
and finally (3) spatially extended SMGs in this work exhibit low IR surface luminosity, matching some expected properties of SMGs in the transition phase to cQs.

\end{enumerate}

The discovery of spatially extended SMGs in this study expands the population of submillimeter-selected galaxies to a lower FIR surface luminosity limit, and therefore provides direct observational constraints on the evolution of dusty starburst galaxies which may be in the quenching phase.
% Our multi-wavelength observations support ``inside-out" quenching as a feasible mechanism to account for both the dissipation of intense star formation in SMGs and the emergence of compact quiescent galaxies at $z\sim 2$.
% Further NIR observations on the optical nebular lines and 4000\,\si{\AA} break, as well as ALMA observations on the molecular gas distribution, would be highly valuable.
% Such observations will provide improved insight into the history of star formation and gas consumption or transportation in these transitional systems, 
Further studies of these (possibly) transitional systems will help to establish the evolutionary picture of massive galaxy at the epoch when the cosmic SFR density peaks \citep{md14}.

%% file: 98_appendix.tex
\section{False detection rate and Completeness of ALMA Detections}
\label{sec:app_1}
\input{98a_alma}

\section{Spatial offset between ALMA and IRAC counterparts}
\label{sec:app_2}
\input{98b_offset}

% \section{The effect of differential magnification}
% \label{sss:05a_iiidifflens}
% \input{98c_difflens}

\section{ALMA, IRAC images and SED plots of all sources}
\label{sec:app_3}
\input{98d_thumbnail}

\clearpage

\input{tables/tb01_observation_log.tex}

% \clearpage

\input{tables/tb02_photometry.tex}

\input{tables/tb03_morphology.tex}

\input{tables/tb04_properties.tex}

%% file: 98a_alma.tex
\begin{figure*}[!tbh]
\centering
\includegraphics[width=0.49\linewidth]{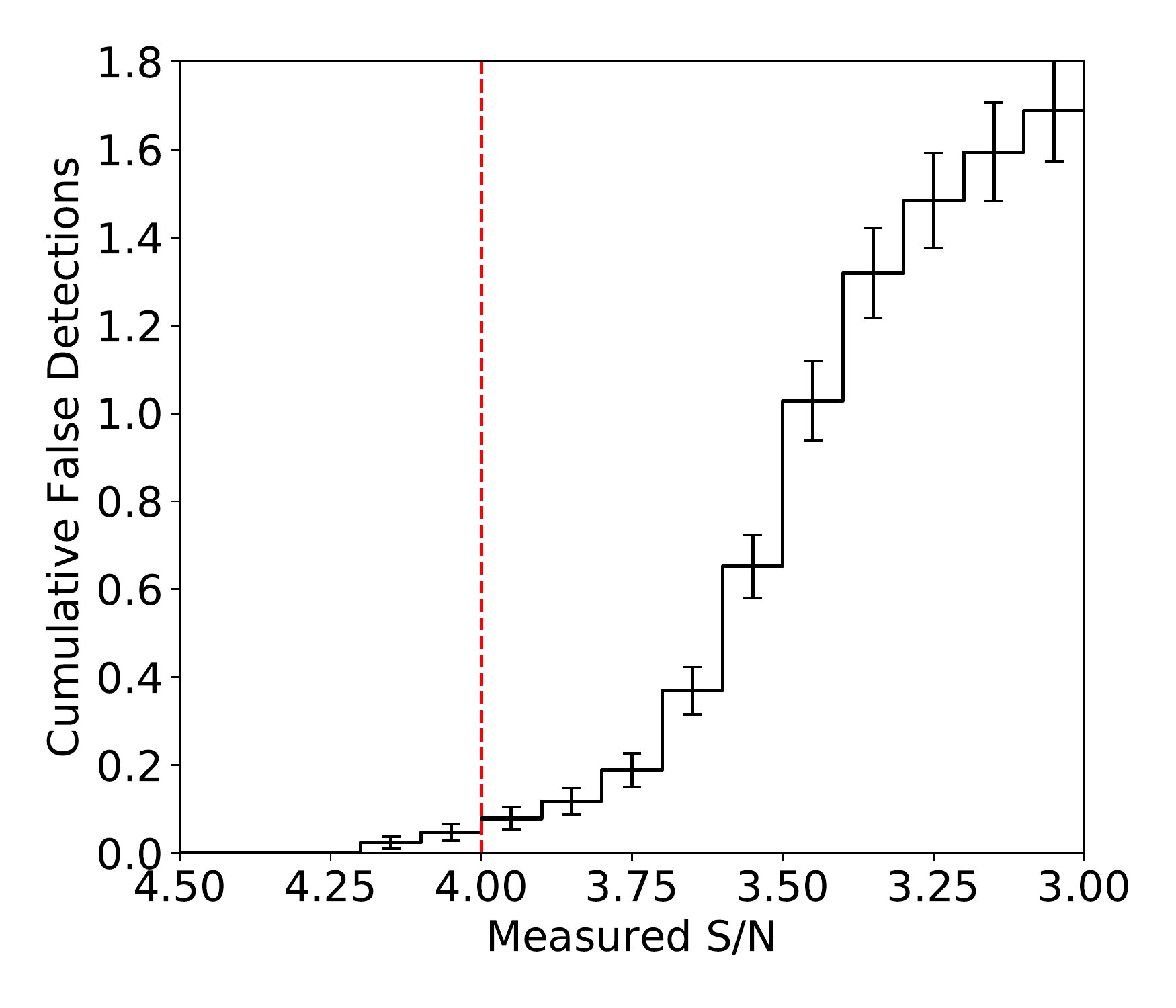}
\includegraphics[width=0.49\linewidth]{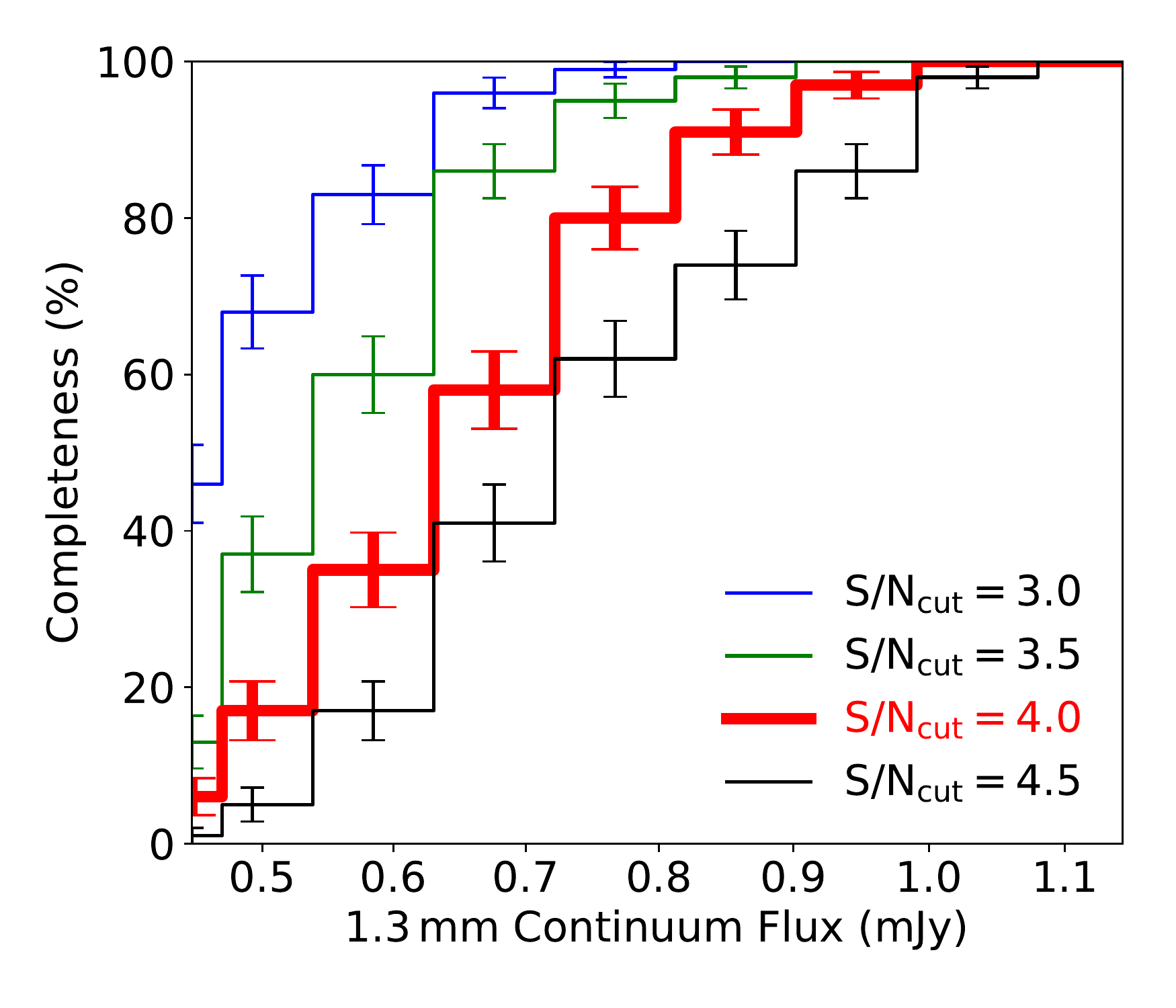}
\vspace{-2mm}
\caption{
\textit{Left:} Cumulative false detections per simulated single-pointing FoV, as a function of measured S/N through \textsc{SExtractor} aperture photometry. We then cut our detection at 4$\sigma$ (red dashed line) to eliminate fake detections.
\textit{Right:} Source detection completeness for simulated point sources, as a function of %median output S/N through aperture photometry (see explanation in Section~\ref{ss:03a_detect}; also converted to 
source flux density assuming a continuum RMS of 0.105\,mJy/beam and primary beam correction factor of $\times\sqrt{2}$). 
We plot the completeness curves at various setting of S/N cut, and highlight S/N$_\mathrm{cut}$=4.0 curve in red solid steps.
}
\label{fig:01_detection}
\end{figure*}

We investigated the false detection rate and completeness of our ALMA detections.
We generated Gaussian white noise maps and smoothed them with a 2D Gaussian kernel to simulate the noise distribution of ALMA continuum images obtained with a 1\arcsec\ beam. 
We then run \textsc{SExtractor} on the mock noise maps with the same settings used for the scientific maps, and treated all the extracted sources as false detections.
This experiment demonstrates that for a typical ALMA Band 6 single-pointing FoV at this \textit{uv}-taper, we expect to detect $\sim1.69$ fake sources at S/N$>$3, $\sim0.84$ fake sources at S/N$>$3.5, but only $\sim0.06$ at S/N$>$4 (Figure~\ref{fig:01_detection}, left).
Therefore, for the purpose of eliminating false detections, we applied an S/N cut of our detections at S/N=4. 
This ensures the total false detection number to be $\lesssim1$ in all of our ALMA maps.

We also studied the point-source detection completeness through similar simulation. 
We generate 100 sources with a given flux density on a Gaussian white noise map, comparable to the combined area of 20 single-pointing FoVs.  
We then blur the image with a 2D Gaussian kernel of FWHM=1\arcsec, simulating the 1\arcsec-tapered continuum images.
This mock image was fed into the same \textsc{SExtractor} routine for source detection, and we counted the number of recovered detections. % with their output S/N through automatic aperture photometry.
Such an experiment is repeated at various input source strength, and the completeness as a function of source flux density is shown in the right panel of Figure~\ref{fig:01_detection}.
% Here the theoretical output S/N is the expected S/N measured on mock images through aperture photometry.
% In real cases, the measured S/N of each detection is fluctuated due to noise, usually deviated from the theoretical S/N. 
% Due to noise, the measured S/N fluctuate among sources with the same input flux density.
% Therefore, we used the median measured S/N for all 100 input sources as the representative ones.
%theoretical one for a given combination of input source and noise strength.
Based on our simulations and a S/N cut of 4.0, our detection is $47\pm 5\%$ complete for point sources at S/N=4.0, corresponding to a flux density of 0.63\,mJy for median continuum RMS (0.105\,mJy) and half-radius primary beam correction ($\times \sqrt{2}$).
At S/N=5.0 (0.79\,mJy for median continuum RMS), the point-source completeness is $80\pm4\%$.
Therefore, our ALMA survey is fairly complete for any point source brighter than 0.9\,mJy.

Note that \citet{franco18} showed the completeness of ALMA detection decreases dramatically for larger galaxy sizes.
With 0\farcs6-tapering, their 1.1\,mm observation is 94\% complete for point sources at 1.2\,mJy but only 9\% for FWHM=0\farcs6 sources. 
We notice that 7 out of 26 sources with our FIR morphology modeling exhibit effective radii $R_\mathrm{e}>0\farcs5$. 
Assuming similar completeness curves as \citet{franco18} while scaled to 0.1\,mJy continuum RMS and 1\arcsec-tapering, our completeness can be lower than 50\% at 1.4\,mJy for these extended sources.

%% file: 98b_offset.tex
\begin{figure*}[!t]
\centering
\includegraphics[width=0.66\linewidth]{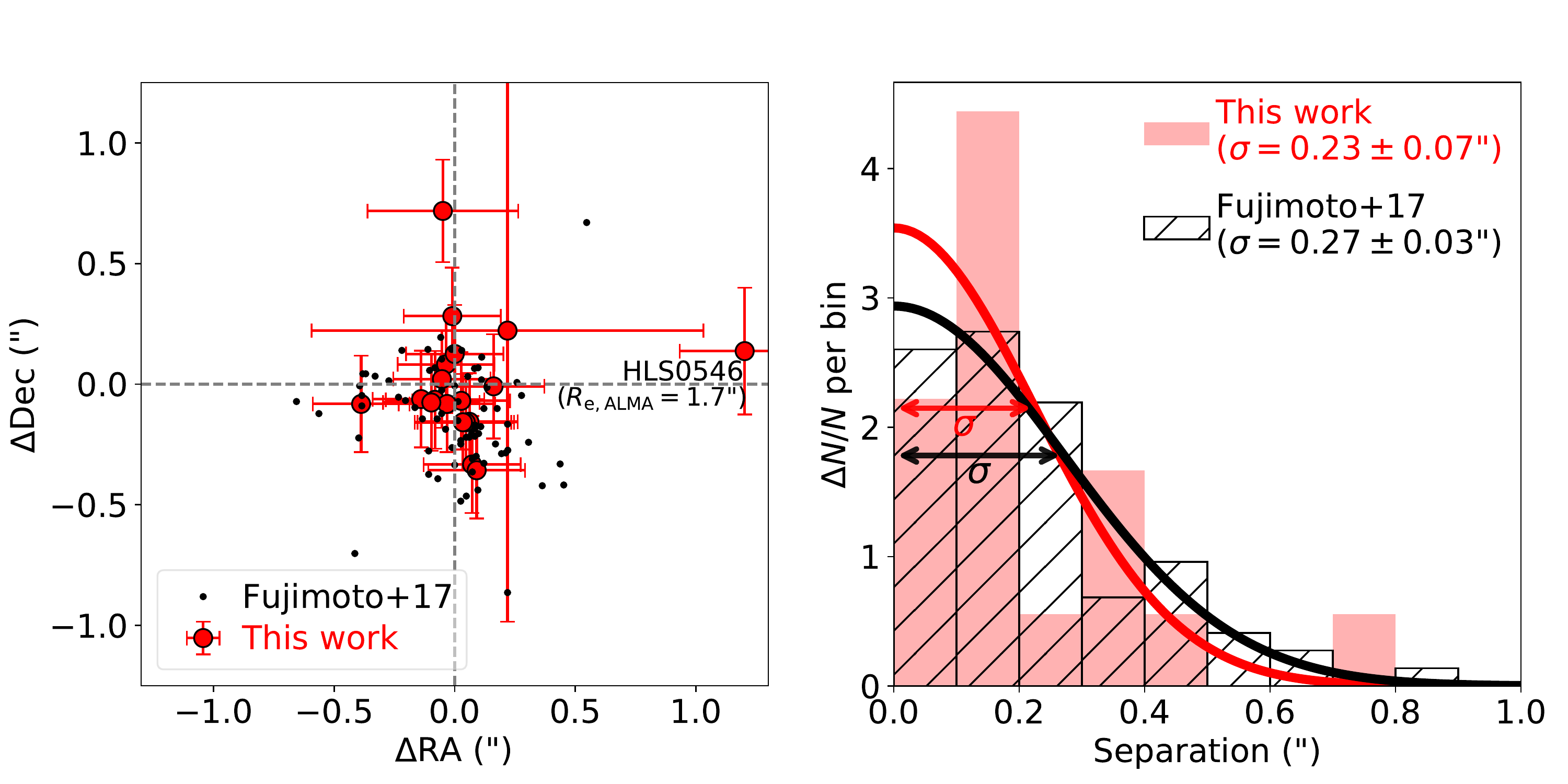}
\includegraphics[width=0.33\linewidth]{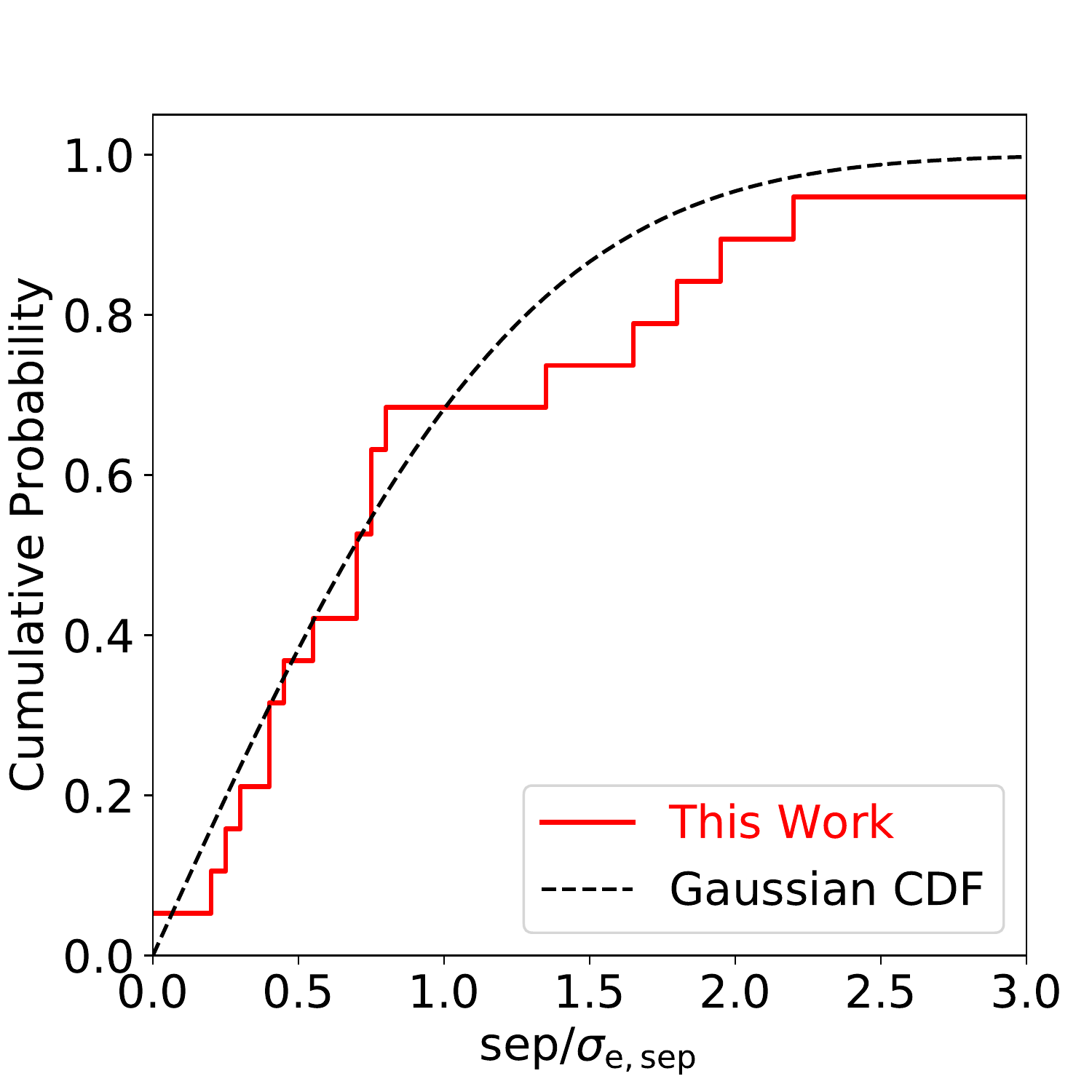}
\caption{\textit{Left}: Spatial offsets between the dust and stellar continua of SMGs. Lensed SMGs in this work are plotted as red circles, and 73 SMGs in CANDELS field are plotted as black dots (ALMA positions by \citealt{fujimoto17}; matched with \hst\ counterparts in \citealt{vdw12}). The only case where offset is larger than 1\arcsec, namely HLS0546, are specially noted.
\textit{Middle}: Histograms of dust/stellar offset distribution in this work (red) and \citet[][black]{fujimoto17}. Best-fit normal distribution profiles are shown as solid lines, with their standard deviations ($\sigma$) labeled.
\textit{Right}: K-S test of the dust/stellar offset distribution in this work (red steps), compared with cumulative distribution function (CDF) of normal distribution (black dashed line). Assuming a typical IRAC position error at 0\farcs20 ($\sim$10\% of PSF FWHM), the hypothesis that the dust and stellar continuum of SMG is co-centered cannot be ruled out ($p$-value as 0.909).
}
% \vspace{-0.2cm}
\label{fig:11_cocen}
\end{figure*}

% In this subsection we study the spatial offset between the stellar and dust continuum of lensed SMGs to investigate the underlying triggering mechanism of these dusty starbursts.

Through our IRAC morphological fitting in Section~\ref{ss:03b_photo}, the sky coordinates of 19 out of 26 ALMA sources were allowed to vary, so we compare these IRAC positions (average of 3.6/4.5\,\micron) with ALMA positions obtained from the \textit{uv}-plane fitting. 
We find that the mean offset between ALMA and IRAC sources is $0\farcs05\pm0\farcs05$ in RA and $0\farcs00\pm 0\farcs04$ in DEC. 
Assuming an uncertainty of 0\farcs2 with IRAC images (i.e., $\sim10\%$ of PSF FWHM), the expected standard error of the dust/stellar offset is $\sim$0\farcs05 if the stellar and dust continua are co-centrally distributed,
matching our result.

We compare our lensed SMGs with unlensed sample compiled in \citet{fujimoto17}. In their work, the authors conducted a large-sample analysis of $\sim$1\,mm ALMA sources on the \textit{uv}-plane, and a significant number of sources are also covered in the \hst\ CANDELS fields \citep{vdw12}, enabling precise measurements of the offset between the dust and stellar continua in SMGs at $z\sim 2.5 \pm 0.5$.
We cross-matched these two catalogs and identified the \hst\ counterparts of 73 SMGs with a maximum allowed separation of 1\arcsec. 
The spatial offset between the dust and stellar continuum center in each source is plotted in the left panel of Figure~\ref{fig:11_cocen}. 
% We obtained a median spatial offset of $0.16\pm0.06$\arcsec\ in our lensed sample, and $0.17\pm0.02$\arcsec\ in the unlensed sample of \citet{fujimoto17}.
The histograms of the dust/stellar offsets in these two SMG samples are shown in the middle panel of Figure~\ref{fig:11_cocen}, and the Gaussian fittings of these histograms suggest a standard deviation between dust and stellar continuum centroids as $0\farcs23\pm0\farcs07$ in our lensed sample, and $0\farcs27\pm0\farcs03$ in the sample of \citet{fujimoto17}.

Such a comparison suggests an intrinsically small spatial offset between the centroids of stellar/dust continua in lensed SMGs, since our offset measurements are amplified by lensing.
Though \hst\ has a higher angular resolution and thus a higher precision in determining the centroid of SMG in the optical/NIR bands, strong dust obscuration could affect the observed morphology of stellar continuum in \hst\ (e.g., \citealt{hodge16}). 
Such an effect can probably bias the direct spatial mapping between observed stellar light in the near-infrared and the intrinsic stellar mass \citep[also suggested by][]{lang19}, result in an underestimate of the uncertainty for stellar centroid determination in \hst\ images \citep[e.g.,][]{chen15}.
% Since IRAC 3.6/4.5\,\micron\ images are much less sensitive to the dust extinction, we are able to determine the stellar centroids of SMGs at a precision no less than \hst, especially when the effect of lensing magnification is also taken into account.

The only SMG showing offset larger than 1\arcsec\ is HLS0546, which is an extended SMG ($R_\mathrm{e,dust}=1\farcs7\pm0\farcs3$) with high S\'{e}rsic index ($n\sim2.3$) for stellar component.
These features may suggest that HLS0546 has already entered a late phase of SMG evolution, showing cuspy stellar component at its center and ongoing star formation at outskirts as discussed in Section~\ref{sss:05b_iiiquench}. 
This can cause the observed offset if the star-forming regions are not symmetrically distributed in the galaxy disk plane.

We further show that such a distribution of dust/stellar offsets may be primarily due to astrometric errors through a K-S test. % to the distribution of these spatial offsets relative to their uncertainties, in comparison with the Gaussian distribution. 
We keep the assumption of the stellar position uncertainty as 0\farcs2 for IRAC images, and the uncertainties of dust continua are determined during the \textit{uv}-plane fitting (median value is 0\farcs04).
We cannot rule out the null hypothesis that the sample of spatial offsets relative to astrometric errors are drawn from a Gaussian distribution (\textit{p}-value as 0.909). 
Assuming a slightly lower IRAC astrometric error (e.g., 0\farcs15) will not change the conclusion.
Since the co-centered distribution of stellar and dust continuum will result in a Gaussian distribution of spatial offsets relative to their uncertainty, this K--S test demonstrates that we cannot rule out the co-centered distributing scenario between the ongoing star formation and the evolved stellar population in $z\sim 2$ SMGs. 
% We therefore conclude that 
% there is no measurable offset between the IRAC and ALMA source positions for the observed lensed SMGs.

%% file: 98d_thumbnail.tex
\begin{figure}[!thb]
\centering
\includegraphics[width=\linewidth]{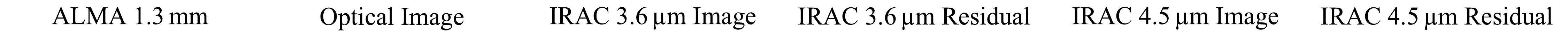}
\includegraphics[width=\linewidth]{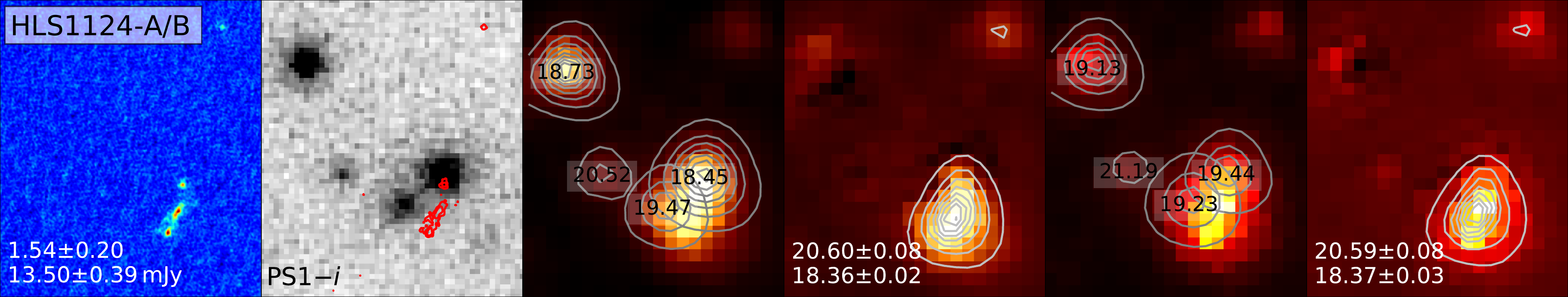}
\includegraphics[width=\linewidth]{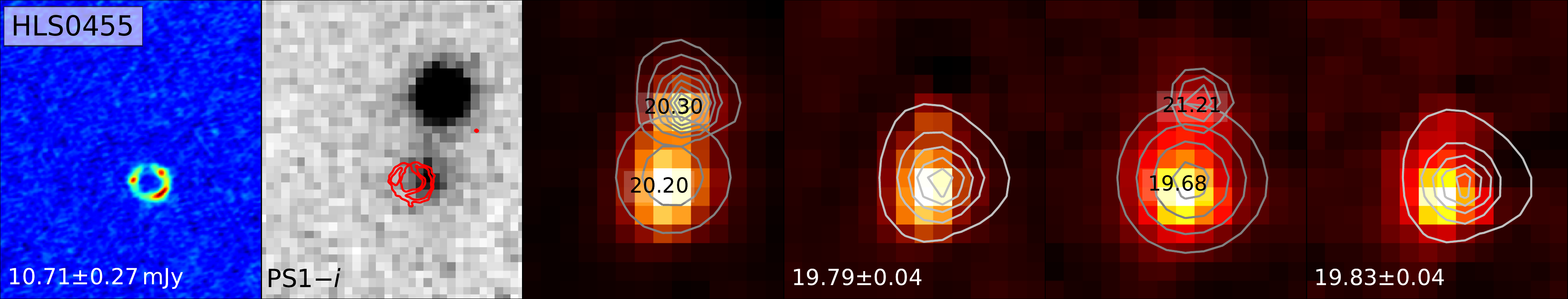}
\includegraphics[width=\linewidth]{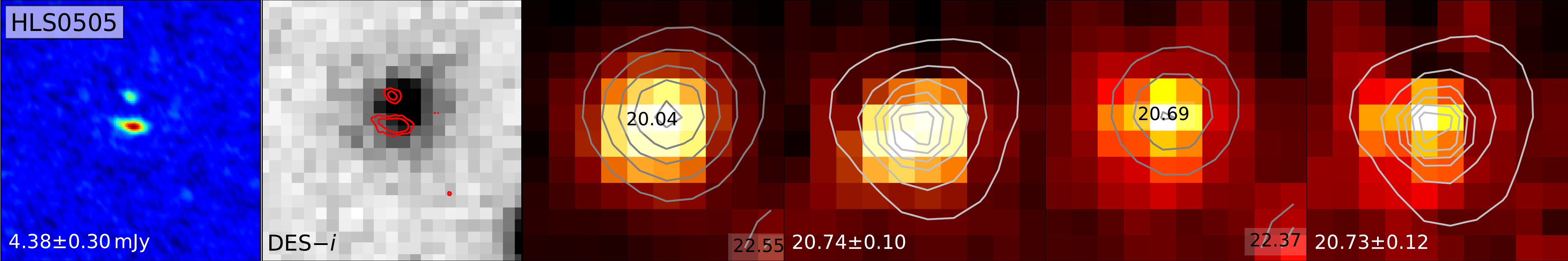}
\includegraphics[width=\linewidth]{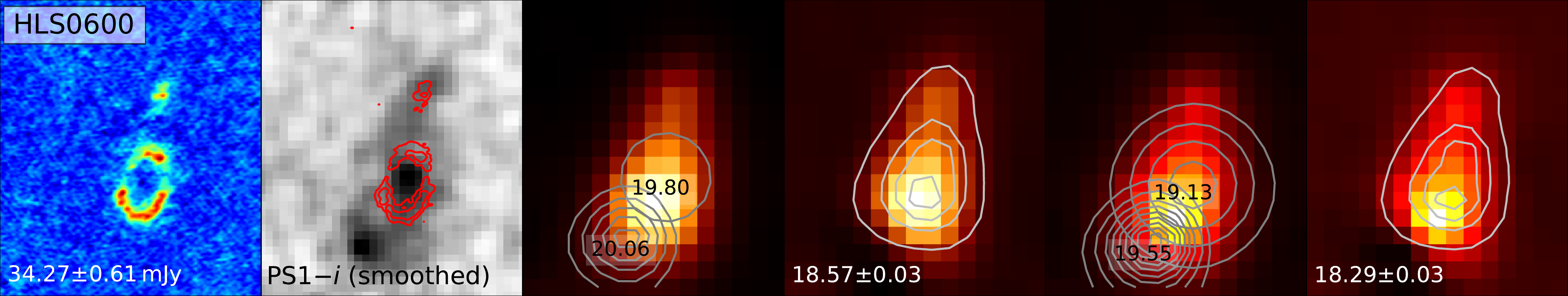}
\caption{Illustration of the IRAC photometry of five highly-blended SMGs with complex morphology.
Note we define two sources in CODEX\,39326 (first row), namely HLS1124-A (the upper-right faint one) and HLS1124-B (the lower bright one with three sub-components in ALMA map), both at $z=1.80$ with CO(6-5) detection (Sun et al., in prep.).
We display their Briggs-weighted ALMA 1.3\,mm continuum images with their flux densities at first column, optical images with ALMA contours (levels: 4, 10$\sigma$) at second column. 
Original IRAC 3.6/4.5\,\micron\ images are displayed at Column 3 and 5, with the best-fit optical source models shown in contours and magnitudes shown in text.
Residual images after nearby source subtraction are shown at Column 4 and 6, with ALMA-model contours and measured magnitudes in text.
}
\label{fig:02_sunfit}
\end{figure}
\begin{figure}[!htp]
\centering
\includegraphics[width=0.49\linewidth]{figures/thumbnail/header.pdf}
\includegraphics[width=0.49\linewidth]{figures/thumbnail/header.pdf}
\includegraphics[width=0.49\linewidth]{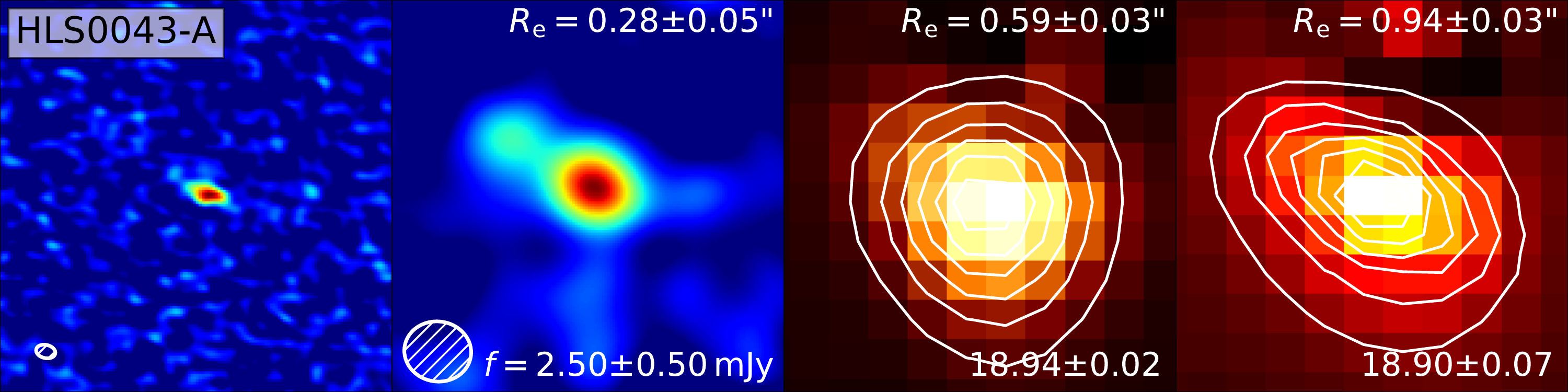}
\includegraphics[width=0.49\linewidth]{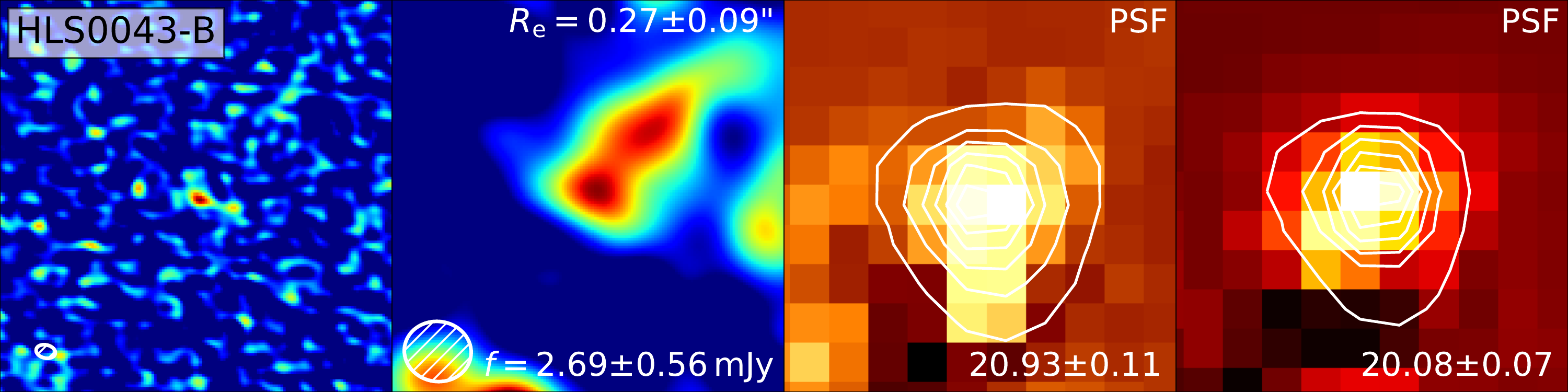}
\includegraphics[width=0.49\linewidth]{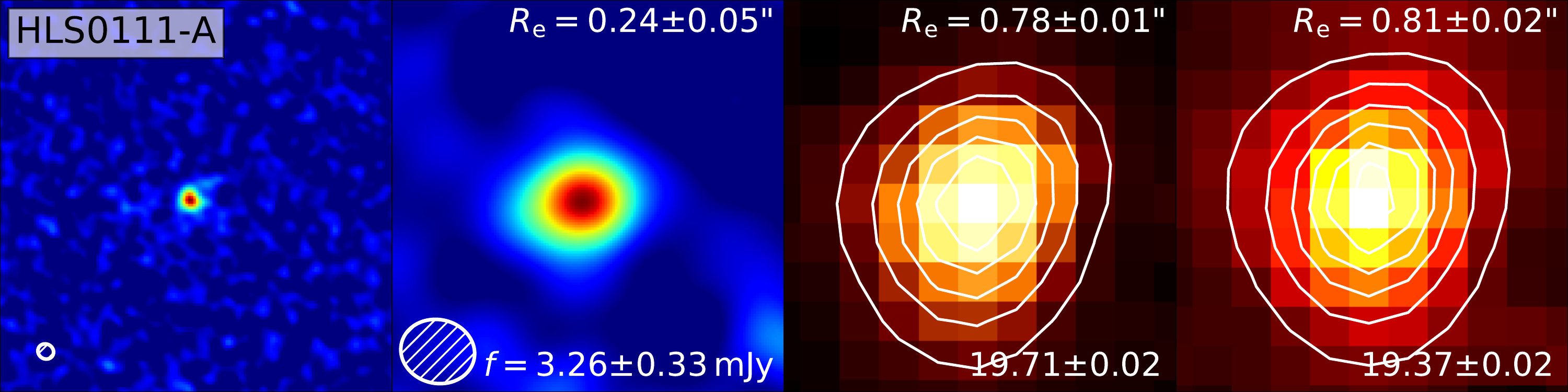}
\includegraphics[width=0.49\linewidth]{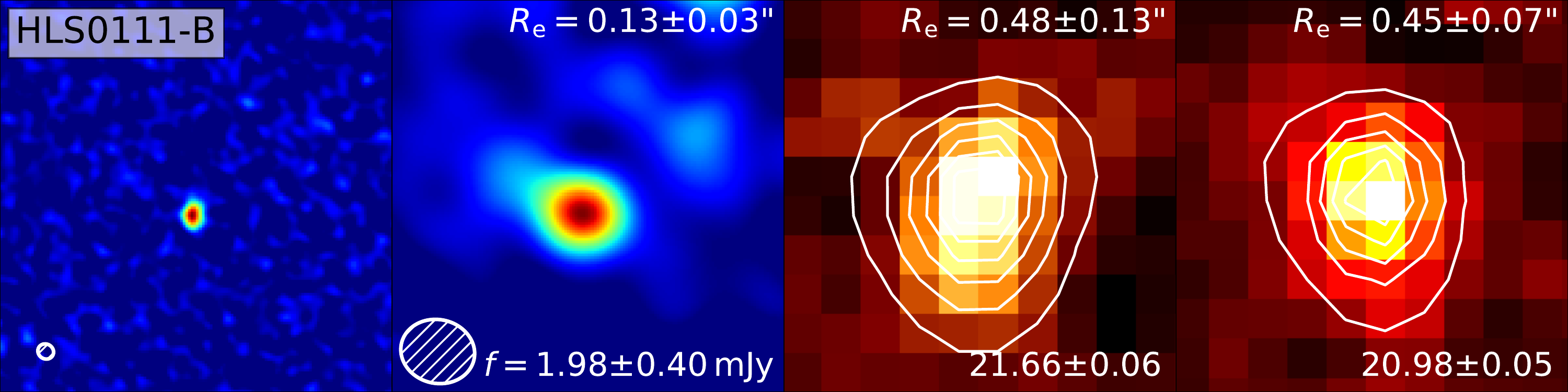}
\includegraphics[width=0.49\linewidth]{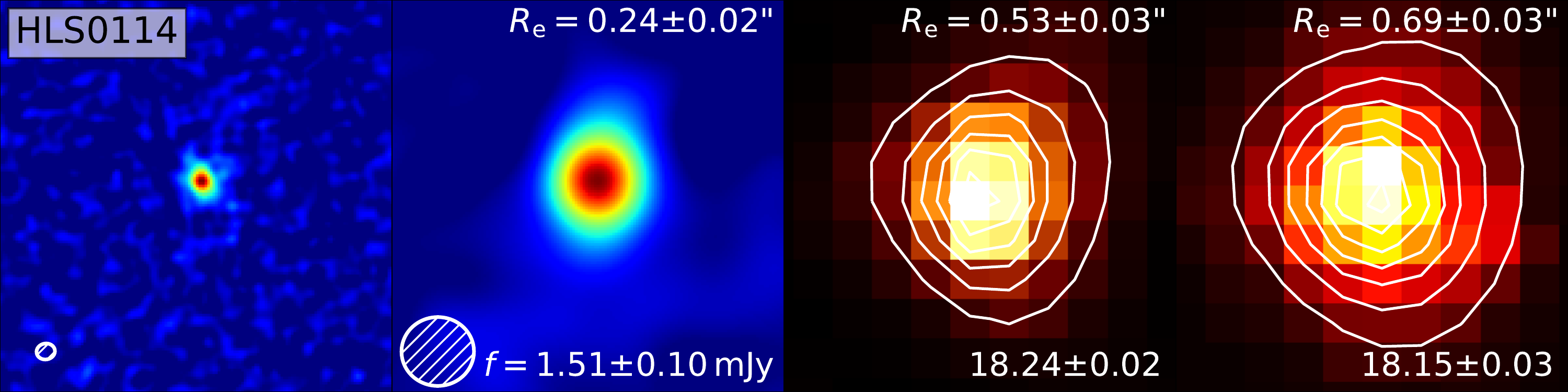}
\includegraphics[width=0.49\linewidth]{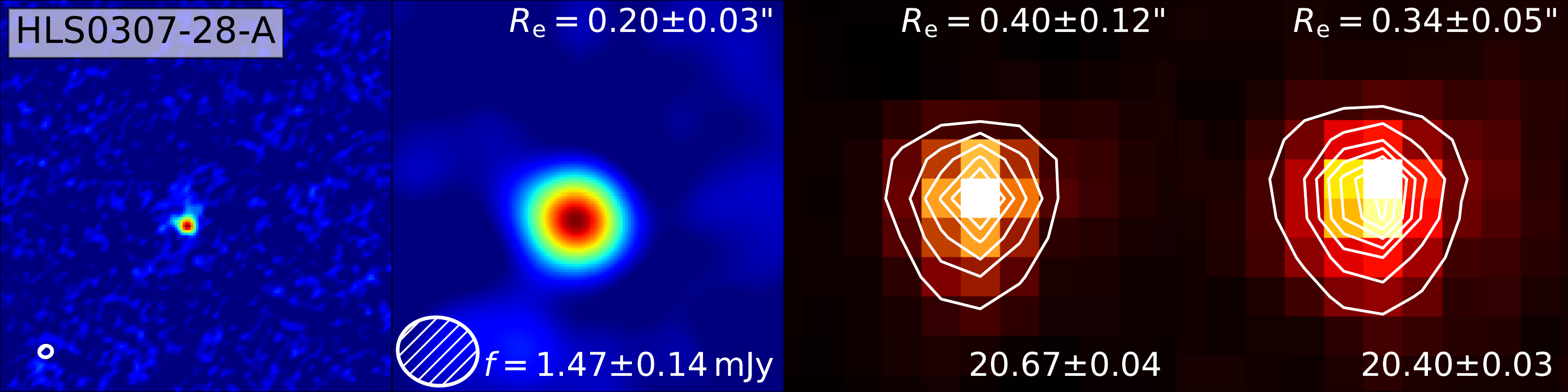}
\includegraphics[width=0.49\linewidth]{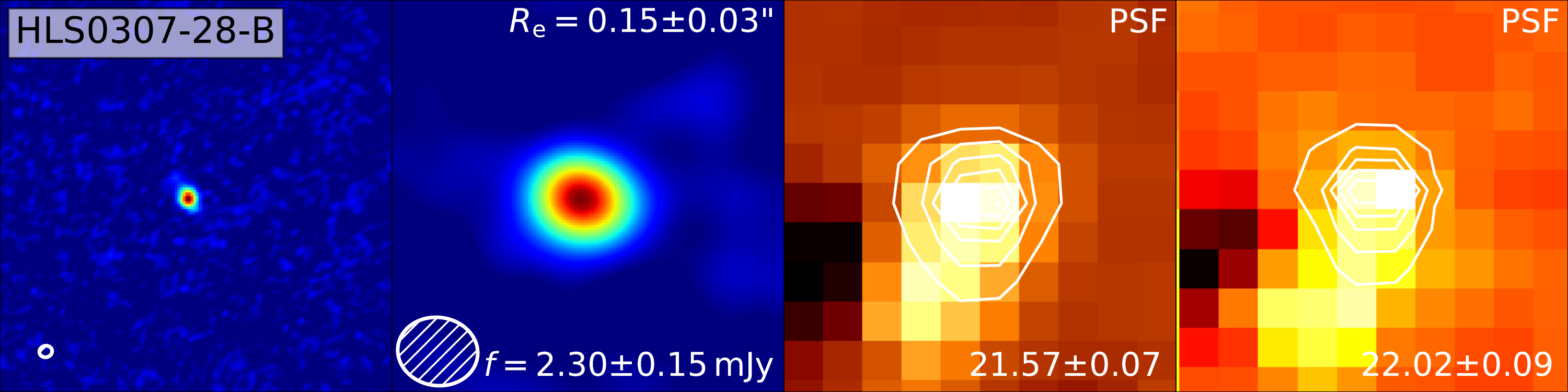}
\includegraphics[width=0.49\linewidth]{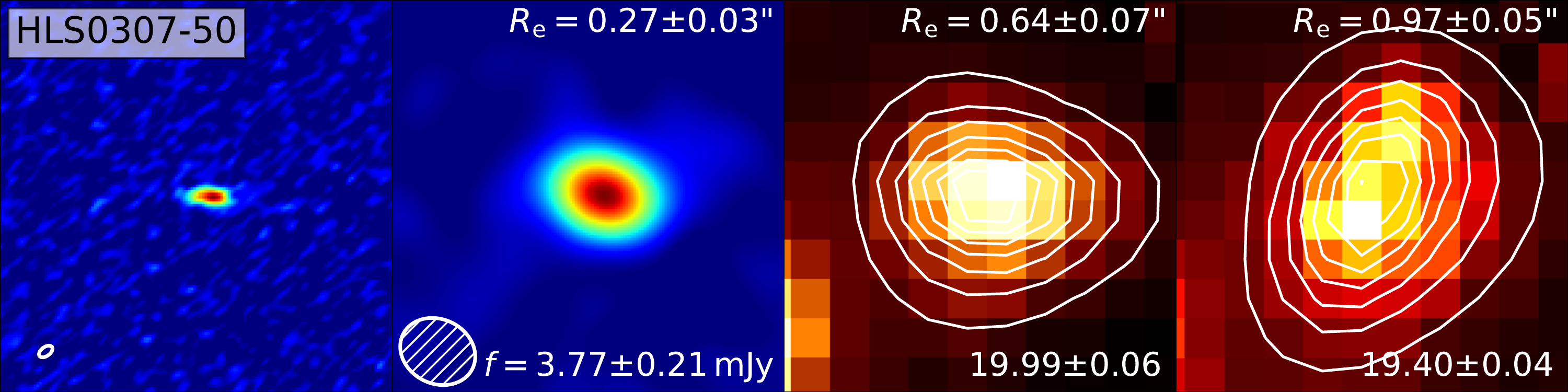}
\includegraphics[width=0.49\linewidth]{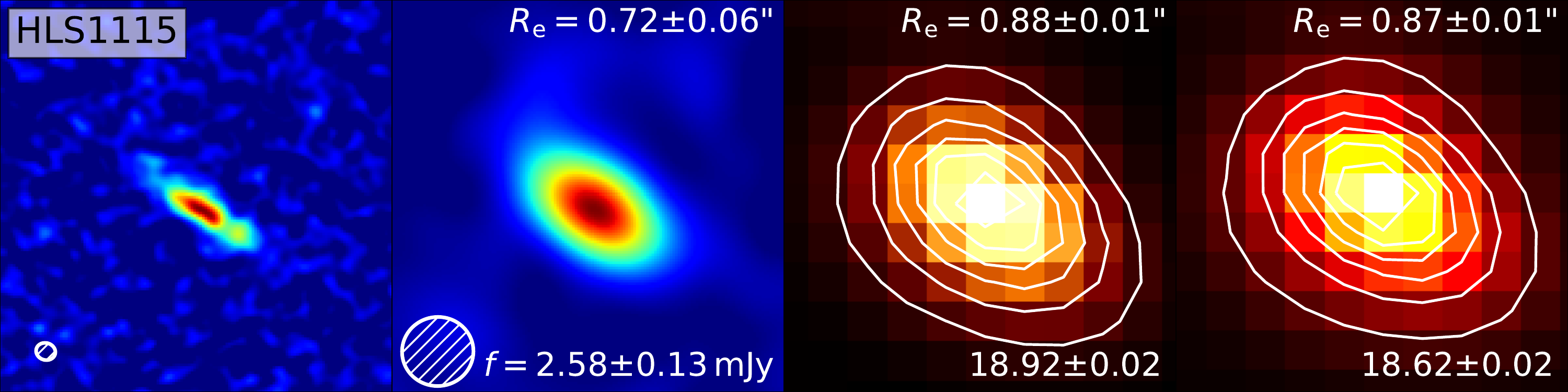}
\includegraphics[width=0.49\linewidth]{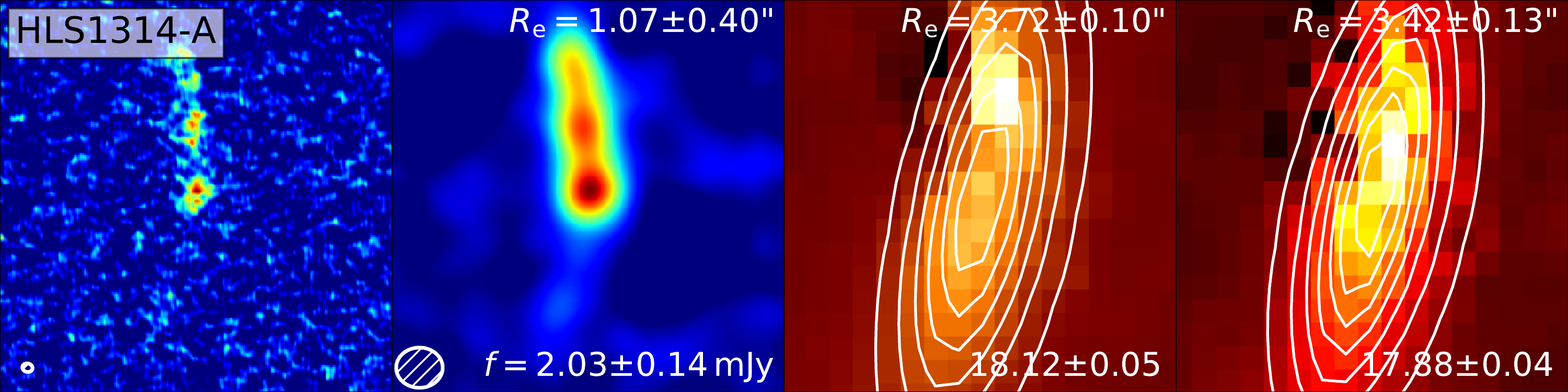}
\includegraphics[width=0.49\linewidth]{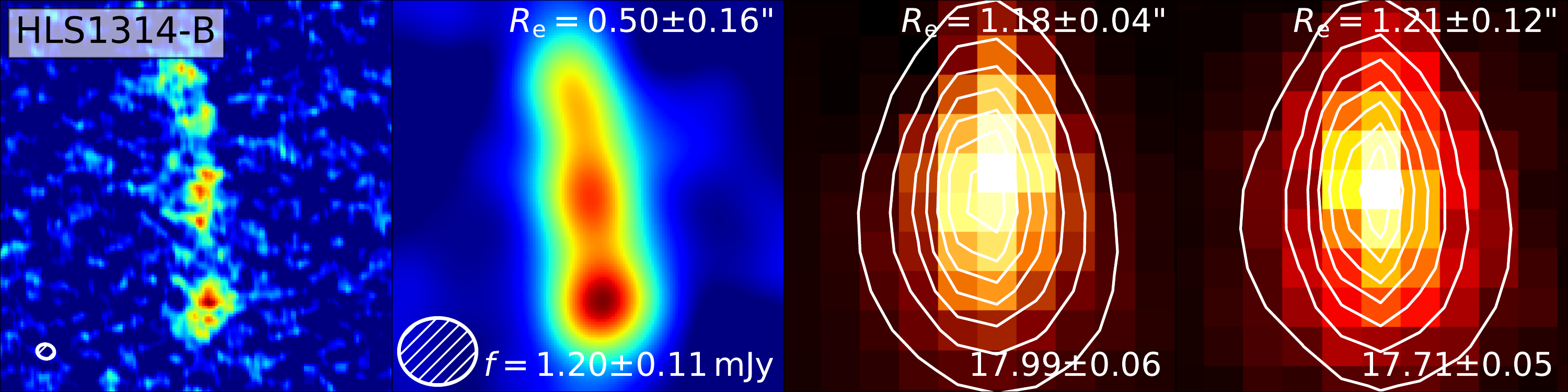}
\includegraphics[width=0.49\linewidth]{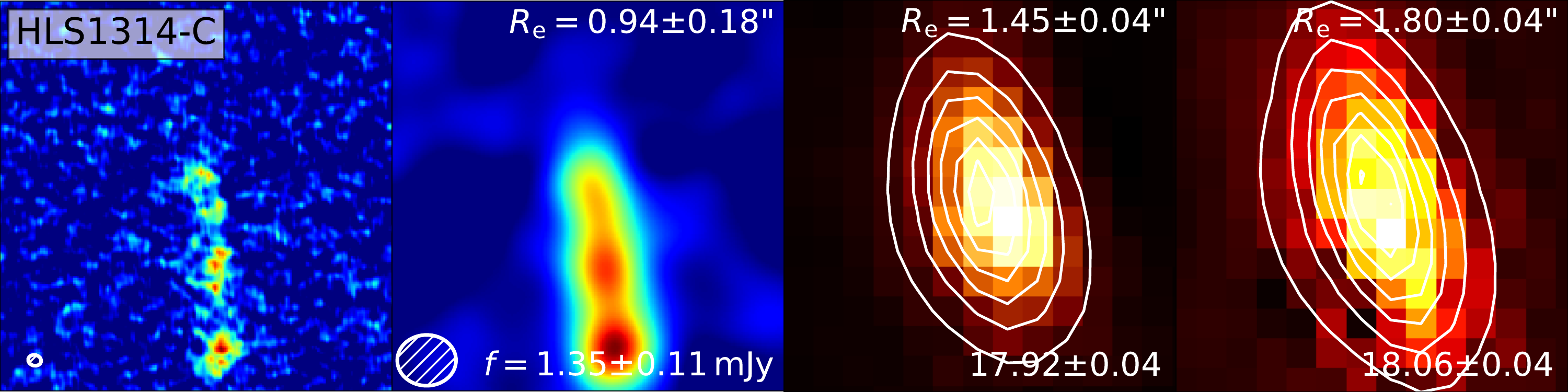}
\includegraphics[width=0.49\linewidth]{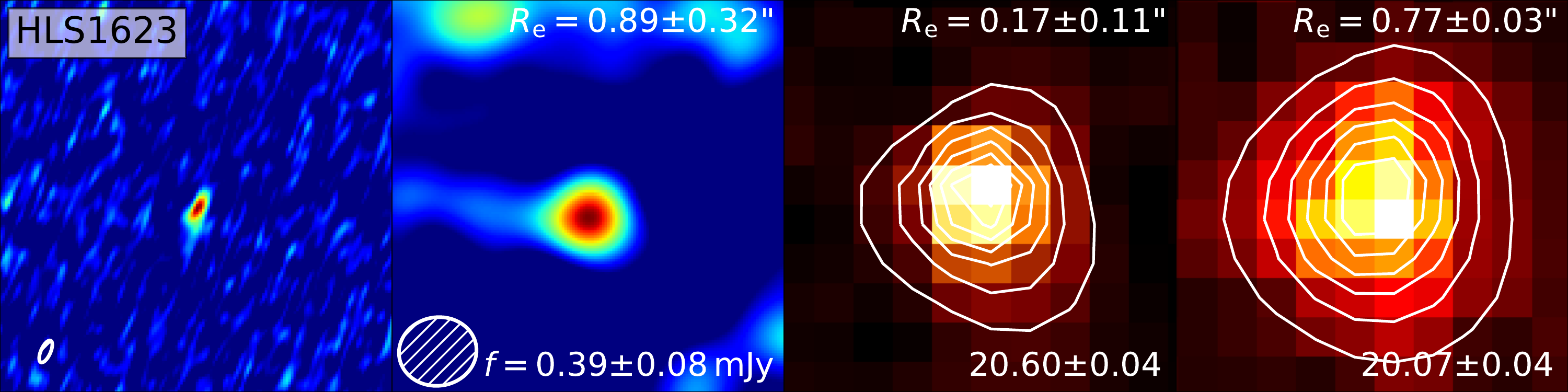}
\includegraphics[width=0.49\linewidth]{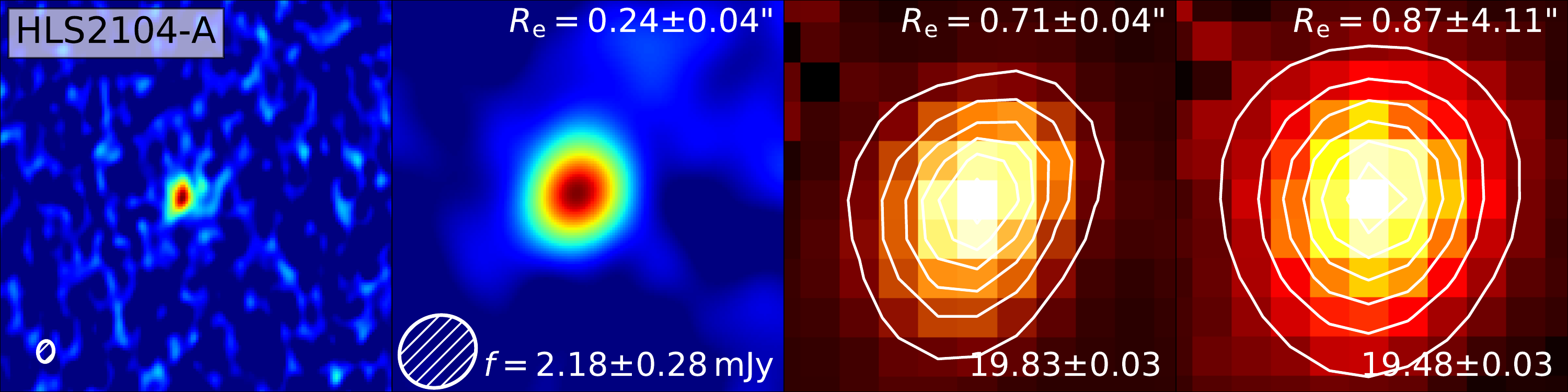}
\includegraphics[width=0.49\linewidth]{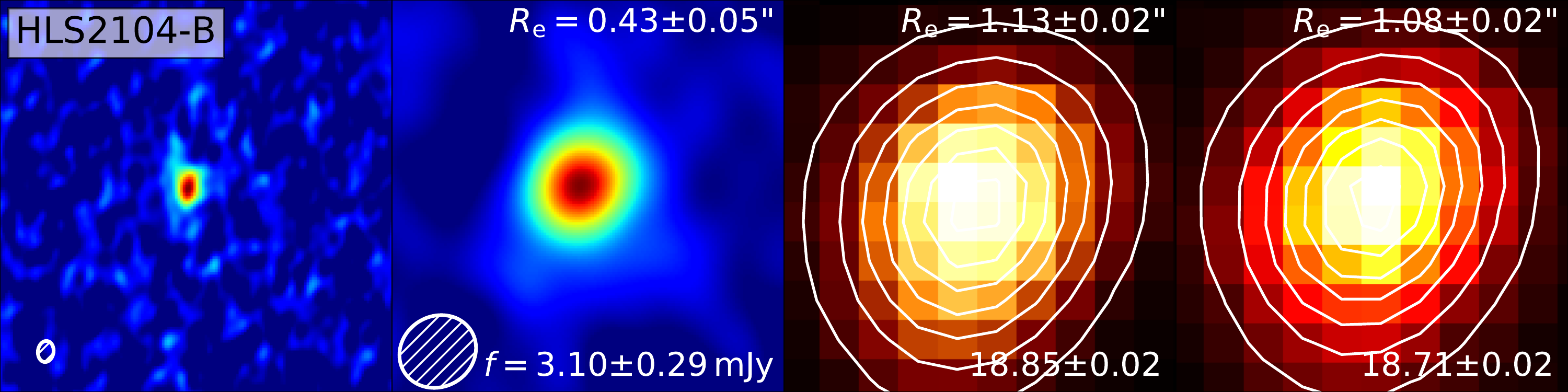}
\includegraphics[width=0.49\linewidth]{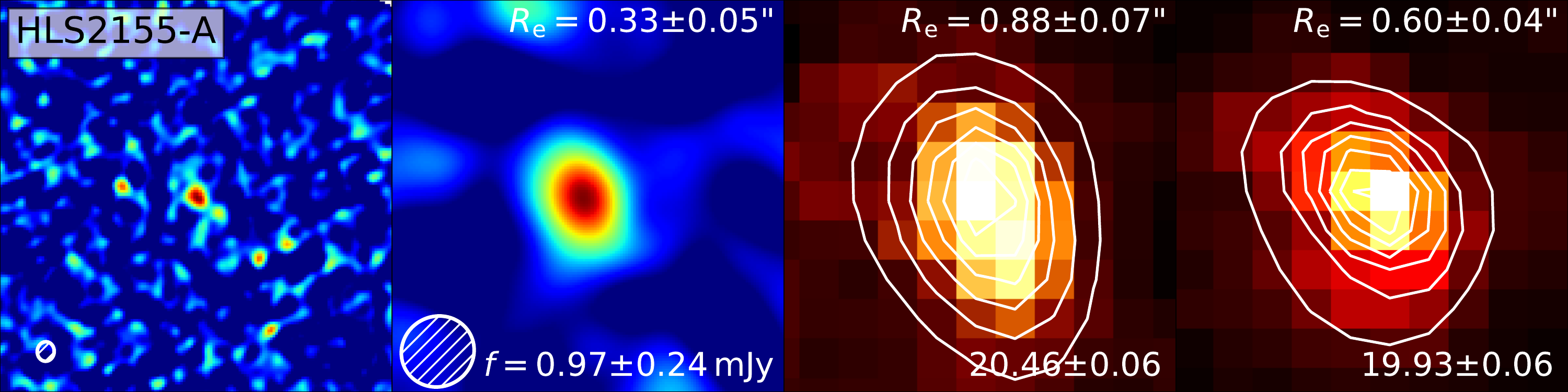}
\includegraphics[width=0.49\linewidth]{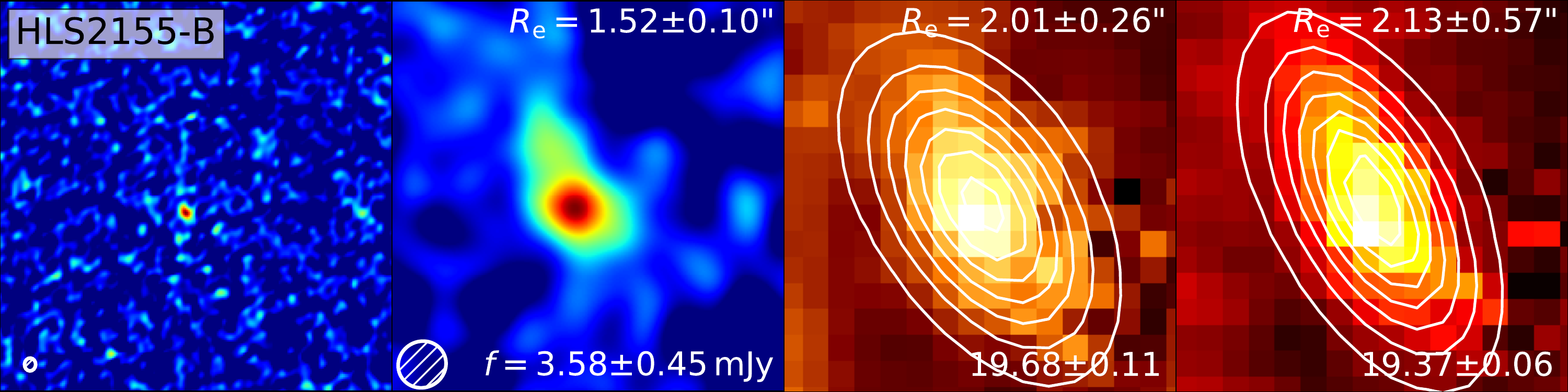}
\includegraphics[width=0.49\linewidth]{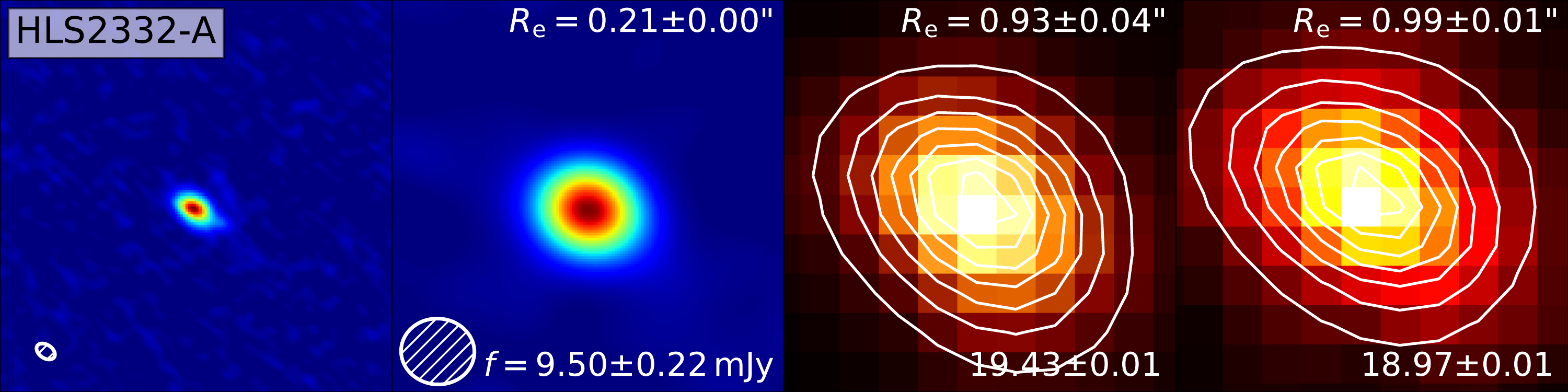}
\includegraphics[width=0.49\linewidth]{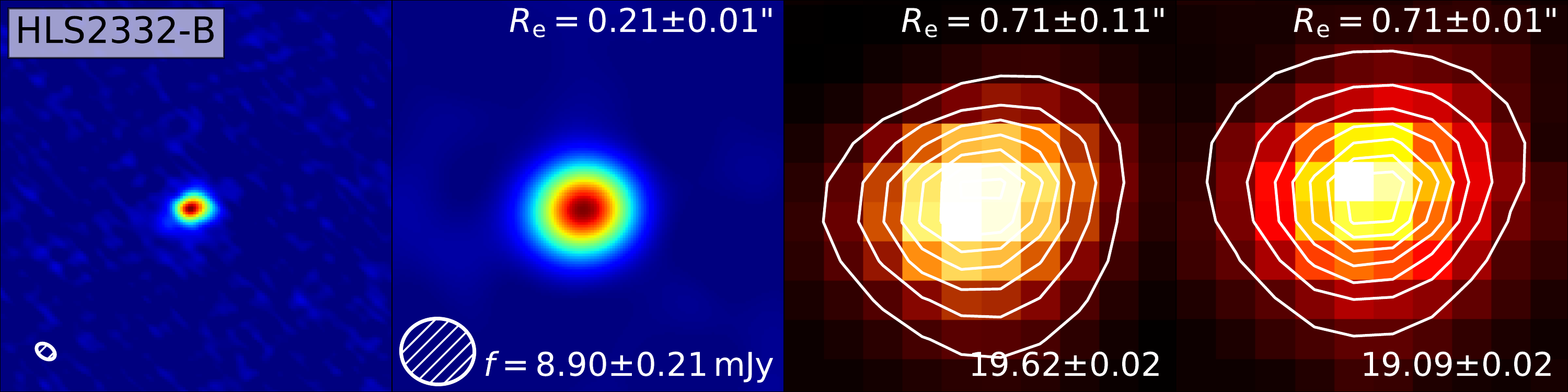}
\includegraphics[width=0.49\linewidth]{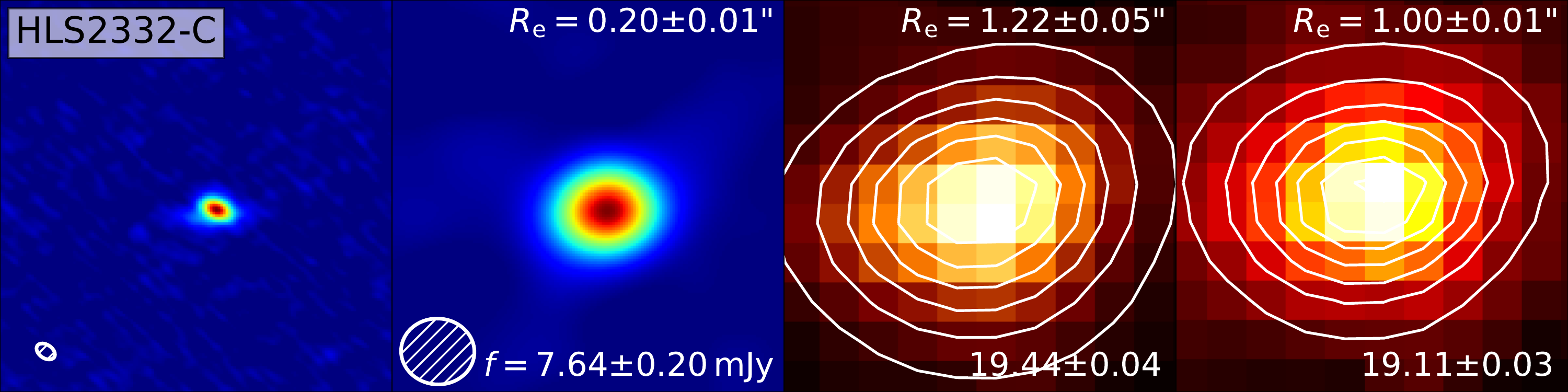}
\caption{Postage stamp images of the remaining 20 cluster-lensed SMGs in this work. 
The layouts of images are the same as those in Figure~\ref{fig:3_cutout}.
Note that we do not subtract other sources when we display the ALMA images of HLS1314-A/B/C, and therefore three components can be identified in their 1\arcsec-tapered ALMA images.
% The galaxies are at the center of each image.
% For each source, we show their ALMA 1.3\,mm map with Briggs weighting in the first column (resolution $\Delta \theta \sim 0.2$\arcsec), and uv-tapered map in the second column ($\Delta \theta \sim 1$\arcsec; $2$\arcsec-tapered for HLS0546 and HLS0840). 
% The synthesized ALMA beam is shown as a hatched ellipse at the lower-left corner, and the source sizes and fluxes are noted in the tapered images.
% IRAC 3.6/4.5\,\micron\ maps after the neighborhood subtraction with \textsc{GALFIT} are shown in the third and forth columns. The best-fit models, convolved with the PSFs, are overlaid as white contours. The source sizes and fluxes/magnitudes are noted in the upper-right and lower-right corners of tapered ALMA images and IRAC images.
}
\label{fig:apd_01}
\end{figure}
\begin{figure*}[!htp]
\centering
\includegraphics[width=0.24\linewidth]{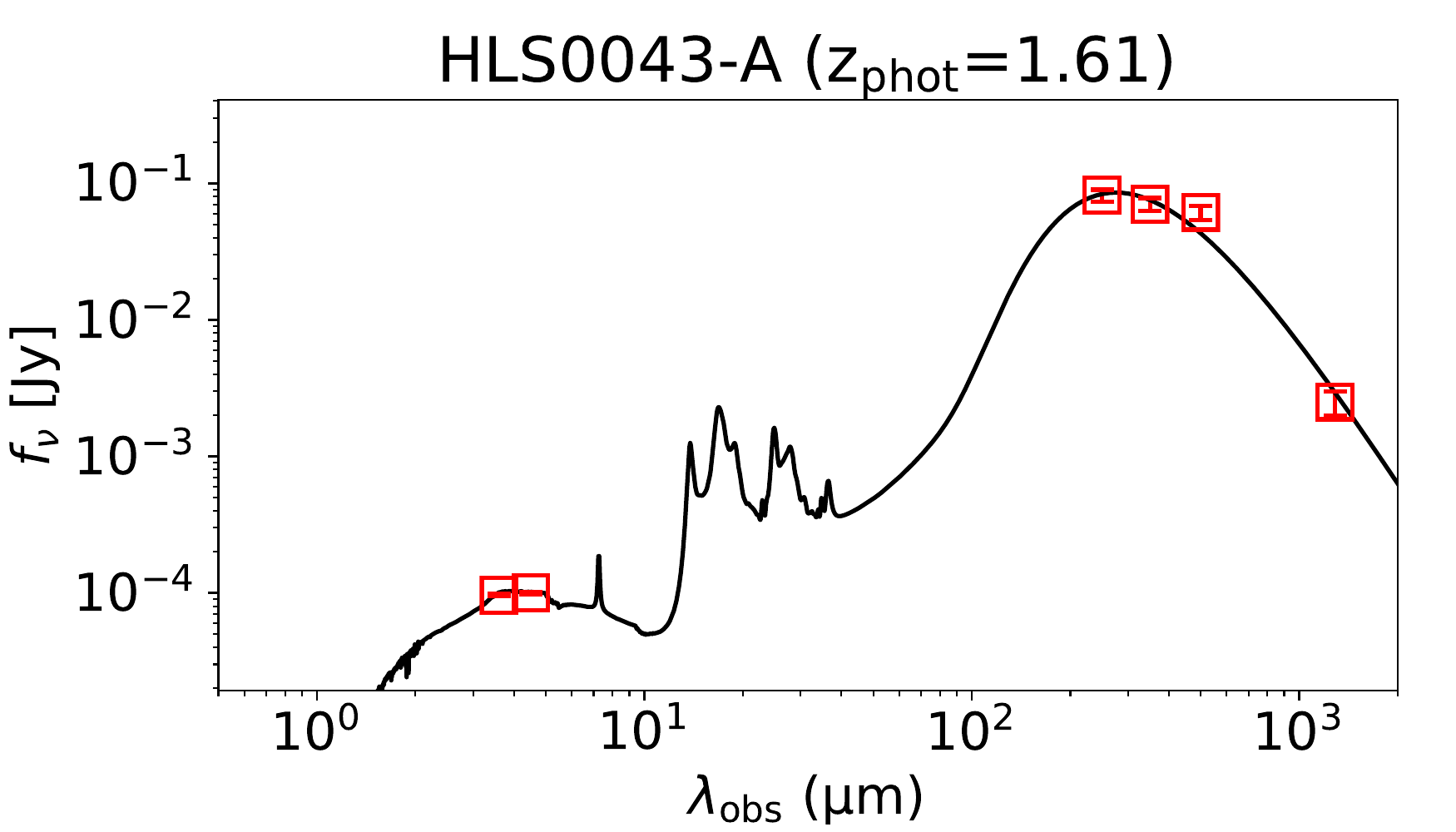}
\includegraphics[width=0.24\linewidth]{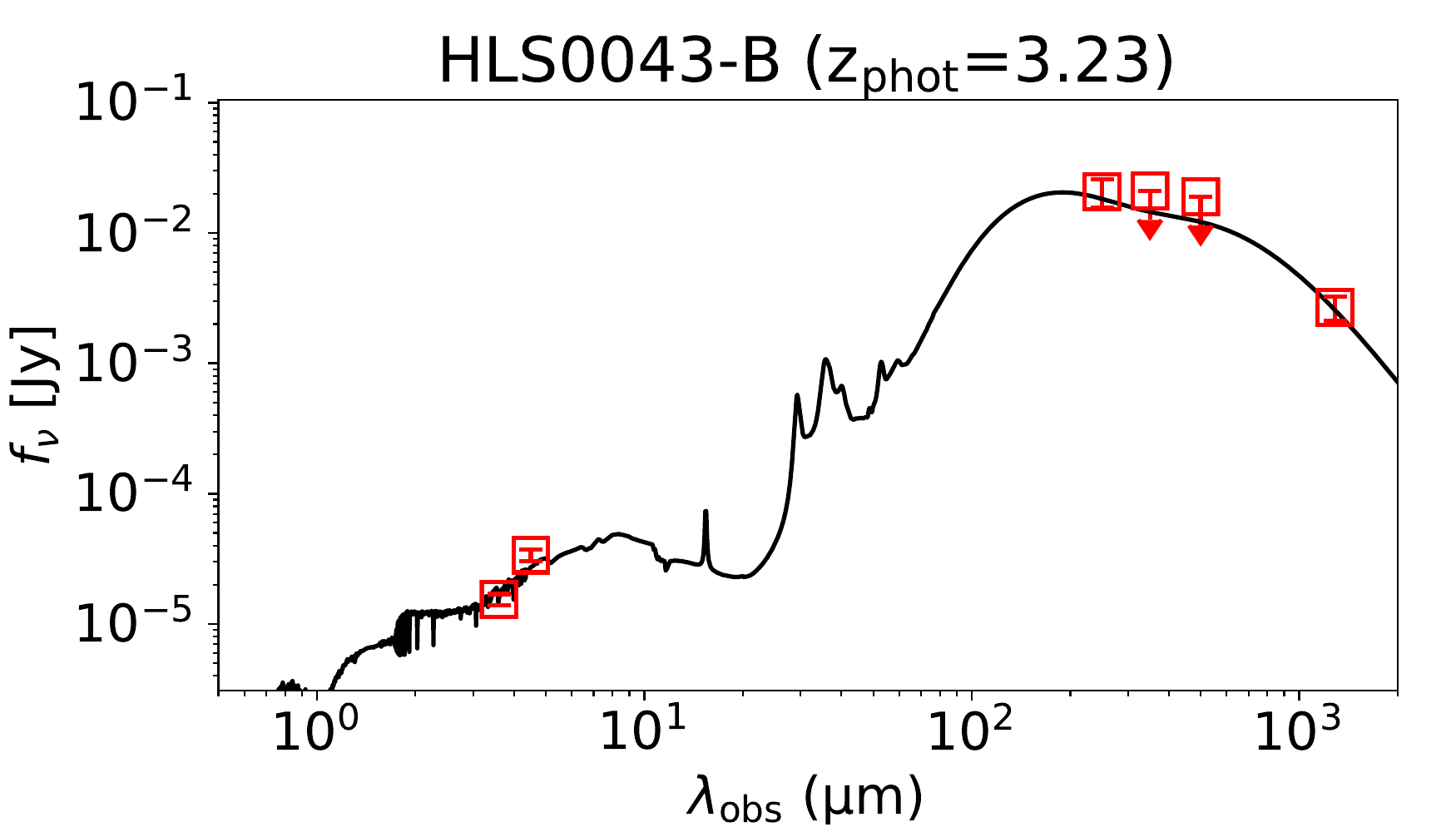}
\includegraphics[width=0.24\linewidth]{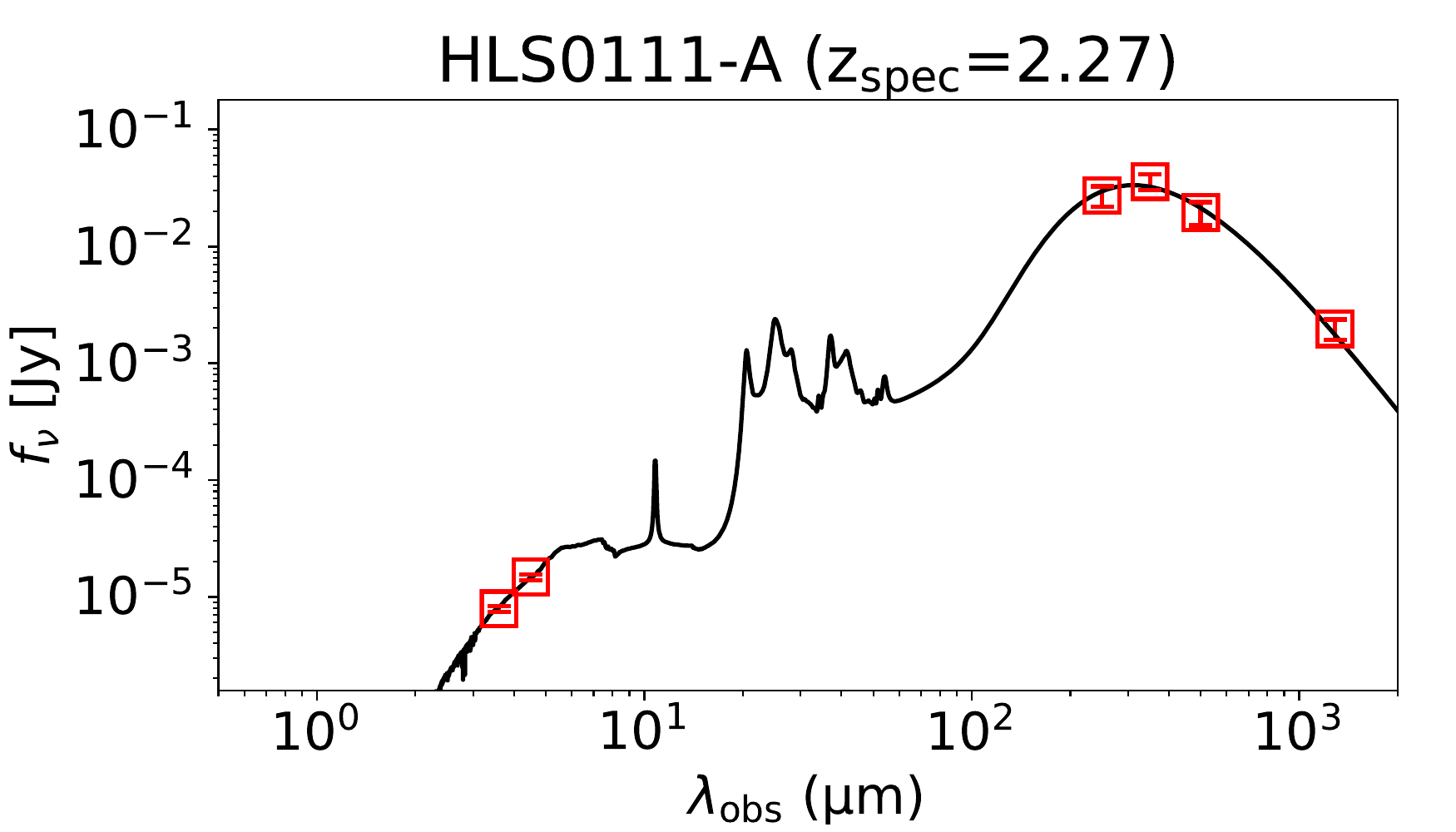}
\includegraphics[width=0.24\linewidth]{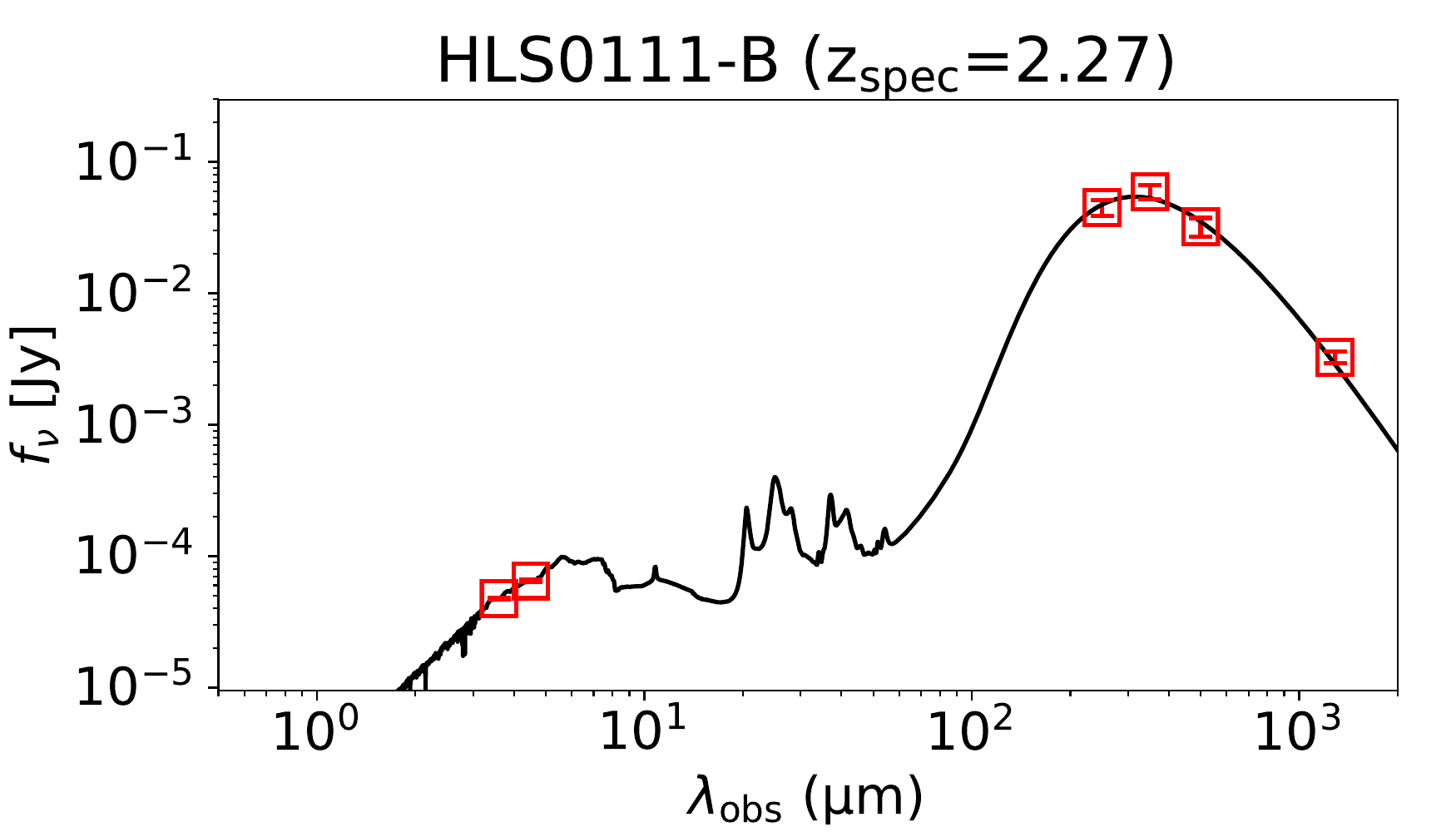}
\includegraphics[width=0.24\linewidth]{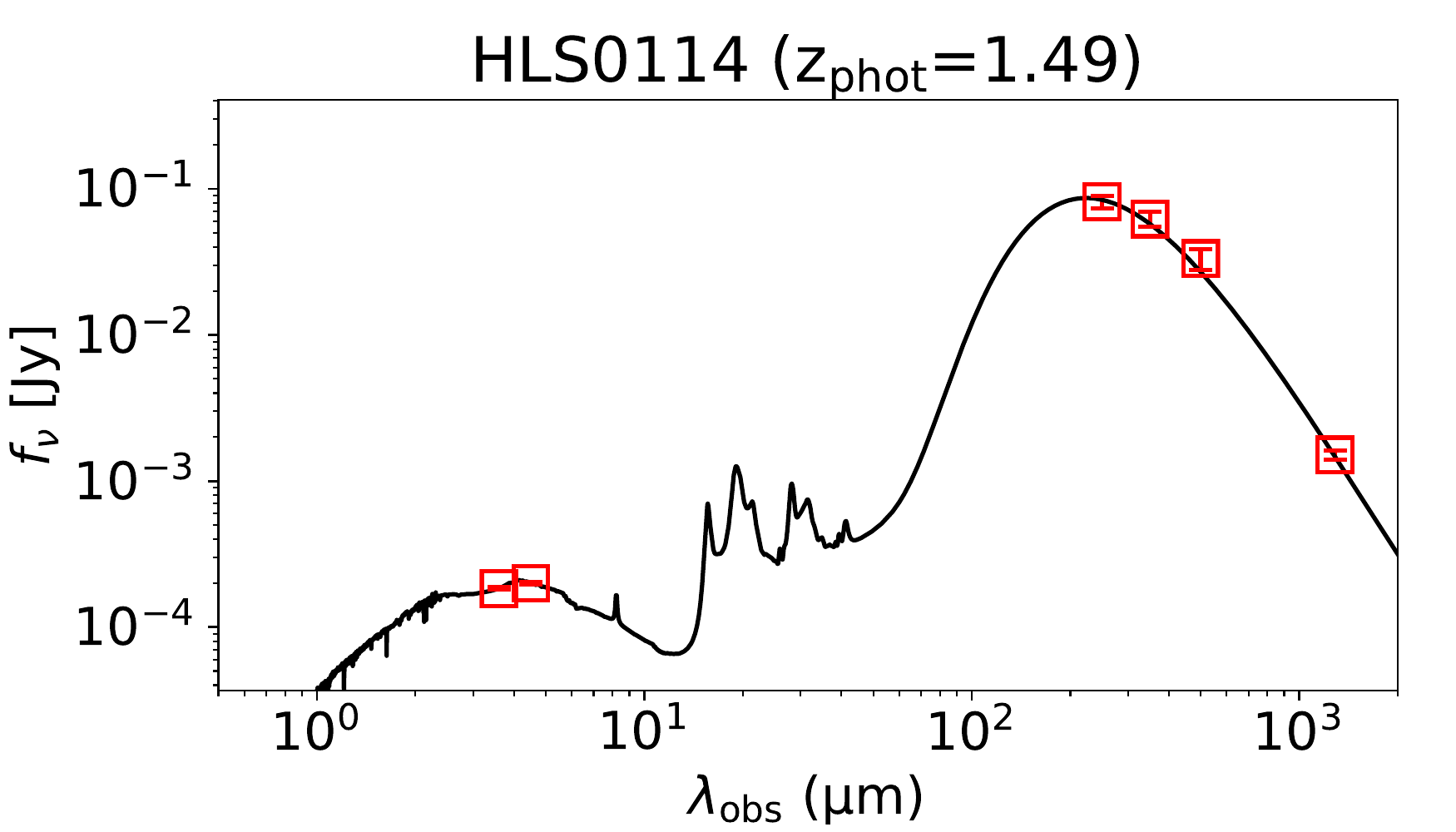}
\includegraphics[width=0.24\linewidth]{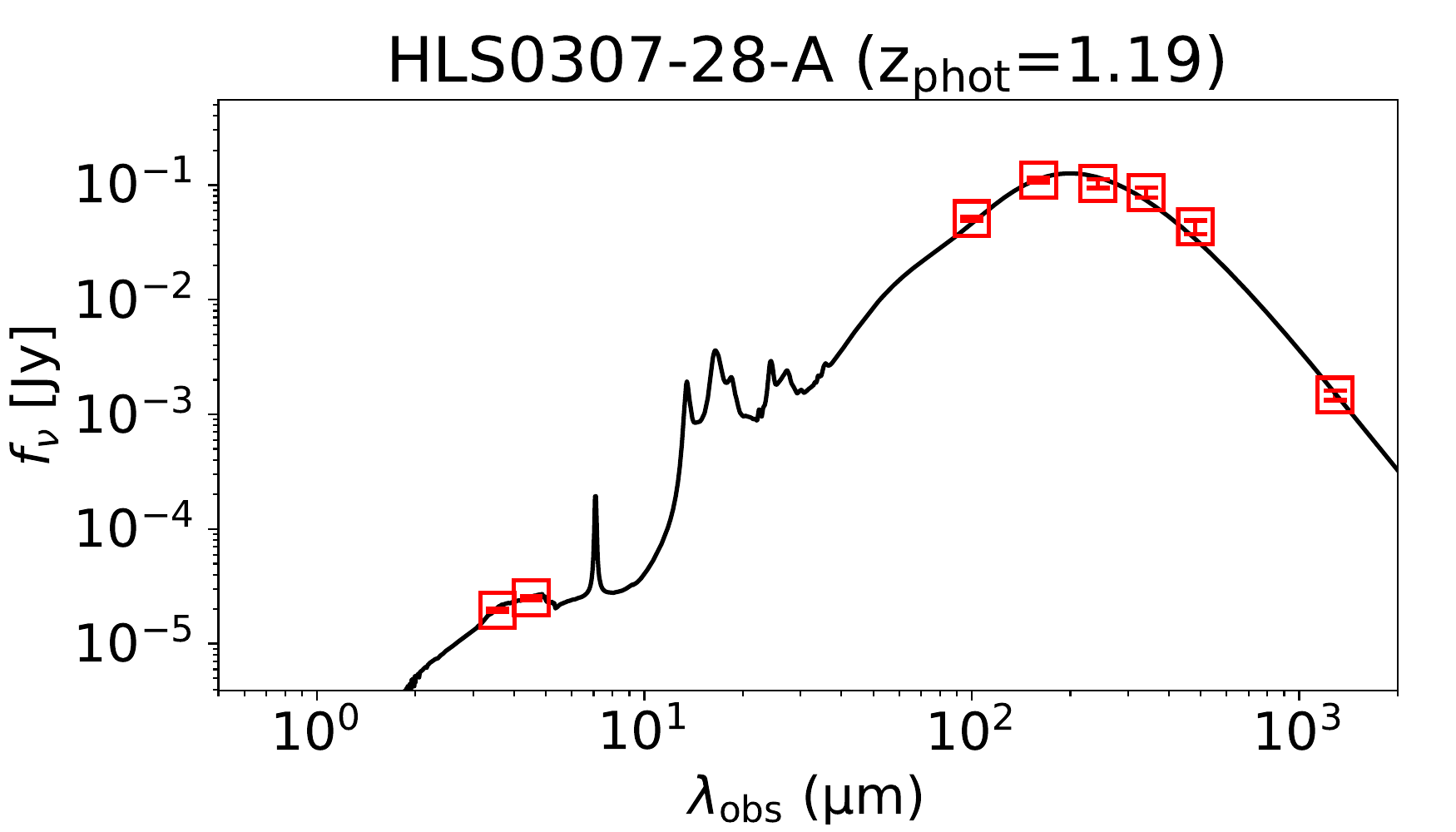}
\includegraphics[width=0.24\linewidth]{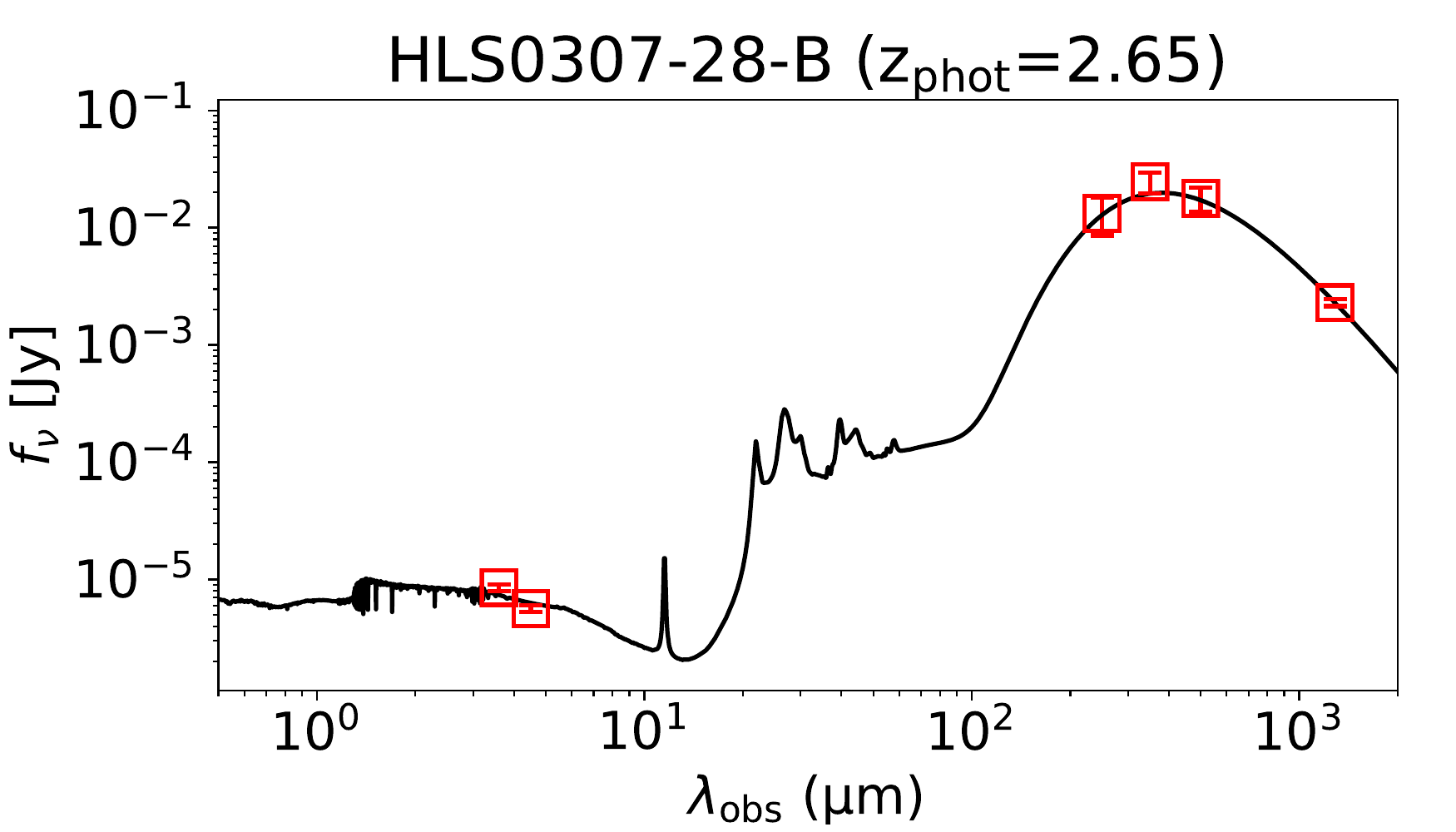}
\includegraphics[width=0.24\linewidth]{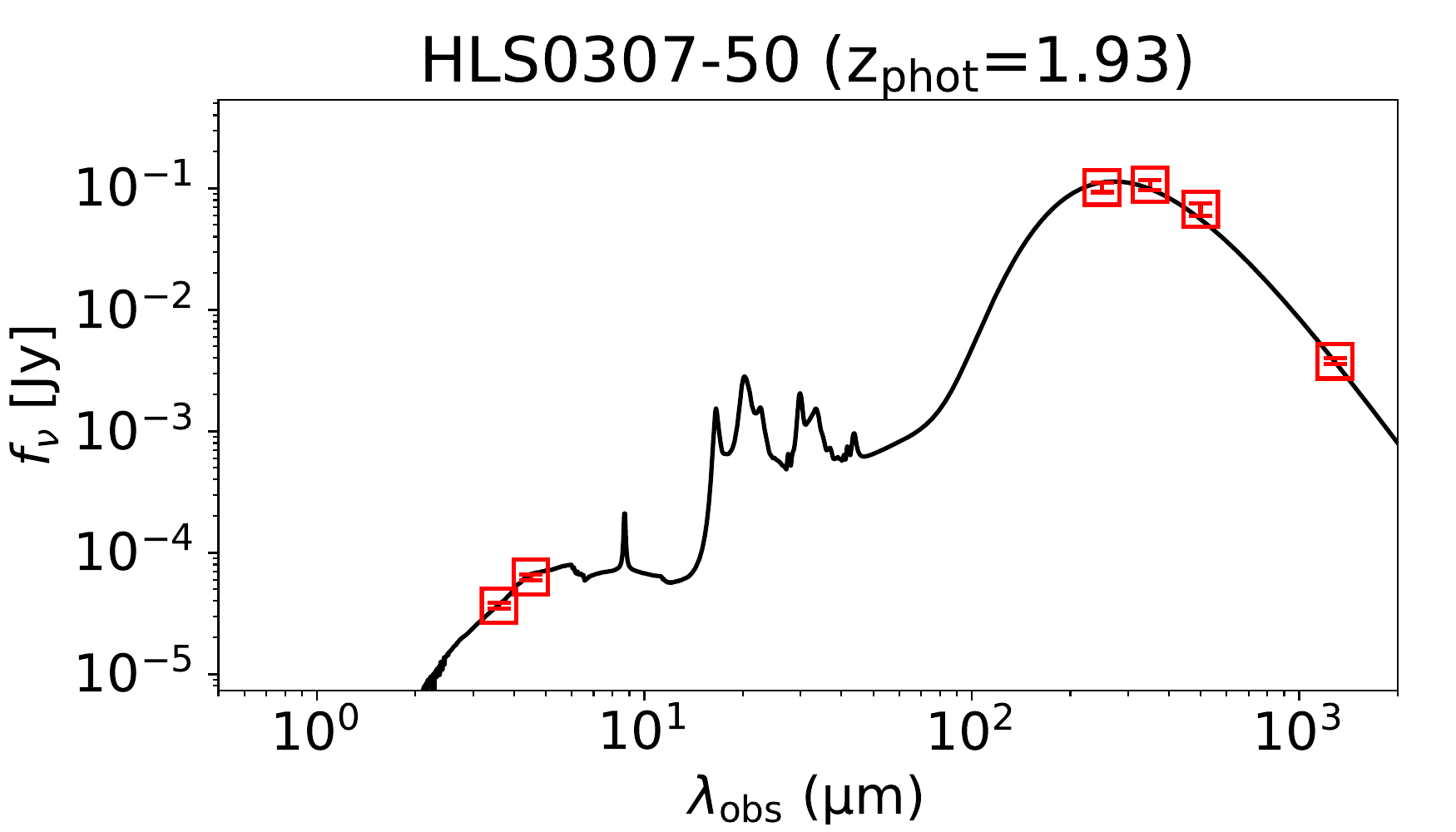}
\includegraphics[width=0.24\linewidth]{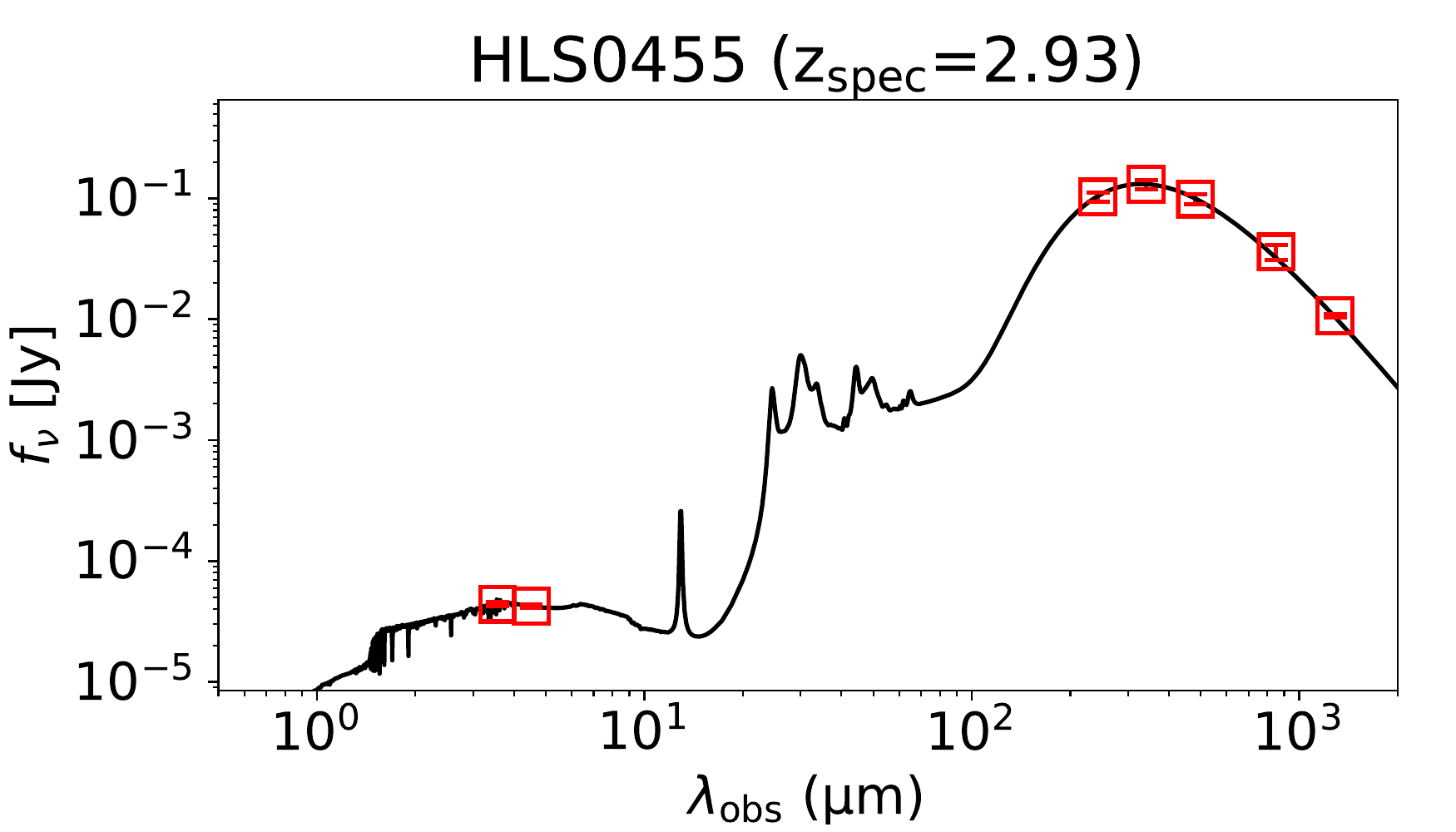}
\includegraphics[width=0.24\linewidth]{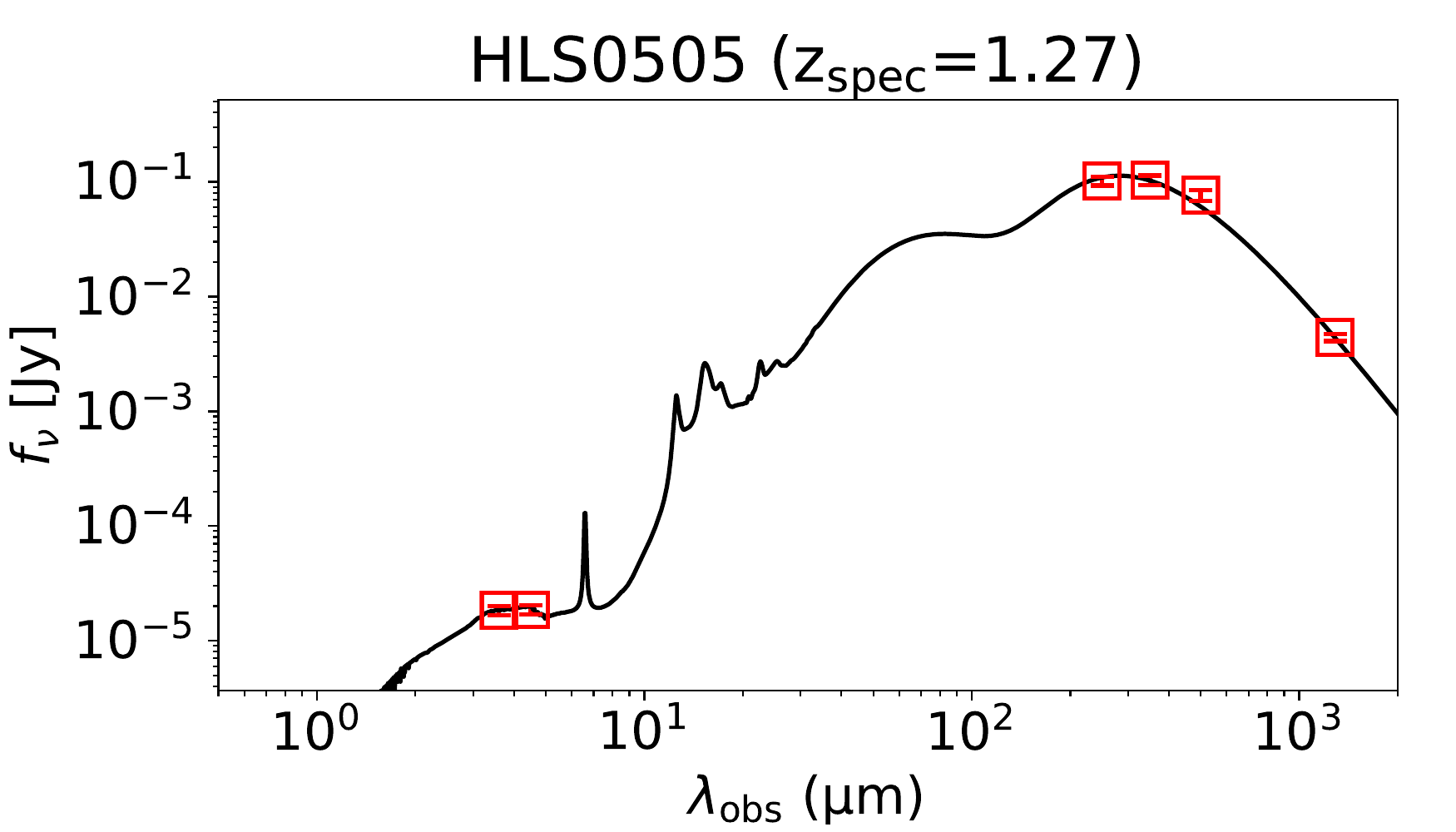}
\includegraphics[width=0.24\linewidth]{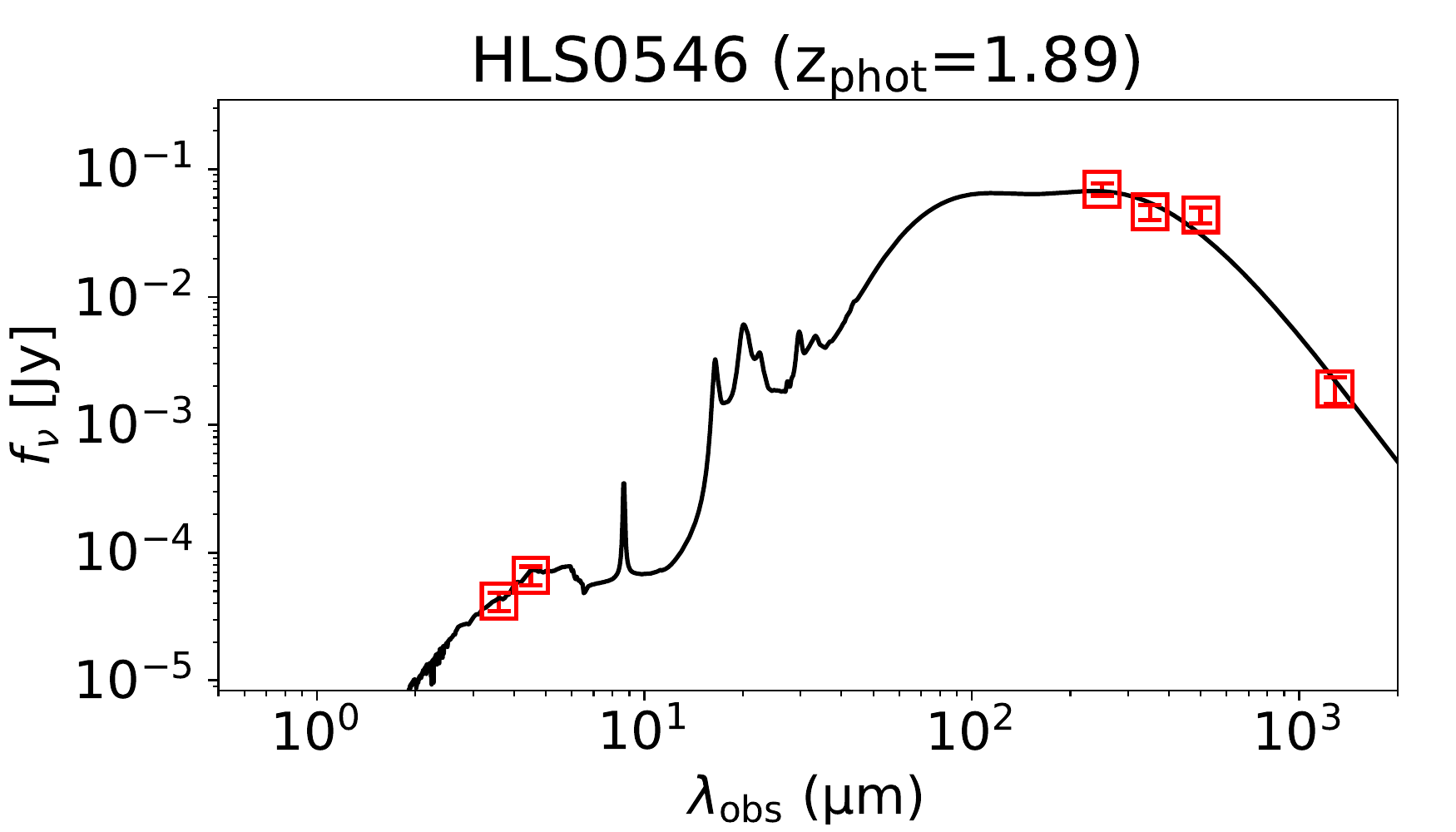}
\includegraphics[width=0.24\linewidth]{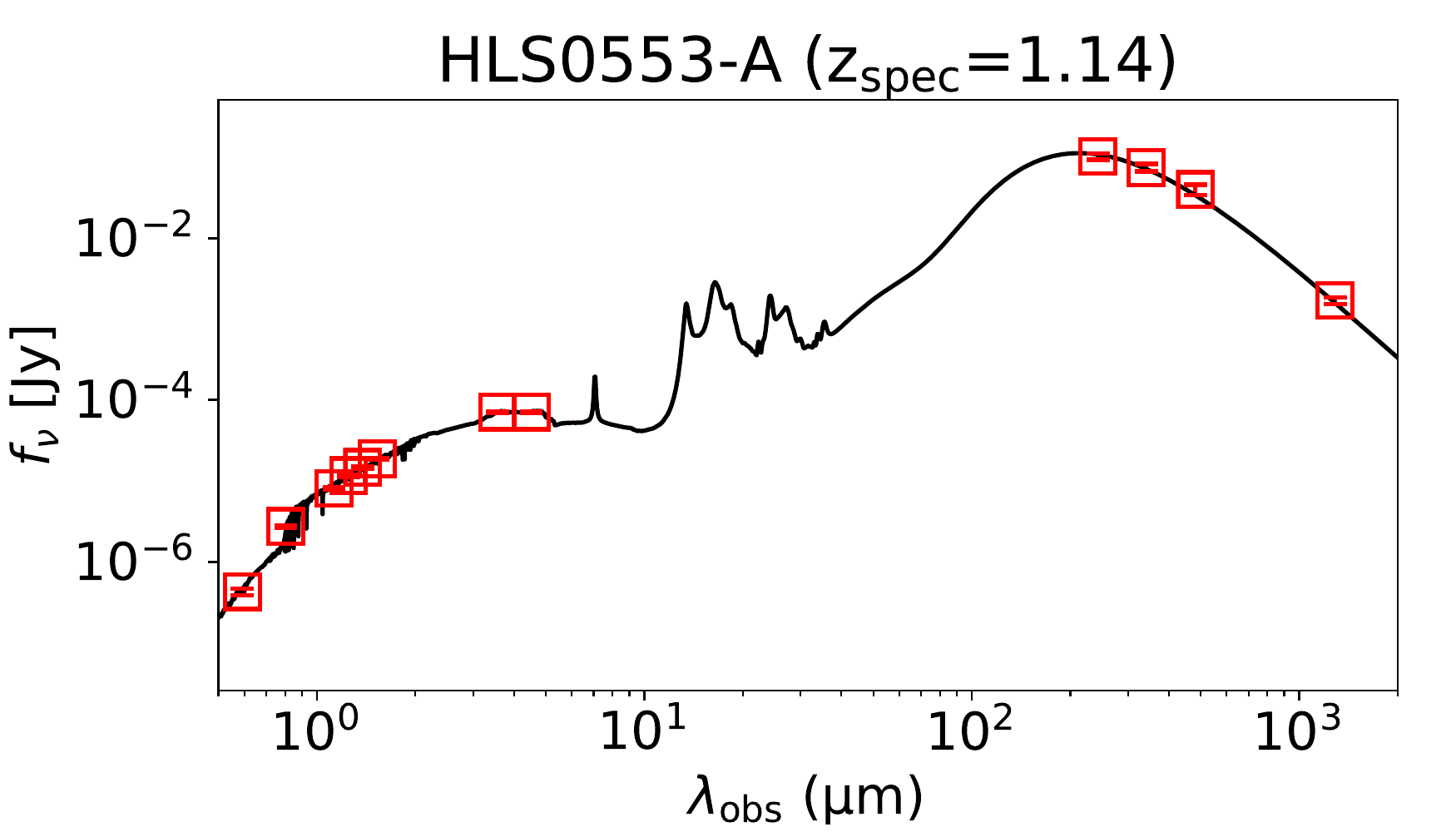}
\includegraphics[width=0.24\linewidth]{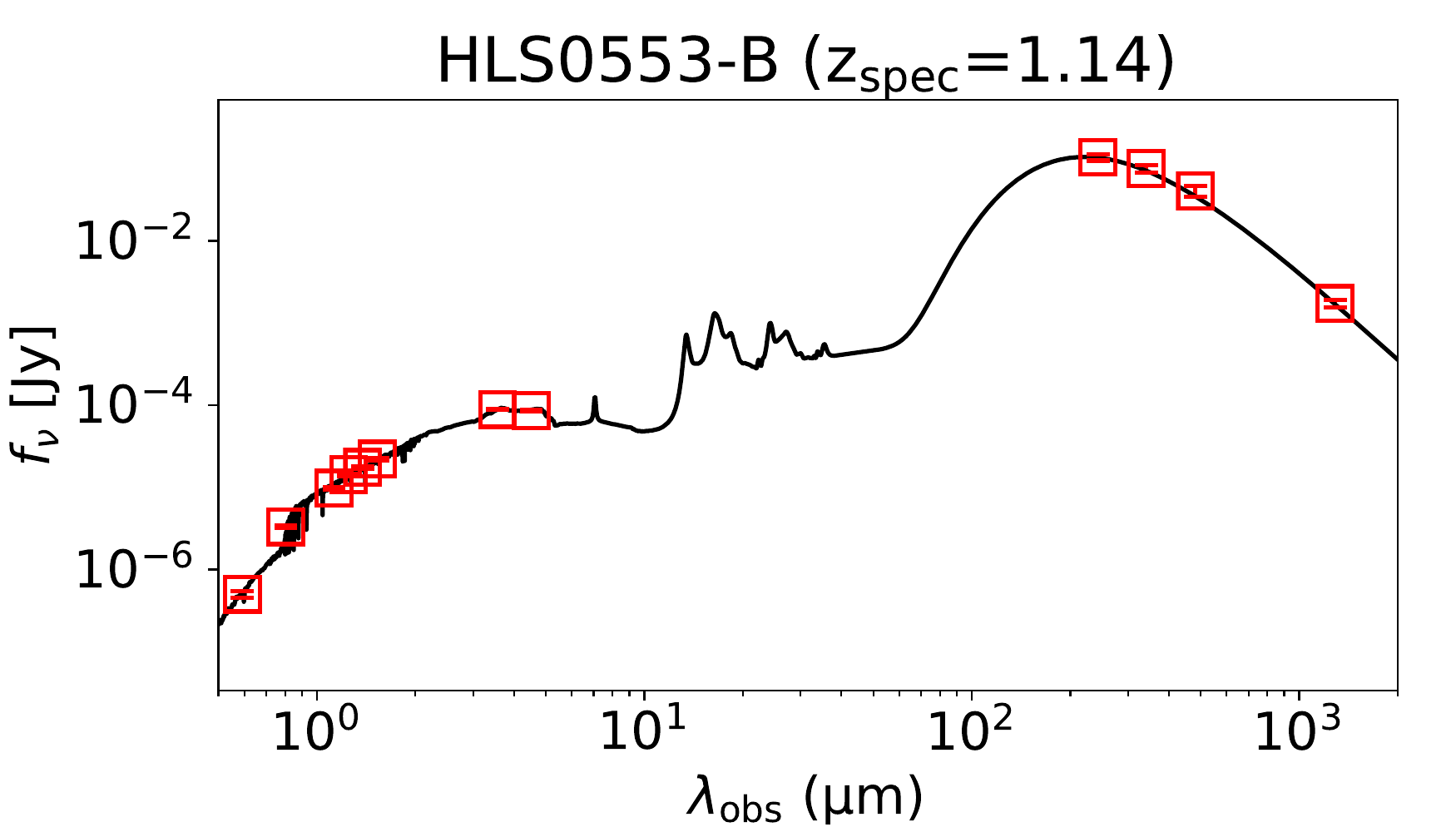}
\includegraphics[width=0.24\linewidth]{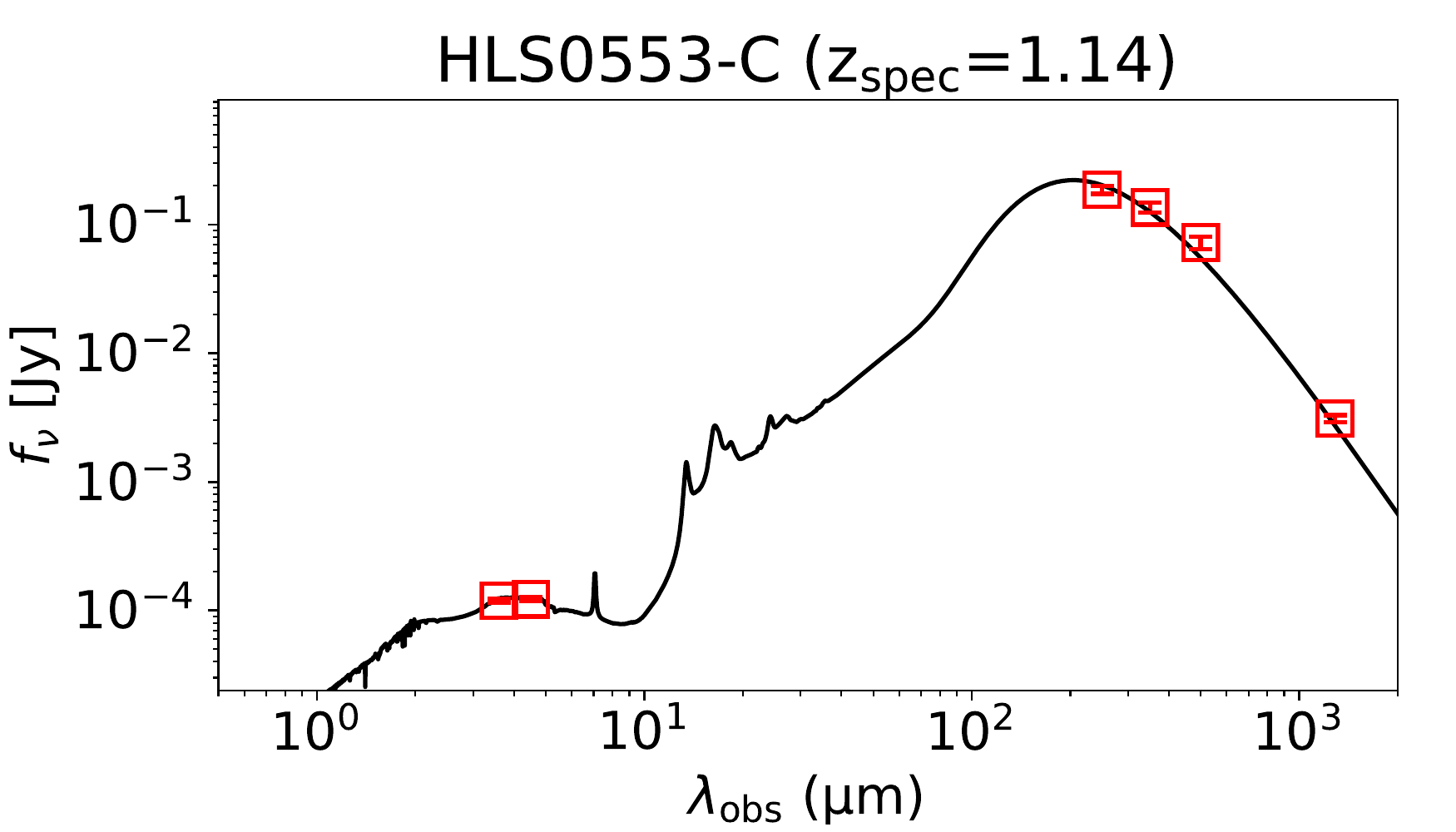}
\includegraphics[width=0.24\linewidth]{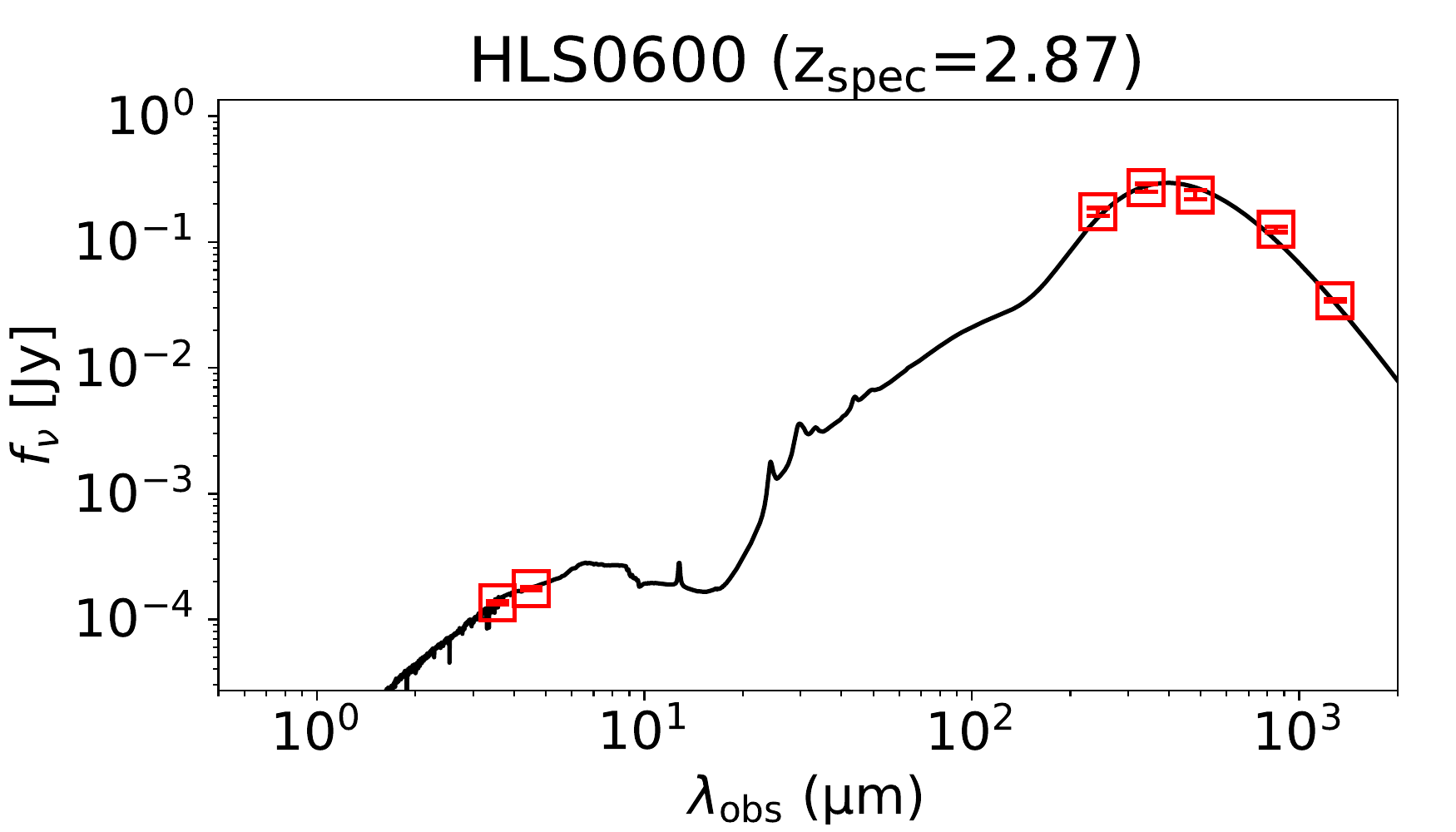}
\includegraphics[width=0.24\linewidth]{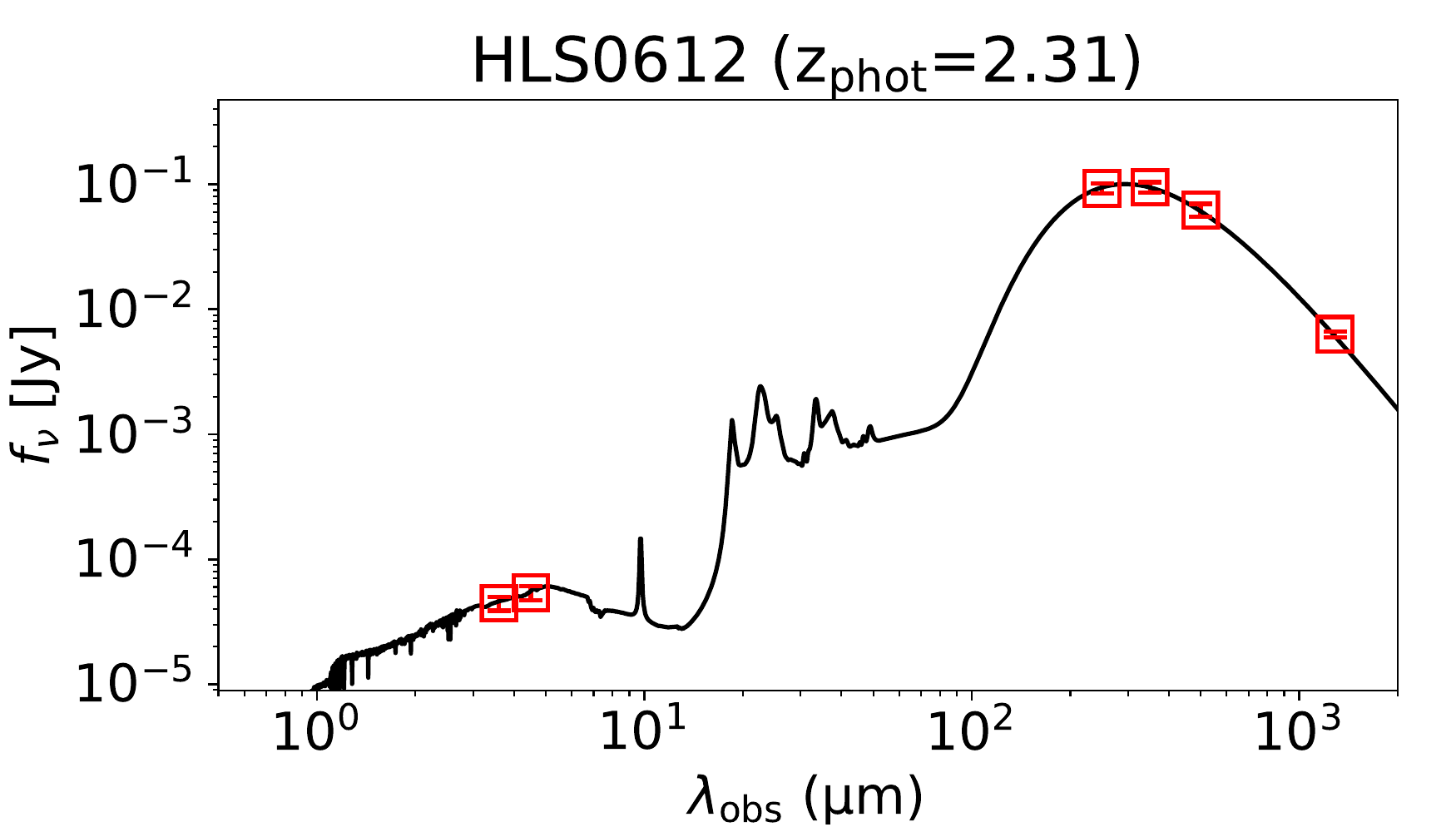}
\includegraphics[width=0.24\linewidth]{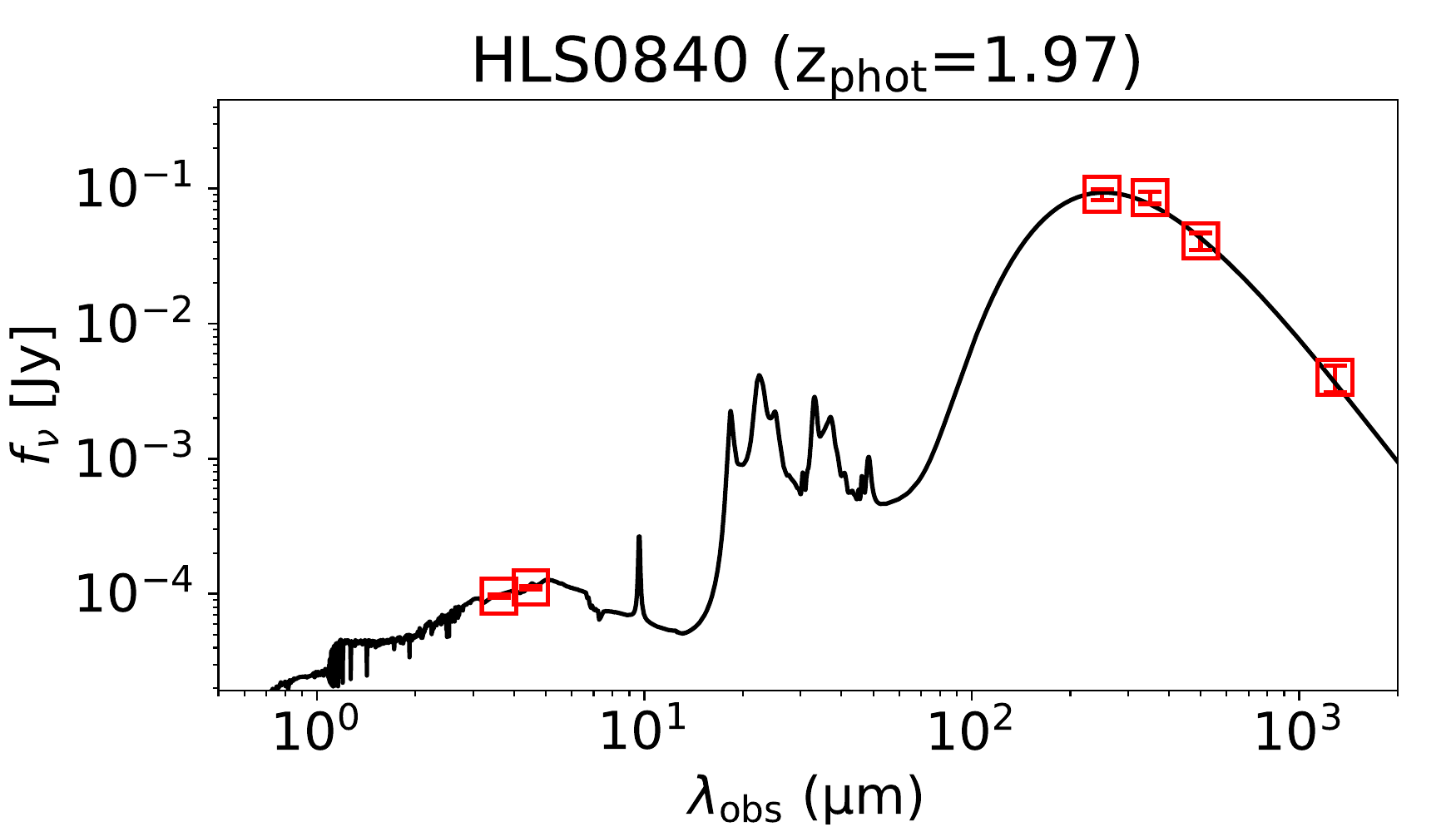}
\includegraphics[width=0.24\linewidth]{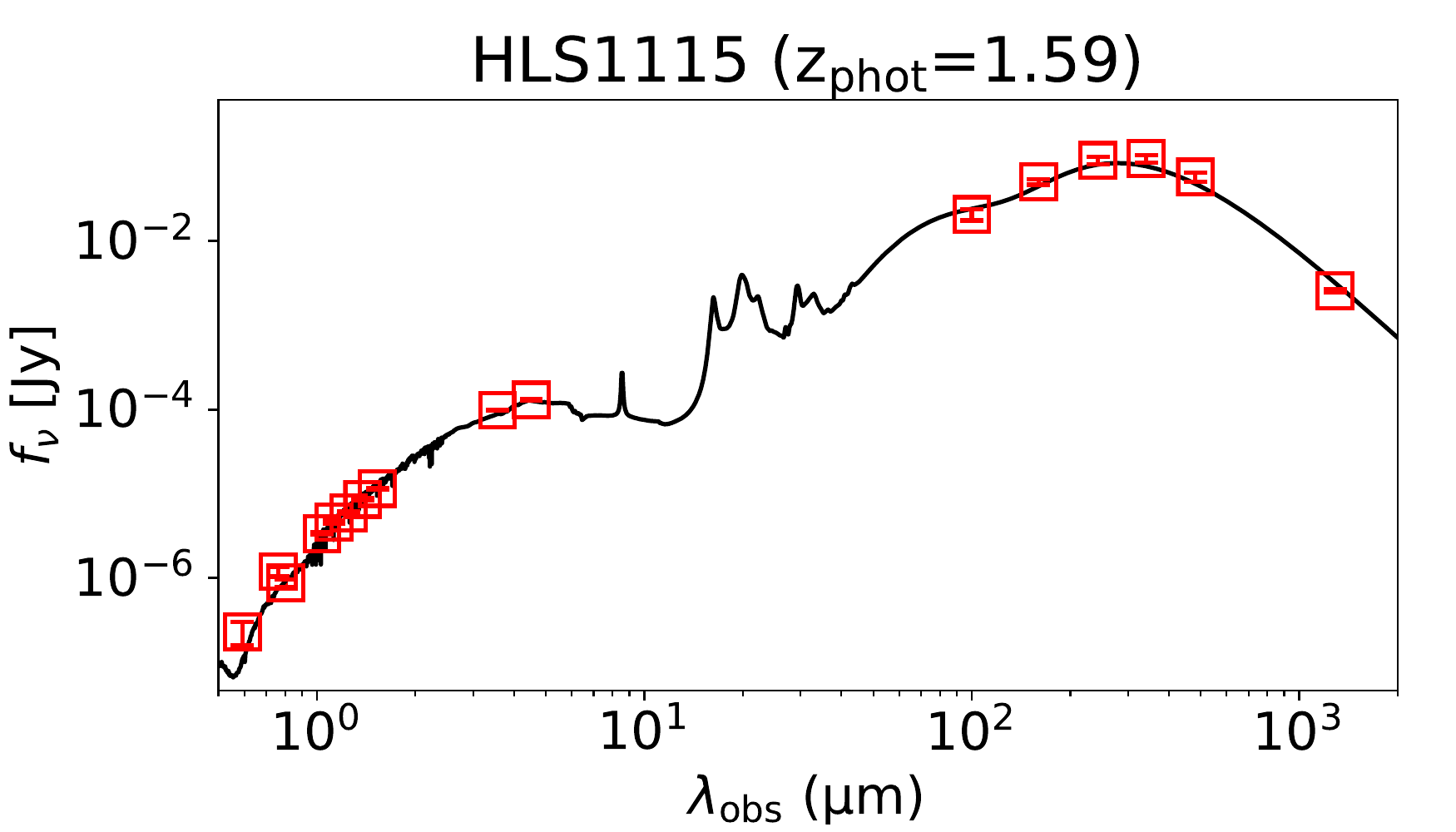}
\includegraphics[width=0.24\linewidth]{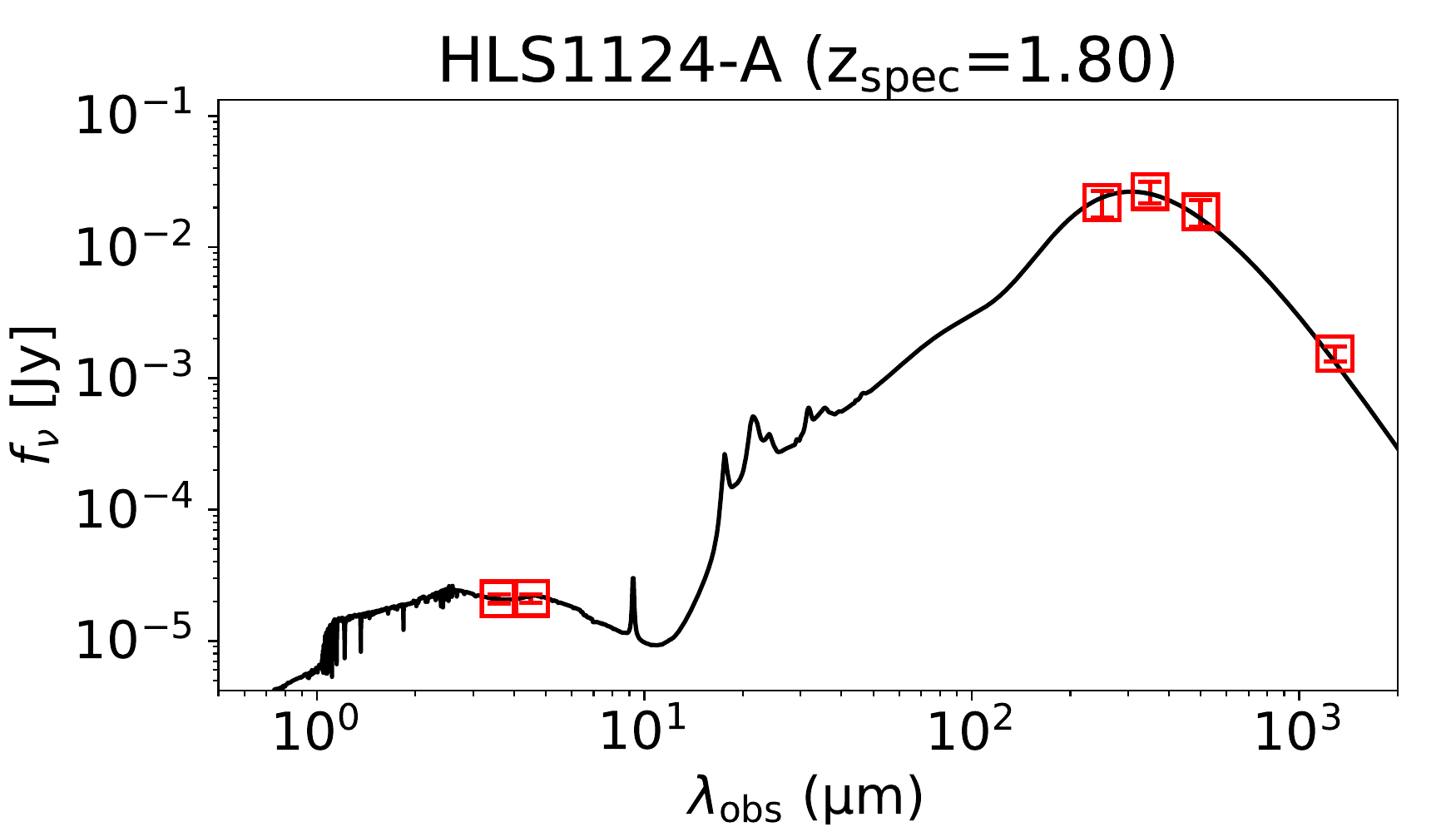}
\includegraphics[width=0.24\linewidth]{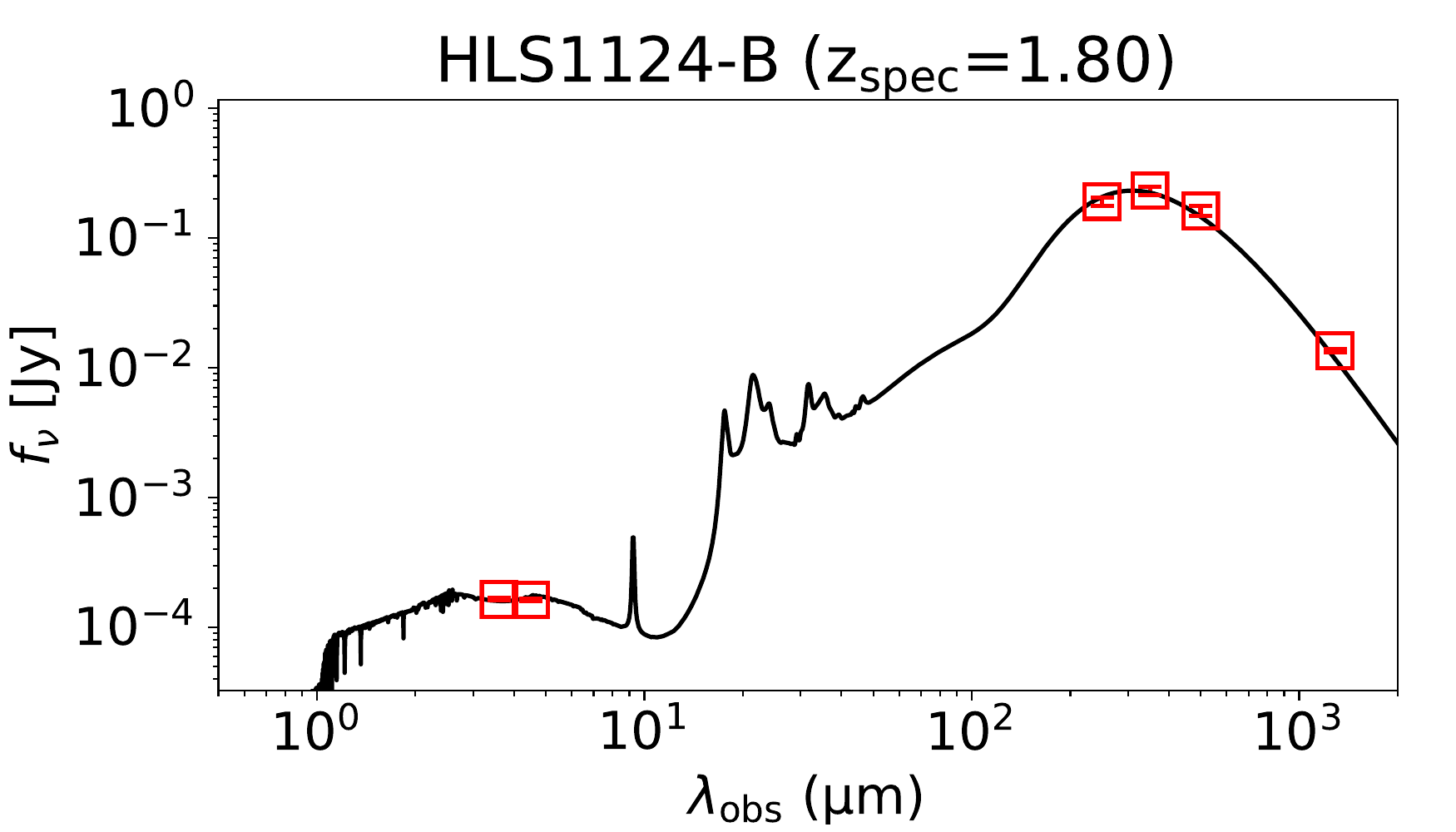}
\includegraphics[width=0.24\linewidth]{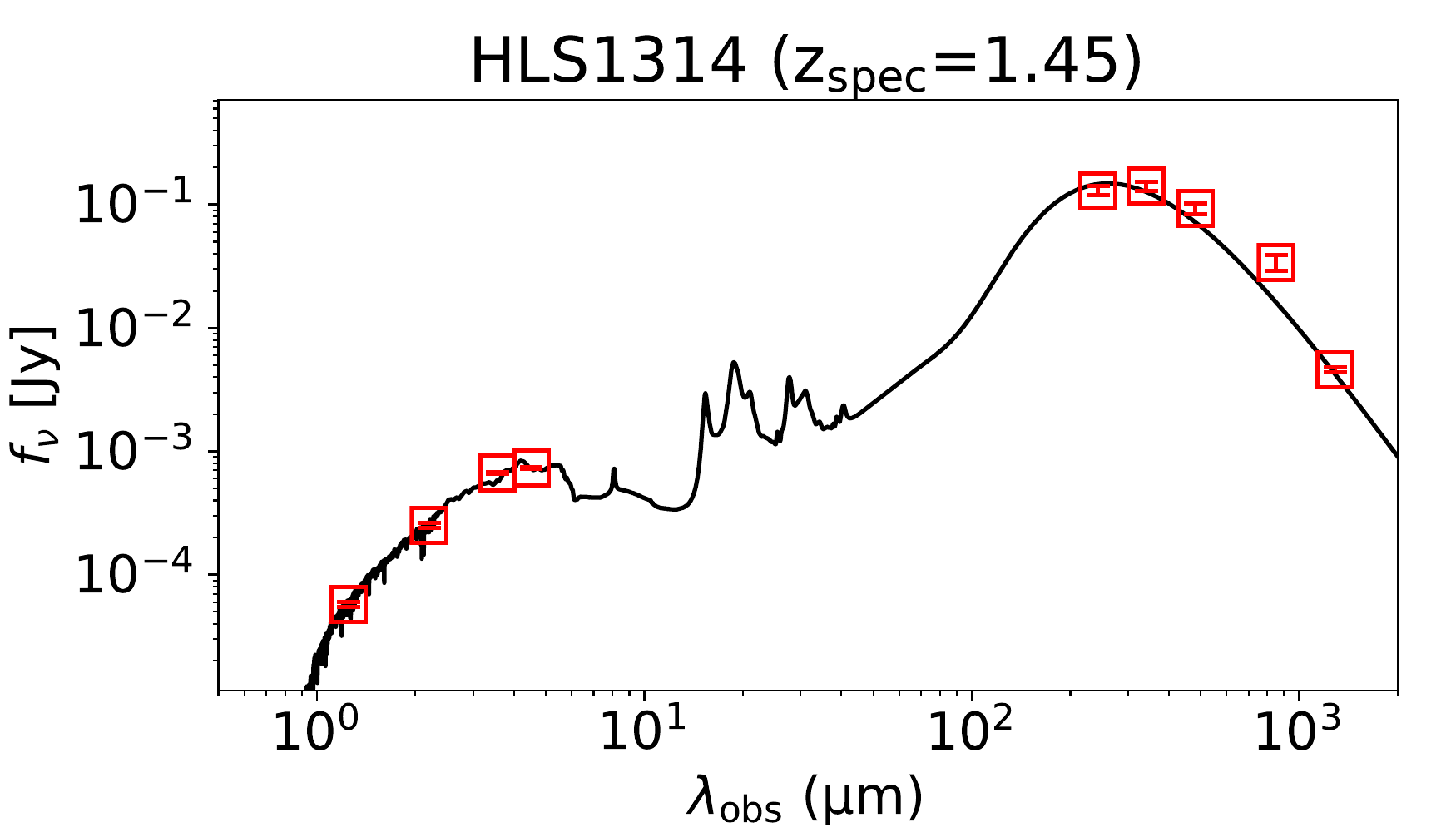}
\includegraphics[width=0.24\linewidth]{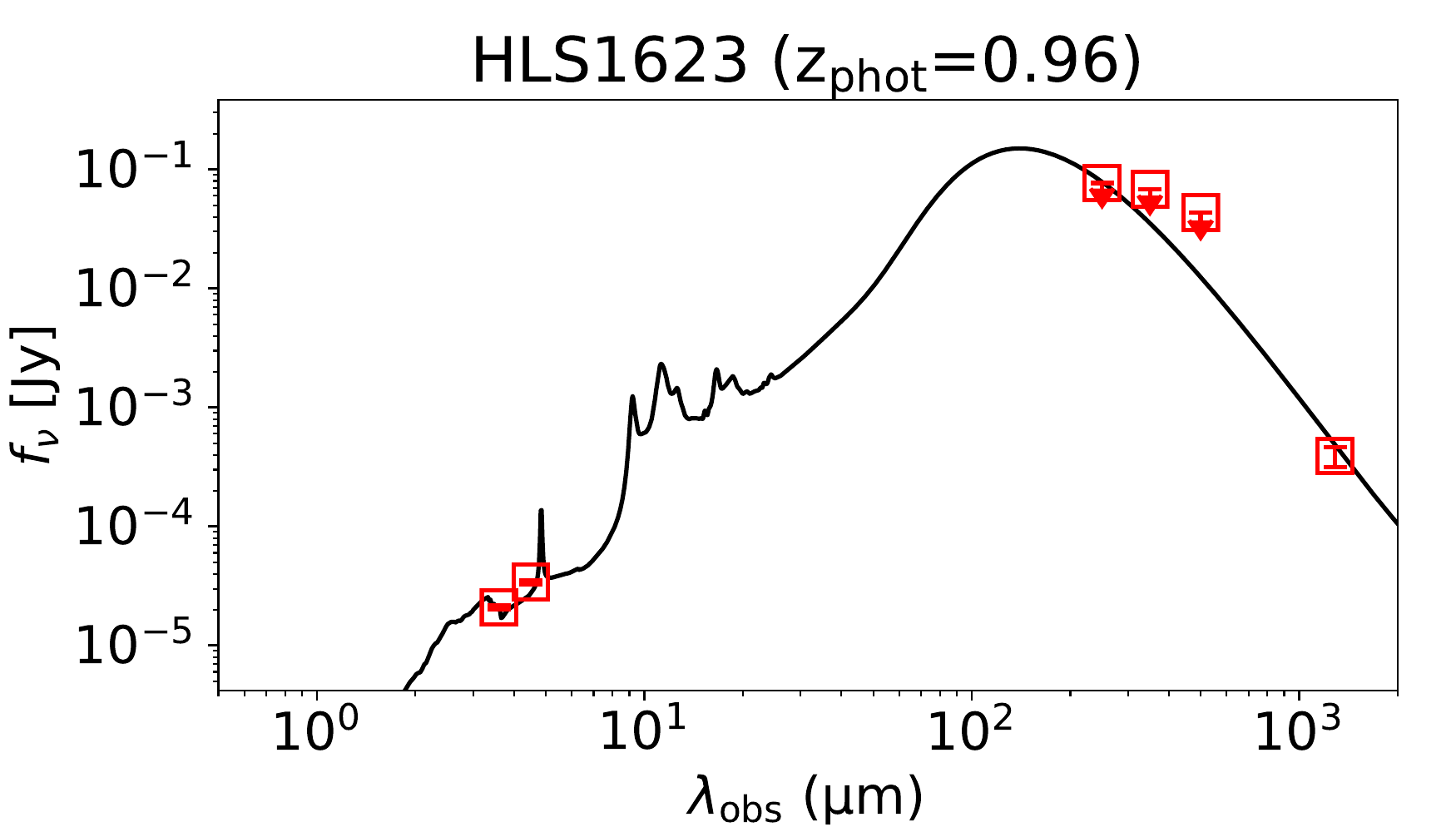}
\includegraphics[width=0.24\linewidth]{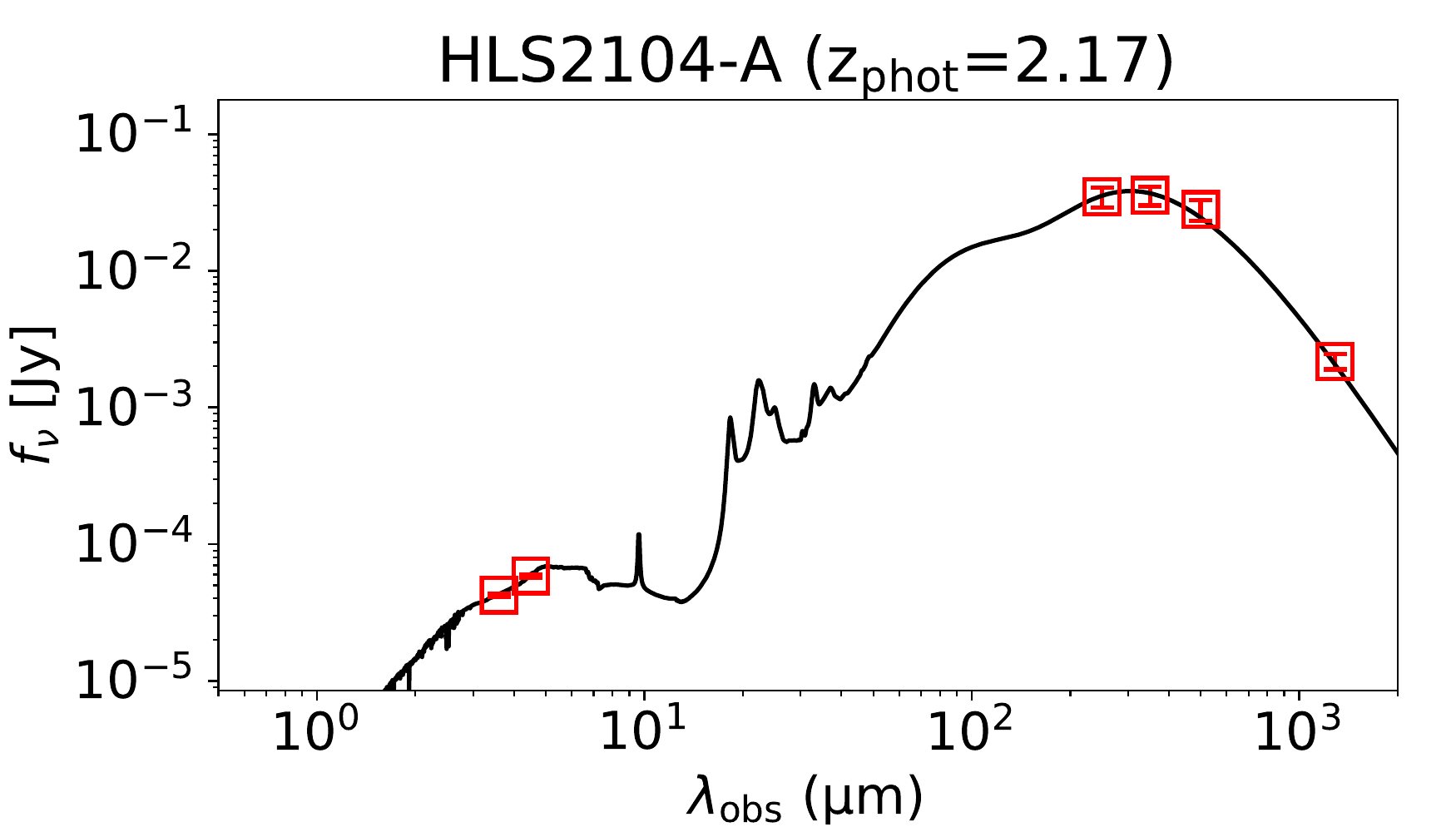}
\includegraphics[width=0.24\linewidth]{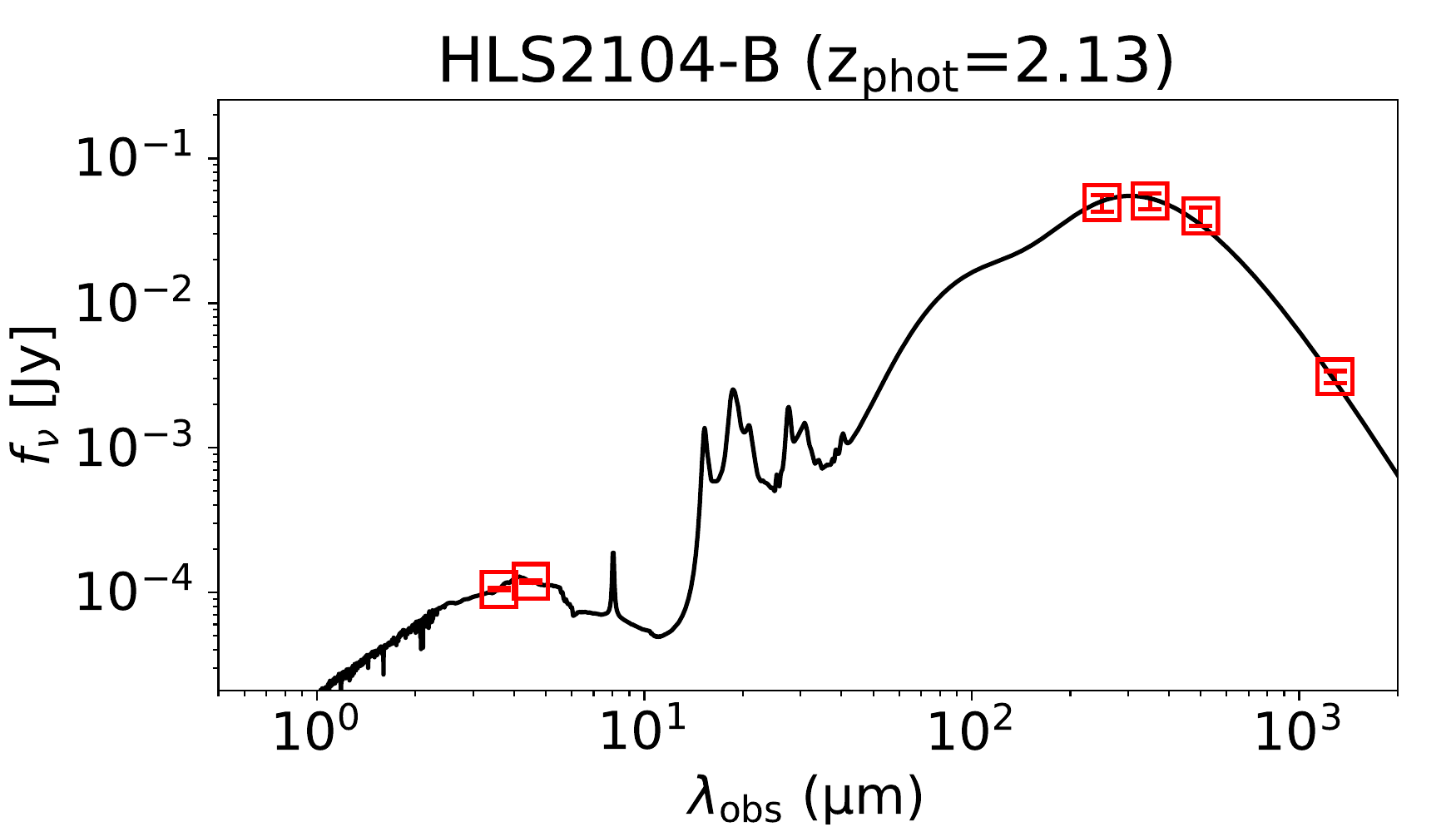}
\includegraphics[width=0.24\linewidth]{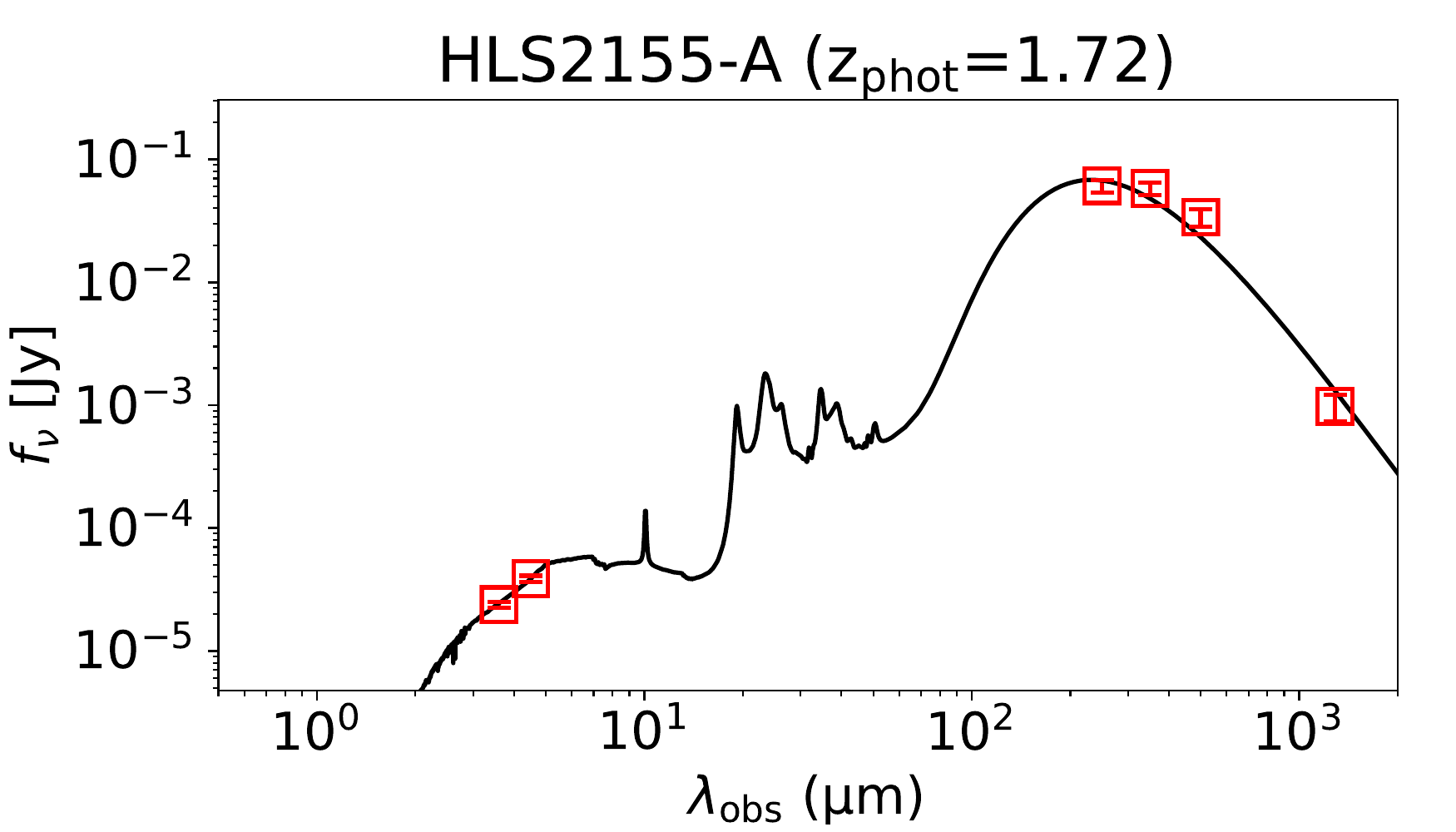}
\includegraphics[width=0.24\linewidth]{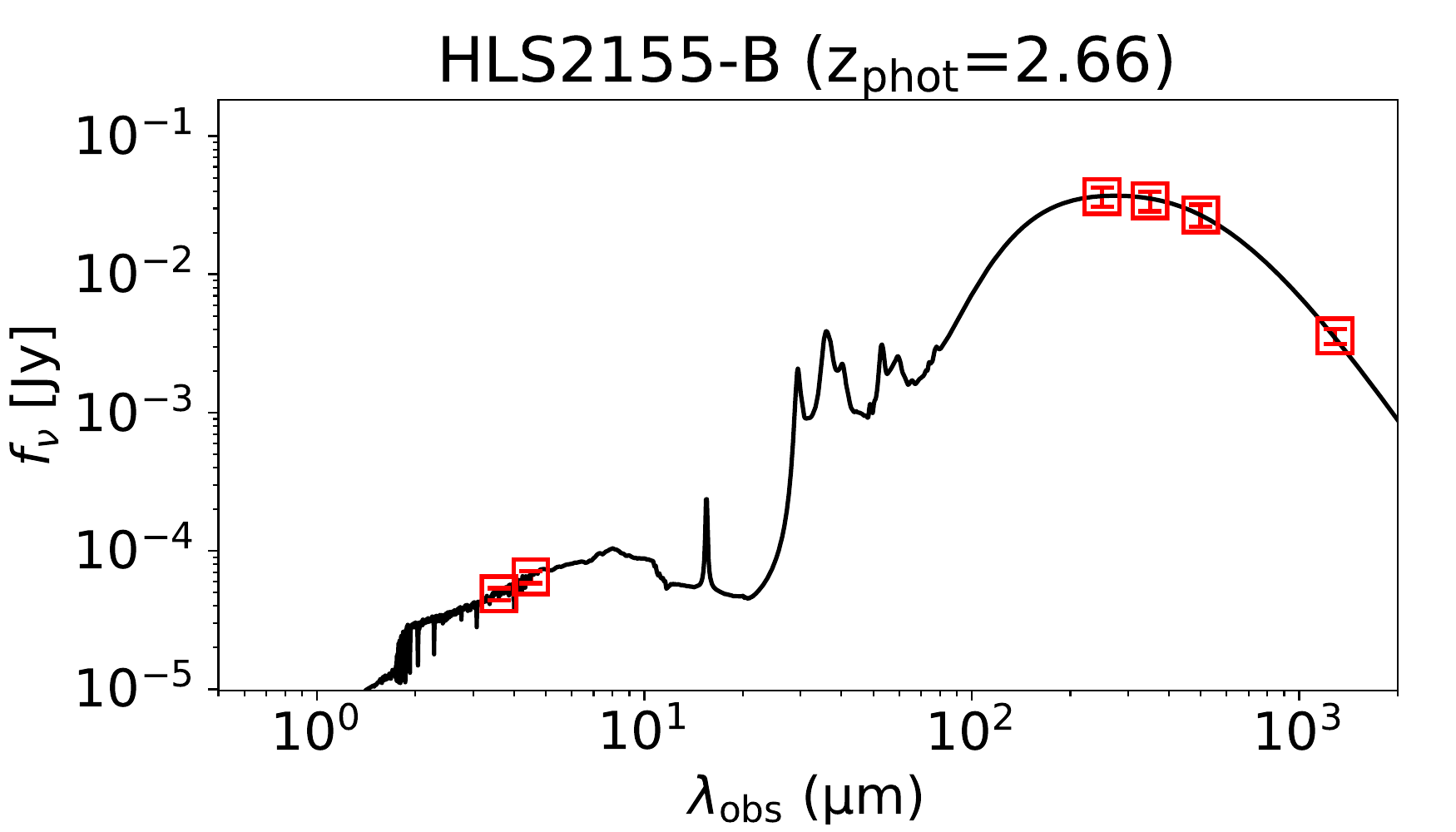}
\includegraphics[width=0.24\linewidth]{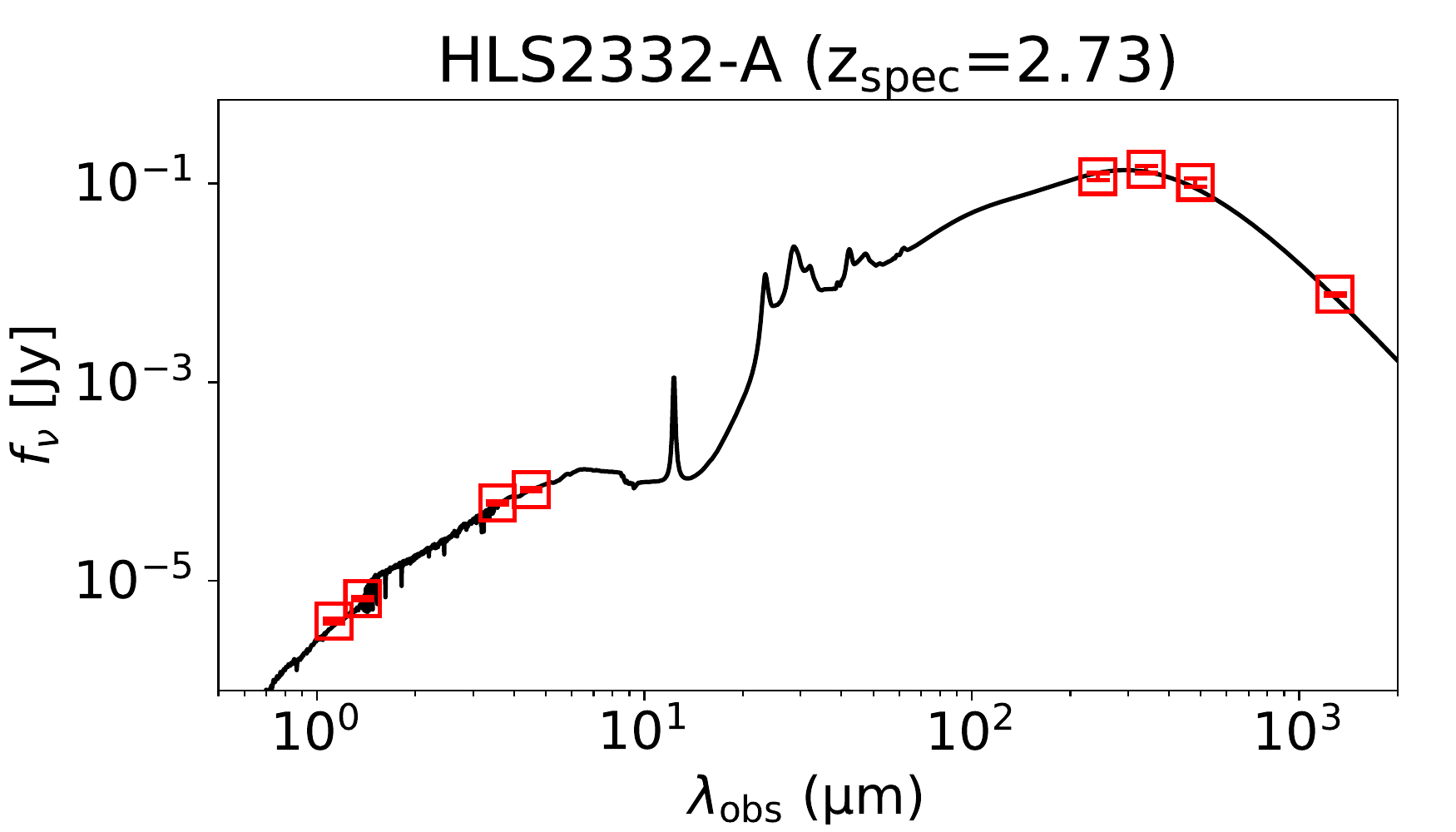}
\includegraphics[width=0.24\linewidth]{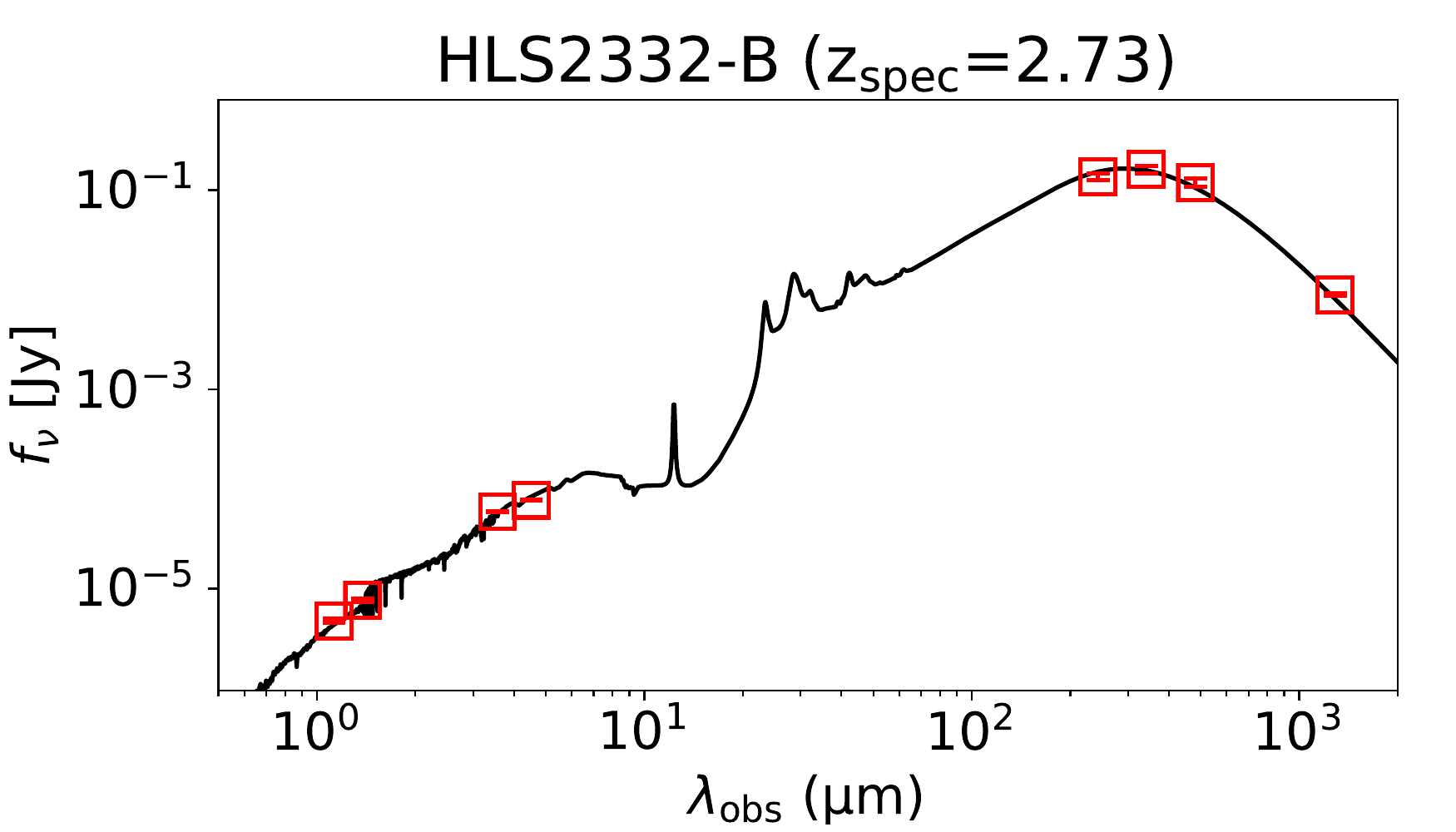}
\includegraphics[width=0.24\linewidth]{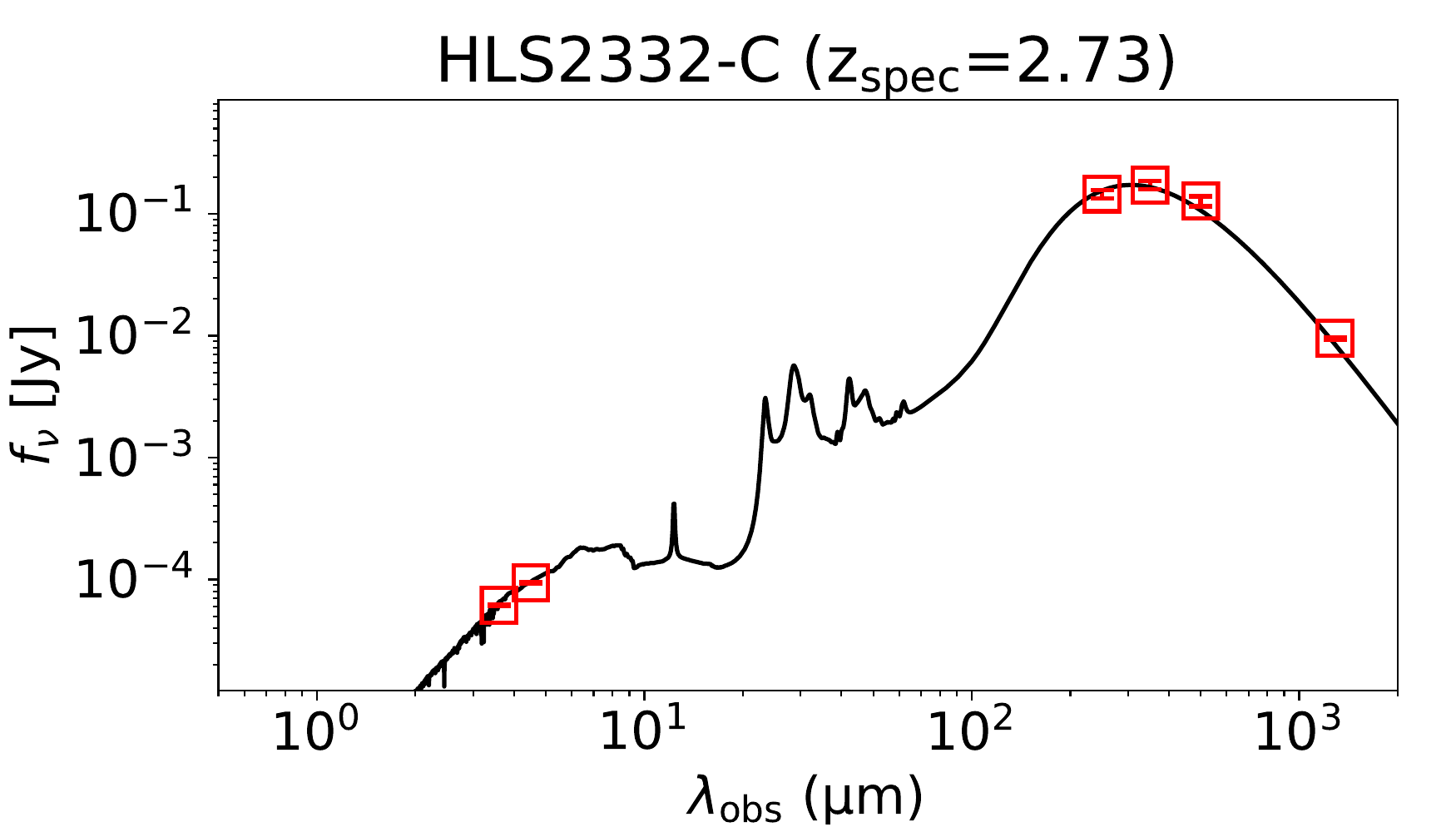}
\caption{\textsc{magphys} best-fit SEDs of all the 29 sources in this work. Source names with their best redshifts are noted above the plots.
The photometric data are plotted as open red squares, and the best-fit SEDs are shown as the black solid lines. 
Key parameters derived from the SED fitting are presented in Table~\ref{tab:04_prop}.
\textred{The measured \herschel\ flux densities of HLS1623 are shown as upper limits, since they are likely partially contributed by other sources not detected by ALMA. }
}
\label{fig:apd_03}
\end{figure*}

Here we show the multi-wavelength images, i.e., ALMA 1.3\,mm (native and tapered) and \spitzer/IRAC 3.6/4.5\,\micron\ images, of five highly blended SMGs in Figure~\ref{fig:02_sunfit} and 20 cluster-lensed SMGs in Figure~\ref{fig:apd_01}.
We also present \textsc{magphys} best-fit SED of all sources in Figure~\ref{fig:apd_03}.

%% file: tables/tb01_observation_log.tex
% \begin{longrotatetable}
\startlongtable 

\begin{deluxetable*}{lrrrrrrcrr} %[p!]
\tablecaption{Summary of \herschel/SPIRE, ALMA and \spitzer/IRAC observations \label{tab:01_log}}
% \tablenum{01}
\tablewidth{0pt}
\tabletypesize{\footnotesize}
\tablehead{
\colhead{Cluster Name} 
& \multicolumn2c{Coordinates\tablenotemark{a}}
& \multicolumn2c{\herschel/SPIRE}
& \multicolumn3c{ALMA Band 6}
& \multicolumn2c{\spitzer/IRAC}
\\
\colhead{} & \colhead{RA} & \colhead{Dec} 
& \colhead{OBSID} & \colhead{$t_\mathrm{obs}$\tablenotemark{b}} 
& \colhead{Pointing\tablenotemark{c}} & \colhead{$t_\mathrm{obs}$\tablenotemark{b}} & \colhead{RMS\tablenotemark{d}}
& \colhead{Program ID} &
\colhead{$5\sigma$ Depth\tablenotemark{e}} \\
% \colhead{}  & \colhead{} & \colhead{} & \nocolhead{} & \colhead{} & \colhead{} & \colhead{} & \colhead{} & \colhead{} & \colhead{3.6/4.5\,\micron} \\
\colhead{}  & \colhead{} 
& \colhead{} & \nocolhead{} & \colhead{(s)} 
& \colhead{} & \colhead{(s)} & \colhead{(\si{mJy.beam^{-1}})}
& \colhead{} &
\colhead{(mag, mag)}
}
% \decimalcolnumbers
\startdata
SPT\,J0114--4123 &   1:14:39.38 &  --41:24:03.1 & 1342247867  & 169 &   single & 907 & 0.038 & 12095 & 22.27, 22.37 \\ 
SPT\,J0307--5042 &   3:07:49.38 &  --50:41:39.1 & 1342240119  & 169 &   single & 302 & 0.111 & 12095 & 22.28, 22.39 \\ 
% 0345-6419\tablenotemark{f} &   3:45:41.33 &  --64:18:05.3 & 1342240116  & 445 &   single & 195 & 0.107 & \nodata & \nodata \\ 
SPT\,J0505--6145 &   5:05:33.14 &  --61:43:55.0 & 1342229231  & 169 &   single & 195 & 0.116 & 80162 & 21.23, 21.60 \\  %\todo{bad cluster name? previously MACSJ0505.5-6144}
% http://simbad.u-strasbg.fr/simbad/sim-id?Ident=%409589854
SPT\,J0546--5345 &   5:46:39.76 &  --53:45:18.2 & 1342240055  & 445 &   single & 302 & 0.104 & 60099, 70149 & 22.78, 22.84 \\ 
SPT\,J0612--4317 &   6:12:03.68 &  --43:17:10.8 & 1342240063  & 169 &   single & 302 & 0.148 & 80012, 90233 & 21.42, 21.78 \\ 
Abell\,2813 &   0:43:35.41 &  --20:42:00.4 & 1342188582  & 5803 &   single & 121 & 0.169 & 60034 & 23.62, 23.65 \\ 
Abell\,3088 &   3:07:08.09 &  --28:40:17.4 & 1342188659  & 5803 &   single & 604 & 0.067 & 80066 & 22.69, 22.65 \\ 
CODEX\,35646 &  16:23:46.51 &   26:34:11.7 & 1342239984  & 169 &   single & 302 & 0.126 & 90218 & 22.28, 22.26 \\ 
CODEX\,39326 &  11:24:02.22 &   24:04:38.4 & 1342256839  & 169 &   single & 532 & 0.101 & 12095 & 22.25, 22.17 \\ 
CODEX\,52909 &  11:53:20.37 &   07:56:00.3 & 1342247962  & 169 &   multiple & 171 & 0.150 & 90218 & 22.38, 22.29 \\ 
MACS\,J0111.5+0855 &   1:11:27.73 &   08:55:28.6 & 1342237548  & 169 &   single & 151 & 0.142 & 90218 & 22.39, 22.33 \\ 
MACS\,J0455.2+0657 &   4:55:17.97 &   07:01:02.6 & 1342229655  & 169 &   single & 433 & 0.097 & 90218 & 22.24, 22.13 \\ 
MACS\,J0553.4--3342 &   5:53:27.79 &  --33:42:35.1 & 1342227700  & 169 &   multiple & 134 & 0.114 & 12005, 12123 & 23.14, 23.43 \\ 
                 &              &              &  &  &  &   &  & 14281, 90218 &   \\ 
MACS\,J0600.1--2008 &   6:00:23.90 &  --20:06:38.0 & 1342230801  & 169 &   single & 423 & 0.106 & 12005, 12123 & 23.10, 23.20 \\ 
                 &              &              &  &  &  &   &  & 90218 &   \\ 
RXC\,J0840.5+0544 &   8:40:32.09 &   05:45:01.1 & 1342230784  & 169 &   single & 302 & 0.195 & 12095 & 22.21, 22.04 \\ 
MACS\,J1115.8+0129 &  11:15:50.76 &   01:30:41.1 & 1342256866  & 1580 &   single & 866 & 0.042 & 80168, 90213 & 23.66, 23.56 \\ 
RXC\,J1314.3--2515 &  13:14:21.43 &  --25:15:47.8 & 1342236193  & 169 &   single & 1512 & 0.043 & 90218 & 22.24, 22.13 \\ 
RXC\,J2104.8+1401 &  21:04:54.78 &   14:01:43.6 & 1342211300  & 169 &   single & 302 & 0.078 & 90218 & 22.40, 22.45 \\ 
RXC\,J2155.6+1231 &  21:55:41.33 &   12:31:50.8 & 1342211302  & 169 &   multiple & 616 & 0.047 & 30344, 90218 & 22.18, 21.91 \\ 
RXC\,J2332.4--5358\tablenotemark{f} &  23:32:26.46 &  --53:58:41.2 & 1342234736  & 169 &   multiple & 605 & 0.088 & 40370, 60099 & 22.64, 21.84 \\ 
  &  &  &  &   &  &  &  & 60194, 80096 &   \\
RXC\,J2332.4--5358\tablenotemark{f} &  \nodata  & \nodata & \nodata & \nodata &  single & 141 & 0.064  & \nodata &  \nodata \\
\enddata
\tablenotetext{a}{Coordinates of the ALMA pointing centers, not the exact positions of the ALMA-detected sources.}
\tablenotetext{b}{Total observation time. For the ALMA observations obtained in the ``single" pointing mode, this is the full on-source integration time. For SPIRE and the ALMA multiple-pointing observation, this indicates the full scan time for the scientific targets.}
\tablenotetext{c}{Mode of ALMA observations (single or multiple pointings).}
\tablenotetext{d}{Continuum RMS noise with a 1\arcsec\ \textit{uv}-tapered beam.}
\tablenotetext{e}{Median $5\sigma$ point-source depth in the final IRAC 3.6/4.5\,\micron\ image products.}
% \tablenotetext{f}{This source is finally identified as a low-redshift ($z<0.1$) LIRG, thus IRAC data reduction is not performed.}
\tablenotetext{f}{Here we distinguish two portions of ALMA observations of R2332 obtained at different modes with slightly different spectral window settings. 
The multiple-pointing observation contains all the three components of this lensed SMGs with CO(7-6) coverage, and the single-pointing one targets pure dust continua of only two components.}
% \tablecomments{This table ``hides'' the third column in the \latex\ when compiled.
% The Distance is also centered on the decimals.  Note that when using decimal
% alignment you need to include the {\tt\string\decimals} command before
% {\tt\string\startdata} and all of the values in that column have to have a
% space before the next ampersand.}
\end{deluxetable*}
% \end{longrotatetable}
% \endlongtable

%% file: tables/tb02_photometry.tex
% \startlongtable

\begin{rotatetable*}
\begin{deluxetable*}{llrrllrrrrr} %[htb!]
\tablecaption{Summary of \spitzer/IRAC, \herschel/SPIRE and ALMA Band-6 photometry \label{tab:02_phot}}
% \tablenum{01}
\tablewidth{0pt}
\tabletypesize{\scriptsize}
\tablehead{
\colhead{Galaxy ID} & \colhead{Short ID} 
& \multicolumn2c{Coordinates\tablenotemark{a}}
% & \colhead{$z$}
& \multicolumn2c{\spitzer/IRAC}
& \multicolumn3c{\herschel/SPIRE}
& \multicolumn2c{ALMA}
\\
\colhead{} & \colhead{} & \colhead{RA} & \colhead{Dec} 
% & \colhead{} 
& \colhead{[3.6\,\micron]} & \colhead{[4.5\,\micron]}
& \colhead{$f_{250}$} & \colhead{$f_{350}$}
& \colhead{$f_{500}$} & \colhead{S/N$_\mathrm{peak}$\tablenotemark{b}} &  \colhead{$f_{1.3\,\mathrm{mm}}$}\\
\colhead{} & \colhead{} & \colhead{(deg)} & \colhead{(deg)} % & \colhead{} 
& \colhead{(mag)} & \colhead{(mag)} 
& \colhead{(mJy)} & \colhead{(mJy)} & \colhead{(mJy)}
& \colhead{} & \colhead{(mJy)} }
% \decimalcolnumbers
\startdata
HLS\,J004335.3--204203 & HLS0043-A    &  10.89702 & --20.70082 & 18.94$\pm$0.02 & 18.90$\pm$0.07 & 81.6$\pm$8.1 & 70.3$\pm$7.7 & 61.2$\pm$7.3 & 9.9 & 2.50$\pm$0.50 \\
HLS\,J004335.1--204159 & HLS0043-B    &  10.89642 & --20.69959 & 20.93$\pm$0.11 & 20.08$\pm$0.07 & 20.7$\pm$5.0 & $<$11.0 & $<$11.0 & 4.7 & 2.69$\pm$0.56 \\
HLS\,J011127.9+085525 & HLS0111-A    &  17.86606 &   8.92353 & 21.66$\pm$0.06 & 20.98$\pm$0.05 & 27.3$\pm$5.4 & 36.0$\pm$5.7 & 19.5$\pm$4.4 & 19.4 & 1.98$\pm$0.40 \\
HLS\,J011127.7+085529 & HLS0111-B    &  17.86547 &   8.92467 & 19.71$\pm$0.02 & 19.37$\pm$0.02 & 45.1$\pm$6.3 & 59.4$\pm$7.1 & 32.2$\pm$5.3 & 19.1 & 3.26$\pm$0.33 \\
HLS\,J011439.3--412404 & HLS0114      &  18.66373 & --41.40105 & 18.24$\pm$0.02 & 18.15$\pm$0.03 & 81.5$\pm$8.1 & 62.3$\pm$7.2 & 33.3$\pm$5.3 & 26.5 & 1.51$\pm$0.10 \\
HLS\,J030707.7--284017 & HLS0307-28-A &  46.78209 & --28.67132 & 20.67$\pm$0.04 & 20.40$\pm$0.03 & 103.0$\pm$9.2 & 85.6$\pm$8.6 & 43.0$\pm$6.0 & 24.6 & 1.47$\pm$0.14 \\
HLS\,J030708.2--284027 & HLS0307-28-B &  46.78405 & --28.67422 & 21.57$\pm$0.07 & 22.02$\pm$0.09 & 13.3$\pm$4.7 & 24.7$\pm$5.0 & 17.8$\pm$4.2 & 36.2 & 2.30$\pm$0.15 \\
HLS\,J030749.6--504138 & HLS0307-50   &  46.95667 & --50.69391 & 19.99$\pm$0.06 & 19.40$\pm$0.04 & 102.1$\pm$9.1 & 107.1$\pm$9.9 & 67.2$\pm$7.7 & 28.7 & 3.77$\pm$0.21 \\
HLS\,J045518.1+070102 & HLS0455      &  73.82525 &   7.01721 & 19.79$\pm$0.04$^\bullet$ & 19.83$\pm$0.04$^\bullet$ & 102.4$\pm$9.1 & 130.5$\pm$11.3 & 98.7$\pm$9.9 & 58.1 & 10.71$\pm$0.27 \\
HLS\,J050532.9--614357 & HLS0505      &  76.38710 & --61.73249 & 20.74$\pm$0.10$^\bullet$ & 20.73$\pm$0.12$^\bullet$ & 101.5$\pm$9.1 & 103.3$\pm$9.7 & 76.3$\pm$8.3 & 32.1 & 4.38$\pm$0.30 \\
HLS\,J054639.2--534520 & HLS0546      &  86.66346 & --53.75560 & 19.85$\pm$0.18 & 19.34$\pm$0.08 & 69.8$\pm$7.5 & 46.3$\pm$6.3 & 44.0$\pm$6.1 & 7.0 & 1.91$\pm$0.45 \\
HLS\,J055327.8--334216 & HLS0553-A    &  88.36579 & --33.70446 & 19.27$\pm$0.02 & 19.27$\pm$0.02 & 102.6$\pm$9.1 & 74.9$\pm$8.0 & 40.0$\pm$5.8 & 24.9 & 1.71$\pm$0.16 \\
HLS\,J055327.6--334244 & HLS0553-B    &  88.36510 & --33.71220 & 19.03$\pm$0.02 & 19.07$\pm$0.02 & 103.7$\pm$9.2 & 75.7$\pm$8.0 & 40.4$\pm$5.8 & 23.2 & 1.72$\pm$0.17 \\
HLS\,J055327.8--334231 & HLS0553-C    &  88.36602 & --33.70849 & 18.71$\pm$0.04 & 18.68$\pm$0.05 & 186.8$\pm$13.3 & 136.4$\pm$11.7 & 72.8$\pm$8.1 & 35.4 & 3.11$\pm$0.19 \\
HLS\,J060023.8--200638 & HLS0600      &  90.09901 & --20.11067 & 18.57$\pm$0.03$^\bullet$ & 18.29$\pm$0.03$^\bullet$ & 173.7$\pm$12.7 & 270.5$\pm$19.7 & 237.8$\pm$19.6 & 77.0 & 34.27$\pm$0.61 \\
HLS\,J061203.5--431712 & HLS0612      &  93.01468 & --43.28678 & 19.78$\pm$0.14 & 19.57$\pm$0.05 & 93.0$\pm$8.7 & 95.0$\pm$9.2 & 62.4$\pm$7.4 & 28.7 & 6.31$\pm$0.37 \\
HLS\,J084032.1+054503 & HLS0840      & 130.13366 &   5.75073 & 18.94$\pm$0.03 & 18.78$\pm$0.03 & 90.5$\pm$8.5 & 85.5$\pm$8.6 & 41.0$\pm$5.9 & 8.0 & 4.00$\pm$0.90 \\
HLS\,J111550.7+013036 & HLS1115      & 168.96117 &   1.50988 & 18.92$\pm$0.02 & 18.62$\pm$0.02 & 91.1$\pm$8.6 & 95.5$\pm$9.2 & 57.6$\pm$7.0 & 31.4 & 2.58$\pm$0.13 \\
HLS\,J112402.2+240447 & HLS1124-A    & 
171.00896 &  24.07963 & 20.60$\pm$0.08$^\bullet$ & 20.59$\pm$0.08$^\bullet$ & 21.8$\pm$5.1 & 26.5$\pm$5.1 & 18.5$\pm$4.3 & 10.9 & 1.54$\pm$0.20 \\
HLS\,J112402.3+240437 & HLS1124-B    & 
171.00967 &  24.07699 & 18.36$\pm$0.02$^\bullet$ & 18.37$\pm$0.03$^\bullet$ & 191.0$\pm$13.6 & 231.6$\pm$17.4 & 161.7$\pm$14.3 & 51.7 & 13.50$\pm$0.39 \\
HLS\,J131421.3--251546 & HLS1314\tablenotemark{c}  & 
198.58876 & --25.26276 & 16.84$\pm$0.01 & 16.74$\pm$0.01 & 130.2$\pm$10.5 & 141.2$\pm$12.0 & 93.1$\pm$9.5 & 26.5 & 4.58$\pm$0.21 \\
HLS\,J162346.5+263412 & HLS1623      & 245.94394 &  26.57011 & 20.60$\pm$0.04 & 20.07$\pm$0.04 & 76.9$\pm$7.8 & 68.0$\pm$7.6 & 43.3$\pm$6.0 & 12.7 & 0.39$\pm$0.07 \\
HLS\,J210454.6+140149 & HLS2104-A    & 316.22738 &  14.03040 & 19.83$\pm$0.03 & 19.48$\pm$0.03 & 34.8$\pm$5.7 & 35.7$\pm$5.6 & 28.1$\pm$5.0 & 15.6 & 2.18$\pm$0.28 \\
HLS\,J210454.7+140141 & HLS2104-B    & 316.22777 &  14.02815 & 18.85$\pm$0.02 & 18.71$\pm$0.02 & 49.4$\pm$6.5 & 50.8$\pm$6.5 & 40.0$\pm$5.8 & 19.9 & 3.10$\pm$0.29 \\
HLS\,J215540.5+123208 & HLS2155-A    & 328.91861 &  12.53547 & 20.46$\pm$0.06 & 19.93$\pm$0.06 & 61.1$\pm$7.1 & 58.1$\pm$7.0 & 33.9$\pm$5.4 & 5.9 & 0.97$\pm$0.24 \\
HLS\,J215540.8+123135 & HLS2155-B    & 328.92018 &  12.52629 & 19.68$\pm$0.11 & 19.37$\pm$0.06 & 36.6$\pm$5.8 & 34.1$\pm$5.5 & 27.0$\pm$4.9 & 9.4 & 3.58$\pm$0.45 \\
HLS\,J233227.2--535844 & HLS2332-A    & 353.11334 & --53.97895 & 19.44$\pm$0.04 & 19.11$\pm$0.03 & 116.8$\pm$9.8 & 138.4$\pm$11.8 & 102.2$\pm$10.2 & 73.6 & 7.64$\pm$0.20 \\
HLS\,J233225.5--535839 & HLS2332-B    & 353.10645 & --53.97746 & 19.62$\pm$0.02 & 19.09$\pm$0.02 & 136.0$\pm$10.8 & 161.2$\pm$13.2 & 119.1$\pm$11.3 & 83.0 & 8.90$\pm$0.21 \\
HLS\,J233229.5--535840 & HLS2332-C    & 353.12295 & --53.97767 & 19.43$\pm$0.01 & 18.97$\pm$0.01 & 145.2$\pm$11.3 & 172.0$\pm$13.8 & 127.1$\pm$11.9 & 83.1 & 9.50$\pm$0.22 \\
\enddata
\tablenotetext{a}{ALMA coordinates of detected source.}
\tablenotetext{b}{Defined as the maximum signal-to-noise ratio of peak flux density ($f_\mathrm{peak}/\sigma_\mathrm{RMS}$) among the four-level continuum images (Section~\ref{ss:02b_alma}).}
\tablenotetext{c}{Combination of all the three clumps (HLS1314-A/B/C) seen in the ALMA map.}
% \tablenotetext{\dagger}{Photometric redshift with a typical uncertainty of $\Delta z \sim 0.2$.}
\tablenotetext{\bullet}{Highly-blended object with complex morphology. IRAC photometry is obtained through a simultaneous MCMC fitting of ALMA and optical source models (see descriptions in Section~\ref{ss:03b_photo} and Figure~\ref{fig:02_sunfit}).}
% \tablecomments{This table ``hides'' the third column in the \latex\ when compiled.
% The Distance is also centered on the decimals.  Note that when using decimal
% alignment you need to include the {\tt\string\decimals} command before
% {\tt\string\startdata} and all of the values in that column have to have a
% space before the next ampersand.}
\end{deluxetable*}
\end{rotatetable*}

%% file: tables/tb03_morphology.tex
\startlongtable

\begin{deluxetable*}{lrrrrrrrrrr} %[htb!]
\tablecaption{Summary of source structural parameters in ALMA Band 6 and \spitzer/IRAC \label{tab:03_irm}}
% \tablenum{01}
\tablewidth{0pt}
\tabletypesize{\small}
\tablehead{
\colhead{ID} 
& \multicolumn2c{ALMA Band 6}
& \multicolumn3c{IRAC 3.6\,\micron}
& \multicolumn3c{IRAC 4.5\,\micron}
& \multicolumn2c{Offset}
\\
\colhead{} 
& \colhead{$R_\mathrm{e}$} & \colhead{$b/a$}
& \colhead{$R_\mathrm{e}$} & \colhead{$b/a$} & \colhead{$n$} 
& \colhead{$R_\mathrm{e}$} & \colhead{$b/a$} & \colhead{$n$} 
& \colhead{$\Delta$RA} & \colhead{$\Delta$Dec} %& \colhead{$\Delta\theta$}
\\
\colhead{}  &
\colhead{(\arcsec)} & \colhead{} 
& \colhead{(\arcsec)} & \colhead{} & \colhead{} 
& \colhead{(\arcsec)} & \colhead{} & \colhead{} 
& \colhead{(\arcsec)} & \colhead{(\arcsec)} %& \colhead{(\arcsec)} 
}
% \decimalcolnumbers
\startdata
HLS0043-A  & 0.28$\pm$0.05 & 0.28$\pm$0.10 & 0.59$\pm$0.03 & 0.92$\pm$0.07 & 1.2$\pm$0.3 & 0.94$\pm$0.03 & 0.14$\pm$0.06 & 0.7$\pm$0.1 & --0.14 & --0.02 \\
HLS0043-B  & 0.27$\pm$0.09 & 0.41$\pm$0.25 & $<$0.40 & \nodata & \nodata & $<$0.40 & \nodata & \nodata & 0.01 & 0.01 \\
HLS0111-A  & 0.24$\pm$0.05 & 0.51$\pm$0.18 & 0.78$\pm$0.01 & 0.67$\pm$0.03 & [0.2] & 0.81$\pm$0.02 & 0.69$\pm$0.02 & 1.0$\pm$0.2 & 0.03 & --0.07 \\
HLS0111-B  & 0.13$\pm$0.03 & 0.29$\pm$0.34 & 0.48$\pm$0.13 & [0.54] & [0.2] & 0.45$\pm$0.07 & 0.56$\pm$0.30 & [0.2] & 0.06 & --0.16 \\
HLS0114    & 0.24$\pm$0.02 & 0.44$\pm$0.08 & 0.53$\pm$0.03 & 0.36$\pm$0.06 & 1.5$\pm$0.4 & 0.69$\pm$0.03 & 0.82$\pm$0.05 & 1.0$\pm$0.3 & --0.04 & 0.08 \\
HLS0307-28-A & 0.20$\pm$0.03 & 0.65$\pm$0.23 & 0.40$\pm$0.12 & 0.32$\pm$0.72 & [0.5] & 0.34$\pm$0.05 & 0.89$\pm$0.24 & [0.5] & 0.07 & --0.33 \\
HLS0307-28-B & 0.15$\pm$0.03 & 0.09$\pm$1.37 & $<$0.40 & \nodata & \nodata & $<$0.40 & \nodata & \nodata & 0.10 & 0.01 \\
HLS0307-50 & 0.27$\pm$0.03 & 0.30$\pm$0.06 & 0.64$\pm$0.07 & 0.35$\pm$0.21 & [0.2] & 0.97$\pm$0.05 & 0.39$\pm$0.07 & [0.2] & --0.14 & --0.06 \\
HLS0546    & 1.68$\pm$0.27 & 0.99$\pm$0.26 & 1.16$\pm$0.26 & 0.05$\pm$0.21 & 1.8$\pm$1.1 & 1.02$\pm$0.21 & 0.04$\pm$0.12 & 2.8$\pm$1.0 & 1.20 & 0.14 \\
HLS0553-A  & 0.10$\pm$0.01 & 0.56$\pm$0.10 & 0.33$\pm$0.02 & 0.62$\pm$0.12 & [1.0] & 0.35$\pm$0.02 & 0.88$\pm$0.08 & [1.0] & 0.05 & --0.16 \\
HLS0553-B  & 0.15$\pm$0.01 & 0.43$\pm$0.05 & 0.76$\pm$0.01 & 0.32$\pm$0.03 & [1.0] & 0.85$\pm$0.03 & 0.54$\pm$0.02 & [1.0] & 0.09 & --0.36 \\
HLS0553-C  & 0.14$\pm$0.01 & 0.58$\pm$0.12 & 0.67$\pm$0.02 & 0.48$\pm$0.03 & [1.0] & 0.64$\pm$0.01 & 0.60$\pm$0.02 & [1.0] & 0.03 & --0.16 \\
HLS0612    & 0.37$\pm$0.02 & 0.83$\pm$0.08 & 0.99$\pm$0.14 & 0.62$\pm$0.07 & 1.7$\pm$0.9 & 0.68$\pm$0.07 & 0.81$\pm$0.10 & 1.0$\pm$0.6 & --0.01 & 0.28 \\
HLS0840    & 3.66$\pm$1.71 & 0.43$\pm$0.30 & 1.13$\pm$0.02 & 0.40$\pm$0.03 & [0.2] & 1.15$\pm$0.03 & 0.49$\pm$0.02 & 0.2$\pm$0.1 & 0.22 & 0.22 \\
HLS1115    & 0.72$\pm$0.06 & 0.21$\pm$0.03 & 0.88$\pm$0.01 & 0.32$\pm$0.01 & 1.0$\pm$0.1 & 0.87$\pm$0.01 & 0.44$\pm$0.01 & 0.8$\pm$0.0 & --0.08 & --0.07 \\
HLS1314-A  & 1.07$\pm$0.40 & 0.27$\pm$0.07 & 3.72$\pm$0.10 & [0.16] & [0.2] & 3.42$\pm$0.13 & 0.14$\pm$0.01 & 0.2$\pm$0.1 & --0.10 & 0.08 \\
HLS1314-B  & 0.50$\pm$0.16 & 0.50$\pm$0.17 & 1.18$\pm$0.04 & [0.17] & 1.2$\pm$0.2 & 1.21$\pm$0.12 & 0.18$\pm$0.02 & 1.1$\pm$0.2 & --0.30 & --0.03 \\
HLS1314-C  & 0.94$\pm$0.18 & 0.22$\pm$0.18 & 1.45$\pm$0.04 & 0.21$\pm$0.03 & 0.9$\pm$0.1 & 1.80$\pm$0.04 & 0.21$\pm$0.02 & 0.8$\pm$0.1 & --0.12 & 0.32 \\
HLS1623    & 0.89$\pm$0.32 & 0.81$\pm$0.04 & 0.17$\pm$0.11 & [0.71] & [0.2] & 0.77$\pm$0.03 & 0.71$\pm$0.06 & [0.2] & --0.05 & 0.72 \\
HLS2104-A  & 0.24$\pm$0.04 & 1.00$\pm$0.24 & 0.71$\pm$0.04 & 0.42$\pm$0.10 & [0.2] & 0.87$\pm$4.11 & 0.80$\pm$0.05 & 0.0$\pm$0.5 & --0.05 & 0.02 \\
HLS2104-B  & 0.43$\pm$0.05 & 0.58$\pm$0.10 & 1.13$\pm$0.02 & 0.75$\pm$0.02 & 0.7$\pm$0.1 & 1.08$\pm$0.02 & 0.73$\pm$0.02 & [1.0] & --0.00 & 0.12 \\
HLS2155-A  & 0.33$\pm$0.05 & 0.16$\pm$0.11 & 0.88$\pm$0.07 & [0.20] & [0.5] & 0.60$\pm$0.04 & [0.20] & [0.5] & 0.00 & --0.01 \\
HLS2155-B  & 1.52$\pm$0.10 & 0.47$\pm$0.05 & 2.01$\pm$0.26 & 0.41$\pm$0.05 & [1.5] & 2.13$\pm$0.57 & 0.30$\pm$0.03 & 2.5$\pm$0.9 & 0.16 & --0.01 \\
HLS2332-A  & 0.21$\pm$0.00 & 0.58$\pm$0.01 & 0.93$\pm$0.04 & 0.42$\pm$0.05 & 0.9$\pm$0.3 & 0.99$\pm$0.01 & 0.48$\pm$0.01 & [1.0] & --0.03 & --0.08 \\
HLS2332-B  & 0.21$\pm$0.01 & 0.59$\pm$0.03 & 0.71$\pm$0.11 & 0.50$\pm$0.10 & 0.2$\pm$0.5 & 0.71$\pm$0.01 & 0.72$\pm$0.01 & [1.0] & --0.10 & --0.08 \\
HLS2332-C  & 0.20$\pm$0.01 & 0.67$\pm$0.04 & 1.22$\pm$0.05 & 0.62$\pm$0.04 & [1.0] & 1.00$\pm$0.01 & 0.60$\pm$0.01 & [1.0] & --0.39 & --0.08 \\
\enddata
% \tablenotetext{a}{ALMA coordinates of detected source.}
% \tablenotetext{\dagger}{Photometric redshift with a typical uncertainty of $\Delta z \sim 0.2$.}
% \tablenotetext{\bullet}{Highly-blended object with complex morphology. IRAC photometry is obtained though a simultaneous MCMC fitting of ALMA and optical source models (see descriptions in Section~\ref{ss:03b_photo} and Figure~\ref{fig:02_sunfit}).}
\tablecomments{Values enclosed with square brackets are fixed during morphological fitting (Section~\ref{ss:03b_photo}).
$R_\mathrm{e}$ is the effective radius, $b/a$ is the ratio between semi-minor and semi-major axis and $n$ is the S\'ersic index.
Spatial offsets in the last two columns are measured between ALMA and IRAC (average of 3.6 and 4.5\,\micron) centroids (see Appendix~\ref{sec:app_2}).
}
% The Distance is also centered on the decimals.  Note that when using decimal
% alignment you need to include the {\tt\string\decimals} command before
% {\tt\string\startdata} and all of the values in that column have to have a
% space before the next ampersand.}
\end{deluxetable*}

%% file: tables/tb04_properties.tex
\startlongtable

\begin{deluxetable*}{lrccrcrcc} %[htb!]
\tablecaption{Summary of source physical properties derived from SED fitting \label{tab:04_prop}}
\tabletypesize{\small}
% \tablenum{01}
\tablewidth{0pt}
\tablehead{
\colhead{ID} &  \colhead{Redshift\tablenotemark{a}} & \colhead{log($M_\ast$)\tablenotemark{b}} & \colhead{log(SFR)\tablenotemark{b}}
& \colhead{log(sSFR)} & \colhead{log($L_\mathrm{IR}$)\tablenotemark{b}\tablenotemark{c}} & \colhead{log($M_\mathrm{dust}$)\tablenotemark{b}} & \colhead{$T_\mathrm{dust}$\tablenotemark{d}} 
& \colhead{$A_V$}
\\
\colhead{}  & \colhead{}  &
\colhead{(\si{\mu^{-1} M_\odot})} & \colhead{(\si{\mu^{-1} M_\odot yr^{-1}})} 
& \colhead{(\si{Gyr^{-1}})} & \colhead{(\si{\mu^{-1} L_\odot})} & \colhead{(\si{\mu^{-1} M_\odot})} 
& \colhead{(K)} & \colhead{(mag)} }
% \decimalcolnumbers
\startdata
HLS0043-A & 1.61$^{+0.79}_{-0.44}$ & 11.77$\pm$0.35 & 2.73$\pm$0.48 & --0.01$\pm$0.58 & 12.87$\pm$0.39 & 9.30$\pm$0.39 & 29.3$\pm$7.0 & 2.3$\pm$1.0 \\
HLS0043-B & 3.23$^{+0.48}_{-0.59}$ & 11.69$\pm$0.29 & 2.66$\pm$0.43 & --0.02$\pm$0.64 & 12.78$\pm$0.23 & 9.01$\pm$0.23 & 32.3$\pm$5.0 & 2.2$\pm$0.7 \\
HLS0111-A & 2.27 & 11.94$\pm$0.21 & 2.55$\pm$0.23 & --0.38$\pm$0.40 & 12.78$\pm$0.10 & 8.96$\pm$0.10 & 31.5$\pm$1.9 & 5.5$\pm$0.5 \\
HLS0111-B & 2.27 & 11.96$\pm$0.26 & 2.84$\pm$0.19 & --0.12$\pm$0.42 & 12.98$\pm$0.10 & 9.20$\pm$0.10 & 31.5$\pm$1.2 & 2.9$\pm$0.5 \\
HLS0114  & 1.49$^{+0.43}_{-0.27}$ & 11.94$\pm$0.32 & 2.71$\pm$0.42 & --0.21$\pm$0.64 & 12.88$\pm$0.27 & 8.92$\pm$0.27 & 35.4$\pm$5.2 & 2.0$\pm$0.9 \\
HLS0307-28-A & 1.19$^{+0.32}_{-0.23}$ & 11.61$\pm$0.39 & 2.59$\pm$0.38 & --0.01$\pm$0.61 & 12.79$\pm$0.27 & 8.97$\pm$0.27 & 33.4$\pm$4.2 & 7.1$\pm$1.6 \\
HLS0307-28-B & 2.65$^{+0.54}_{-0.30}$ & 10.70$\pm$0.32 & 2.44$\pm$0.14 & 0.74$\pm$0.29 & 12.63$\pm$0.18 & 9.15$\pm$0.18 & 29.6$\pm$3.7 & 2.0$\pm$0.9 \\
HLS0307-50 & 1.93$^{+0.53}_{-0.39}$ & 12.25$\pm$0.33 & 2.86$\pm$0.38 & --0.38$\pm$0.55 & 13.13$\pm$0.25 & 9.34$\pm$0.25 & 33.5$\pm$5.3 & 5.0$\pm$0.9 \\
HLS0455  & 2.93 & 11.82$\pm$0.21 & 3.43$\pm$0.06 & 0.62$\pm$0.28 & 13.57$\pm$0.06 & 9.62$\pm$0.06 & 34.4$\pm$0.6 & 2.2$\pm$0.4 \\
HLS0505  & 1.27$^{+1.09}_{-0.22}$ & 11.31$\pm$0.44 & 2.61$\pm$0.41 & 0.44$\pm$0.52 & 12.74$\pm$0.40 & 9.62$\pm$0.40 & 24.6$\pm$7.1 & 5.1$\pm$1.6 \\
HLS0546  & 1.89$^{+0.50}_{-0.42}$ & 11.85$\pm$0.41 & 2.74$\pm$0.36 & --0.09$\pm$0.65 & 12.93$\pm$0.26 & 9.02$\pm$0.26 & 33.9$\pm$5.8 & 3.5$\pm$1.1 \\
HLS0553-A & 1.14 & 11.45$\pm$0.03 & 2.68$\pm$0.10 & 0.22$\pm$0.10 & 12.85$\pm$0.09 & 8.99$\pm$0.09 & 31.2$\pm$1.5 & 3.8$\pm$0.2 \\
HLS0553-B & 1.14 & 11.54$\pm$0.07 & 2.66$\pm$0.15 & 0.07$\pm$0.17 & 12.82$\pm$0.14 & 9.00$\pm$0.14 & 31.1$\pm$1.5 & 3.7$\pm$0.3 \\
HLS0553-C & 1.14 & 11.74$\pm$0.32 & 2.91$\pm$0.18 & 0.18$\pm$0.47 & 13.05$\pm$0.10 & 9.25$\pm$0.10 & 31.4$\pm$1.1 & 4.2$\pm$0.9 \\
HLS0600  & 2.87 & 12.51$\pm$0.27 & 3.72$\pm$0.12 & 0.22$\pm$0.38 & 13.84$\pm$0.07 & 10.29$\pm$0.07 & 29.8$\pm$0.3 & 2.7$\pm$0.5 \\
HLS0612  & 2.31$^{+0.59}_{-0.55}$ & 11.76$\pm$0.40 & 3.11$\pm$0.29 & 0.37$\pm$0.48 & 13.23$\pm$0.27 & 9.53$\pm$0.27 & 31.8$\pm$5.5 & 3.0$\pm$0.9 \\
HLS0840  & 1.97$^{+0.52}_{-0.58}$ & 11.84$\pm$0.36 & 2.97$\pm$0.36 & 0.12$\pm$0.53 & 13.09$\pm$0.30 & 9.26$\pm$0.30 & 33.3$\pm$6.4 & 2.2$\pm$0.9 \\
HLS1115  & 1.59$^{+0.05}_{-0.01}$ & 12.15$\pm$0.01 & 2.83$\pm$0.01 & --0.32$\pm$0.04 & 12.92$\pm$0.03 & 9.36$\pm$0.03 & 31.2$\pm$0.6 & 3.1$\pm$0.1 \\
HLS1124-A & 1.80 & 10.80$\pm$0.30 & 2.36$\pm$0.14 & 0.57$\pm$0.40 & 12.44$\pm$0.13 & 9.03$\pm$0.13 & 27.0$\pm$1.5 & 1.7$\pm$0.6 \\
HLS1124-B & 1.80 & 11.70$\pm$0.30 & 3.27$\pm$0.12 & 0.57$\pm$0.40 & 13.35$\pm$0.12 & 9.98$\pm$0.12 & 27.0$\pm$0.4 & 1.5$\pm$0.4 \\
HLS1314\tablenotemark{e}  & 1.45 & 12.73$\pm$0.06 & 2.44$\pm$0.21 & --1.32$\pm$0.20 & 12.94$\pm$0.07 & 9.50$\pm$0.07 & 28.4$\pm$0.7 & 2.3$\pm$0.1 \\
HLS1623  & 0.96$^{+0.17}_{-0.45}$ & 11.84$\pm$0.48 & 2.26$\pm$0.41 & --0.60$\pm$0.53 & 12.68$\pm$0.34 & 8.42$\pm$0.34 & 22.0$\pm$3.8 & 9.2$\pm$1.4 \\
HLS2104-A & 2.17$^{+0.71}_{-0.54}$ & 11.73$\pm$0.33 & 2.60$\pm$0.41 & --0.08$\pm$0.56 & 12.77$\pm$0.31 & 9.10$\pm$0.31 & 31.0$\pm$6.3 & 2.3$\pm$0.9 \\
HLS2104-B & 2.13$^{+0.62}_{-0.63}$ & 11.87$\pm$0.30 & 2.79$\pm$0.44 & --0.07$\pm$0.56 & 12.91$\pm$0.33 & 9.28$\pm$0.33 & 30.7$\pm$6.2 & 1.5$\pm$0.8 \\
HLS2155-A & 1.72$^{+0.29}_{-0.30}$ & 11.92$\pm$0.35 & 2.63$\pm$0.33 & --0.31$\pm$0.60 & 12.87$\pm$0.20 & 8.81$\pm$0.20 & 37.0$\pm$4.7 & 5.2$\pm$1.1 \\
HLS2155-B & 2.66$^{+0.94}_{-0.73}$ & 11.72$\pm$0.31 & 2.92$\pm$0.37 & 0.20$\pm$0.45 & 12.99$\pm$0.32 & 9.25$\pm$0.32 & 31.9$\pm$7.4 & 1.6$\pm$0.7 \\
HLS2332-A & 2.73 & 11.91$\pm$0.09 & 3.61$\pm$0.12 & 0.72$\pm$0.17 & 13.64$\pm$0.12 & 9.44$\pm$0.12 & 36.7$\pm$0.7 & 3.0$\pm$0.2 \\
HLS2332-B & 2.73 & 11.73$\pm$0.12 & 3.64$\pm$0.04 & 0.93$\pm$0.15 & 13.70$\pm$0.07 & 9.51$\pm$0.07 & 36.7$\pm$0.6 & 2.9$\pm$0.1 \\
HLS2332-C & 2.73 & 12.49$\pm$0.24 & 3.51$\pm$0.13 & 0.03$\pm$0.35 & 13.64$\pm$0.07 & 9.53$\pm$0.07 & 36.7$\pm$0.6 & 3.8$\pm$0.4 \\
\enddata
\tablenotetext{a}{Photometric redshifts are presented as the median of their likelihood distributions with 1$\sigma$ confidence range.}
\tablenotetext{b}{These quantities are not corrected for lensing magnification.}
\tablenotetext{c}{Defined as the total IR luminosity integrated from 8 to 1000\,\micron.}
\tablenotetext{d}{Modeled with MBB spectrum with fixed dust emissivity at $\beta = 1.8$.} %for spectroscopically confirmed sources 
\tablenotetext{e}{\textred{Combination of all the three clumps (HLS1314-A/B/C) seen in the ALMA map (See Section~\ref{ss:04a_sed}).}}
% \tablenotetext{\dagger}{Photometric redshift with a typical uncertainty of $\Delta z \sim 0.2$.}
% \tablenotetext{\bullet}{Highly-blended object with complex morphology. IRAC photometry is obtained though a simultaneous MCMC fitting of ALMA and optical source models (see descriptions in Section~\ref{ss:03b_photo} and Figure~\ref{fig:02_sunfit}).}
% The Distance is also centered on the decimals.  Note that when using decimal
% alignment you need to include the {\tt\string\decimals} command before
% {\tt\string\startdata} and all of the values in that column have to have a
% space before the next ampersand.}
\end{deluxetable*}